\documentclass{lmcs}
\pdfoutput=1

\usepackage{lastpage}
\lmcsdoi{22}{3}{5}
\lmcsheading{}{\pageref{LastPage}}{}{}%
{Aug.~30,~2024}{Aug.~06,~2026}{}

\usepackage[utf8]{inputenc}
\usepackage[T1]{fontenc}
\usepackage{amssymb}
\usepackage{bm}
\usepackage{hyperref}
\usepackage[capitalise]{cleveref}

\newcommand{\Con}{\mathsf{Con}}
\newcommand{\norm}[1]{\lVert #1 \rVert}
\newcommand{\newR}{\mathcal{R}}
\newcommand{\newU}{\mathcal{U}}
\DeclareMathOperator{\Ext}{Ext}
\DeclareMathOperator{\ext}{ext}
\DeclareMathOperator{\ndepth}{nd}
\newcommand{\Rn}{\mathcal{R}}
\newcommand{\N}{\mathbb{N}}
\newcommand{\Z}{\mathbb{Z}}
\newcommand{\Q}{\mathbb{Q}}
\newcommand{\card}[1]{\left|#1\right|}
\newcommand{\FO}{\mathsf{FO}}
\newcommand{\FOord}[1][\null]
{
    \ifthenelse{\equal{#1}{\null}}
    {
	\mathsf{FO}[\mathsf{<}]
    }
    {
	\mathsf{FO}_{#1}[\mathsf{<}]
    }
}
\newcommand{\FOplus}[1][\null]
{
    \ifthenelse{\equal{#1}{\null}}
    {
	\mathsf{FO}[\mathsf{+}]
    }
    {
	\mathsf{FO}_{#1}[\mathsf{+}]
    }
}
\newcommand{\FOarb}{\mathsf{FO}[\mathsf{arb}]}
\newcommand{\MODpred}{\mathsf{MOD}}
\newcommand{\TCO}{\mathsf{TC}^0}
\newcommand{\ACO}{\mathsf{AC}^0}
\newcommand{\ACCO}{\mathsf{ACC}^0}
\newcommand{\NCI}{\mathsf{NC}^1}
\newcommand{\reduce}{\leq_{\mathrm{cd}}}
\newcommand{\displaypunct}[1]{\,\,\text{#1}}
\newcommand{\call}{\text{call}}
\newcommand{\return}{\text{ret}}
\newcommand{\internal}{\text{int}}
\newcommand{\leftmult}[1]{\mathrm{left}_{#1}}
\newcommand{\rightmult}[1]{\mathrm{right}_{#1}}
\newcommand{\emptyword}{\varepsilon}
\newcommand{\ExtMon}{\mathcal{O}}
\newcommand{\id}{\mathrm{id}}
\newcommand{\length}[1]{\mathopen{}\left|#1\right|\mathclose{}}
\newcommand{\set}[1]{\{#1\}}
\newcommand{\bigset}[1]{\bigl\{#1\bigr\}}
\newcommand{\biggset}[1]{\biggl\{#1\biggr\}}
\newcommand{\bigmid}{\mathrel{\big|}}
\newcommand{\biggmid}{\mathrel{\bigg|}}
\newcommand{\powerset}[1]{\mathfrak{P}(#1)}
\newcommand{\DLang}[1]{L\mathopen{}\left(#1\right)\mathclose{}}
\newcommand{\quotient}{\mathclose{}/\mathopen{}}
\newcommand{\restr}[2]{{#1|}_{#2}}
\newcommand{\eqACO}{=_{\mathrm{cd}}}
\newcommand{\GreenJ}{\mathrel{\mathfrak{J}}}
\newcommand{\GreenJleq}{\leq_{\mathfrak{J}}}
\newcommand{\GreenJlt}{<_{\mathfrak{J}}}
\newcommand{\GreenR}{\mathrel{\mathfrak{R}}}
\newcommand{\GreenRleq}{\leq_{\mathfrak{R}}}
\newcommand{\GreenL}{\mathrel{\mathfrak{L}}}
\newcommand{\GreenLleq}{\leq_{\mathfrak{L}}}
\newcommand{\GreenH}{\mathrel{\mathfrak{H}}}
\newcommand{\EQ}{\mathrm{EQUALITY}}
\newcommand{\SM}{\mathcal{L}}
\newcommand{\MOD}{\mathrm{MOD}}
\DeclareMathOperator{\eval}{eval}
\newcommand{\match}{\leftrightsquigarrow}
\newcommand{\matching}[1]{M^\triangle\left(#1\right)}

\Crefname{lemma}{Lemma}{Lemmata}
\Crefname{equation}{Line}{Lines}\crefrangelabelformat{equation}{(#3#1#4)--(#5#2#6)}
\Crefname{inequality}{Inequality}{Inequalities}\creflabelformat{inequality}{(#2#1#3)}
\Crefname{equivalence}{Equivalence}{Equivalences}\creflabelformat{equivalence}{(#2#1#3)}
\Crefname{property}{Property}{Properties}\creflabelformat{property}{(#2#1#3)}


\begin{document}

\bibliographystyle{alphaurl}

\title{The \texorpdfstring{$\ACO$}{AC0}-Complexity Of Visibly Pushdown Languages}
\titlecomment{This is the full version of the extended abstract that appeared in
the proceedings of STACS 2024.}
\thanks{The first author was supported by the Agence Nationale de la Recherche grant no. ANR-17-CE40-0010.}

\author{Stefan G\"oller}[a]
\author{Nathan Grosshans\lmcsorcid{0000-0003-3400-1098}}[b]
\address{School of Electrical Engineering and Computer Science, Universit\"at Kassel, Germany}
\email{stefan.goeller@uni-kassel.de}
\address{Independent Scholar, Paris Region, France}
\email{nathan.grosshans@melix.org}

\begin{abstract}
	We study the question of which visibly pushdown languages (VPLs) are 
	in the complexity class $\ACO$ and how to effectively decide this 
	question.
	Our contribution is to introduce a particular subclass of 
	one-turn VPLs, called {\em intermediate VPLs},
	for which the raised question is entirely unclear: to the 
	best of our knowledge
	our research community is unaware of 
	containment or non-containment in $\ACO$ for {\em any} 
	language in our newly introduced class.
	Our main result states that there is an algorithm that, 
	given a visibly pushdown automaton,
	correctly outputs exactly one of the following: that its
	language $L$ is in $\ACO$, some $m\geq 2$
	such that $L$ is $\ACCO(m)$-hard (implying that $L$ is not in $\ACO$), or
	a finite disjoint union of intermediate VPLs 
	that $L$ is constant-depth
	equivalent to.
	In the latter of the three cases one can moreover effectively 
	compute $k,l\in\N_{>0}$ with $k\not=l$ such that 
	the concrete intermediate 
	VPL $L(S\rightarrow \varepsilon\mid a c^{k-1} S b_1\mid ac^{l-1}Sb_2)$
	is constant-depth reducible to the language $L$.
	Due to their particular nature we conjecture that either all 
	intermediate VPLs are in $\ACO$ or all are not.
As a corollary of our main result we obtain that in case the input language is a visibly counter language our algorithm 
can effectively determine if it is in $\ACO$ --- hence our main result generalizes 
	a result by Krebs et al.
	stating that it is decidable if a given visibly
	counter language is in $\ACO$ (when restricted to well-matched words).

	For our proofs we revisit so-called $\Ext$-algebras
	(introduced by Czarnetzki et al.), 
	which are closely related to forest algebras 
	(introduced by Boja{\'n}czyk and Walukiewicz),
	and use Green's relations.
\end{abstract}

\maketitle

\section{Introduction}\label{Section Introduction}
This paper studies the circuit complexity of formal word languages.
It is well-known that the regular word languages are characterized as
the languages recognizable by finite monoids. 
When restricting the finite monoids to be aperiodic Sch\"utzenberger 
proved that one obtains precisely the star-free regular languages~\cite{Schutzenberger65}.
In terms of logic, these correspond to the languages definable in first-order logic $\FOord$
by a result of McNaughton and Papert~\cite{Books/Straubing-1994}.
The more general class of regular languages expressible in $\FOarb$, i.e. first-order logic with arbitrary
numerical predicates, coincides with the regular languages in $\ACO$~\cite{GL84,Immerman87}.
These can be characterized algebraically as the regular languages whose
syntactic morphism is 
quasi-aperiodic~\cite{BCST92}.
The latter algebraic characterization also shows that 
it is decidable if a regular language is in $\ACO$.

Generalizing regular languages, Mehlhorn introduced input-driven languages~\cite{Mehlhorn80}.
They are described by pushdown automata whose input alphabet is
partitioned into letters that are either of type call, internal, or return.
Rediscovered by Alur and Madhusudan in 2004~\cite{AlurM04}
under the name of {\em visibly pushdown languages (VPLs)}, it was shown
that they enjoy many of the 
desirable effective closure properties of the regular languages.
For instance, the visibly pushdown languages form an effective Boolean algebra.
Algebraically, VPLs were characterized by Alur et al.~\cite{AKMV05} 
by congruences on well-matched words of finite index. 
Extending upon these, Czarnetzki et al. introduced so-called $\Ext$-algebras~\cite{Czarnetzki-Krebs-Lange-2018};
these involve pairs of monoids $(R,O)$ where $O$ is a submonoid of $R^R$.
Being tailored towards recognizing word languages, $\Ext$-algebras are closely connected
to forest algebras, introduced by Boja{\'n}czyk and Walukiewicz~\cite{BW08}:
in~\cite{Czarnetzki-Krebs-Lange-2018} it is shown that a language of well-matched
words is visibly pushdown if, and only if, its syntactic $\Ext$-algebra is finite.
While context-free languages are generally in $\mathsf{LOGCFL}=\mathsf{SAC}^1$, 
the visibly pushdown languages, as the regular languages, 
 are known to be in $\NCI$~\cite{Dymond88}.
By a famous result of Barrington~\cite{Barrington89}, there already exist regular
languages that are $\NCI$-hard.

\medskip
\noindent
{\bf Related work.}
Visibly pushdown languages (VPLs) were introduced~\cite{AlurM04}
via deterministic visibly pushdown automata (DVPA for short).
Inspired by forest algebras~\cite{BW08} the paper~\cite{Czarnetzki-Krebs-Lange-2018} 
introduces $\Ext$-algebras.
Unfortunately, the definition of $\Ext$-algebra 
morphisms in~\cite{Czarnetzki-Krebs-Lange-2018} is incorrect in that it
provably does not lead to freeness.

The regular languages that are in $\ACO$ were effectively characterized by 
Barrington et al.~\cite{BCST92}.
By an automata-theoretic approach, Krebs et al.~\cite{Krebs-Lange-Ludwig-2015}
effectively characterized the visibly counter languages that are in $\ACO$.
These are particular VPLs that are essentially accepted 
 by visibly pushdown automata that use only one stack symbol.
In his PhD thesis~\cite{Ludwigthesis} Ludwig already considers the question
of which VPLs
 are in $\ACO$. Unfortunately, his conjectural characterization
contains several serious flaws.

\medskip
\noindent
{\bf Our contribution.}
We reintroduce $\Ext$-algebras, fix the notion of $\Ext$-algebra morphisms and
define the languages they recognize. We also reintroduce the syntactic
$\Ext$-algebra of languages of well-matched words.
We rigorously prove classical results like freeness and
minimality of syntactic $\Ext$-algebras with respect to recognition.
We prove that a language of well-matched words is a VPL if, and only if, 
it is recognizable by a finite $\Ext$-algebra.
While these results essentially revisit the constructions 
of~\cite{Czarnetzki-Krebs-Lange-2018}, 
we use $\Ext$-algebras as a technical tool for studying the complexity 
of visibly pushdown languages.

Fix a visibly pushdown alphabet $\Sigma$, i.e.\ $\Sigma$ is partitioned into $\Sigma_\call$
(call letters), $\Sigma_\internal$ (internal letters), and $\Sigma_\return$ (return letters).
Denoting $\Delta(u)$ as the difference between the number of occurrences of
call and return letters in $u\in\Sigma^*$,
a word $w\in\Sigma^*$ is {\em well-matched} if $\Delta(w)=0$ and $\Delta(u)\geq 0$ for all
prefixes $u$ of $w$. A {\em context} is a pair $(u,v)$ such that $uv$ is well-matched ---
contexts have a natural composition operation: $(u,v)\circ(u',v')=(uu',v'v)$.

A set of contexts $\Rn$ is {\em length-synchronous} if 
$|u|/|v|=|u'|/|v'|$ for all $(u,v),(u',v')\in\Rn$ with $\Delta(u),\Delta(u')>0$
and {\em weakly length-synchronous} if
$u=u'$ implies $|v|=|v'|$ and $v=v'$ implies $|u|=|u'|$ for all $(u,v),(u',v')\in\Rn$
with $\Delta(u),\Delta(u')>0$.
Any language $L$ of well-matched words induces a congruence $\equiv_L$ on contexts:
$(u,v)\equiv_L(u',v')$ if $xuwvy\in L\Leftrightarrow xu'wv'y\in L$ for all contexts
$(x,y)$ and all well-matched words $w$.
We introduce the notion of quasi-counterfreeness: a language
is {\em quasi-counterfree} if for all contexts 
$\sigma\in\Sigma^k\times\Sigma^l$ with $k,l\in\N$ arbitrary at least one of the
following holds: (1) there exists some $n\in\N$ such that
$\sigma^n\equiv_L\sigma^{n+1}$ or (2) no context in $\Sigma^k\times\Sigma^l$ is
$\equiv_L$-equivalent to $\sigma\circ\sigma$.
Finally, we introduce our central class of {\em intermediate VPLs}: 
a VPL is {\em intermediate} if it is quasi-counterfree and generated by a
context-free grammar containing the production $S\rightarrow_G \varepsilon$, where $S$ is the
start nonterminal and whose other productions are of
the form $T\rightarrow_G u T'v$ where $uv$ is well-matched, 
$u\in(\Sigma_\internal^*\Sigma_\call\Sigma_\internal^*)^+$ and 
$v\in(\Sigma_\internal^*\Sigma_\return\Sigma_\internal^*)^+$,
such that the set of contexts $\set{(u,v)\mid S\Rightarrow_G^*uSv}$ is weakly length-synchronous
but not length-synchronous.
Note that intermediate VPLs are particular one-turn visibly pushdown languages, that is,
visibly pushdown languages that are subsets of 
$(\Sigma\setminus\Sigma_\return)^*(\Sigma\setminus\Sigma_\call)^*$.
As an example, for all $k,l\geq 1$ with $k\not=l$, a concrete intermediate 
VPL, denoted by
$\SM_{k,l}$, is the one that is generated by the context-free grammar 
$S\rightarrow \varepsilon\mid a c^{k-1} S b_1\mid ac^{l-1}Sb_2$:
here $a$ is a call letter, $c$ is an internal letter and $b_1$ and $b_2$ are return
letters.

As far as we know the techniques known to our community 
do not directly suffice to show
whether {\em at all} there is some intermediate VPL that is 
{\em provably in $\ACO$ 
or provably not in $\ACO$} --- analogous remarks apply to $\ACCO$.
 Our main result states that there is an algorithm 
that, given a DVPA $A$ correctly outputs exactly one of the
following: $L(A)\in\ACO$, some $m\geq 2$ such
that $L(A)$ is $\ACCO(m)$-hard (implying that 
$L(A)\not\in\ACO$), or a non-empty disjoint finite union of
intermediate VPLs that $L(A)$ is  constant-depth equivalent to.
In the latter of the three cases one can moreover effectively compute $k,l\in\N_{>0}$ with $k\not=l$ such that 
the above-mentioned $\SM_{k,l}$ is constant-depth reducible to $L(A)$.
We conjecture that either all intermediate VPLs are in $\ACO$ or all are not:
	note that together with our main result this conjecture implies the existence
	of an algorithm that can effectively determine if a given visibly pushdown language
	is in $\ACO$.
As a corollary of our main result we obtain that in case the input language is a visibly counter language our algorithm 
can effectively determine if it is in $\ACO$, hence our main result generalizes 
	a result by Krebs et al.
	stating that it is decidable if a given visibly
	counter lanugage is in $\ACO$ (when restricted to well-matched words).

For our main result we extensively study $\Ext$-algebras, the
syntactic morphisms of VPLs, and make use of Green's relations.

\medskip
\noindent
{\bf Organization.} Our paper is organized as follows.
We introduce notation 
and give an overview of our main result in Section~\ref{Section Preliminaries}.
In Section~\ref{Section Ext Algebras} we first recall general algebraic concepts 
and then revisit $\Ext$-algebras and their correspondence to visibly pushdown languages.
Section~\ref{Section Notions} introduces
central notions like length-synchronicity and 
weak length-synchronicity for $\Ext$-algebra morphisms
and visibly pushdown languages.
The proof of our main result is content of 
Section~\ref{Section Proof Strategy}.
In Section~\ref{Section Effectiveness} we concern ourselves
with the computability of the syntactic $\Ext$-algebra as well as decidability of
quasi-aperiodicity and (weak) length-synchronicity.
We conclude in Section~\ref{Conclusion}.

\section{Preliminaries}\label{Section Preliminaries}

By $\N$ we denote the non-negative integers and by
$\N_{>0}$ the positive integers.
For integers $i,j\in\Z$ we denote by $[i,j]$ the set $\set{i,\ldots,j}$.
For any function $f\colon X\rightarrow Y$ and any subset $Z\subseteq X$ we denote
by $\restr{f}Z\colon Z\rightarrow Y$ the restriction of $f$ to domain $Z$, i.e.
$\restr{f}Z(z)=f(z)$ for all $z\in Z$.

For all words $w=w_1\cdots w_n$, where $w_i\in\Sigma$ for all $i\in[1,n]$, and for
subsets $\Gamma\subseteq\Sigma$, let 
$\card{w}_\Gamma=|\set{i\in[1,|w|]\mid w_i\in\Gamma}|$
denote the number of occurrences of letters in $\Gamma$.
For all $a\in\Sigma$ we write $\card{w}_a$ to denote
$\card{w}_{\set{a}}$.
Given a finite set $I$ and languages $L_i \subseteq \Sigma^*$ for all
$i \in I$, we call \emph{disjoint union of the $L_i$'s} and denote by
$\biguplus_{i \in I} L_i$ the union $\bigcup_{i \in I} \$^{\sigma(i)} L_i$ for
an arbitrary ordering bijection $\sigma\colon I \to [1, \card{I}]$ and a letter
$\$ \notin \Sigma$.
We define the languages 
\begin{itemize}
	\item $\EQ=\bigset{w\in\set{0,1}^* \bigmid |w|_0=|w|_1}$, and
	\item $\MOD_m=\bigset{w\in\set{0,1}^* \bigmid m\text{ divides } |w|_{1}}$
for each $m\geq 2$.
\end{itemize}
\noindent
A \emph{visibly pushdown alphabet} is a finite alphabet
$\Sigma=\Sigma_\call
\cup\Sigma_\internal
\cup\Sigma_\return$,
where the alphabets $\Sigma_\call,\Sigma_\internal$, and $\Sigma_\return$ are 
pairwise disjoint.

\begin{samepage}
\begin{defi}
The set of \emph{well-matched words over a visibly pushdown alphabet $\Sigma$}, denoted by
    $\Sigma^\triangle$, is the smallest set satisfying the following:
    \begin{itemize}
	\item
	    $\varepsilon \in \Sigma^\triangle$ and $c \in \Sigma^\triangle$ for
	    all $c \in \Sigma_\internal$,
	\item
	    $a w b \in \Sigma^\triangle$ for all $w \in \Sigma^\triangle$,
	    $a \in \Sigma_\call$ and $b \in \Sigma_\return$, and
	\item
	    $u v \in \Sigma^\triangle$ for all
	    $u, v \in \Sigma^\triangle \setminus \set{\emptyword}$.
    \end{itemize}
\end{defi}
\end{samepage}
\noindent
	A well-matched word $w\in\Sigma^\triangle$ 
	is {\em one-turn} if 
	$w\in(\Sigma\setminus\Sigma_\return)^*(\Sigma\setminus\Sigma_\call)^*$.
	A language $L\subseteq\Sigma^\triangle$ is {\em one-turn} if it contains only one-turn words.
    Let $\Sigma$ be a visibly pushdown alphabet.
    We define $\Delta\colon \Sigma^* \to \Z$ to be the height monoid morphism
    such that $\Delta(w) = \card{w}_{\Sigma_\call}-\card{w}_{\Sigma_\return}$
    for all $w \in \Sigma^*$.

A {\em context} is a pair $(u,v)\in\Sigma^*\times\Sigma^*$ such that
$uv\in\Sigma^\triangle$.
The \emph{composition} of two contexts $(u,v),(x,y)\in\Con(\Sigma)$
is defined as $(u,v)\circ(x,y)=(ux,yv)$.  
For $\sigma\in\Con(\Sigma)$ by $\sigma^k$ we denote the $k$-fold composition
$\sigma\circ\dots\circ\sigma$.
For any context $(u,v)\in\Con(\Sigma)$ and well-matched
word $w\in\Sigma^\triangle$ we define
$(u,v)w=uwv$.
An equivalence relation $\equiv$ on $\Con(\Sigma)$ is a
\emph{congruence} if for all $\chi,\chi',\sigma,\tau\in\Con(\Sigma)$
we have that $\sigma\equiv\tau$ implies 
$\chi\circ\sigma\circ\chi'\equiv\chi\circ\tau\circ\chi'$.
Given a congruence $\equiv$ over $\Con(\Sigma)$ we denote by
$\lbrack \sigma\rbrack_{\equiv}$
the equivalence class of $\sigma$.
Given a language of well-matched words $L\subseteq\Sigma^\triangle$ 
we write $\sigma\equiv_L\tau$ if for all $\chi\in\Con(\Sigma)$
and all $w\in\Sigma^\triangle$ we have
$(\chi\circ\sigma)w\in L$ if, and only if, 
$(\chi\circ\tau)w\in L$.
Clearly, $\equiv_L$ is a congruence.

A {\em context-free grammar} is a tuple $G=(V,\Sigma,P,S)$,
where $V$ is a finite set of \emph{nonterminals}, 
$\Sigma$ is a non-empty finite \emph{alphabet},
$P\subseteq V\times(V\cup\Sigma)^*$ is a finite set of
\emph{productions}, and $S\in V$ is the {\em start nonterminal}.
We write $T\rightarrow_Gy$ whenever $(T,y)\in P$.
The binary relation $\Rightarrow_G$ over 
$(V\cup\Sigma)^*$ is defined as $u\Rightarrow_G v$ if there exists a production 
$T\rightarrow_Gy$ and $x,z\in(V\cup\Sigma)^*$ such that $u=xTz$ and $v=xyz$.
By $L(G)=\set{w\in\Sigma^*\mid S\Rightarrow_G^*w}$ we denote
the \emph{language} of $G$, where 
$\Rightarrow_G^*$ is the reflexive transitive
closure of $\Rightarrow_G$.

In the following we introduce deterministic visibly pushdown
automata, remarking that nondeterministic visibly pushdown automata are determinizable~\cite{AlurM04}.
\begin{defi}
	A \emph{deterministic visibly pushdown automaton} (DVPA) is a tuple
	$A = (Q, \Sigma, \Gamma, \delta, \allowbreak q_0, F, \bot)$, where
    \begin{itemize}
	\item
	    $Q$ is a finite set of \emph{states},
	\item
	    $\Sigma$ is a visibly pushdown alphabet, the \emph{input alphabet},
	\item
		$\Gamma$ is a finite alphabet, the \emph{stack alphabet},
	\item
	    $q_0 \in Q$ is the \emph{initial state},
	\item
	    $F \subseteq Q$ is the set of \emph{final states},
	\item
		$\bot \in \Gamma$ is the \emph{bottom-of-stack symbol}, and
	\item
	    $\delta\colon Q \times \Sigma \times \Gamma \to
			  Q \times \bigl(\set{\emptyword} \cup \Gamma \cup
					 (\Gamma \setminus \set{\bot}) \Gamma
				   \bigr)$
	    is the \emph{transition function} such that for all
	    $q \in Q, a \in \Sigma, \alpha \in \Gamma$:
    \begin{itemize}
		\item
		    if $a \in \Sigma_\call$, then
		    $\delta(q, a, \alpha) \in
			    Q \times (\Gamma \setminus \set{\bot}) \alpha$,
		\item
		    if $a \in \Sigma_\return$, then
		    $\delta(q, a, \alpha) \in Q \times \set{\emptyword}$, and
		\item
		    if $a \in \Sigma_\internal$, then
		    $\delta(q, a, \alpha) \in Q \times \set{\alpha}$.
	    \end{itemize}
    \end{itemize}
\end{defi}
\begin{samepage}
    We define the \emph{extended transition function}
    $\widehat{\delta}\colon Q \times \Sigma^* \times \Gamma^* \to Q \times \Gamma^*$ inductively as
    \begin{itemize}
	\item
	    $\widehat{\delta}(q, \emptyword, \beta) = (q, \beta)$
	    for all $q \in Q$ and $\beta\in \Gamma^*$,
	\item
	    $\widehat{\delta}(q, w, \emptyword) = (q, \emptyword)$
	    for all $q \in Q$ and $w \in \Sigma^+$, and
    \item
	    $\widehat{\delta}(q, a w, \alpha \beta) =
	     \widehat{\delta}(p, w, \gamma \beta)$,
	    where $\delta(q, a, \alpha) = (p, \gamma)$ for all $q \in Q$,
	    $a \in \Sigma$, $w \in \Sigma^*$, $\alpha \in \Gamma$ and
	    $\beta\in \Gamma^*$.
   \end{itemize}
\end{samepage}
    The {\em language} accepted by $A$ is the language
	$\DLang{A} = \set{w \in \Sigma^* \mid
			 \widehat{\delta}(q_0, w, \bot) \in F \times \set{\bot}}$.
	We call such a language a \emph{visibly pushdown language} (VPL).
	We remark that visibly pushdown languages are always subsets of $\Sigma^\triangle$.

We refer to~\cite{Har78} for further details on formal language theory.

	\paragraph*{Semi-linear sets.}
Given $d \in \N_{>0}$, for
$\vec{x}=(x_1,\ldots,x_d),\vec{y}=(y_1,\ldots,y_d)\in\N^d$ we define 
$\vec{x}+\vec{y}=(x_1+y_1,\ldots,x_d+y_d)$.
We define the {\em norm} of a vector $\vec{x}\in\N^d$ as
$\norm{\vec{x}}=\max\set{x_i\mid i\in[1,d]}$.
For $X,Y\subseteq\N^d$ define $X+Y=\set{\vec{x}+\vec{y}\mid\vec{x}\in X,\vec{y}\in Y}$.
For $\vec{x}=(x_1,\ldots,x_d)\in\N^d$ and $n\in\N$ we
define $n\vec{x}=(nx_1,\ldots,nx_d)$ and $\N\vec{x}=\set{n\vec{x}\mid n\in\N}$.
A set $X\subseteq\N^d$ is {\em linear} if $X=\vec{y}+\sum_{i=1}^k\N\vec{x}_i$
for $k\in\N$ and $y, x_1, \ldots, x_k \in \N^d$ and it is {\em semilinear} if
$X$ is a finite union of linear sets.

\subsection{Complexity and logic}
We assume familiarity with standard circuit complexity,
we refer to \cite{Vollmerbook,Juknabook} for an introduction
to the topic.
Recall the following Boolean functions: the AND-function, the OR-function, 
the majority function (that outputs $1$ if the majority of its inputs are $1$s),
and the $\text{mod}_m$ function (that outputs $1$ if the number of its inputs
that are $1$s is divisible by $m$) for all $m\geq 2$.

A circuit family $(C_n)_{n\in\N}$ {\em decides} a binary language 
$L\subseteq\set{0,1}^*$ if $C_n$ is a circuit with $n$ inputs
such that $L\cap\set{0,1}^n=\bigset{x_1\dots x_n\in\set{0,1}^n \bigmid C_n(x_1,\dots,x_n)=1}$
for all $n\in\N$.
In this paper, we will consider circuits deciding languages over arbitrary
finite alphabets: to do this, we just consider implicitly that any language over
an arbitrary finite alphabet comes with a fixed binary encoding that encodes
each letter as a block of bits of fixed size.
By $\reduce$ we mean {\em constant-depth truth table reducibility}
(or just constant-depth reducibility) as
introduced in~\cite{CSV84}. Formally for two languages
$K \subseteq \Gamma^*$ and $L \subseteq \Sigma^*$ for finite
alphabets $\Sigma, \Gamma$, we write $K\reduce L$ in case there is a 
polynomial $p$, a constant $d\in\N$,
and circuit family $(C_n)_{n\in\N}$ deciding $K$
such that each circuit $C_n$ satisfies the following:
it is of depth at most $d$ and size at most $p(n)$
and its non-input gates are either AND-labeled, OR-labeled, or so-called oracle
gates, labeled by $L$, that are gates deciding $L \cap \Sigma^m$, where
$m\leq p(n)$, such that there is no path from the output of an oracle gate to an
input of another oracle gate.
We write $K \eqACO L$ if $K \reduce L$ and $L \reduce K$; we also say that
$K$ and $L$ are {\em constant-depth equivalent}.
We say a language $L$ is {\em hard} for a complexity class $\mathsf{C}$ (or
just {\em $\mathsf{C}$-hard})
if $L' \reduce L$ for all $L'\in\mathsf{C}$.
We say $L$ is {\em $\mathsf{C}$-complete} if $L$ is $\mathsf{C}$-hard and
$L\in\mathsf{C}$.
The following complexity classes are relevant in this paper:
\begin{itemize}
    \item
	$\ACO$ is the class of all languages decided by circuit families with
	NOT gates, AND, OR gates of unbounded fan-in, constant depth and polynomial size;
		
    \item
	    $\ACCO(m)$ is the class of all languages decided by circuit
		families with NOT gates, AND, OR and $\text{mod}_m$ gates  
		of unbounded fan-in, constant depth and polynomial size;
    \item
	    $\ACCO=\bigcup_{m\geq 2}\ACCO(m)$ is 
		the class of all languages decided by circuit
		families with NOT gates, AND, OR and modular gates 
		(for some fixed $m$) 
		of unbounded fan-in, constant depth and polynomial size;
    \item
	$\TCO$ is the class of all languages decided by circuit families with
	NOT gates, AND, OR and majority gates of unbounded fan-in, constant depth and
	polynomial size;
    \item
	$\NCI$ is the class of all languages decided by circuit
	families with NOT gates, AND, OR gates of bounded fan-in, logarithmic depth and
	polynomial size.
\end{itemize}
\noindent
We also consider the framework of first order logic over finite
words. (See~\cite{Books/Immerman-1999, Books/Straubing-1994} for a proper
introduction to the topic.)
A \emph{numerical predicate of arity $r \in \N_{>0}$} is a symbol of arity $r$
associated to a subset of ${\N_{>0}}^r$.
Given a class $\mathcal{C}$ of numerical predicates and a finite
alphabet $\Sigma$, we call $\FO_\Sigma[\mathcal{C}]$-formula a first order formula over
finite words using the alphabet $\Sigma$ and numerical predicates from the class
$\mathcal{C}$. On occasions, we might also consider
$\FO_{\Sigma, \match}[\mathcal{C}]$-formulas that in comparison to the previous
ones can use an additional binary predicate $\match$ and are interpreted on
structures $(w, M)$ with $w \in \Sigma^*$ and $M \subseteq [1, \length{w}]^2$,
where everything is interpreted as for $\FO_\Sigma[\mathcal{C}]$-formulas on $w$
excepted for $\match$ that is interpreted by $M$. 
Given a class $\mathcal{C}$ of numerical predicates, by $\FO[\mathcal{C}]$ we
denote the class of all languages over any finite alphabet $\Sigma$ defined by a
$\FO_\Sigma[\mathcal{C}]$-sentence. A classical result at the interplay of
circuit complexity and logic is that $\ACO = \FOarb$, where $\mathsf{arb}$
denotes the class of all numerical predicates
(see~\cite[Theorem~IX.2.1]{Books/Straubing-1994}
or~\cite[Corollary~5.32]{Books/Immerman-1999}). The other numerical predicates
that we will encounter in this paper are $<$, $+$ and $\MODpred_m$ for all
$m \in \N_{>0}$ (gathered together in the set
$\MODpred = \set{\MODpred_m\mid m>0}$),
where $\MODpred_m$ holds for those positions that are divisible
by $m$.

\subsection{Main result}
The notion of length-synchronicity and weak length-synchronicity will be a central
notion in our main result.
In the following $\Sigma$ always denotes a visibly pushdown alphabet.

\begin{samepage}
	\begin{defi}[(Weak) Length-Synchronicity]\label{def synchronicity}
		Let $\newR\subseteq\Con(\Sigma)$ be a set of contexts.
\begin{itemize} 
    \item $\newR$ is {\em length-synchronous} if
	$\length{u}/\length{v}=\length{u'}/\length{v'}$ for all $(u,v),(u',v')\in \newR$ with $\Delta(u),\Delta(u')>0$.
		\item $\newR$ is {\em weakly length-synchronous} if 
		    $u=u'$ implies $\length{v}=\length{v'}$ and $v=v'$ 
		implies $|u|=|u'|$ for all $(u,v),(u',v')\in 
		\newR$
		with $\Delta(u),\Delta(u')>0$.
		\end{itemize}
\end{defi}
\end{samepage}
	\noindent
	Note that a set of contexts $\newR$ is weakly length-synchronous if 
	$\newR$ is length-synchronous. Indeed, if, say $(u,v),(u,v')\in\newR$,
	where $\length{v}\not=\length{v'}$ and $\Delta(u)>0$, then
	$\length{u},\length{v},\length{v'}>0$  
	and so the quotients $\frac{\length{u}}{\length{v}}$ and
	$\frac{\length{u}}{\length{v'}}$ are distinct,
	thus violating length-synchronicity of $\newR$.

\begin{defi}[Quasi-Counterfree]\label{def vertically}
	
		A VPL $L\subseteq\Sigma^\triangle$ 
		is {\em quasi-counterfree} 
if for all contexts 
$\sigma\in\Sigma^k\times\Sigma^l$ with $k,l\in\N$ arbitrary at least one of the
following holds: (1) there exists some $n\in\N$ such that
$\sigma^n\equiv_L\sigma^{n+1}$ or (2) no context in $\Sigma^k\times\Sigma^l$ is
$\equiv_L$-equivalent to $\sigma\circ\sigma$.

\end{defi}
	We will later show that quasi-counterfreeness of a VPL
$L\subseteq\Sigma^\triangle$ is equivalent to the condition that 
every set of contexts in $\Con(\Sigma)$ forming a non-trivial group
when considering the associated equivalence classes with respect to $\equiv_L$
contains two contexts one of which is in $\Sigma^k\times\Sigma^l$
the other of which is in $\Sigma^{k'}\times\Sigma^{l'}$ where moreover
$(k,l)\not=(k',l')$, or equivalently there
is no $k,l\in\N$ such that there is a subset of 
$\Con(\Sigma)\cap\Sigma^k\times\Sigma^l$
that forms a non-trivial group 
when considering the associated equivalence classes with respect
to $\equiv_L$ (Proposition~\ref{prop quasi-counterfree}).

\begin{samepage}
\begin{exa}\label{example SM}
	
	Consider the visibly pushdown alphabet $\Sigma$, 
	where
	$\Sigma_\call=\set{a}$, $\Sigma_\internal=\set{c}$ and $\Sigma_\return=\set{b_1,b_2}$.
    For all $k, l \in \N_{>0}$ satisfying $k \neq l$, consider the language
$\SM_{k, l}$ generated by the context-free grammar
	$
	S\rightarrow ac^{k - 1} S b_1 \mid ac^{l - 1} S b_2 \mid \emptyword
	\displaypunct{.}$
	We have that the set of contexts
	$\set{(u,v)\in\Con(\Sigma)\mid S\Rightarrow_G^*uSv}$
	is weakly length-synchronous since both the relation and its reverse is a partial function --- however, it is not length-synchronous.
It is also not hard to see that $\SM_{k,l}$ is quasi-counterfree. Indeed, given
$(u, v) \in \Con(\Sigma)$, if $u$ contains the letter $b_1$ or $b_2$, or $v$
contains the letter $a$ or the letter $c$ along with the letter $b_1$ or $b_2$,
we have that $(\chi\circ(u,v)^2)w \notin \SM_{k,l}$ for all $\chi\in\Con(\Sigma)$
and all $w\in\Sigma^\triangle$, so $(u, v)^2 \equiv_{\SM_{k,l}} (u, v)^3$. If
$u$ and $v$ happen to contain the letter $c$ and only this letter, we can
argue similarly.
In the cases remaining, $u$ contains only letters from $\set{a,c}$ and $v$
letters from $\set{b_1,b_2}$.
	
\end{exa}
\end{samepage}

We say a context-free grammar $G=(V,\Sigma,P,S)$ is \emph{vertically visibly pushdown} if 
the underlying alphabet $\Sigma$ is a visibly pushdown alphabet,
$S\rightarrow_G\varepsilon$, 
and all other productions of $G$ are of the form 
			$T\rightarrow_G uT'v$,
			where $uv\in\Sigma^\triangle$ is one-turn
			such that $u\in
			(\Sigma_\internal^*\Sigma_\call\Sigma_\internal^*)^+$
			and $v\in(\Sigma_\internal^*\Sigma_\return\Sigma_\internal^*)^+$.
			Note that each grammar generating $\SM_{k,l}$ 
			in Example~\ref{example SM} is
			vertically visibly pushdown.
The following remark is obvious.
\begin{rem}\label{remark simple grammar}
The languages generated by vertically visibly pushdown grammars are one-turn VPLs.
\end{rem}

\begin{defi}[Intermediate VPL]
	\label{def intermediate}
	A VPL $L$ is intermediate if
it is quasi-counterfree and
	$L=L(G)$
	for some vertically visibly pushdown grammar $G$
for which \[\newR(G)=\set{(u,v)\in\Con(\Sigma)\mid
	    S\Rightarrow_G^* uSv}\] is 
			weakly length-synchronous but not 
			length-synchronous.
\end{defi}

Thus the languages $\SM_{k,l}$ from Example~\ref{example SM}
are all intermediate VPLs.
Loosely speaking, they are the simplest intermediate VPLs.
We remark that every intermediate language is in $\TCO$.
We have the following conjecture.
\begin{conj}\label{conjecture main}
There is no intermediate VPL that is in $\ACCO$ or $\TCO$-hard.
\end{conj}
In fact, the authors are not even aware of any intermediate VPL that is provably not in $\ACO$.
An indication for the inadequacy of known techniques to prove 
the latter is that the {\em robustness}~\cite{Juknabook} of
intermediate VPLs can be proven to be constant. Further techniques, 
based for instance on the {\em switching lemma}~\cite{Hastad86} or 
on the {\em polynomial method}~\cite{Beigel93} also do 
not seem to be applicable.

Our main result is the following theorem.

\begin{samepage}
\begin{thm} \label{thm main}
	There is an algorithm that, given a DVPA $A$, correctly outputs either
	\begin{itemize}
		\item $L(A)\in\ACO$,
		\item $m\geq 2$ such that
			$L$ is $\ACCO(m)$-hard
			(hence implying that $L(A)\not\in\ACO$),
		\item vertically visibly pushdown grammars
			$G_1,\dots,G_m$ each generating intermediate
			VPLs such that
			$L\eqACO\biguplus_{i\in[1,m]}L(G_i)$. 
			In this case one can moreover effectively compute 
$k,l\in\N$ with $k\not=l$ such that $\SM_{k,l}\reduce L(A)$.
	\end{itemize}
	\end{thm}
\end{samepage}
\noindent
Theorem~\ref{thm main} and the following conjecture imply the existence of an algorithm
that decides if a given visibly pushdown language is in $\ACO$.
\begin{conj}
Either all intermediate VPLs are in $\ACO$ or all are not.
\end{conj}

\subsection{Corollary for visibly counter languages}
A {\em visibly counter automaton} with threshold $m$
($m$-VCA) over a visibly pushdown alphabet $\Sigma$
is a tuple $A=(Q,\Sigma,q_0,F,\delta_0,\dots,\delta_m)$,
where $Q$ is a finite set of {\em states},
$q_0$ is the {\em initial state}, $F\subseteq Q$ is a set of
{\em final states}, $m\geq 0$ is a {\em threshold},
and $\delta_i\colon Q\times\Sigma\rightarrow Q$ is a {\em transition function}
for each $i\in[0,m]$.

A {\em configuration of $A$} is an element of $Q\times\N$.
For any two configurations $(q,n),(q',n')$ and any $x\in\Sigma$ we define 
$(q,n)\xrightarrow{x}_A(q',n')$ 
if $q'=\delta_{\min(n,m)}(q,x)$ and $n'=n+\Delta(x)$.
The relation $\xrightarrow{x}_A$ is naturally extended to $\xrightarrow{w}_A$
for $w\in\Sigma^*$.
By $L(A)=\set{w\in\Sigma^\triangle\mid \exists q\in F : (q_0,0)\xrightarrow{w}_A(q,0)}$
we denote the {\em language} (of well-matched words) of $A$.
We remark that the language of any $m$-VCA is a visibly 
pushdown language.
We also remark that the languages of $m$-VCAs are defined to be sets
of well-matched words as in~\cite{BLS06}, 
whereas in~\cite{Krebs-Lange-Ludwig-2015} 
the well-matched requirement is not present.

The following corollary of Theorem~\ref{thm main} implies 
the main result of \cite{Krebs-Lange-Ludwig-2015}
when restricted to well-matched words.

\begin{cor}\label{corollary OCA}
	There is an algorithm that, given an $m$-VCA $A$, 
	correctly outputs either that $L(A)$ is in $\ACO$ or some $m\geq 2$
	such that $L(A)$ is $\ACCO(m)$-hard
	(hence implying $\MOD_m\reduce L(A)$ and thus $L(A)\not\in\ACO$).
\end{cor}

For the proof of Corollary~\ref{corollary OCA} we 
refer to Section~\ref{S Corollary}.

\section{Language-theoretic and algebraic foundations and Ext-Algebras}\label{Section Ext Algebras}

\subsection{Basic algebraic automata theory}\label{Algebraic Foundations}

For a thorough introduction to algebraic automata theory, we refer the reader to
the two classical references of the domain by
Eilenberg~\cite{Books/Eilenberg-1974, Books/Eilenberg-1976} and
Pin~\cite{Books/Pin-1986}, but also to the following central 
reference in automata
theory~\cite[Chapter~1]{Books/HAT}.

A {\em semigroup} is a pair $(M,\cdot)$, where $M$ is a non-empty set and
$\cdot$ is a binary operation on $M$ that is associative, i.e.\
$x\cdot (y\cdot z)=(x\cdot y)\cdot z$ for all $x,y,z\in M$. Usually, when the
operation is clear from the context, we write it multiplicatively and write just
$M$ instead of $(M,\cdot)$.
The semigroup $M$ is {\em trivial} if $\card{M}=1$, and {\em non-trivial}
otherwise.
A {\em subsemigroup} of $M$ is a semigroup $N$ such that $N$ is a subset of $M$
and the operation of $N$ is the restriction of the operation of $M$ to $N$.
We often just write $xy$ to denote $x \cdot y$.
An {\em idempotent} of a semigroup $M$ is an element $x\in M$ satisfying $x=xx$.
The {\em idempotent power} of a finite semigroup $M$ is the smallest positive
integer $\omega$ such that $x^\omega$ is an idempotent for all $x\in M$.
The {\em zero} of a semigroup $M$ is the unique element $x\in M$ (if it exists) 
satisfying $xy=yx=x$ for all $y\in M$.
A {\em monoid} is a semigroup $M$ with a neutral element, that is, an element
$e \in M$ such that $x \cdot e = e \cdot x = x$ for all $x\in M$. We usually
denote the neutral element of a monoid $M$ by $1_M$.
A {\em submonoid} of $M$ is a monoid $N$ that is a subsemigroup of $M$
containing $1_M$ (which is thus also the neutral element of $N$).
Consider some monoid $M$. 
A {\em congruence} on $M$ is an equivalence relation $\sim$ on $M$ that
satisfies $v x z\sim vyz$ for all $v,z\in M$ and all $x,y\in M$ with $x\sim y$.
We denote by $[x]_\sim$ the equivalence class of $x\in M$.
The {\em quotient} of $M$ with respect to a congruence $\equiv$ 
is the monoid $M\quotient\equiv$ with base set
$M\quotient\equiv = \set{[m]_\equiv\mid m\in M}$ and operation given by
$[x]_\equiv\cdot[y]_\equiv=[xy]_\equiv$ for all $x,y\in M$.

A {\em group} is a monoid $M$ in which for all $x\in M$ there exists an inverse,
that is, an element $x' \in M$ such that $xx'=x'x=1_M$. Each element in a group
$M$ has a unique inverse, so we denote by $x^{-1}$ the unique inverse of an
$x \in M$.
A {\em subgroup} of a group $M$ is a submonoid of $M$ that is a group.
Given a semigroup $M$, a set $S \subseteq M$ and a subsemigroup $N$ of $M$, whenever
$N \subseteq S$, $N$ is said to be \emph{contained} in $S$.
A semigroup $M$ is {\em aperiodic} if it does not contain any non-trivial group.
It is well-known that a finite semigroup $M$ is aperiodic if, and only if, given
$\omega$ the idempotent power of $M$, it holds that $x^\omega=x^{\omega+1}$ for
all $x\in M$ if, and only if, there exists $k \in \N_{>0}$ such that
$x^k = x^{k + 1}$ for all $x \in M$.

A {\em morphism} from a monoid $M$ to a monoid $N$ is a mapping
$\varphi\colon M\to N$ such that $\varphi(1_M)=1_N$ and
$\varphi(xy)=\varphi(x)\varphi(y)$ for all $x,y\in M$.
If $M=\Sigma^*$ and $N=\Gamma^*$ where $\Sigma$ and $\Gamma$ are finite
alphabets, we call $\varphi$ {\em length-multiplying} whenever there exists
$k \in \N$ such that $\varphi(\Sigma) \subseteq \Gamma^k$.
Let $\varphi\colon \Sigma^*\to M$ be a morphism, where $\Sigma$ is a finite
alphabet and $M$ is finite.
Then there exists $l\in\N_{>0}$ such that
$\varphi(\Sigma^l)=\varphi(\Sigma^{2l})$: this implies that $\varphi(\Sigma^l)$
is a semigroup. The smallest such $l$ is called the {\em stability index of the
morphism $\varphi$}. 
It is easily shown that if $\varphi(\Sigma^n)$ contains a non-trivial group for
some $n\in\N$, then so does $\varphi(\Sigma^l)$.
We say that $h$ is {\em quasi-aperiodic} if $\varphi(\Sigma^n)$ does not contain
any non-trivial group for all $n\in\N$, which is equivalent to the fact that
$\varphi(\Sigma^l)$ is aperiodic. (See~\cite{BCST92, Books/Straubing-1994} for
the original definition and~\cite{Straubing-2002} for the definition using the
stability index, though it has been only formulated for surjective morphisms.)

A language $L$ over a finite alphabet $\Sigma$ is \emph{recognized} by a monoid
$M$ if there is a morphism $\varphi\colon \Sigma^* \to M$ and $F \subseteq M$
such that $L = \varphi^{-1}(F)$.
The {\em syntactic monoid} of a language $L\subseteq\Sigma^*$ 
is the quotient of $\Sigma^*$ by the congruence $\sim_L$ (called the
\emph{syntactic congruence} of $L$) defined by
$x \sim_L y$ for $x, y \in \Sigma^*$ whenever for all $u, v \in \Sigma^*$,
$uxv \in L \Leftrightarrow uyv \in L$. The syntactic monoid of $L$ recognizes
$L$ via the {\em syntactic morphism} of $L$ sending any word $w \in \Sigma^*$ to
$[w]_{\sim_L}$.
A fundamental and well-known result is that a language $L$ is regular if, and
only if, it is recognized by a finite monoid if, and only if, its syntactic
monoid is finite.

\subsection{Ext-algebras}

This section builds on~\cite{Czarnetzki-Krebs-Lange-2018}, but identifies an
inaccuracy in the definition of $\Ext$-algebra morphisms to establish freeness.

Let $(M,\cdot,1_M)$ be a monoid.
For each $m \in M$, we shall respectively denote
by $\leftmult{m}$ and $\rightmult{m}$ the \emph{left-multiplication map}
$x \mapsto m \cdot x$ and the \emph{right-multiplication map}
$x \mapsto x \cdot m$.

\begin{defi}
    An \emph{$\Ext$-algebra} $(R, O, \cdot, \circ)$ consists of a monoid
    $(R, \cdot, 1_R)$ and a monoid $(O, \circ, 1_O)$ that is a submonoid of
    $(R^R, \circ)$ containing the maps $\leftmult{r}$ and $\rightmult{r}$ for
    each $r \in R$.
\end{defi}

We note that $(O,\circ)$ completely determines $(R,\cdot)$, indeed
$xy=\rightmult{y}(x)$.

    \begin{samepage}
\begin{defi}\label{def morphism}
    Let $(R, O)$ and $(S, P)$ be $\Ext$-algebras.
    An \emph{$\Ext$-algebra morphism from $(R, O)$ to $(S, P)$} is a pair
    $(\varphi, \psi)$ of monoid morphisms $\varphi\colon R \to S$ and
    $\psi\colon O \to P$ such that:
    \begin{itemize}
	\item
	    for all $e \in O$ and $r \in R$ we have
	    $\psi(e)(\varphi(r)) = \varphi(e(r))$;
	\item
	    for all $r \in R$ we have
	    $\psi(\leftmult{r}) = \leftmult{\varphi(r)}$ and
	    $\psi(\rightmult{r}) = \rightmult{\varphi(r)}$.
    \end{itemize}
    We write $(\varphi, \psi)\colon (R, O) \to (S, P)$.
    The morphism $(\varphi, \psi)$ is called \emph{surjective} (respectively
    \emph{bijective}) if both $\varphi$ and $\psi$ are \emph{surjective}
    (respectively \emph{bijective}).
\end{defi}
    \end{samepage}

When it is clear from the context, we shall write \emph{morphism} to mean
\emph{$\Ext$-algebra morphism}.

\begin{rem}
    In the above definition, $\varphi$ is totally determined by $\psi$, because
    for each $r \in R$, we have
	$\varphi(r) = \varphi(\leftmult{r}(1_R)) =
	\psi(\leftmult{r})(\varphi(1_R)) = \psi(\leftmult{r})(1_S)$.
\end{rem}

    \begin{samepage}
\begin{defi}
    Let $(R, O)$ and $(S, P)$ be $\Ext$-algebras.
    Then
    \begin{itemize}
	\item
	    $(R, O)$ is a \emph{sub-$\Ext$-algebra} of $(S, P)$ whenever $R$ is
	    a submonoid of $S$ and there exists a submonoid $O'$ of $P$ such
	    that $O = \set{\restr{e}{R} \mid e \in O'}$, so that we may denote
	    $O$ by $\restr{O'}{R}$.
	\item
	    $(R, O)$ is a \emph{quotient} of $(S, P)$ whenever there exists a
	    surjective morphism from $(S, P)$ to $(R, O)$.
	\item
	    $(R, O)$ \emph{divides} $(S, P)$ whenever $(R, O)$ is a quotient of
	    a sub-$\Ext$-algebra of $(S, P)$.
    \end{itemize}
\end{defi}
    \end{samepage}

\noindent
For the rest of this section, let us fix some visibly pushdown alphabet $\Sigma$.

\begin{defi}
	For all $(u, v) \in\Con(\Sigma)$,
    consider the function $\ext_{u, v}\colon \Sigma^\triangle \to \Sigma^*$ such
    that $\ext_{u, v}(x) = u x v$ for all $x \in \Sigma^\triangle$.
	We call \[\ext_{x_1,y_1}\circ\cdots\circ\ext_{x_m,y_m}\]
	a factorization of $\ext_{u,v}$
    if $u=x_1\cdots x_m$ and $v=y_m\cdots y_1$.
\end{defi}
The following lemma states that each $\ext_{u,v}$ has a unique factorization 
when restricting the $(x_i,y_i)$ to be from $\Sigma^\triangle\times\Sigma^\triangle$
or from $\Sigma_\call\times\Sigma_\return$ and minimizing the number
of $(x_i,y_i)\in\Sigma^\triangle\times\Sigma^\triangle$:
we obtain its so-called {\em stair factorization}.
\begin{lem}\label{lemma ext}
	For all $\ext_{u,v}$ there exists a unique factorization 
    \[
	\ext_{u,v}
	\quad=\quad
	\ext_{x_1,y_1} \circ \ext_{a_1,b_1} \circ \quad\cdots\quad \circ
	\ext_{x_{h-1},y_{h-1}} \circ \ext_{a_{h-1},b_{h-1}} \circ \ext_{x_h,y_h}
    \]
	satisfying $h\geq 1$, $x_i,y_i\in\Sigma^\triangle$ for all $i\in[1,h]$
	and $a_i\in\Sigma_\call$ and $b_i\in\Sigma_\return$ for all 
	$i \in [1,h - 1]$.
    In particular, $\ext_{u,v}$ is in fact a function from $\Sigma^\triangle$ to
    $\Sigma^\triangle$.
\end{lem}
\begin{proof}
    We show additionally that the required factorization must satisfy
    $h=\Delta(u)+1$.
    We prove the statement by induction on $|uv|$.
    In case $|uv|\leq 1$, then either 
    $\ext_{u,v}=\ext_{\emptyword,\emptyword}$,
    or there is some $c\in\Sigma_\internal$ such that
    $\ext_{u,v}=\ext_{\emptyword,c}$
    or $\ext_{u,v}=\ext_{c,\emptyword}$.
    In any of these cases, we uniquely factorize $\ext_{u, v}$ as
    $\ext_{x_1, y_1}$ with $x_1 = u$ and $y_1 = v$.

    Let us consider the case when $|uv|\geq 2$ and let $h=\Delta(u)+1$.
    Note that since $uv\in\Sigma^\triangle$ we
    have $u\in\Sigma^\triangle$ if, and only if, $v\in\Sigma^\triangle$.
    In case $u,v\in\Sigma^\triangle$ we have $\Delta(u)=0$, hence
    the only factorization
    of the desired form is indeed $\ext_{u,v}=\ext_{x_1,y_1}$,
    where $x_1=u$ and $y_1=v$.
    Let us finally treat the case when $u,v\notin\Sigma^\triangle$,
    thus $\Delta(u)\geq 1$ and hence $h\geq 2$.
    Let $x$ be the maximal prefix of $u$ satisfying $x\in\Sigma^\triangle$
    and let $y$ be the maximal suffix of $v$ satisfying $y\in\Sigma^\triangle$.
    Due to maximality of $x$ and $y$ there must exist $a\in\Sigma_\call$,
    $b\in\Sigma_\return$, and $u',v'\in\Sigma^*$ such that $u=xau'$, $v=v'by$
    and $u'v'\in\Sigma^\triangle$ with $\Delta(u')=\Delta(u)-1=h-2$.
    Let
    $\ext_{x_1,y_1} \circ \ext_{a_1,b_1} \circ \cdots \circ
     \ext_{x_{h-2},y_{h-2}} \circ \ext_{a_{h-2},b_{h-2}} \circ
     \ext_{x_{h-1},y_{h-1}}$
    be the unique factorization of the desired form for $\ext_{u',v'}$ by
    induction hypothesis.
    We claim that 
    \[
	\ext_{x,y} \circ \ext_{a,b} \circ
	\ext_{x_1,y_1} \circ \ext_{a_1,b_1} \circ \cdots \circ
	\ext_{x_{h-2},y_{h-2}} \circ \ext_{a_{h-2},b_{h-2}} \circ
	\ext_{x_{h-1},y_{h-1}}
    \]
    is the unique factorization of the desired form for $\ext_{u,v}$. 
    Indeed, since $\Delta(u) \geq 1$ any potential factorization of the desired
    form for $\ext_{u,v}$ must be of the form
    $\ext_{x',y'}\circ\ext_{a',b'}\circ\pi$, where $x'$ is a prefix of $u$
    satisfying $x'\in\Sigma^\triangle$, $y'$ is a suffix of $v$ satisfying
    $y'\in\Sigma^\triangle$, $a'\in\Sigma_\call$, and $b'\in\Sigma_\return$. 
    In particular $x'$ is a prefix of $x$ and $y'$ is suffix of $y$. 
    In case $x'=x$ and $y'=y$ it follows $a'=a$ and $b'=b$ and
    uniqueness follows from induction hypothesis.
    It remains to consider the case when $x'$ is a strict prefix of $x$ or
    $y'$ is a strict suffix of $y$. We only treat the former case.
    It must hold $x=x'a's$ for some $s\in\Sigma^+$ such that
    $a's\in\Sigma^\triangle$. But then $\pi$ is a factorization for
    $\ext_{su',v'z}$ for some $z\in\Sigma^*$ which is a contradiction since
    $\Delta(s)=-1$ due to $a's\in\Sigma^\triangle$.
\end{proof}

In the following we will denote the unique factorization provided by Lemma~\ref{lemma ext}
as the {\em stair factorization} of $\ext_{u,v}$.
Consider now the set $\ExtMon(\Sigma^\triangle)$ of all functions $\ext_{u, v}$ 
for $(u, v) \in \Con(\Sigma)$: it is a subset of
$(\Sigma^\triangle)^{\Sigma^\triangle}$ closed under composition. 
Thus,
$(\ExtMon(\Sigma^\triangle), \circ)$ is a submonoid of
$((\Sigma^\triangle)^{\Sigma^\triangle}, \circ)$.
Since for all $w\in\Sigma^\triangle$ we have 
$\leftmult{w}=\ext_{w,\emptyword}$ and $\rightmult{w}=\ext_{\emptyword,w}$,
the set $\ExtMon(\Sigma^\triangle)$ contains the functions
$\leftmult{w}$ and $\rightmult{w}$ for all $w \in \Sigma^\triangle$.
Hence, $(\Sigma^\triangle, \ExtMon(\Sigma^\triangle), \cdot, \circ)$ is an
$\Ext$-algebra.
The following important proposition establishes freeness
of $(\Sigma^\triangle,\ExtMon(\Sigma^\triangle))$.

\begin{samepage}
\begin{prop}\label{ptn:Freeness_Ext-algebras}
    Let $(R, O)$ be an $\Ext$-algebra and consider two functions
    $\varphi\colon \Sigma_\internal \to R$ and
    $\psi\colon\set{\ext_{a, b}\mid a \in \Sigma_\call, b \in \Sigma_\return} \to
     O$.
    Then there exists a unique $\Ext$-algebra morphism
    $(\overline{\varphi}, \overline{\psi})$ from
    $(\Sigma^\triangle, \ExtMon(\Sigma^\triangle))$ to $(R, O)$ satisfying
    $\overline{\varphi}(c) = \varphi(c)$ for each $c \in \Sigma_\internal$ and
    $\overline{\psi}(\ext_{a, b}) = \psi(\ext_{a, b})$ for each
    $a \in \Sigma_\call, b \in \Sigma_\return$.
\end{prop}
\end{samepage}
\begin{proof}
We define $\overline{\varphi}$ based on a refinement
of the structural definition of well-matched words.
For each $w\in\Sigma^\triangle$ we inductively define:
\[
    \overline{\varphi}(w) = 
    \begin{cases}
	1_R & \text{if $w=\emptyword$ (type 1)}\\
	\varphi(c) & \text{if $w=c\in\Sigma_\internal$ (type 2)}\\
	\psi(\ext_{a,b})(\overline{\varphi}(x)) & 
	\text{if $w=axb$ for $a\in\Sigma_\call$, $b\in\Sigma_\return$ and 
	$x\in\Sigma^\triangle$ (type 3)}\\
	\overline{\varphi}(x)\overline{\varphi}(y) & \text{if $w=xy$ for
	$x,y\in\Sigma^\triangle\setminus\set{\emptyword}$, where $|x|$ is minimal 
	(type 4)}
    \end{cases}
\]
Observe that the four above types give unique decompositions.
For proving that $\overline{\varphi}$ is indeed a monoid morphism 
one proves that for all $w,v\in\Sigma^\triangle$ we 
have $\overline{\varphi}(wv)=\overline{\varphi}(w)\overline{\varphi}(v)$
by structural induction on $w$ given by the four types.
The case $v=\emptyword$ is direct, we only treat the case 
$v\in\Sigma^\triangle\setminus\set{\emptyword}$ in the following.
If $w$ is of type 1
we have $\overline{\varphi}(wv)=\overline{\varphi}(v)=1_R\cdot
\overline{\varphi}(v)=\overline{\varphi}(w)\overline{\varphi}(v)$.
If $w$ is of type 2 or 3, then $wv$ is of type 4 and $w$ is the shortest prefix
of $wv$ with $w\in\Sigma^\triangle\setminus\set{\emptyword}$, hence
$\overline{\varphi}(wv)=\overline{\varphi}(w)\overline{\varphi}(v)$.
If $w$ is of type 4, then $w=xy$ for some
$x,y\in\Sigma^\triangle\setminus\set{\varepsilon}$, where $x$ is of minimal
length.
Then $wv$ is of type 4, where $wv=x(yv)$ and $x$ is the shortest prefix of $wv$
with $x\in\Sigma^\triangle\setminus\set{\emptyword}$.
Hence
$\overline{\varphi}(wv)=\overline{\varphi}(x)\overline{\varphi}(yv)=
\overline{\varphi}(x)\overline{\varphi}(y)\overline{\varphi}(v)=
\overline{\varphi}(xy)\overline{\varphi}(v)=\overline{\varphi}(w)\overline{\varphi}(v)$,
where the first equality follows by definition of $\overline{\varphi}$
and the second and third equality follow from the induction hypothesis.
Given any $\ext_{u,v}\in\ExtMon(\Sigma^\triangle)$ 
let 
\[
    \ext_{u,v}
    \quad=\quad
    \ext_{x_1,y_1} \circ \ext_{a_1,b_1} \circ \quad\cdots\quad \circ
    \ext_{x_{h-1},y_{h-1}} \circ \ext_{a_{h-1},b_{h-1}} \circ \ext_{x_h,y_h}
\]
be the unique stair factorization given by Lemma~\ref{lemma ext}.
We define
\[
    \overline{\psi}(\ext_{u,v})=
    \bigcirc_{i=1}^{h-1}
    \bigl(
	\leftmult{\overline{\varphi}(x_i)} \circ
	\rightmult{\overline{\varphi}(y_i)} \circ
	\psi(\ext_{a_i,b_i})
    \bigr) \circ
    \leftmult{\overline{\varphi}(x_h)} \circ \rightmult{\overline{\varphi}(y_h)}
    \displaypunct{.}
\]
For showing that $\overline{\psi}$ is indeed a monoid morphism,
one proves
$\overline{\psi}(\ext_{uu',v'v}) =
 \overline{\psi}(\ext_{u,v}) \circ \overline{\psi}(\ext_{u',v'})$
for all $\ext_{u, v}, \ext_{u', v'} \in \ExtMon(\Sigma^\triangle)$ by observing
simply that the unique stair factorization of $\ext_{u u', v' v}$ is obtained by
composing the unique stair factorizations of $\ext_{u, v}$ and $\ext_{u', v'}$.

We now show that $(\overline{\varphi},\overline{\psi})$ is in fact an
$\Ext$-algebra morphism.
The discussion above first shows that both $\overline{\varphi}\colon\Sigma^\triangle\rightarrow R$
and $\overline{\psi}\colon\ExtMon(\Sigma^\triangle)\rightarrow O$
are monoid morphisms.
Next, let us prove that for all $\ext_{u,v}\in\ExtMon(\Sigma^\triangle)$ and 
$w\in\Sigma^\triangle$ we have
$\overline{\psi}(\ext_{u,v})(\overline{\varphi}(w))=\overline{\varphi}(\ext_{u,v}(w))$.
Let
\[
    \ext_{u,v}
    \quad=\quad
    \ext_{x_1,y_1} \circ \ext_{a_1,b_1} \circ \quad\cdots\quad \circ
    \ext_{x_{h-1},y_{h-1}} \circ \ext_{a_{h-1},b_{h-1}} \circ \ext_{x_h,y_h}
\]
be the unique stair factorization of $\ext_{u,v}$ provided by
Lemma~\ref{lemma ext}.
If $h = 1$, then
\[
    \overline{\psi}(\ext_{u,v})(\overline{\varphi}(w)) =
    \leftmult{\overline{\varphi}(x_h)} \circ
    \rightmult{\overline{\varphi}(y_h)}(\overline{\varphi}(w)) =
    \overline{\varphi}(x_hwy_h) =
    \overline{\varphi}(\ext_{u, v}(w))
    \displaypunct{.}
\]
Otherwise, we have 
\begin{align*}
    & \overline{\psi}(\ext_{u,v})(\overline{\varphi}(w))\\
    =&
    \bigcirc_{i=1}^{h-1}
    \bigl(
	\leftmult{\overline{\varphi}(x_i)} \circ
	\rightmult{\overline{\varphi}(y_i)} \circ
	\psi(\ext_{a_i,b_i})
    \bigr) \circ
    \leftmult{\overline{\varphi}(x_h)}\circ\rightmult{\overline{\varphi}(y_h)}
    (\overline{\varphi}(w))\\
    =&
    \bigcirc_{i=1}^{h-1}
    \bigl(
	\leftmult{\overline{\varphi}(x_i)} \circ
	\rightmult{\overline{\varphi}(y_i)} \circ
	\psi(\ext_{a_i,b_i})
    \bigr)
    \bigl(\overline{\varphi}(x_hwy_h)\bigr)\\
    =&
    \bigcirc_{i=1}^{h-2}
    \bigl(
	\leftmult{\overline{\varphi}(x_i)} \circ
	\rightmult{\overline{\varphi}(y_i)} \circ
	\psi(\ext_{a_i,b_i})
    \bigr) \circ
    \\
    &
    \leftmult{\overline{\varphi}(x_{h-1})}\circ\rightmult{\overline{\varphi}(y_{h-1})}\circ\psi(\ext_{a_{h-1},b_{h-1}})
    \bigl(\overline{\varphi}(x_hwy_h)\bigr)\\
    =&
    \bigcirc_{i=1}^{h-2}
    \bigl(
	\leftmult{\overline{\varphi}(x_i)} \circ
	\rightmult{\overline{\varphi}(y_i)} \circ
	\psi(\ext_{a_i,b_i})
    \bigr) \circ
    \\
    &
    \leftmult{\overline{\varphi}(x_{h-1})}\circ\rightmult{\overline{\varphi}(y_{h-1})}
    \bigl(\overline{\varphi}(a_{h-1}x_hwy_hb_{h-1})\bigr)\\
    =&
    \bigcirc_{i=1}^{h-2}
    \bigl(
	\leftmult{\overline{\varphi}(x_i)} \circ
	\rightmult{\overline{\varphi}(y_i)} \circ
	\psi(\ext_{a_i,b_i})
    \bigr)
    \bigl(\overline{\varphi}(x_{h-1}a_{h-1}x_hwy_hb_{h-1}y_{h-1})\bigr)\\
    =& \cdots\\
    =& \overline{\varphi}(x_1a_1\cdots x_{h-1}a_{h-1}x_hwy_hb_{h-1}y_{h-1}\cdots b_1y_1)\\
    =& \overline{\varphi}(\ext_{u,v}(w))
    \displaypunct{.}
\end{align*}

Let us prove that for all $w\in\Sigma^\triangle$
we have $\overline{\psi}(\leftmult{w})=\leftmult{\overline{\varphi}(w)}$.
Noting that the unique stair factorization of $\leftmult{w}$ is $\ext_{w,\emptyword}$
we obtain
\[
    \overline{\psi}(\leftmult{w})=
    \overline{\psi}(\ext_{w,\emptyword})=
    \leftmult{\overline{\varphi}(w)}\circ\rightmult{\overline{\varphi}(\emptyword)}=
    \leftmult{\overline{\varphi}(w)}\circ\rightmult{1_R}=
    \leftmult{\overline{\varphi}(w)}\circ 1_O
    = \leftmult{\overline{\varphi}(w)}
    \displaypunct{.}
\]
One proves $\overline{\psi}(\rightmult{w})=\rightmult{\overline{\varphi}(w)}$
for all $w \in \Sigma^\triangle$ analogously.

Therefore, $(\overline{\varphi}, \overline{\psi})$ is an $\Ext$-algebra morphism
and it is the unique one satisfying $\overline{\varphi}(c) = \varphi(c)$ for
each $c \in \Sigma_\internal$ and
$\overline{\psi}(\ext_{a, b}) = \psi(\ext_{a, b})$ for each
$a \in \Sigma_\call, b \in \Sigma_\return$.
Take indeed any such $\Ext$-algebra morphism $(\varphi', \psi')$: using the
properties of $\Ext$-algebra morphisms, it is straightforward to prove that then
$\overline{\varphi}(w) = \varphi'(w)$ for all $w \in \Sigma^\triangle$ by
structural induction on $w$ and then to prove that
$\overline{\psi}(\ext_{u, v}) = \psi'(\ext_{u, v})$ for all
$\ext_{u, v} \in \ExtMon(\Sigma^\triangle)$ by using the unique stair
factorization of $\ext_{u,v}$ provided by Lemma~\ref{lemma ext}.
\end{proof}

\begin{rem}
    The second condition in Definition~\ref{def morphism}, i.e.\ for 
    all $r\in R$ we have $\psi(\leftmult{r})=\leftmult{\varphi(r)}$
    and $\psi(\rightmult{r})=\rightmult{\varphi(r)}$,
    does not appear 
    in the definition of $\Ext$-algebra morphisms given
    in~\cite{Czarnetzki-Krebs-Lange-2018}. But this is actually problematic,
    because then Proposition~\ref{ptn:Freeness_Ext-algebras}
    would not hold in general.

    Indeed, consider for instance the visibly pushdown alphabet $\Gamma$ where
    $\Gamma_\call = \set{a}$, $\Gamma_\internal = \emptyset$ and
    $\Gamma_\return = \set{b}$, where $R$ the is semi-lattice on two elements
    $\set{0, 1}$ such that
    $1 \cdot 1 = 1$ and $0 \cdot 1 = 1 \cdot 0 = 0 \cdot 0 = 0$; and moreover
    $O$ is defined as  $\set{\id, \mathbf{0}, \mathbf{1}}$ with
    $\mathbf{0}(0) = \mathbf{0}(1) = 0$ and $\mathbf{1}(0) = \mathbf{1}(1) = 1$.
    Then $(R, O)$ is an $\Ext$-algebra.
    Let us define the function $\varphi\colon \Gamma^\triangle \to R$ by
    $\varphi(w) = 1$ for all $w \in \Gamma^\triangle$ and the function
    $\psi\colon \ExtMon(\Gamma^\triangle) \to O$ by
    $\psi(\ext_{a^n, b^n}) = \id$ for all $n \in \N$ and
    $\psi(\ext_{u, v})  = \mathbf{1}$ for all $u, v \in \Gamma^*$ with
    $u v \in \Gamma^\triangle$ and
    $(u \in a \Gamma^* b \Gamma^*$ or $v \in \Gamma^* a \Gamma^* b)$.
    The pair $(\varphi, \psi)$ forms a morphism from
    $(\Gamma^\triangle, \ExtMon(\Gamma^\triangle))$ to $(R, O)$, but it is not
    the only one sending $\ext_{a, b}$ to $\id$, because we could also take
    $\psi$ to send all elements of $\ExtMon(\Gamma^\triangle)$ to $\id$.
\end{rem}

\begin{defi}
    A language $L \subseteq \Sigma^\triangle$ is \emph{recognized} by an
    $\Ext$-algebra $(R, O)$ whenever there exists a morphism 
    $(\varphi, \psi)\colon
     (\Sigma^\triangle, \ExtMon(\Sigma^\triangle)) \to (R, O)$
    such that $L = \varphi^{-1}(F)$ for some $F \subseteq R$.
\end{defi}

\begin{exa}\label{example L2}
	
    Consider the language
    $L = \SM_{1,2}=L(S\rightarrow a S b_1 \mid ac S b_2 \mid \emptyword)$ 
    from Example~\ref{example SM} over the visibly pushdown
    alphabet $\Gamma$, where $\Gamma_\internal=\set{c}$, $\Gamma_\call=\set{a}$ and 
    $\Gamma_\return=\set{b_1,b_2}$.
    Set
    $R_L=\set{\lbrack acb_1\rbrack_{\sim_L}, \lbrack\emptyword\rbrack_{\sim_L},
	      \lbrack c\rbrack_{\sim_L},\lbrack cab_1\rbrack_{\sim_L},
	      \lbrack ab_1\rbrack_{\sim_L}}$, 
    and, given for all $(u,v)\in\Con(\Sigma)$ the function
    $f_{u,v} \in (R_L)^{R_L}$ satisfying
    $f_{u,v}(\lbrack x\rbrack_{\sim_L}])=\lbrack uxv\rbrack_{\sim_L}$ for all
    $x\in\Sigma^\triangle$, set
    \[
	O_L =
	\set{f_{acb_1,\emptyword}, f_{\emptyword,\emptyword},
	     f_{c,\emptyword}, f_{\emptyword,c}, f_{ab_1,\emptyword},
	     f_{\emptyword,ab_1}, f_{cab_1,\emptyword},
	     f_{a,b_2}, f_{ca,b_2}, f_{ca,ab_1b_2}, f_{ca,b_1},
	     f_{a,ab_1b_2}, f_{a,b_1}}
	\displaypunct{.}
    \]
    For instance, note that 
    $\lbrack ab_1\rbrack_{\sim_L}=\lbrack acb_2\rbrack_{\sim_L}$,
    that $\lbrack acb_1\rbrack_{\sim_L}$ is the zero 
    of $R_L$ and that $f_{acb_1,\varepsilon}$ is the zero of $O_L$.
    Then $(R_L, O_L)$ is an $\Ext$-algebra recognizing $L$ thanks to the
    morphism
    $(\varphi_L,\psi_L)\colon 
     (\Gamma^\triangle,\ExtMon(\Gamma^\triangle))\to (R_L,O_L)$
    satisfying $\varphi_L(c)=\lbrack c\rbrack_{\sim_L}$, 
    $\psi_L(\ext_{a,b_1})=f_{a,b_1}$ and
    $\psi_L(\ext_{a,b_2})=f_{a,b_2}$.
    Note that $L=\varphi_L^{-1}(\set{\lbrack\emptyword\rbrack_{\sim_L},
    \lbrack ab_1\rbrack_{\sim_L}})$.
    Finally, note that for instance 
    $\psi_L(\ext_{ca,ab_1b_2})=f_{ca,ab_1b_2}
    \not=f_{a,ab_1b_2}=\psi_L(\ext_{a,ab_1b_2})$
    since we have
    \[\psi_L(\ext_{a,b_2})\circ
    \psi_L(\ext_{ca,ab_1b_2})(\lbrack c\rbrack_{\sim_L})=
    \lbrack acacab_1b_2b_2\rbrack_{\sim_L}
    =\lbrack ab_1\rbrack_{\sim_L}\]
    whereas 
    \[\psi_L(\ext_{a,b_2})\circ
    \psi_L(\ext_{a,ab_1b_2})(\lbrack c\rbrack_{\sim_L})=
    \lbrack aacab_1b_2b_2\rbrack_{\sim_L}=\lbrack acb_1\rbrack_L\displaypunct{.}\]
	
\end{exa}

\begin{defi}
    Let $(R, O)$ be an $\Ext$-algebra. An \emph{equivalence relation} on
	$(R, O)$ is an equivalence relation $\sim$ on $R$.
 We say an equivalence relation $\sim$ is a \emph{congruence} on $(R, O)$ 
	whenever
	for all $e \in O$ and for all $x, y \in R$ we have that
	$x \sim y$ implies $e(x) \sim e(y)$.
	In case $\sim$ is a congruence we denote by $(R,O)\quotient\sim$
	the pair $(R\quotient\sim,O')$, where
\[
	O' = \set{e' \in (R \quotient \sim)^{R \quotient \sim} \mid
		  \exists e \in O\ \forall x \in R\colon e'([x]_\sim) = [e(x)]_\sim}.
    \]
\end{defi}

The following lemma
actually shows that $(R,O)\quotient\sim$ is again an
$\Ext$-algebra, that we call {\em the quotient of $(R,O)$ by} $\sim$.

\begin{lem}
\label{lem:Ext_algebras_congruence_quotient}
    Let $(R, O)$ be an $\Ext$-algebra and $\sim$ be a congruence on
	$(R, O)$. 
	Then $(R \quotient \sim, O')$, with
    \[
	O' = \set{e' \in (R \quotient \sim)^{R \quotient \sim} \mid
		  \exists e \in O\ \forall x \in R\colon e'([x]_\sim) = [e(x)]_\sim}
    \]
	a submonoid of $(R \quotient \sim)^{R \quotient \sim}$,
	is an $\Ext$-algebra and the pair $(\varphi, \psi)$ of functions
    $\varphi\colon R \to R \quotient \sim$ and $\psi\colon O \to O'$ satisfying
    $\varphi(r) = [r]_\sim$ for all $r \in R$ and
    $\psi(e)([r]_\sim) = [e(r)]_\sim$ for all $e \in O$ and $r \in R$ is a
    surjective morphism from $(R, O)$ to $(R \quotient \sim, O')$.
\end{lem}

\begin{proof}
    Let $u, v \in R$ such that $u \sim v$.
    Take any $x, y \in R$: we have that
    \[
	x u y = \rightmult{y} \circ \leftmult{x}(u) \sim
	\rightmult{y} \circ \leftmult{x}(v) = x v y
    \]
	by definition of congruence. Thus, $\sim$ is a congruence on $R$. This implies that
    $R \quotient \sim$ is a monoid.

    Let $e', f' \in O'$: this means there exist $e, f \in O$ such that
    $e'([r]_\sim) = [e(r)]_\sim$ and $f'([r]_\sim) = [f(r)]_\sim$ for all
    $r \in R$. Given any $r \in R$, we thus have
    \[
	e' \circ f'([r]_\sim) = e'([f(r)]_\sim) = [e(f(r))]_\sim =
	[e \circ f(r)]_\sim
	\displaypunct{,}
    \]
    so that $e' \circ f' \in O'$. Therefore, $O'$ is a submonoid of
	$(R \quotient \sim)^{R \quotient \sim}$ 
	that contains the functions $\leftmult{[r]_{\sim}}$
	and $\rightmult{[r]_{\sim}}$ for all $[r]_\sim\in R\quotient\sim$.
	Thus, $(R\quotient\sim,O')$ is an $\Ext$-algebra.

    Now define the functions
    $\varphi\colon R \to R \quotient \sim$ and $\psi\colon O \to O'$ by
    respectively $\varphi(r) = [r]_\sim$ for all $r \in R$ and
    $\psi(e) = e'$ with $e' \in O'$ such that $e'([r]_\sim) = [e(r)]_\sim$ for
    all $r \in R$: this is well-defined because $\sim$ is a
    congruence on $(R, O)$.
    Since $\sim$ is a congruence on $R$, $\varphi$ is a surjective monoid
    morphism.
    Further, let $e, f \in O$. We have
    \begin{align*}
	\psi(e) \circ \psi(f) ([r]_\sim)
	& = \psi(e)([f(r)]_\sim)\\
	& = [e(f(r))]_\sim\\
	& = [e \circ f(r)]_\sim\\
	& = \psi(e \circ f)([r]_\sim)
    \end{align*}
    for all $r \in R$, so that $\psi(e) \circ \psi(f) = \psi(e \circ f)$.
    Therefore, as $\psi(1_O)([r]_\sim) = [1_O(r)]_\sim = [r]_\sim$ for all
    $r \in R$, it follows that $\psi$ is also a monoid morphism, that is
    obviously surjective.
    By construction, we do of course have that
    \[
	\psi(e)(\varphi(r)) = \psi(e)([r]_\sim) = [e(r)]_\sim = \varphi(e(r))
    \]
    for all $e \in O$ and $r \in R$.
    Moreover, for all $r \in R$, it holds that
    \[
	\psi(\leftmult{r})([x]_\sim) = [\leftmult{r}(x)]_\sim = [r x]_\sim =
	[r]_\sim [x]_\sim = \leftmult{\varphi(r)}([x]_\sim)
    \]
    for all $x \in R$, so that $\psi(\leftmult{r}) = \leftmult{\varphi(r)}$.
    Similarly, we can prove that $\psi(\rightmult{r}) = \rightmult{\varphi(r)}$
    for all $r \in R$.
    Thus, $(\varphi, \psi)$ is a surjective morphism from $(R, O)$ to
    $(R \quotient \sim, O')$.
\end{proof}

The lemma also proves that the pair $(\varphi, \psi)$ of functions
$\varphi\colon R \to R \quotient \sim$ and $\psi\colon O \to O'$ satisfying
$\varphi(r) = [r]_\sim$ for all $r \in R$ and $\psi(e)([r]_\sim) = [e(r)]_\sim$
for all $e \in O$ and $r \in R$ is a surjective morphism from $(R, O)$ to
$(R,O)\quotient\sim$.
We also call this pair $(\varphi,\psi)$ the {\em morphism associated to the
congruence} $\sim$.

\begin{defi}\label{def syntactic}
    The \emph{syntactic congruence of a language $L\subseteq\Sigma^\triangle$}
    is the congruence $\sim_L$ on
    $(\Sigma^\triangle, \ExtMon(\Sigma^\triangle))$ defined by $u \sim_L v$ for
    $u, v \in \Sigma^\triangle$ whenever 
	
$xuy\in L \Leftrightarrow xvy\in L$ for all $(x,y)\in\Con(\Sigma)$.
    We define the \emph{syntactic $\Ext$-algebra of $L$} to be
    $(R_L,O_L)=(\Sigma^\triangle, \ExtMon(\Sigma^\triangle)) \quotient\sim_L$
    and the \emph{syntactic morphism of $L$} to be the morphism
    $(\varphi_L,\psi_L)$ associated to $\sim_L$.
\end{defi}

Note that the syntactic $\Ext$-algebra $(R_L,O_L)$ of $L$ recognizes $L$ via the
syntactic morphism $(\varphi_L,\psi_L)$.
Indeed, for all $u, v \in \Sigma^\triangle$, we have that if $u \sim_L v$, then
$u \in L \Leftrightarrow v \in L$. This implies that
$L = \varphi_L^{-1}(\varphi_L(L))$.
For instance, it can be proven that the $\Ext$-algebra recognizing the language
$\SM_{1,2}$ in Example~\ref{example L2} is in fact a certain presentation of the
syntactic $\Ext$-algebra of $\SM_{1,2}$.    

The next lemma states that all languages recognized by an $\Ext$-algebra are
also recognized by the $\Ext$-algebras it divides.

\begin{lem}\label{lem:Division_and_recognisability}
	Let $(R, O)$ and $(S, P)$ be two $\Ext$-algebras such that $(R, O)$ divides
    $(S, P)$.
	Then any language $L \subseteq \Sigma^\triangle$ recognized
	by $(R, O)$ is also recognized by $(S, P)$.
\end{lem}

\begin{proof}
    Let $L \subseteq \Sigma^\triangle$ be a language recognized by $(R, O)$.
    This means that there exists a morphism
    $(\varphi, \psi)\colon
     (\Sigma^\triangle, \ExtMon(\Sigma^\triangle)) \to (R, O)$
    such that $L = \varphi^{-1}(F)$ for some $F \subseteq R$.
    We will prove the lemma by combining the following
    two points: 
    \begin{enumerate}[(1)]
	\item
	    if $(R,O)$ is a sub-$\Ext$-algebra of $(S,P)$, then so does $(S,P)$
	    recognize $L$, and 
	\item
	    if $(R,O)$ is a quotient of $(S,P)$, then so does $(S,P)$ recognize
	    $L$.
    \end{enumerate}
	\noindent
    For Point (1), assume that $(R, O)$ is a sub-$\Ext$-algebra of $(S, P)$.
    This means that $R$ is a submonoid of $S$ and that there exists a submonoid
    $O'$ of $P$ satisfying $O = \restr{O'}{R}$. Take an arbitrary function
    $\sigma\colon O \to P$ such that $\restr{\sigma(e)}{R} = e$ for all
    $e \in O$.
    Let us consider the unique morphism
    $(\varphi', \psi')\colon
     (\Sigma^\triangle, \ExtMon(\Sigma^\triangle)) \to (S, P)$
    such that $\varphi'(c) = \varphi(c)$ for all $c \in \Sigma_\internal$
    and $\psi'(\ext_{a, b}) = \sigma(\psi(\ext_{a, b}))$ for all
    $a \in \Sigma_\call, b \in \Sigma_\return$, given to us by
    Proposition~\ref{ptn:Freeness_Ext-algebras}.
    We can prove by induction on $w$ that $\varphi'(w) = \varphi(w)$ for all
    $w \in \Sigma^\triangle$:
    \begin{itemize}
	\item
		$w = \emptyword$. Then $\varphi'(w) = 1_S = 1_R = \varphi(w)$.
	\item
	    $w = c$ for some $c \in \Sigma_\internal$.
	    Then $\varphi'(w) = \varphi'(c) = \varphi(c) = \varphi(w)$.
	\item
	    $w = a w' b$ for some $a \in \Sigma_\call$, $b \in \Sigma_\return$
	    and $w' \in \Sigma^\triangle$.
	    Then
	    \begin{align*}
		\varphi'(w)
		& = \varphi'(\ext_{a, b}(w'))\\
		& = \psi'(\ext_{a, b})(\varphi'(w'))\\
		& \stackrel{\text{IH}}{=}\sigma(\psi(\ext_{a, b}))(\varphi(w'))\\
		    & = \restr{\sigma(\psi(\ext_{a, b}))}{R}(\varphi(w'))\\
		& = \psi(\ext_{a, b})(\varphi(w'))\\
		& = \varphi(\ext_{a, b}(w'))\\
		& = \varphi(w)
		\displaypunct{.}
	    \end{align*}
	\item
	    $w = u v$ for some
	    $u, v \in \Sigma^\triangle \setminus \set{\emptyword}$.
	    Then
	    \[
		    \varphi'(w) = \varphi'(u) \varphi'(v) \stackrel{\text{IH}}{=}
		\varphi(u) \varphi(v) = \varphi(w)
		\displaypunct{.}
	    \]
    \end{itemize}
    Thus, $\varphi'^{-1}(F) = L$, which implies that $(S, P)$ recognizes $L$.

    For Point (2), assume that $(R, O)$ is a quotient of $(S, P)$. 
    This means that there exists a surjective morphism
    $(\alpha, \beta)\colon (S, P) \to (R, O)$.
    Let us define an arbitrary function $\rho\colon \Sigma_\internal \to S$
    such that $\rho(c) \in \alpha^{-1}(\varphi(c))$ for all
    $c \in \Sigma_\internal$ as well as an arbitrary function
    $\sigma\colon\set{\ext_{a, b}\mid a \in \Sigma_\call, b \in \Sigma_\return} \to P$
    such that $\sigma(\ext_{a, b}) \in \beta^{-1}(\psi(\ext_{a, b}))$ for all
    $a \in \Sigma_\call, b \in \Sigma_\return$.
    Now, take the unique morphism
    $(\varphi', \psi')\colon
     (\Sigma^\triangle, \ExtMon(\Sigma^\triangle)) \to (S, P)$
    given by Proposition~\ref{ptn:Freeness_Ext-algebras} for $\rho$ and
    $\sigma$: we claim that it is such that $\alpha(\varphi'(w)) = \varphi(w)$
    for all $w \in \Sigma^\triangle$.
    We can prove it by induction on $w$:
    \begin{itemize}
	\item
	    $w = \emptyword$. Then
	    $\alpha(\varphi'(w)) = \alpha(1_S) = 1_R = \varphi(w)$.
	\item
	    $w = c$ for some $c \in \Sigma_\internal$.
	    Then
	    $\alpha(\varphi'(w)) = \alpha(\rho(c)) = \varphi(c) = \varphi(w)$.
	\item
	    $w = a w' b$ for some $a \in \Sigma_\call$, $b \in \Sigma_\return$
	    and $w' \in \Sigma^\triangle$.
	    Then
	    \begin{align*}
		\alpha(\varphi'(w))
		& = \alpha\bigl(\varphi'(\ext_{a, b}(w'))\bigr)\\
		& = \alpha\bigl(\psi'(\ext_{a, b})(\varphi'(w'))\bigr)\\
		& = \beta(\psi'(\ext_{a, b}))\bigl(\alpha(\varphi'(w'))\bigr)\\
		    & \stackrel{\text{IH}}{=}
		    \beta(\sigma(\ext_{a, b}))(\varphi(w'))\\
		& = \psi(\ext_{a, b})(\varphi(w'))\\
		& = \varphi(\ext_{a, b}(w'))\\
		& = \varphi(w)
		\displaypunct{.}
	    \end{align*}
	\item
	    $w = u v$ for some
	    $u, v \in \Sigma^\triangle \setminus \set{\emptyword}$.
	    Then
	    \[
		\alpha(\varphi'(w)) = \alpha(\varphi'(u)) \alpha(\varphi'(v)) 
\stackrel{\text{IH}}{=} \varphi(u) \varphi(v) = \varphi(w)
		\displaypunct{.}
	    \]
    \end{itemize}
    Therefore, $\varphi'^{-1}(\alpha^{-1}(F)) = L$, which implies that $(S, P)$
    recognizes $L$.
\end{proof}

Next, we show that any language recognized by an $\Ext$-algebra is
also recognized by one of its sub-$\Ext$-algebras via a surjective morphism.

\begin{lem}\label{lem:Surjective_recognition}
    Let
    $(\varphi, \psi)\colon (\Sigma^\triangle, \ExtMon(\Sigma^\triangle)) \to
     (R, O)$
    be a morphism and let $L \subseteq \Sigma^\triangle$ be a language
    recognized by $(R, O)$ via $(\varphi, \psi)$.
    Then
    $\bigl(\varphi(\Sigma^\triangle),
	   \restr{\psi(\ExtMon(\Sigma^\triangle))}{\varphi(\Sigma^\triangle)}
     \bigr)$
    is a sub-$\Ext$-algebra of $(R, O)$ recognizing $L$ via the surjective
    morphism $(\varphi, \psi')$ where
    $\psi'(\ext_{u, v}) = \restr{\psi(\ext_{u, v})}{\varphi(\Sigma^\triangle)}$
    for all $\ext_{u, v} \in \ExtMon(\Sigma^\triangle)$.
\end{lem}

\begin{proof}
    Since $(R, O)$ recognizes $L$ via $(\varphi, \psi)$, this means that there
    exists $F \subseteq R$ such that $\varphi^{-1}(F) = L$.
    We have that $\varphi(\Sigma^\triangle)$ is a submonoid of $R$ and
    $\psi(\ExtMon(\Sigma^\triangle))$ is a submonoid of $O$.
    Observe that for all $e \in \psi(\ExtMon(\Sigma^\triangle))$ and
    $r \in \varphi(\Sigma^\triangle)$, we have
    \[
	e(r) = \psi(\ext_{u, v})(\varphi(w)) = \varphi(u w v) \in
	\varphi(\Sigma^\triangle)
    \]
    because $\psi(\ext_{u,v}) = e$ for
    $\ext_{u, v} \in \ExtMon(\Sigma^\triangle)$ and $r = \varphi(w)$ for
    $w \in \Sigma^\triangle$.
    Moreover, for all $e, f \in \psi(\ExtMon(\Sigma^\triangle))$, it holds that
    $\restr{e}{\varphi(\Sigma^\triangle)} \circ
     \restr{f}{\varphi(\Sigma^\triangle)} =
     \restr{(e \circ f)}{\varphi(\Sigma^\triangle)}$.
    Therefore,
    $\restr{\psi(\ExtMon(\Sigma^\triangle))}{\varphi(\Sigma^\triangle)}$ is a
    submonoid of $\varphi(\Sigma^\triangle)^{\varphi(\Sigma^\triangle)}$.
    In addition, for each $r \in \varphi(\Sigma^\triangle)$, we have that
    $r = \varphi(w)$ for some $w \in \Sigma^\triangle$ and thus that
    \[
	\leftmult{r} = \leftmult{\varphi(w)} =
	\restr{\psi(\leftmult{w})}{\varphi(\Sigma^\triangle)} =
	\restr{\psi(\ext_{w, \emptyword})}{\varphi(\Sigma^\triangle)}
    \]
    as well as
    $\rightmult{r} =
     \restr{\psi(\ext_{\emptyword, w})}{\varphi(\Sigma^\triangle)}$.
    Thus, 
    $\bigl(\varphi(\Sigma^\triangle),
	   \restr{\psi(\ExtMon(\Sigma^\triangle))}{\varphi(\Sigma^\triangle)}
     \bigr)$
    is a sub-$\Ext$-algebra of $(R, O)$.

    It is clear that $\varphi$ is a surjective monoid morphism from
    $\Sigma^\triangle$ to $\varphi(\Sigma^\triangle)$.
    Further,
    \begin{align*}
	\psi'(\ext_{u, v}) \circ \psi'(\ext_{u', v'})
	& = \restr{\psi(\ext_{u, v})}{\varphi(\Sigma^\triangle)} \circ
	    \restr{\psi(\ext_{u', v'})}{\varphi(\Sigma^\triangle)}\\
	& = \bigl(\psi(\ext_{u, v}) \circ \psi(\ext_{u', v'})\bigr)
	    \big|_{\varphi(\Sigma^\triangle)}\\
	& = \restr{\psi(\ext_{u, v} \circ \ext_{u', v'})}
		  {\varphi(\Sigma^\triangle)}\\
	& = \psi'(\ext_{u, v} \circ \ext_{u', v'})
    \end{align*}
    for all $\ext_{u, v}, \ext_{u', v'} \in \ExtMon(\Sigma^\triangle)$, hence
    since
    $\psi'(\ext_{\emptyword, \emptyword}) =
     \restr{1_O}{\varphi(\Sigma^\triangle)}$,
    it follows that $\psi'$ is a surjective monoid morphism from
    $\ExtMon(\Sigma^\triangle)$ to
    $\restr{\psi(\ExtMon(\Sigma^\triangle))}{\varphi(\Sigma^\triangle)}$.
    Moreover, we have
    \begin{itemize}
	\item
	    $\psi'(\ext_{u, v})(\varphi(w)) =
	     \restr{\psi(\ext_{u, v})}{\varphi(\Sigma^\triangle)}(\varphi(w)) =
	     \psi(\ext_{u, v})(\varphi(w)) = \varphi(\ext_{u, v}(w))$
	    for all $\ext_{u, v} \in \ExtMon(\Sigma^\triangle)$ and
	    $w \in \Sigma^\triangle$;
	\item
	    $\psi'(\leftmult{w}) =
	     \restr{\psi(\ext_{w, \emptyword})}{\varphi(\Sigma^\triangle)} =
	     \leftmult{\varphi(w)}$
	    and $\psi'(\rightmult{w}) = \rightmult{\varphi(w)}$ for all
	    $w \in \Sigma^\triangle$.
    \end{itemize}
    Therefore, $(\varphi, \psi')$ is a surjective morphism recognizing $L$.
\end{proof}

The following lemma states that a language is recognized by an $\Ext$-algebra via a
surjective morphism if, and only if, the syntactic morphism of the 
language factors through the former morphism.

\begin{lem}\label{lem:Factorization_surjective_recognition}
    Let
    $(\varphi, \psi)\colon (\Sigma^\triangle, \ExtMon(\Sigma^\triangle)) \to
     (R, O)$
    be a surjective morphism, let $L \subseteq \Sigma^\triangle$ and let
    $(\varphi_L, \psi_L)\colon (\Sigma^\triangle, \ExtMon(\Sigma^\triangle)) \to
     (R_L, O_L)$
    be the syntactic morphism of $L$.
    Then $(R, O)$ recognizes $L$ via $(\varphi, \psi)$ if and only if there is a
    surjective morphism $(\alpha, \beta)\colon (R, O) \to (R_L, O_L)$ such that
    $\varphi_L = \alpha \circ \varphi$ (we say that $(\varphi_L, \psi_L)$
    \emph{factors through} $(\varphi, \psi)$).
\end{lem}

\begin{proof}
    Assume first that there is a surjective morphism
    $(\alpha, \beta)\colon (R, O) \to (R_L, O_L)$ such that
    $\varphi_L = \alpha \circ \varphi$.
    Then
    \[
	\varphi^{-1}\bigl(\alpha^{-1}(\varphi(L))\bigr) =
	(\alpha \circ \varphi)^{-1}\bigl(\varphi_L(L)) =
	\varphi_L^{-1}(\varphi_L(L)) =
	L
	\displaypunct{,}
    \]
    hence $(R, O)$ recognizes $L$ via $(\varphi, \psi)$.

    Assume now that $(R, O)$ recognizes $L$ via $(\varphi, \psi)$. This means
    that there exists $F \subseteq R$ satisfying $\varphi^{-1}(F) = L$.
    Take $w, w' \in \Sigma^\triangle$ such that $\varphi(w) = \varphi(w')$.
    Then, given any $e \in \ExtMon(\Sigma^\triangle)$, we have that
    \[
	\varphi(e(w)) = \psi(e)(\varphi(w)) = \psi(e)(\varphi(w')) =
	\varphi(e(w'))
	\displaypunct{.}
    \]
    Therefore, since $\varphi^{-1}(F) = L$, it holds that $w \sim_L w'$, that
    is, $\varphi_L(w) = \varphi_L(w')$.

    Take $\ext_{u, v}, \ext_{u', v'} \in \ExtMon(\Sigma^\triangle)$ such that
    $\psi(\ext_{u, v}) = \psi(\ext_{u', v'})$. Then, for each
    $w \in \Sigma^\triangle$, we have that
    \[
	\varphi(\ext_{u, v}(w)) =
	\psi(\ext_{u, v})(\varphi(w)) =
	\psi(\ext_{u', v'})(\varphi(w)) =
	\varphi(\ext_{u', v'}(w))
	\displaypunct{,}
    \]
    that is, $\ext_{u, v}(w) \sim_L \ext_{u', v'}(w)$.
    Hence, $\psi_L(\ext_{u, v}) = \psi_L(\ext_{u', v'})$.

    We can now define the functions $\alpha\colon R \to R_L$ and
    $\beta\colon O \to O_L$ such that $\alpha(\varphi(w)) = \varphi_L(w)$ for
    all $w \in \Sigma^\triangle$ and
    $\beta(\psi(\ext_{u, v})) = \psi_L(\ext_{u, v})$ for all
    $\ext_{u, v} \in \ExtMon(\Sigma^\triangle)$: those are well-defined by
    surjectivity of $(\varphi, \psi)$ and what we have proven just above.
    Since $(\varphi_L, \psi_L)$ is a surjective morphism from
    $(\Sigma^\triangle, \ExtMon(\Sigma^\triangle))$ to $(R_L, O_L)$, we can
    easily prove that $(\alpha, \beta)$ is a surjective morphism from $(R, O)$
    to $(R_L, O_L)$ that does of course satisfy
    $\varphi_L = \alpha \circ \varphi$.
\end{proof}

The following proposition shows that the syntactic $\Ext$-algebra of a given
language of well-matched words is the least $\Ext$-algebra recognizing this
language.

\begin{prop}\label{prop:syntactic:recognizes}
    An $\Ext$-algebra $(R, O)$ recognizes a language
    $L \subseteq \Sigma^\triangle$ if, and only if, its syntactic $\Ext$-algebra
    $(R_L,O_L)$ divides $(R, O)$.
\end{prop}

\begin{proof}
    Let $(R, O)$ be an $\Ext$-algebra and let $L \subseteq \Sigma^\triangle$ be
    a language. Consider also its syntactic $\Ext$-algebra $(R_L,O_L)$ and its
    syntactic morphism $(\varphi_L, \psi_L)$.

    \subparagraph{Implication from right to left.}
    Assume that the syntactic $\Ext$-algebra $(R_L, O_L)$ of $L$ divides
    $(R, O)$. We have that $(R_L,O_L)$ recognizes $L$ and we then use
    Lemma~\ref{lem:Division_and_recognisability} to conclude that $(R, O)$ does
    also recognize $L$.

    \subparagraph{Implication from left to right.}
    Assume that $(R, O)$ recognizes $L$ through a morphism
    $(\varphi, \psi)\colon
     (\Sigma^\triangle, \ExtMon(\Sigma^\triangle)) \to (R, O)$.
    By Lemma~\ref{lem:Surjective_recognition}, we have that
    $\bigl(\varphi(\Sigma^\triangle),
	   \restr{\psi(\ExtMon(\Sigma^\triangle))}{\varphi(\Sigma^\triangle)}
     \bigr) =
     (R', O')$
    is a sub-$\Ext$-algebra of $(R, O)$ recognizing $L$ via the surjective
    morphism $(\varphi, \psi')$ where
    $\psi'(\ext_{u, v}) = \restr{\psi(\ext_{u, v})}{\varphi(\Sigma^\triangle)}$
    for all $\ext_{u, v} \in \ExtMon(\Sigma^\triangle)$.
    Then, by Lemma~\ref{lem:Factorization_surjective_recognition}, there exists
    a surjective morphism $(\alpha, \beta)\colon (R', O') \to (R_L, O_L)$ such
    that $\varphi_L = \alpha \circ \varphi$.
    Thus, we have that $(R_L, O_L)$ divides $(R, O)$.
\end{proof}

We say that an $\Ext$-algebra $(R, O)$ is \emph{finite} whenever $R$ is finite
(which is the case if and only if $O$ is finite).
The following theorem establishes the equivalence between visibly pushdown
languages and languages recognizable by finite $\Ext$-algebras.
Its proof provides effective translations from DVPAs to $\Ext$-algebras and vice
versa.

\begin{thm}\label{thm:vpl:ext}
    A language $L \subseteq \Sigma^\triangle$ is a VPL if, and only if, it is recognized by a finite
    $\Ext$-algebra.
\end{thm}
\begin{proof}
    Let $L \subseteq \Sigma^\triangle$ be a language.
    Before we prove the theorem we have the following claim,
    which can be easily proven by induction on $|u|$ and structural induction on 
    $w$, respectively.

    \subparagraph{Claim.}
	Let $A = (Q, \Sigma, \Gamma, \delta, q_0, F, \bot)$ be a DVPA.
    We denote by $\pi_Q$ the projection of $Q \times \Gamma^*$ on $Q$ and by
    $\pi_{\Gamma^*}$ the projection of $Q \times \Gamma^*$ on $\Gamma^*$.
    It holds that $\DLang{A} \subseteq \Sigma^\triangle$ and additionally we
    have that
    \[
	\widehat{\delta}(q, u v, \sigma) =
	\widehat{\delta}\bigl(\pi_Q(\widehat{\delta}(q, u, \sigma)), v,
			      \pi_{\Gamma^*}(\widehat{\delta}(q, u, \sigma))
			\bigr)
    \]
    and
    \[
	\widehat{\delta}(q, w, \alpha \sigma) =
	\bigl(\pi_Q(\widehat{\delta}(q, w, \alpha)), \alpha \sigma\bigr)
    \]
    for all $q \in Q$, $u, v \in \Sigma^*$, $\sigma \in \Gamma^*$,
    $w \in \Sigma^\triangle$ and $\alpha \in \Gamma$.

    \subparagraph{Implication from left to right.}
	Assume that $L$ is a VPL. This means there exists a DVPA
	$A = (Q, \Sigma, \Gamma, \delta, q_0, F, \bot)$ such that $\DLang{A} = L$.
	Consider the operation $*$ on $R=Q^{Q \times \Gamma}$ defined so that for all
    $f, g \in R$, we have $f * g(q, \alpha) = g(f(q, \alpha), \alpha)$ for all
    $q \in Q$ and $\alpha \in \Gamma$.
    Observe that for all $f, g, h \in Q^{Q \times \Gamma}$, we have
    \[
	(f * g) * h(q, \alpha) = h(f * g(q, \alpha), \alpha) =
	h(g(f(q, \alpha), \alpha), \alpha) = g * h(f(q, \alpha), \alpha) =
	f * (g * h)(q, \alpha)
    \]
    for all $q \in Q$ and $\alpha \in \Gamma$. Thus $*$ is associative and since
    it also has $i \in R$ with $i(q, \alpha) = q$ for all $q \in Q$ and
    $\alpha \in \Gamma$ as an identity, we have that $R = Q^{Q \times \Gamma}$
    with operation $*$ forms a monoid.
	Take $O$ to be the monoid $R^R$ (for composition). 
	Since $O$ clearly contains the functions $\leftmult{r}$ and $\rightmult{r}$ for all $r\in R$,
	it follows that $(R, O)$ is a finite $\Ext$-algebra.
    We now prove that it recognizes $L$.
    For each $w \in \Sigma^\triangle$, define $f_w \in R$ by
    $f_w(q, \alpha) = \pi_Q(\widehat{\delta}(q, w, \alpha))$ for all $q \in Q$
    and $\alpha \in \Gamma$.
    Let us consider the unique morphism
    $(\varphi, \psi)\colon
     (\Sigma^\triangle, \ExtMon(\Sigma^\triangle)) \to (R, O)$,
    given by Proposition~\ref{ptn:Freeness_Ext-algebras}, such that for each
    $c \in \Sigma_\internal$, we have $\varphi(c) = f_c$ and for each
    $a \in \Sigma_\call, b \in \Sigma_\return$, we have that $\psi(\ext_{a, b})$
    sends any $f \in R$ to $g \in R$ satisfying that
	$g(q, \alpha) = \pi_Q\bigl(\delta(f(p, \beta), b, \beta)\bigr)$ with
    $\delta(q, a, \alpha) = (p, \beta \alpha)$ for all $q \in Q$ and
    $\alpha \in \Gamma$.
    We claim that for all $w \in \Sigma^\triangle$, we have that
    $\varphi(w) = f_w$. We prove it by induction on $w$.
    \begin{itemize}
	\item
	    $w = \emptyword$. Then $\varphi(w) = i = f_w$.
	\item
	    $w = c$ for some $c \in \Sigma_\internal$.
	    Then $\varphi(w) = f_c = f_w$.
	\item
	    $w = a w' b$ for some $a \in \Sigma_\call$, $b \in \Sigma_\return$
	    and $w' \in \Sigma^\triangle$.
	    Then
	    \[
		\varphi(w) = \varphi(\ext_{a, b}(w')) =
		    \psi(\ext_{a, b})(\varphi(w')) \stackrel{\text{IH}}{ = } \psi(\ext_{a, b})(f_{w'})
		\displaypunct{.}
	    \]
	    So $\varphi(w) = g$ such that for all $q \in Q$ and
		    $\alpha \in \Gamma$, if we set $\delta(q, a, \alpha) = (p, \beta\alpha)$,
		    we have, recalling that $\widehat{\delta}$ extends $\delta$,
	    \begin{align*}
		g(q, \alpha)
		& = \pi_Q\bigl(\delta(f_{w'}(p, \beta), b, \beta)\bigr)\\
		    &  = \pi_Q\Bigl(\delta\bigl(\pi_Q(\widehat{\delta}(p,w',\beta)), b, \beta
		    \bigr)\Bigr) \\
		    &  = \pi_Q\Bigl(\widehat{\delta}\bigl(\pi_Q(\widehat{\delta}(p,w',\beta)), b, \beta
		    \bigr)\Bigr) \\
		    & = \pi_Q\Bigl(\widehat{\delta}
			       \bigl(\pi_Q(\widehat{\delta}(q, a w', \alpha)),
				     b,
				     \pi_{\Gamma^*}(\widehat{\delta}(q, a w',
								     \alpha))
			       \bigr)\Bigr)\\
				& = \pi_Q(\widehat{\delta}(q, a w' b, \alpha))\\
		& = f_{a w' b}(q, \alpha)\\
		    & = f_w(q,\alpha)
		\displaypunct{.}
	    \end{align*}
	    Thus $\varphi(w) = f_w$.
	\item
	    $w = u v$ for some
	    $u, v \in \Sigma^\triangle \setminus \set{\emptyword}$.
		    Then $\varphi(w) = \varphi(u) * \varphi(v) \stackrel{\text{IH}}{=} f_u * f_v$.
	    But
	    \begin{align*}
		f_u * f_v(q, \alpha)
		& = f_v(f_u(q, \alpha), \alpha)\\
		& = f_v\bigl(\pi_Q(\widehat{\delta}(q, u, \alpha)),
			     \alpha\bigr)\\
		& = \pi_Q\Bigl(\widehat{\delta}
			       \bigl(\pi_Q(\widehat{\delta}(q, u, \alpha)), v,
				     \alpha\bigr)\Bigr)\\
		& = \pi_Q\Bigl(\widehat{\delta}
			       \bigl(\pi_Q(\widehat{\delta}(q, u, \alpha)), v,
				     \pi_{\Gamma^*}(\widehat{\delta}(q, u,
								     \alpha))
			       \bigr)\Bigr)\\
		& = \pi_Q(\widehat{\delta}(q, u v, \alpha))\\
		& = f_{u v}(q, \alpha)
	    \end{align*}
	    for all $q \in Q$ and $\alpha \in \Gamma$.
	    Therefore $\varphi(w) = f_w$.
    \end{itemize}
    Finally, set $P = \set{f \in R \mid f(q_0, \bot) \in F}$. It holds that
    \begin{align*}
	\varphi^{-1}(P)
	    & = \set{w \in \Sigma^\triangle \mid f_w(q_0, \bot) \in F}\\
	& = \set{w \in \Sigma^\triangle \mid
		 \pi_Q(\widehat{\delta}(q_0, w, \bot)) \in F}\\
	& = \set{w \in \Sigma^\triangle \mid
		 \widehat{\delta}(q_0, w, \bot) \in F \times \set{\bot}}\\
	& = \DLang{A}\\
	    & =  L
	\displaypunct{.}
    \end{align*}
    Therefore, $(R, O)$ recognizes $L$.

    \subparagraph{Implication from right to left.}
    Assume there exists a finite $\Ext$-algebra $(R, O)$ that recognizes $L$.
    This means that there exists a morphism
    $(\varphi, \psi)\colon
     (\Sigma^\triangle, \ExtMon(\Sigma^\triangle)) \to (R, O)$
    such that $L = \varphi^{-1}(F)$ for some $F \subseteq R$.
    Let us now define the DVPA
    \[
	    A = (Q, \Sigma, \Gamma, \delta, 1, F, 
	    \bot),
    \]
    where $Q=R$,
    $1=1_R$,
$\Gamma=R \times \Sigma_\call \cup \set{\bot}$,
    and
    \[
	\delta(r, a, \alpha) =
	\begin{cases}
	    (1, (r, a) \alpha) & \text{if $a \in \Sigma_\call$}\\
	    (s \psi(\ext_{b, a})(r), \emptyword) &
		\text{if $a \in \Sigma_\return$ and
		      $\alpha = (s, b) \in R \times \Sigma_\call$}\\
	    (r, \emptyword) &
		\text{if $a \in \Sigma_\return$ and $\alpha = \bot$}\\
	    (r \varphi(c), \alpha) & \text{if $a \in \Sigma_\internal$}
	\end{cases}
    \]
    for all $r \in R$, $a \in \Sigma$ and $\alpha \in \Gamma$.
    We prove that $\widehat{\delta}(r, w, \sigma) = (r \varphi(w), \sigma)$ for
    all $r \in R$, $w \in \Sigma^\triangle$ and $\sigma \in \Gamma^* \bot$ by
    induction on $w$.
    \begin{itemize}
	\item
	    $w = \emptyword$. Then
	    $\widehat{\delta}(r, w, \sigma) = (r, \sigma) =
	     (r \varphi(w), \sigma)$.
	\item
	    $w = c$ for some $c \in \Sigma_\internal$.
	    Then
	    $\widehat{\delta}(r, w, \sigma) = (r \varphi(c), \sigma) =
	     (r \varphi(w), \sigma)$.
	\item
	    $w = a w' b$ for some $a \in \Sigma_\call$, $b \in \Sigma_\return$
	    and $w' \in \Sigma^\triangle$.
	    Then
	    \begin{align*}
		\widehat{\delta}(r, w, \sigma)
		& = \widehat{\delta}(1, w' b, (r, a) \sigma)\\
		& = \widehat{\delta}\bigl(\pi_Q(1, w', (r, a) \sigma), b,
					  \pi_{\Gamma^*}(1, w', (r, a) \sigma)
				    \bigr)\\
		& \stackrel{\text{IH}}{=} \widehat{\delta}(\varphi(w'), b, (r, a) \sigma)\\
		& = \bigl(r \psi(\ext_{a, b})(\varphi(w')), \sigma\bigr)\\
		& = (r \varphi(w), \sigma)
		\displaypunct{.}
	    \end{align*}
	\item
	    $w = u v$ for some
	    $u, v \in \Sigma^\triangle \setminus \set{\emptyword}$.
	    Then
	    \begin{align*}
		\widehat{\delta}(r, w, \sigma)
		    & = \widehat{\delta}\bigl(\pi_Q(\widehat{\delta}(r, u, \sigma)), v,
					  \pi_{\Gamma^*}(\widehat{\delta}(r, u, \sigma))\bigr)\\
		    & \stackrel{\text{IH}}{=} \widehat{\delta}(r \varphi(u), v, \sigma)\\
		& \stackrel{\text{IH}}{=} (r \varphi(u) \varphi(v), \sigma)\\
		& = (r \varphi(w), \sigma)
		\displaypunct{.}
	    \end{align*}
    \end{itemize}
    Hence,
    \begin{align*}
	\DLang{A}
	& = \set{w \in \Sigma^\triangle \mid
		 \widehat{\delta}(1, w, \bot) \in F \times \set{\bot}}\\
	& = \set{w \in \Sigma^\triangle \mid
		 \pi_Q(\widehat{\delta}(1, w, \bot)) \in F}\\
	& = \set{w \in \Sigma^\triangle \mid \varphi(w) \in F}\\
	& = \varphi^{-1}(F) = L
	\displaypunct{.}
    \end{align*}
    Therefore, $L$ is a VPL.
\end{proof}

\section{(Weak) length-synchronicity,
nesting depth, and quasi-aperiodicity}\label{Section Notions}

For the rest of this section let us fix a visibly pushdown alphabet $\Sigma$, a
finite $\Ext$-algebra $(R, O)$ and consider a morphism
$(\varphi, \psi)\colon (\Sigma^\triangle, \ExtMon(\Sigma^\triangle)) \to
 (R, O)$.
 Suitably adjusting the pumping lemma for context-free languages
we introduce a pumping lemma for $\Ext$-algebra morphisms 
in Section~\ref{S Pumping}.
In Section~\ref{S length synchronicity} we extend the notions
of weak length-synchronicity and length-synchronicity 
to $\Ext$-algebra morphisms and to visibly pushdown languages.  
It is shown that for languages generated by vertically
visibly pushdown grammars, (weak) length-synchronicity of the word relation of 
the generating grammar coincides with 
(weak) length-synchronicity of the language.
We concern ourselves with the nesting depth
of visibly pushdown languages in Section~\ref{section nesting}.
Finally in Section~\ref{section quasi} we introduce 
quasi-aperiodicity of $\Ext$-algebra morphisms 
and prove that a VPL is quasi-counterfree if, and only if, 
its syntactic morphism is quasi-aperiodic.

 \subsection{A pumping lemma for Ext-algebra morphisms}\label{S Pumping}
The following is an adaption of the pumping lemma for 
context-free languages to $\Ext$-algebra morphisms.
It states that if $uv\in\Sigma^\triangle$ and $u$ (resp.\
$v$) contains a well-matched factor that is sufficiently long, we can pump
certain infixes of $u$ (resp. $v$): thus, one can find longer and longer
words $u_1,u_2, \ldots$ (resp.\ $v_1,v_2,\ldots)$ such that
$u_1 v, u_2 v, \ldots \in \Sigma^\triangle$ (resp.\
$u v_1, u v_2, \ldots \in \Sigma^\triangle$) and the morphism $\psi$ sends
$\ext_{u,v}$ to the same element in $O$ as $\ext_{u_i,v}$ (resp.\ as
$\ext_{u,v_i}$).

\begin{samepage}
\begin{lem}[Pumping Lemma]\label{lem:Ext_pumping}
There exists $n \in \N_{>0}$ such that for all
    $\ext_{u, v} \in \ExtMon(\Sigma^\triangle)$ we have:
    \begin{itemize}
	\item
	    If there exists a factor $w \in \Sigma^\triangle$ of $u$ satisfying
	    $\length{w} \geq n$, then $u = s x z y t$ with
	    $s, x, z, y, t \in \Sigma^*$ such that $\length{x y} \geq 1$,
	    $\length{x z y} \leq n$ and for all $i \in \N$,
	    $s x^i z y^i t v \in \Sigma^\triangle$ and
	    $\psi(\ext_{u, v}) = \psi(\ext_{s x^i z y^i t, v})$.
	\item
	    If there exists a factor $w \in \Sigma^\triangle$ of $v$ satisfying
	    $\length{w} \geq n$, then $v = s x z y t$ with
	    $s, x, z, y, t \in \Sigma^*$ such that $\length{x y} \geq 1$,
	    $\length{x z y} \leq n$ and for all $i \in \N$,
	    $u s x^i z y^i t \in \Sigma^\triangle$ and
	    $\psi(\ext_{u, v}) = \psi(\ext_{u, s x^i z y^i t})$.
    \end{itemize}
\end{lem}
\end{samepage}

\begin{proof}
   For each $r \in R$, let $n_r \in \N_{>0}$ be the pumping constant
	for the context-free language $\varphi^{-1}(r)$: it is a VPL and hence a
    context-free language by Theorem~\ref{thm:vpl:ext}.
	We set $n = \max_{r \in R} n_r$.
    Let $\ext_{u, v} \in \ExtMon(\Sigma^\triangle)$ be such that there exists a
    factor $w \in \Sigma^\triangle$ of $u$ satisfying $\length{w} \geq n$.
    Let
    \[
	\ext_{u,v}
	\quad=\quad
	\ext_{x_1,y_1} \circ \ext_{a_1,b_1} \circ \quad\cdots\quad \circ
	\ext_{x_{h-1},y_{h-1}} \circ \ext_{a_{h-1},b_{h-1}} \circ \ext_{x_h,y_h}
    \]
    be the stair factorization of $\ext_{u,v}$ provided by
    Lemma~\ref{lemma ext}. Since no factor of $u$ spanning one of the $a_j$'s
    in the factorization can be well-matched, there must exist some
    $j \in [1, h]$ satisfying $\length{x_j} \geq n$, so that if we set
    $u' = x_1 a_1 \cdots x_{j - 1} a_{j - 1}$,
    $v' = b_{j - 1} y_{j - 1} \cdots b_1 y_1$,
    $u'' = a_j x_{j + 1} \cdots a_{h - 1} x_h$ and
    $v'' = y_h b_{h - 1} \cdots y_{j + 1} b_j y_j$, we have
    $u'v', u''v'' \in \Sigma^\triangle$ and
    $\ext_{u, v} =
     \ext_{u', v'} \circ \ext_{x_j, \emptyword} \circ \ext_{u'', v''}$.
    By the pumping lemma for context-free languages we have $x_j = x' x z y y'$ with
    $x', x, z, y, y' \in \Sigma^*$ such that $\length{x y} \geq 1$,
    $\length{x z y} \leq n$ and for all $i \in \N$,
    $x' x^i z y^i y' \in \Sigma^\triangle$ and
    $\varphi(x_j) = \varphi(x' x^i z y^i y')$.
    This implies that if we set $s = u' x'$ and $t = y' u''$, then
	for all $i \in \N$, we have
    $s x^i z y^i t v =
     \ext_{u', v'} \circ \ext_{x' x^i z y^i y', \emptyword} \circ
     \ext_{u'', v''}(\emptyword) \in \Sigma^\triangle$
    and
    \begin{align*}
	\psi(\ext_{u, v}) & =
	\psi(\ext_{u', v'}) \circ \leftmult{\varphi(x_j)} \circ
	\psi(\ext_{u'', v''})\\
	& = \psi(\ext_{u', v'}) \circ \leftmult{\varphi(x' x^i z y^i y')} \circ
	    \psi(\ext_{u'', v''})\\
	& = \psi(\ext_{u', v'}) \circ \psi(\ext_{x' x^i z y^i y', \emptyword})
	    \circ \psi(\ext_{u'', v''})\\
	& = \psi(\ext_{s x^i z y^i t, v})
	\displaypunct{.}
    \end{align*}

    We handle the case where for 
    $\ext_{u, v} \in \ExtMon(\Sigma^\triangle)$ there exists
a factor $w \in \Sigma^\triangle$ of $v$
such that $\length{w} \geq n$ symmetrically.
\end{proof}

	\subsection{Weak length-synchronicity and length-synchronicity}
	\label{S length synchronicity}

In this section we introduce the notions of weak length-synchronicity and
length-synchronicity for $\Ext$-algebra morphisms and visibly pushdown
languages. Before we do that, let us give some motivation how 
$\TCO$-hardness can be proven if the syntactic morphism
maps certain $\ext_{u,v},\ext_{u',v}$ with $|u|\not=|u'|$ to particular
idempotents. For these we require the following notion of reachability.

For $F \subseteq R$ we say that an element $r\in R$
is \emph{$F$-reaching} if $e(r) \in F$ for some $e\in O$.
We say $e\in O$ is \emph{$F$-reaching} if $e(r)$ is $F$-reaching for some $r\in R$.
Although we will mainly study $F$-reaching elements over finite $\Ext$-algebras
we remark that the notion of $F$-reachability is defined over any $\Ext$-algebra,
in particular over $(\Sigma^\triangle,\ExtMon(\Sigma^\triangle))$.

Fix any VPL $L$, its syntactic $\Ext$-algebra $(R_L,O_L)$ along
with its syntactic morphism $(\varphi_L,\psi_L)$.\label{proofsketch EQ}
	     Assume there exists some idempotent $e\in O_L$ that is 
	     $\varphi(L)$-reaching.
We claim that if $\psi_L(\ext_{u,v})=\psi_L(\ext_{u',v})=e$
	     and $\Delta(u),\Delta(u')>0$ for some 
	     $\ext_{u,v},\ext_{u',v}\in\ExtMon(\Sigma^\triangle)$
	     with $|u|\not=|u'|$, then $L$ is $\TCO$-hard.
	     We remark that we must have $\Delta(u)=\Delta(u')$.
	     Exploiting the fact that $|u|\not=|u'|$ we give a
	     constant-depth reduction from 
	     the $\TCO$-complete language $\EQ$ to $L$.
	     As $\psi_L(\ext_{u,v})$ is $\varphi(L)$-reaching, 
	     we can fix some $x,y,z\in\Sigma^*$ such 
	     that $xuyvz\in L$.
	     Given a word $w\in\set{0,1}^*$ of length $2n$ (binary words of odd length can directly be rejected) 
	     we map it to $xh(w)yv^{n\cdot(|u|+|u'|)}z$,
	     where $h\colon\set{0,1}^*\rightarrow\Sigma^*$ is the 
	     length-multiplying morphism 
	     (i.e. $\exists l\in\N:h(0),h(1)\in\Sigma^l$)
	     satisfiying
$h(0)=u^{|u'|}$ and $h(1)=u'^{|u|}$.
We have $w\in\EQ$ if, and only if, 
$|w|_0=|w|_1=n$ if, and only if, 
$\Delta(h(w))=n\cdot(|u|+|u'|)\cdot\Delta(u)=
-n\cdot(|u|+|u'|)\cdot\Delta(v)$ if, and only if, 
$h(w)v^{n\cdot(|u|+|u'|)}\in\Sigma^\triangle$.
Hence, since $\psi_L(\ext_{u^s,v^s})=\psi_L(\ext_{(u')^t,v^t})=e$ for all $s,t\geq 1$
it follows that $w\in\EQ$ if, and only, if $xh(w)yv^{n\cdot(|u|+|u'|)}z\in\Sigma^\triangle$
if, and only if, $xh(w)yv^{n\cdot(|u|+|u'|)}z\in L$.

Dually, one can show that $L$ is $\TCO$-hard in 
case $\psi_L(\ext_{u,v})=\psi_L(\ext_{u,v'})=e$
and $\Delta(u)>0$ for some $\ext_{u,v},\ext_{u,v'}\in\ExtMon(\Sigma^\triangle)$
with $|v|\not=|v'|$.

The following definition of weak length-synchronicity captures the situation when 
such idempotents do not exist --- it adapts the notion of
weak length-synchronicity of sets of contexts, given in
Definition~\ref{def synchronicity}, to morphisms and VPLs, respectively.
Recall that $\mathcal{R}\subseteq\Con(\Sigma)$ is defined
to be weakly length-synchronous if 
$u=u'$ implies $\length{v}=\length{v'}$ and 
$v=v'$ implies $\length{u}=\length{u'}$
for all $(u,v),(u',v')\in \mathcal{R}$ satisfying $\Delta(u),\Delta(u')>0$.

\begin{defi}
	For all $e\in O$ define the sets of contexts $\newR_e$ and
	$\newU_e$ as follows:
	\[
	    \newR_e=\set{(u,v)\in\Con(\Sigma)\mid \psi(\ext_{u,v})=e}
	    \quad\text{ and }\quad
	    \newU_e=\set{(u,v)\in\Con(\Sigma)\mid e\circ\psi(\ext_{u,v})=e}
	\]
\end{defi}

\begin{defi}[Weak Length-Synchronicity]
	The morphism 
$(\varphi, \psi)\colon (\Sigma^\triangle, \ExtMon(\Sigma^\triangle)) \to
     (R, O)$
	is $F$-weakly-length-synchronous (where $F\subseteq R$) if
for all $F$-reaching idempotents $e\in O$ 
the set of contexts $\newR_e$ is weakly length-synchronous.
	We call a VPL $L\subseteq\Sigma^\triangle$ {\em weakly length-synchronous} if
	its syntactic morphism $(\varphi_L,\psi_L)$ is $\varphi_L(L)$-weakly-length-synchronous.
\end{defi}

In fact, $F$-weak-length-synchronicity actually implies weak
length-synchronicity of the set of contexts associated to any subsemigroup of
$F$-reaching elements.

\begin{lem}
\label{lem:Weak_length-synchronicity_subsemigroups}
    For all $F \subseteq R$ and subsemigroup $P$ of $O$, if all elements in $P$
    are $F$-reaching and $(\varphi,\psi)$ is $F$-weakly-length-synchronous,
    then $\bigcup_{e \in P} \newR_e$ is weakly length-synchronous.
\end{lem}

\begin{proof}
    Let $F \subseteq R$ and $P$ be a subsemigroup of $O$.
    Assume all elements in $P$ are $F$-reaching and $(\varphi,\psi)$ is
    $F$-weakly-length-synchronous.

    Let $(u, v), (u', v) \in \Con(\Sigma)$ be such that $\Delta(u), \Delta(u') > 0$
    and $\psi(\ext_{u, v}), \psi(\ext_{u', v}) \in P$.
    Set $e = \psi(\ext_{u, v})$ and $f = \psi(\ext_{u', v})$. By hypothesis,
    given $\omega \in \N_{>0}$ the idempotent power of $O$, we have
    $(e^\omega f^\omega)^\omega \in P$, hence $(e^\omega f^\omega)^\omega$ is an
    $F$-reaching idempotent and thus $\newR_{(e^\omega f^\omega)^\omega}$
    is weakly length-synchronous.
    But
    \[
	\psi(\ext_{(u^{2\cdot\omega}u'^\omega)^\omega, v^{3\cdot\omega^2}}) =
	(e^\omega f^\omega)^\omega =
	\psi(\ext_{(u^{\omega}u'^{2\cdot\omega})^\omega, v^{3\cdot\omega^2}})
    \]
    so since
    $\Delta((u^{2\cdot\omega}u'^\omega)^\omega) =
    \Delta((u^{\omega}u'^{2\cdot\omega})^\omega) > 0$,
    we obtain
    \[
	    \arraycolsep=1.4pt\def\arraystretch{1.7}
    \begin{array}{lrcl}
	    &\quad{\omega \cdot (2\cdot\omega\cdot |u|+ \omega\cdot |u'|)}& =&
	     \omega\cdot(\omega\cdot|u|+2\cdot\omega\cdot
	    |u'|)\\
	    \Longleftrightarrow\quad & {2\cdot|u|+|u'|} &=& {|u|+2\cdot|u'|}\\
	    \Longleftrightarrow\quad & |u| &=& |u'|
	     	\displaypunct{.}
    \end{array}
    \]
    In the same way, one can prove that for all
    $(u, v), (u, v') \in \Con(\Sigma)$ such that $\Delta(u) > 0$
    and $\psi(\ext_{u, v}), \psi(\ext_{u, v'}) \in P$, we have
    $\length{v} = \length{v'}$.
    
    Therefore, $\bigcup_{e \in P} \newR_e$ is weakly length-synchronous.
\end{proof}

Instead of considering those pairs $(u,v)$ such that $\ext_{u,v}$ is 
being mapped to an $F$-reaching idempotent, the following characterization of weak 
length-synchronicity 
consider pairs $(u,v)$ such that
$\ext_{u,v}$ is being mapped to an element that behaves neutrally with respect
to right multiplication with an $F$-reaching element that is not necessarily
idempotent.
\begin{samepage}
\begin{prop}\label{ptn:Weak_length-synchronicity_equivalence}
    For all $F\subseteq R$ 
	we have that $(\varphi,\psi)$ is $F$-weakly-length-synchronous
	if, and only if, for all $F$-reaching $e\in O$ 
	the set of contexts $\newU_e$ is weakly length-synchrononous.
\end{prop}
\end{samepage}

\begin{proof}
    Let $F \subseteq R$. 

    If $\newU_e=\set{(u,v)\in\Con(\Sigma)\mid e\circ\psi(\ext_{u,v})=e}$ is
    weakly length-synchronous for all $F$-reaching $e\in O$, then in particular 
    the set of contexts
    $\newR_e=\set{(u,v)\in\Con(\Sigma)\mid \psi(\ext_{u,v})=e}$ is weakly
    length-synchronous for all $F$-reaching idempotents $e\in O$.

    Conversely, assume that $(\varphi,\psi)$ is $F$-weakly-length-synchronous.
    Fix any $F$-reaching $e\in O$.
    We need to prove that $\newU_e$ is weakly length-synchronous. It is clear
    that $P = \set{f \in O \mid e \circ f = e}$ forms a subsemigroup
    of $O$ whose elements are all $F$-reaching, by $F$-reachability of $e$.
    Therefore, by Lemma~\ref{lem:Weak_length-synchronicity_subsemigroups}, 
    $\bigcup_{f \in P} \newR_f = \newU_e$ is weakly length-synchronous.
\end{proof}

Using Lemma~\ref{lem:Ext_pumping} and
Proposition~\ref{ptn:Weak_length-synchronicity_equivalence}, the following
proposition follows immediately.

\begin{prop}\label{ptn:hill_small}
	Let $n$ be the pumping constant from Lemma~\ref{lem:Ext_pumping},
	let $F\subseteq R$, let $e\in O$ be $F$-reaching,
	and let $\ext_{u,v}$ be such that $\Delta(u)>0$ and
	$e\circ\psi(\ext_{u,v})=e$.
	If $(\varphi,\psi)$ is $F$-weakly-length-synchronous, then
	the stair factorization 
	\[\ext_{u,v}=\ext_{x_1,y_1}\circ\ext_{a_1,b_1}\circ\dots
	\circ\ext_{x_{h-1},y_{h-1}}\circ\ext_{a_{h-1},b_{h-1}}\circ\ext_{x_h,y_h}\]
	satisfies $|x_i|,|y_i|\leq n$ for all $i\in[1,h]$,
\end{prop}

As above, the following definition adapts the notion
of length-synchronicity of sets of contexts, given in 
Definition~\ref{def synchronicity}, to $\Ext$-algebra morphisms 
and VPLs, respectively.

\begin{samepage}
\begin{defi}[Length-Synchronicity]
The morphism 
$(\varphi, \psi)\colon (\Sigma^\triangle, \ExtMon(\Sigma^\triangle)) \to
     (R, O)$
	is $F$-length-synchronous (where $F\subseteq R$) if
for all $F$-reaching idempotents $e\in O$ 
	    the set of contexts
	$\newR_e$ is length-synchronous.
	We call a VPL $L\subseteq\Sigma^\triangle$ {\em length-synchronous} if
	its syntactic morphism $(\varphi_L,\psi_L)$ is $\varphi_L(L)$-length-synchronous.
\end{defi}
\end{samepage}

\begin{exa}
	Consider our running example $
	\SM_{1,2}=L(S\rightarrow a S b_1 \mid ac S b_2 \mid \emptyword)$.
	Recall that the monoid $O_{\SM_{1,2}}$ of the syntactic $\Ext$-algebra 
	$(R_{\SM_{1,2}},O_{\SM_{1,2}})$ 
	and syntactic morphism $(\varphi_{\SM_{1,2}},\psi_{\SM_{1,2}})$ of
	$\SM_{1,2}$, 
	given in Example~\ref{example L2},
	has the idempotents $(\varepsilon,\varepsilon)$, $(acb_1,\emptyword)$ and 
	$(a,b_1)$. Also recall that $\varphi_{\SM_{1,2}}(\SM_{1,2})
	=\set{\varepsilon,ab_1}$.
	Since $\psi_{\SM_{1,2}}^{-1}((\emptyword,\emptyword))=
	\set{\ext_{\emptyword,\emptyword}}$
	and $(acb_1,\emptyword)$ is a zero we have that 
	$O_{\SM_{1,2}}$'s only idempotent that is $\set{\varepsilon,ab_1}$-reaching 
	and
	whose pre-image under $\psi_{\SM_{1,2}}$ contains at least 
	one $\ext_{u,v}$ with $\Delta(u)>0$ is
	the idempotent $(a,b_1)$.
	However, both $\ext_{a,b_1}$ and $\ext_{ac,b_2}$, where 
	$\Delta(a)=\Delta(ac)=1>0$, are sent to the 
	idempotent $(a,b_1)=(a,b_2)\circ(c,\varepsilon)$.
	Since $|a|/|b_1|=1\not=2=|ac|/|b_2|$, we have that $\SM_{1,2}$ is not length-synchronous.
	On the other hand, note that if any $\ext_{u,v}$ and $\ext_{u',v}$ 
	(resp. $\ext_{u,v}$ and $\ext_{u,v'}$) are sent
	to $(a,b_1)$ then $u=u'$ and thus $|u|=|u'|$ 
	(resp. $v=v'$ and thus $|v|=|v'|$). Hence, $\SM_{1,2}$ is weakly
	length-synchronous.
\end{exa}

As above, $F$-length-synchronicity actually implies length-synchronicity of the
set of contexts associated to any subsemigroup of $F$-reaching elements.

\begin{lem}
\label{lem:Length-synchronicity_subsemigroups}
    For all $F \subseteq R$ and subsemigroup $P$ of $O$, if all elements in $P$
    are $F$-reaching and $(\varphi,\psi)$ is $F$-length-synchronous, then
    $\bigcup_{e \in P} \newR_e$ is length-synchronous.
\end{lem}

\begin{proof}
    Let $F \subseteq R$ and $P$ be a subsemigroup of $O$.
    Assume all elements in $P$ are $F$-reaching and $(\varphi,\psi)$ is
    $F$-length-synchronous.

    Let $(u, v), (u', v') \in \Con(\Sigma)$ be such that
    $\Delta(u), \Delta(u') > 0$ and
    $\psi(\ext_{u, v}), \psi(\ext_{u', v'}) \in P$.
    Set $e = \psi(\ext_{u, v})$ and $f = \psi(\ext_{u', v'})$. By hypothesis,
    given $\omega \in \N_{>0}$ the idempotent power of $O$, we have
    $(e^\omega f^\omega)^\omega \in P$, hence $(e^\omega f^\omega)^\omega$ is an
    $F$-reaching idempotent and thus $\newR_{(e^\omega f^\omega)^\omega}$
    is length-synchronous.
    But
    \[
	\psi(\ext_{(u^{2 \cdot \omega} u'^\omega)^\omega,
		   (v'^\omega v^{2 \cdot \omega})^\omega}) =
	(e^\omega f^\omega)^\omega =
	\psi(\ext_{(u^\omega u'^{2 \cdot \omega})^\omega,
		   (v'^{2 \cdot \omega} v^\omega)^\omega})
    \]
    so since
    $\Delta((u^{2\cdot\omega}u'^\omega)^\omega),
     \Delta((u^{\omega}u'^{2\cdot\omega})^\omega) > 0$,
    setting
    $(x, y) = ((u^{2 \cdot \omega} u'^\omega)^\omega,
	       (v'^\omega v^{2 \cdot \omega})^\omega)$
    and
    $(x', y') = ((u^\omega u'^{2 \cdot \omega})^\omega,
		 (v'^{2 \cdot \omega} v^\omega)^\omega)$,
    we obtain (using that for $a,b,c,d>0$ we have that $\frac{a}{b}=\frac{c}{d}$
    implies $\frac{a}{b}=\frac{c}{d}=\frac{a+c}{b+d}$ and, if additionally
    $a>c$, it implies $\frac{a}{b}=\frac{c}{d}=\frac{a-c}{b-d}$)
    \begin{eqnarray}
	    \frac{|x|}{|y|}=\frac{|x'|}{|y'|} & \quad\Longrightarrow\quad &
	    \frac{|x|}{|y|}=\frac{|x'|}{|y'|}=\frac{|x|+|x'|}{|y|+|y'|}=
	    \frac{\omega^2\cdot(|u|+|u'|)}{\omega^2\cdot(|v|+|v'|)}\nonumber\\
	    &\quad\Longrightarrow\quad&
	    \frac{|x|-\omega^2\cdot(|u|+|u'|)}{|y|-\omega^2\cdot(|v|+|v'|)}=
	    \frac{|x'|-\omega^2\cdot(|u|+|u'|)}{|y'|-\omega^2\cdot(|v|+|v'|)}
	    \label{E square}\\
	    &\quad\Longrightarrow\quad&
	    \frac{|u|}{|v|}=
	    \frac{\omega^2\cdot|u|}{\omega^2\cdot|v|}
	    \stackrel{(\ref{E square})}{=}
	    \frac{\omega^2\cdot|u'|}{\omega^2\cdot|v'|}=
	    \frac{|u'|}{|v'|}\nonumber
	    \displaypunct{.}
    \end{eqnarray}

    Therefore, $\bigcup_{e \in P} \newR_e$ is length-synchronous.
\end{proof}

The two following propositions characterize length-synchronicity of
$\Ext$-algebra morphisms, which will be of
particular importance when approximating the matching relation of a
length-synchronous VPL in terms of $\FO[+]$.
This will be an important ingredient to proving that VPLs that both are
length-synchronous and have a quasi-aperiodic syntactic morphism (a notion to be
defined in Subsection~\ref{section quasi}) are in $\FOplus$ and thus in $\ACO$.

\begin{prop}\label{prop length synchronous}
    For all $F \subseteq R$, we have that $(\varphi,\psi)$ is
    $F$-length-synchronous if, and only if, for all $F$-reaching $e\in O$ the
    set of contexts $\newU_e$ is length-synchronous.
\end{prop}
\begin{proof}
    Let $F \subseteq R$. 

    If $\newU_e=\set{(u,v)\in\Con(\Sigma)\mid e\circ\psi(\ext_{u,v})=e}$ is
    length-synchronous for all $F$-reaching $e\in O$, then in particular the
    set of contexts $\newR_e=\set{(u,v)\in\Con(\Sigma)\mid \psi(\ext_{u,v})=e}$
    is length-synchronous for all $F$-reaching idempotents $e\in O$.

    Conversely, assume that $(\varphi,\psi)$ is $F$-length-synchronous.
    Fix any $F$-reaching $e\in O$.
    We need to prove that $\newU_e$ is length-synchronous. It is clear that
    $P = \set{f \in O \mid e \circ f = e}$ forms a subsemigroup of $O$ whose
    elements are all $F$-reaching, by $F$-reachability of $e$.
    Therefore, by Lemma~\ref{lem:Length-synchronicity_subsemigroups}, 
    $\bigcup_{f \in P} \newR_f = \newU_e$ is length-synchronous.
\end{proof}

\begin{samepage}
\begin{prop}\label{ptn:Length-synchronicity_equivalences}
	Let $F \subseteq R$ and assume $(\varphi,\psi)$ is
	$F$-weakly-length-synchronous.
	Then for all $F$-reaching $e\in O$ 
	the following two statements are equivalent.
    \begin{enumerate}
	\item\label{it:length-synchronous_height}
		The set of contexts $\newU_e$
		    is length-synchronous.
	    \item There exist $\alpha \in \Q_{>0}$,
	    $\beta \in \N$, $\gamma \in \N_{>0}$ such that for all
	    $(u,v)\in\newU_e$ with $\Delta(u) > 0$ we have:
	    \begin{enumerate}
		\item
		    $\frac{\length{u}}{\length{v}} = \alpha$.
		\item
		    For all $u', v' \in \Sigma^+$ with $u'$ prefix of $u$ and
		    $v'$ suffix of $v$ such that
		    $\frac{\length{u'}}{\length{v'}} = \alpha$, we have that
		    $-\Delta(v') - \beta \leq \Delta(u') \leq
		     -\Delta(v') + \beta$.
		\item
		    For all factors $u' \in \Sigma^*$ of $u$ such that
		    $\length{u'} = \gamma$, we have $\Delta(u') \geq 1$.
		\item
		    For all factors $v' \in \Sigma^*$ of $v$ such that
		    $\length{v'} = \gamma$, we have $\Delta(v') \leq -1$.
	    \end{enumerate}
    \end{enumerate}
\end{prop}
\end{samepage}

\begin{proof}
The implication from Point 2 to Point 1 is trivial
	since Point 2 (a) implies Point 1.

	Let us now prove that Point 1 implies Point 2. 
	Fix any $e\in O$ that is $F$-reaching
	and assume that $\newU_e$ is length-synchronous.
	Point 2 (a) follows immediately from length-synchronicity
	of $\newU_e$.
	We can hence write $\alpha=\frac{A}{B}$
	for some $A,B\in\N_{>0}$.

	For proving Point 2 (b),
we define $\beta=
(n+1)\cdot(|O|+\max(A,B)+1)$,
	where $n$ is the constant taken from Lemma~\ref{lem:Ext_pumping}.
	Let $(u,v)\in\newU_e$ with $\Delta(u) > 0$
	and let 
	\[
\ext_{u,v}=\ext_{x_1,y_1}\circ\ext_{a_1,b_1}\circ
\dots
\circ\ext_{x_{h-1},y_{h-1}}\circ\ext_{a_{h-1},b_{h-1}}\circ\ext_{x_h,y_h}
	\]
	be the stair factorization of $\ext_{u,v}$ according
	to Lemma~\ref{lemma ext}.
	Since our morphism $(\varphi,\psi)$ is
	$F$-weakly-length-synchronous by
	assumption, we have $|x_i|,|y_i|\leq n$
	by Lemma~\ref{ptn:hill_small}.
		Let $u'\in\Sigma^*$ be a prefix of $u$ and $v'$ be a suffix of
	$v$ such that $\frac{|u'|}{|v'|}=\alpha$.
	If $(u',v')=(u,v)$ we are done since then
	$\Delta(u')=-\Delta(v')$. 
	Thus, it remains to consider the case when $u'$ is a strict 
	prefix of $u$ and $v'$ is a strict suffix of $v$:
	indeed, due to $\frac{|u|}{|v|}=\frac{|u'|}{|v'|}=\alpha$ 
	we have that $u'$ is a strict prefix of $u$ 
	if, and only if, $v'$ is a strict suffix of $v$.

	Let $j\in[1,h]$ be maximal such that
	$x_1\cdots a_{j-1}x_j$ is a prefix of $u'$
	and $y_jb_{j-1}\cdots y_1$ is a suffix of $v'$.
	If $j = 1$ we are done, since then
	$\min\set{\length{u'}, \length{v'}} \leq n$, so that
	$|\Delta(u')+\Delta(v')| \leq \length{u'} + \length{v'} \leq
	 n + n \cdot \max(A, B) \leq \beta$.
	So assume now that $j > 1$, which implies that $\Delta(u'') > 0$.
	Note that $j<h$ since $(u',v')\not=(u,v)$.
	Hence there exist unique words $s,t\in\Sigma^*$ such that
	$u'=u''s$
	and $v'=tv''$,
	where $u''=x_1\cdots a_{j-1}x_j$ and 
	$v''=y_jb_{j-1}\cdots y_1$.
	By maximality of $j$ we have $\min\set{|s|,|t|}\leq n$.
	Setting $f=\psi(\ext_{u'',v''})$
	and $g=\psi(\ext_{a_jx_{j+1}\cdots a_{h-1}x_h,y_h b_{h-1}
	\cdots y_{j+1}b_j})$ 
	we have $\psi(\ext_{u,v})=f\circ g$.
	We claim that there exist
	$\ext_{x_g,y_g}\in\ExtMon(\Sigma^\triangle)$
	such that $\psi(\ext_{x_g,y_g})=g$ and $|x_g|,|y_g|\leq |O|\cdot (n+1)$:
	indeed, by the pigeonhole principle and Lemma~\ref{ptn:hill_small} (as
	$e \circ f \circ g = e$ and $\Delta(u'') > 0$), any
	$\ext_{x,y}\in\ExtMon(\Sigma^\triangle)$ such that $\psi(\ext_{x,y})=g$
	and $\max(|x|,|y|)>|O|\cdot (n+1)$ must have a stair factorization
	according to Lemma~\ref{lemma ext} with an $h > \card{O}$ and can thus
	be factorized as
	$\ext_{x,y}=\ext_{x',y'}\circ\ext_{x'',y''}\circ\ext_{x''',y'''}$ such
	that $\psi(\ext_{x,y})=\psi(\ext_{x',y'})\circ\psi(\ext_{x''',y'''})$,
	where moreover $(x'',y'')\in\Sigma^+\times\Sigma^+$.
	Thus, $\psi(\ext_{u''x_g,y_gv''})=\psi(\ext_{u,v})$
	and therefore $(u''x_g,y_gv'')\in\newU_e$ with $\Delta(u''x_g) > 0$.
	It follows $\alpha=\frac{|u''x_g|}{|y_gv''|}=
	\frac{|u'|-|s|+|x_g|}{|y_g|+|v'|-|t|}$, or equivalently,
	using $\frac{|u'|}{|v'|}=\alpha$:
	\begin{eqnarray}
		|s|&=&|u'|+|x_g|+\alpha(|t|-|y_g|-|v'|)=
		|x_g|+\alpha(|t|-|y_g|)\label{eq s}\\
		|t|&=&\frac{|s|-|u'|-|x_g|}{\alpha}+|y_g|+|v'|=
		\frac{|s|-|x_g|}{\alpha}+|y_g|\label{eq t}
		\displaypunct{.}
	\end{eqnarray}
	Finally, we obtain
	\begin{eqnarray*}
		|\Delta(u')+\Delta(v')| & = 
		&|\Delta(u''s)+\Delta(tv'')|\\
		&\stackrel{\Delta(u'')=-\Delta(v'')}{=}&|\Delta(s)+\Delta(t)|\\
		&\leq & |s|+|t|\\
		&=& \min(|s|,|t|)+\max(|s|,|t|)\\
		&\leq& n+\max(|s|,|t|)\\
		&\stackrel{(\ref{eq s}),(\ref{eq t})}{\leq}& 
		n+\max\left(|x_g|+\alpha(n-|y_g|),
		\frac{n-|x_g|}{\alpha}+|y_g|\right)\\
		&\leq& n+|O|\cdot (n+1)+n\cdot\max(A,B)\\
		&\leq&(n+1)\cdot(|O|+\max(A,B)+1)\\
		&=&\beta
		\displaypunct{.}
	\end{eqnarray*}
	This proves Point 2 (b).

	For Point 2 (c) and Point 2 (d) 
	we set 
$\gamma=(\lceil\frac{n}{2}\rceil+1)\cdot(n+1)+n$
and remark that $\gamma$ does not depend on $u$ nor $v$.
	We only prove Point 2 (c), the proof of Point 2 (d) is analogous.
	As above, let 
	\[
\ext_{u,v}=\ext_{x_1,y_1}\circ\ext_{a_1,b_1}\circ
\dots
\circ\ext_{x_{h-1},y_{h-1}}\circ\ext_{a_{h-1},b_{h-1}}\circ\ext_{x_h,y_h}
	\]
	be the stair factorization of $\ext_{u,v}$ according
	to Lemma~\ref{lemma ext}.
	Let $u'$ with $|u'|\geq\gamma$ be a factor of $u$ and hence
	of $x_1a_1x_2\cdots x_{h-1}a_{h-1}x_h$.
	By definition of stair factorization we have
	$\Delta(x_i)=0$ for all $i\in[1,h]$ and $\Delta(a_i)=1$
	for all $i\in[1,h-1]$.
	Let $w$ be the longest prefix of $u'$ such that 
	$\Delta(w)=\min\set{\Delta(x)\mid\text{$x$ is a prefix of $u'$}}$.
	Since $|x_1|,|y_1|,\dots,|x_h|,|y_h|\leq n$, it
	immediately follows $\Delta(w)\geq -\frac{n}{2}$ and $|w|\leq n$.
By the same reason, every prefix of the form $ws$ of $u'$ 
	satisfies $\Delta(ws)\geq \Delta(w)+\frac{|s|}{n+1}$.
	Thus we have 
    \[
	\Delta(u')\geq \Delta(w)+\frac{|u'|-|w|}{n+1}
	\geq
	-\frac{n}{2}+
		\frac{((\lceil\frac{n}{2}\rceil+1)\cdot(n+1)+n)-n}{n+1}
		\geq 1
		\displaypunct{.}
		\qedhere
    \]
\end{proof}

The following proposition relates, for languages $L$ generated by vertically 
visibly pushdown grammars $G$, (weak) length-synchronicity of $L$ 
with (weak) length-synchronicity of $\newR(G)$.

\begin{prop}\label{prop correspondence}
	Let $L=L(G)$ for some vertically visibly pushdown grammar $G=(V,\Sigma,P,S)$.
	Moreover, let $\newR(G)=\set{(u,v)\in\Con(\Sigma)\mid S\Rightarrow_G^*uSv}$.
	Then the following equivalences hold:
	\begin{enumerate}
		\item $L(G)$ is length-synchronous if, and only if,
			$\newR(G)$ is length-synchronous.
		\item $L(G)$ is weakly length-synchronous if, and only if,
			$\newR(G)$ is weakly length-synchronous.
	\end{enumerate}
\end{prop}
\begin{proof}
	Let $L=L(G)$ for a vertically visibly pushdown grammar 
	$G=(V,\Sigma,P,S)$.
Moreover, let 
$(\varphi_L,\psi_L)\colon (\ExtMon(\Sigma),\Sigma^\triangle)\to(R_L,O_L)$ be the syntactic morphism.
	For all $e\in O_L$ recall the set of contexts 
	$\newR_e=\set{(u,v)\in\Con(\Sigma)\mid\psi_L(\ext_{u,v})=e}$.

	Let $F=\set{e\in O_L\mid\exists\ext_{u,v}\in\psi_L^{-1}(e) : (u,v)
	\in\Rn(G)}$.
	Observe that $F$ is a submonoid of $O_L$ all of whose
	elements are $\varphi_L(L)$-reaching since $G$ is vertically 
	visibly pushdown. Also observe that
	$\Rn(G)=\bigcup_{f \in F} \newR_f$.
	We claim that since $G$ is vertically visibly pushdown, there exists
	a constant $C>0$ such that 
$\frac{1}{C}\leq\frac{|x|}{|y|}\leq C$
	for all 
	$(x,y)\in\Rn(G)\setminus\set{(\varepsilon,\varepsilon)}$:
	indeed, one can take $C=\frac{A}{B}$ where
	$A = \max\bigset{\max\set{\length{u}, \length{v}} \bigmid
			T\rightarrow_GuT'v}$
	and
	$B = \min\bigset{\min\set{\length{u}, \length{v}} \bigmid
			T\rightarrow_GuT'v}$
	since whenever $T \Rightarrow_G^* x T' y$ thanks to a derivation
	comprising $k \in \N_{>0}$ steps, we have
	\[
	    \frac{1}{C} = \frac{k \cdot B}{k \cdot A} \leq
	    \frac{\length{x}}{\length{y}} \leq \frac{k \cdot A}{k \cdot B} = C
	    \displaypunct{.}
	\]

	Next, we prove that for all $\varphi_L(L)$-reaching idempotents
	$e\in O_L$ there exist $g,h\in O_L$ such that
	$g\circ e\circ h\in F$.
	Fix any such $\varphi_L(L)$-reaching idempotent $e\in O_L$.
	Without loss of generality let us assume  that $e$ is not the identity in $O_L$
	(indeed, if $e$ is the identity in $O_L$, then we are done 
	since we can then choose $g=h=e\in F$).
	Thus there exist $g\in O_L$ and $r\in R_L$ such that
	$(g\circ e)(r)\in\varphi_L(L)$.
	Moreover, let $\ext_{u',v'}\in\psi_L^{-1}(g),\ext_{u,v}\in\psi_L^{-1}(e)$,
	and $w\in\varphi_L^{-1}(r)$. Observe that we must have $(u,v)\not=(\varepsilon,\varepsilon)$
	since $\psi_L(\ext_{\varepsilon,\varepsilon})$ is the identity in $O_L$. 
	Since $e$ is an idempotent we have
	that $u'u^nwv^nv'\in L$ for all $n\geq 1$.
	Fix a sufficiently large $N\geq 1$ such that 
	$\frac{|u'|+|u|}{(N-1)|u|+|w|+N|v|+|v'|}\leq\frac{1}{C}$
	and $\frac{|u'|+N|u|+|w|+(N-1)|v|}{|v|+|v'|}\geq C$, which exists due to
	$(u,v)\not=(\varepsilon,\varepsilon)$.
	Since $u'u^Nwv^Nv'\in L$ there exists $(x,y)\in\Rn(G)\setminus\set{(\varepsilon,\varepsilon)}$
	such that $xy=u'u^Nwv^Nv'$, $S\Rightarrow_G^* xSy$, $|x|\geq|u'u|$, and $|y|\geq |vv'|$.
	Let $(x',y')\in\Sigma^*\times\Sigma^*$ be such that 
	$(x,y)=(u'ux',y'vv')$.
	As $(u',v'),(u,v),(x,y)\in\Con(\Sigma)$,
	$x'\in(\Sigma\setminus\Sigma_\return)^*$,
	and $y'\in(\Sigma\setminus\Sigma_\call)^*$, we can conclude
	$(x',y')\in\Con(\Sigma)$.
	That is, $(x,y)=(u',v')\circ(u,v)\circ(x',y')\in\Rn(G)$.
	Hence,
	\[
	g\circ e\circ\psi_L(\ext_{x',y'})=\psi_L(\ext_{u'ux',y'vv'})=\psi_L(\ext_{x,y})\in F.
	\]

		We are now ready to prove Point 1.
	For the first direction, let us assume that $L(G)$ is length-synchronous.
		Recalling that $F$ is a submonoid of $O_L$ all of whose
		elements are $\varphi_L(L)$-reaching, we obtain that
	$R(G)=\bigcup_{f \in F} \newR_f$ is length-synchronous
	by Lemma~\ref{lem:Length-synchronicity_subsemigroups}.

	Conversely, let us assume that 
	$\Rn(G)$ is length-synchronous.
	Assume by contradiction that $L(G)$ is not length-synchronous. Hence 
	$\newR_e$ is not length-synchronous for some 
	$\varphi_L(L)$-reaching idempotent $e\in O_L$, i.e.
	$\psi_L(\ext_{u,v})=\psi_L(\ext_{u',v'})=e$ for some
	$\ext_{u,v},\ext_{u',v'}\in\ExtMon(\Sigma^\triangle)$ such that
	$\Delta(u),\Delta(u')>0$ and $\frac{|u|}{|v|}\not=\frac{|u'|}{|v'|}$. 
	Without loss of generality we may assume that $|v|=|v'|$
	(indeed, if $|v|\not=|v'|$, then 
	$\ext_{u^{|v'|},v^{|v'|}},\ext_{(u')^{|v|},(v')^{|v|}}$
	satisfies the desired property).
	As a consequence we have $|u|\not=|u'|$, say
	$|u|<|u'|$ without loss of generality.
	Since $e$ is a $\varphi_L(L)$-reaching idempotent, as argued above,
	there exist $g,h\in O_L$ such that $g\circ e\circ h=f'$
	for some $f'\in F$.
	Let us fix
	$\ext_{x_h,y_h}\in\psi_L^{-1}(h)$ and
	$\ext_{x_g,y_g}\in\psi_L^{-1}(g)$.
	We have
	$\psi_L(\ext_{x_gux_h,y_hvy_g}) = f' =
	 \psi_L(\ext_{x_gu'x_h,y_hv'y_g})$.
Since $|x_g u x_h|<|x_g u' x_h|$ and
$|y_h v y_g|=|y_h v' y_g|$ it follows that
$\newR_{f'}$ is not length-synchronous, contradicting
our assumption that 
$R(G)=\bigcup_{f \in F} \newR_f$ is length-synchronous.

	Let us next prove Point 2. 

	Let us first assume that 
	$L(G)$ is weakly length-synchronous.
	Again, since $F$ is a submonoid of $O_L$ all of whose elements
	are $\varphi_L(L)$-reaching, we obtain
	that $R(G)=\bigcup_{f \in F} \newR_f$
	is weakly length-synchronous by 
Lemma~\ref{lem:Weak_length-synchronicity_subsemigroups}.

	Conversely, assume $\Rn(G)$ is weakly 
	length-synchronous.
	Assume by contradiction that $\newR_e$ is not weakly 
	length-synchronous
	for some $\varphi_L(L)$-reaching idempotent $e\in O_L$.
	Thus, 
	$\psi_L(\ext_{u,v})=\psi_L(\ext_{u',v'})=e$
	for some $\ext_{u,v},\ext_{u',v'}\in\ExtMon(\Sigma^\triangle)$
	such that 
$\Delta(u),\Delta(u')>0$ and moreover either
	$u=u'$ and $|v|\not=|v'|$ or $v=v'$ and $|u|\not=|u'|$.
	Without loss of generality let us assume that
	$u=u'$ and $|v|\not=|v'|$.
	As mentioned above, there exist $g,h\in O_L$
	such that $g\circ e\circ h=f'$ for some $f'\in F$. 
	Fix some $\ext_{x_g,y_g}\in\psi_L^{-1}(g)$
	and some $\ext_{x_h,y_h}\in\psi_L^{-1}(h)$.
Analogously, as argued above, we have
	$\psi_L(\ext_{x_gux_h,y_hvy_g})=\psi_L(\ext_{x_gux_h,y_hv'y_g})=f'$,
	$\Delta(x_gux_h)>0$, and $|y_hvy_g|\not=|y_hv'y_g|$, 
	implying that $\newR_{f'}$ is not weakly length-synchronous,
	a contradiction to our assumption that
	$\Rn(G)=\bigcup_{f\in F} \newR_f$ is weakly
	length-synchronous.
\end{proof}

\subsection{The nesting depth of visibly pushdown languages}\label{section nesting}
Another central notion is the nesting depth of well-matched words,
which is the Horton-Strahler number~\cite{ELS14} of the underlying trees.

\begin{defi}
    The \emph{nesting depth} of 
	well-matched words is given by the function 
	$\ndepth\colon \Sigma^\triangle \to \N$ defined inductively as follows:
	    \begin{itemize}
	\item
	    $\ndepth(\emptyword) = 0$;
	\item
	    $\ndepth(c) = 0$ for each $c \in \Sigma_\internal$;
	\item
	    $\ndepth(u v) = \max\set{\ndepth(u), \ndepth(v)}$ for each
	    $u \in
	     \Sigma_\call \Sigma^\triangle \Sigma_\return \cup \Sigma_\internal$
	    and $v \in \Sigma^\triangle \setminus \set{\emptyword}$;
	\item
	    $\ndepth(a w b) =
	     \begin{cases}
		\ndepth(w) + 1 & \text{if $w = u v$ with $u, v \in \Sigma^\triangle$
				   and $\ndepth(w) = \ndepth(u) = \ndepth(v)$}\\
		\ndepth(w) & \text{otherwise}
	     \end{cases}$
	    for each $a \in \Sigma_\call$, $b \in \Sigma_\return$ and
	    $w \in \Sigma^\triangle$.
    \end{itemize}
\end{defi}
\noindent
An important property of weakly length-synchronous VPLs is that their words have
bounded nesting depth.

\begin{prop}\label{ptn:Bounded_nesting-depth}
    For each weakly length-synchronous VPL $L \subseteq \Sigma^\triangle$ there
    exists a constant $d \in \N$ such that
    $L \subseteq \set{w \in \Sigma^\triangle \mid \ndepth(w) \leq d}$.
\end{prop}

Proposition~\ref{ptn:Bounded_nesting-depth} is proved in several steps.
For these we introduce a factorization that can be seen as one that
witnesses the nesting depth of a word.

\begin{defi}
    A \emph{nesting-maximal stair factorization} of $w\in\Sigma^\triangle$ 
	with $\ndepth(w)\geq 1$ is a factorization of $w$ as
    \[
	w =
	\ext_{x_1,y_1} \circ \ext_{a_1,b_1} \circ \cdots \circ
	\ext_{x_{k},y_{k}} \circ \ext_{a_{k},b_{k}} (w')
    \]
    such that $k\geq 0$, $x_i,y_i\in\Sigma^\triangle$, $a_i\in\Sigma_\call$, and
	$b_i\in\Sigma_\return$ for all $i \in [1,l]$, and 
	$w' \in \Sigma_\internal^*$ satisfying
    that for all $i \in [1, k]$ we have
    \[
		\ndepth(\ext_{x_i,y_i}(w_i))=\ndepth(w_i),
    \]
	where $w_i=\ext_{a_i,b_i}\circ\ext_{x_{i+1},y_{i+1}}\circ\cdots\circ\ext_{a_{k},b_{k}}(w')$.
 \end{defi}

\begin{lem}\label{lem:Maximal_stair_factorization}
    All words $w \in \Sigma^\triangle$ have a nesting-maximal stair factorization.
\end{lem}

\begin{proof}
    The proof goes by structural induction on $w$.
    \begin{itemize}
	\item
	    $w = \emptyword$. Then we are done because $w$ contains only
	    internal letters.
	\item
	    $w = c$ for a $c \in \Sigma_\internal$. Then we are again done
	    because $w$ contains only internal letters.
	\item
	    $w = a w' b$ for $a \in \Sigma_\call$, $b \in \Sigma_\return$ and
	    $w' \in \Sigma^\triangle$.
	    By using the inductive hypothesis, $w'$ has a nesting-maximal stair
	    factorization
		    $\ext_{x_1,y_1}\circ\ext_{a_1, b_1} \circ \cdots \circ
		    \ext_{x_{k},y_k}\circ\ext_{a_{k}, b_{k}} (w'')$.
	    It directly follows that
		    $\ext_{a,b}\circ\ext_{x_1,y_1}\circ\ext_{a_1, b_1} 
		    \circ \cdots \circ
		    \ext_{x_{k},y_k}\circ\ext_{a_{k}, b_{k}} (w'')$
	    is a nesting-maximal stair factorization of $w$.
	\item
	    $w = u v$ for
	    $u, v \in \Sigma^\triangle \setminus \set{\emptyword}$.
	    Then $w$ can be decomposed as $z_1 \cdots z_m$ with
	    $z_1, \ldots, z_m \in
	     \Sigma_\call \Sigma^\triangle \Sigma_\return \cup \Sigma_\internal$
	    and $m \in \N, m \geq 2$.
	    In this case, either $z_i \in \Sigma_\internal$ for all
	    $i \in [1, m]$ and thus we are done because $w$ contains only
	    internal letters, or there exists some $i \in [1, m]$ such that
	    $z_i \in \Sigma_\call \Sigma^\triangle \Sigma_\return$ and has
	    maximal nesting depth, i.e.\ $\ndepth(w) = \ndepth(z_i)$.
	    In this second subcase, we have that $z_i = a z_i' b$ with
	    $a \in \Sigma_\call$, $b \in \Sigma_\return$ and
	    $z_i' \in \Sigma^\triangle$.
	    By using the inductive hypothesis, $z_i'$ has a nesting-maximal stair
	    factorization
		    $\ext_{x_1,y_1}\circ\ext_{ a_1, b_1}\circ 
		    \cdots \circ \ext_{x_k,y_k}\circ\ext_{a_k,b_k} (w'')$.
	    Therefore,
	    \[
		\ext_{z_1 \cdots z_{i - 1}, z_{i + 1} \cdots z_k}\circ
		    \ext_{a,b}\circ\ext_{x_1,y_1}\circ\ext_{a_1,b_1}\circ
		    \dots\ext_{x_k,y_k}\circ\ext_{a_k,b_k}(w'')
	    \]
	    is a nesting-maximal stair factorization of $w$.
	    \qedhere
    \end{itemize}
\end{proof}
\noindent
The following lemma will be a useful tool for proofs by induction on the nesting
depth of well-matched words.

\begin{samepage}
	\begin{lem}{\label{L nesting induction}}
	Let $u=a_1vb_1\in\Sigma^\triangle$ for some $a_1\in\Sigma_\call$,
	$b_1\in\Sigma_\return$, and $v\in\Sigma^\triangle$ such that $\ndepth(u)=d>0$.
	Moreover, let $u=\ext_{x_1,y_1}\circ\ext_{a_1,b_1}\circ\cdots\circ
	\ext_{x_{k},y_{k}}\circ\ext_{a_{k},b_{k}}(u')$
	be a nesting-maximal stair factorization of $u$ 
	(i.e. $x_1=y_1=\varepsilon$).
	Then there exists $h\in[1,k]$ such that,
	setting $u_i=\ext_{a_i,b_i}\circ\ext_{x_{i+1},y_{i+1}}
			\circ\cdots\circ\
			\ext_{a_k,b_k}(u')$
	for all $i \in [1, k]$ and $u_{k + 1} = u'$,
	we have
	\begin{enumerate}
		\item $\ndepth(u)=\ndepth(u_h)=d$,
		\item $\ndepth(u_{h+1})=d-1$, and 
		\item $\ndepth(x_1),\ndepth(y_1),\dots,\ndepth(x_h),\ndepth(y_h)<d$.
	\end{enumerate}
\end{lem}
\end{samepage}
\begin{proof}
	Let $u_j=\ext_{a_j,b_j}\circ\ext_{x_{j+1},y_{j+1}}
			\circ\cdots\circ\
			\ext_{a_{k},b_{k}}(u')$
for all $j\in[1,k]$.
	Note that we have $\ndepth(u)=\ndepth(u_1)=d>0$ by assumption.
	Moreover, $\ndepth(u_{j})\geq\ndepth(u_{j+1})$ for all $j\in[1,k-1]$
	by definition of nesting depth.
	Thus, since $\ndepth(u_k) = 1 > 0 = \ndepth(u_{k + 1})$, it follows that 
	\[h=\min\set{j\in[1,k]\mid\ndepth(u_j)>\ndepth(u_{j+1})}\] is
	well-defined and 
	$\ndepth(u)=\ndepth(u_1)=\ndepth(u_h)=d$, thus showing Point 1.
	Since 
	\[d=\ndepth(u_h)\leq\ndepth(u_{h+1})+1\]
	and $\ndepth(u_{h+1})<\ndepth(u_h)=d$
	it follows $\ndepth(u_{h+1})=d-1$, thus showing Point 2.
	To prove Point 3, assume by contradiction that
	$\ndepth(x_j)\geq d$ or $\ndepth(y_j)\geq d$ for some $j\in[1,h]$.
	Without loss of generality assume $\ndepth(x_j)\geq d$.
	Since $x_1=y_1=\varepsilon$ and $d>0$ we must have $j\in[2,h]$.
	It follows \[\ndepth(u)\geq\ndepth(u_{j-1})=\ndepth(a_{j-1}x_ju_jy_jb_{j-1})\geq
	\min(\ndepth(x_j),\ndepth(u_j))+1\geq d+1>d=\ndepth(u)\displaypunct{,}\]
	which is a contradiction.
\end{proof}

We are now ready to prove Proposition~\ref{ptn:Bounded_nesting-depth}.

\begin{proof}[Proof of Proposition~\ref{ptn:Bounded_nesting-depth}]
	Let $L\subseteq\Sigma^\triangle$ be a weakly length-synchronous VPL.
	We claim that $\ndepth(L)\leq n+1$, where $n$ is the 
	pumping constant from Lemma~\ref{lem:Ext_pumping}.
	Assume by contradiction that $\ndepth(u)=d$ for some 
	$u\in L$ and some $d>n+1$.
	Let \[u=\ext_{x_1,y_1}\circ\ext_{a_1,b_1}\circ\dots\circ
	\ext_{x_{k},y_{k}}\circ\ext_{a_k,b_k}(u')\] be a nesting-maximal stair
	factorization of $u$ according to 
	Lemma~\ref{lem:Maximal_stair_factorization}.
	According to Lemma~\ref{L nesting induction} there exists
	$i\in[1,k]$ such that, setting
	\[u_j=\ext_{a_j,b_j}\circ\ext_{x_{j+1},y_{j+1}}\circ\cdots\circ
	\ext_{a_k,b_k}(u')\] for all $j \in [1, k]$ and $u_{k + 1} = u'$, we have
	$\ndepth(u)=\ndepth(u_i)=d$ and $\ndepth(u_{i+1})=d-1$.
	Since $d - 1 > n > 0$, we must have $i + 1 \leq k$, so that
	$u_i=a_ix_{i+1}u_{i+1}y_{i+1}b_i$ with
	$u_{i+1} \in \Sigma_\call \Sigma^\triangle \Sigma_\return$ and
	$\ndepth(x_{i+1}u_{i+1}y_{i+1})=\ndepth(u_{i+1})=d-1$.
	Hence it follows that $\ndepth(x_{i+1})=d-1$ or $\ndepth(y_{i+1})=d-1$.
	Without loss of generality let us assume $\ndepth(y_{i+1})=d-1>n$.
	A simple induction shows that $|x|\geq 2^{\ndepth(x)}-1\geq\ndepth(x)$
	for all $x\in\Sigma^\triangle$.
	Thus, we have $|y_{i+1}|\geq\ndepth(y_{i+1})>n$, contradicting 
	Proposition~\ref{ptn:hill_small}.
	\end{proof}

\subsection{Quasi-aperiodicity and its correspondence with 
quasi-counterfreeness\label{section quasi}}

Let us revisit the notion of
quasi-aperiodicity. 
Towards characterizing the circuit complexity of visibly
pushdown languages it has already been defined 
for visibly pushdown languages in~\cite{Ludwigthesis}.
Let us define
$\ExtMon(\Sigma^\triangle)^{k,l} =
 \bigset{\ext_{u,v} \in \ExtMon(\Sigma^\triangle) \bigmid \length{u} = k, \length{v} = l}$
for all $k,l\in\N$.
We say the morphism
$(\varphi,\psi)\colon(\Sigma^\triangle, \ExtMon(\Sigma^\triangle))\to (R,O)$
is 
{\em quasi-aperiodic} if all semigroups
contained in the set 
$\psi(\ExtMon(\Sigma^\triangle)^{k,l})$ are aperiodic for all $k,l\in\N$.

The following proposition relates 
quasi-counterfreeness of a VPL with 
 quasi-aperiodicity of its syntactic morphism.

\begin{prop}\label{prop quasi-counterfree}
	A VPL $L\subseteq\Sigma^\triangle$ is quasi-counterfree
	if, and only if, its syntactic morphism is quasi-aperiodic.
\end{prop}

\begin{proof}
	Recall that 
	$(R_L,O_L)=(\Sigma^\triangle,\ExtMon(\Sigma^\triangle))\quotient\sim_L$
	by Definition~\ref{def syntactic}.
	Hence for all $(u,v),\allowbreak(u',v')\in\Con(\Sigma)$ we have
	$(u,v)\equiv_L(u',v')$ if, and only if, 
	$\psi_L(\ext_{u,v})=\psi_L(\ext_{u',v'})$.

	First, let us assume that 
$(\varphi_L,\psi_L)\colon (\Sigma^\triangle,\ExtMon(\Sigma^\triangle))\rightarrow(R_L,O_L)$ is quasi-aperiodic.
	Assume by contradiction that $L$ is not
	quasi-counterfree. 
	Thus, there exists some $\sigma=(u,v)\in\Con(\Sigma)$ such that
	$\sigma^n\not\equiv_L\sigma^{n+1}$ for all $n\in\N$
	and 
	$\tau\equiv_L\sigma\circ\sigma$
	for some $\tau=(x,y)\in\Con(\Sigma)\cap\Sigma^{|u|}\times
	\Sigma^{|v|}$. The latter can equivalently be
	rephrased as $\psi_L(\ext_{x,y})=\psi_L(\ext_{u^2,v^2})$.
	By choice of $\sigma$ and the fact that
	$\equiv_L$ has finite index there exist $s\geq 1$ and $t\geq 2$ such that $\sigma^{s+i}\equiv_L\sigma^{s+i+t}$
	for all $i\geq 0$ and 
	$\sigma^{s+i}\not\equiv_L\sigma^{s+j}$ for all $i,j\in[0,t-1]$ with 
	$i\not=j$.
	It follows that the set
	$G=\psi_L(\set{\ext_{u^{s+i},v^{s+i}}\mid i\in[0,t-1]})$
	is a non-trivial group
	with identity
	$g_0=\psi_L(\ext_{u^{s+k},v^{s+k}})\in\psi_L(\ExtMon(\Sigma^\triangle)^{(s+k)|u|,(s+k)|v|})$,
	where $k$ is the unique integer in $[0,t-1]$ such that
	$s+k$ is divisible by $t$.
	Also observe that $G$ is a cyclic group that is generated by 
	$g_1=\psi_L(\ext_{u^{s+k+1},v^{s+k+1}})$.
	That is $G=\set{g_0,g_1,\dots,g_{t-1}}$, where $g_i=g_1^i$ for
	all $i\in[2,t-1]$.
	But note that due to $\psi_L(\ext_{x,y})=\psi_L(\ext_{u^2,v^2})$
	we obtain 
	$g_1=\psi_L(\ext_{u^{s+k+1},v^{s+k+1}})=\psi_L(\ext_{xu^{s+k-1},v^{s+k-1}y})$
	and thus $g_0,g_1\in\psi_L(\ExtMon(\Sigma^\triangle)^{(s+k)|u|,(s+k)|v|})$
	due to $|x|=|u|$ and $|y|=|v|$.
	Since $g_i=g_0^{t-1-i}g_1^i$ for all $i\in[0,t-1]$
	we obtain that $G=\set{g_0,g_1,\dots,g_{t-1}}$ is contained in
	$\psi_L(\ExtMon(\Sigma^\triangle)^{(t-1)(s+k)|u|,(t-1)(s+k)|v|})$,
	hereby contradicting quasi-aperiodicity of $(\varphi_L,\psi_L)$.

	For the converse direction, let us assume that $L$ is
	quasi-counterfree. Assume by contradiction that $(\varphi_L,\psi_L)$ is
	not quasi-aperiodic. That is, for some $k,l\in\N$ the set 
	$\psi_L(\ExtMon(\Sigma^\triangle)^{k,l})$
	contains a non-trivial group $G$. 
	Let $g_0\in G$ be the identity of $G$.
	Fix some $g\in G$ with $g\not=g_0$ and some 
	$\ext_{u,v}\in\ExtMon(\Sigma^\triangle)^{k,l}$ such that $\psi_L(\ext_{u,v})=g$.
	Let $\sigma=(u,v)\in\Con(\Sigma)\cap\Sigma^k\times\Sigma^l$.
	Since $g$ is not the identity in $G$ we have that 
	$\psi_L(\ext_{u^n,u^n})\not=\psi_L(\ext_{u^{n+1},v^{n+1}})$ for all $n\in\N$,
	equivalently $\sigma^n\not\equiv_L\sigma^{n+1}$ for 
	all $n\in\N$.
	But since moreover $\psi_L(\ext_{u^n,v^n})$ is in $G$ and
	thus in $\psi_L(\ExtMon(\Sigma^\triangle)^{k,l})$ for all $n\in\N$
	it follows that $\sigma\circ\sigma\equiv_L\tau$ for some 
	$\tau\in\Con(\Sigma)\cap\Sigma^{k}\times\Sigma^{l}$.
	We thus obtain a contradiction to our assumption that $L$ is 
	quasi-counterfree.
\end{proof}

\section{Proof of the main theorem}\label{Section Proof Strategy}

Before giving an overview of the proof of Theorem~\ref{thm main}
we will state a proposition saying that the syntactic $\Ext$-algebra
and the syntactic morphism of a given visibly pushdown language $L$ is 
computable
and that it is decidable if $L$ is quasi-aperiodic, length-synchronous, 
and weakly length-synchronous, respectively. 
Its proof is the subject of Section~\ref{Section Effectiveness}.
\begin{samepage}
\begin{prop}\label{prop effectivity}
	The following computability and decidability results hold:
	\begin{enumerate}
		\item Given a DVPA $A$, one can effectively compute the syntactic
    $\Ext$-algebra of $L = L(A)$, its syntactic morphism
    $(\varphi_L, \psi_L)$ and $\varphi_L(L)$.
\item 
   Given a morphism 
    $(\varphi,\psi)\colon(\Sigma^\triangle,\ExtMon(\Sigma^\triangle))\to (R,O)$
    for a visibly pushdown alphabet $\Sigma$ and a finite
    $\Ext$-algebra $(R, O)$, 
	all of the following are decidable for $(\varphi,\psi)$:
			\begin{enumerate}
				\item Quasi-aperiodicity. In case
					$(\varphi,\psi)$ is not 
					quasi-aperiodic, one can effectively 
			compute $k,l\in\N$ such that 
			$\psi(\ExtMon(\Sigma^\triangle)^{k,l})$ is
			not aperiodic.
				\item $F$-length-synchronicity for a given 
					$F\subseteq R$. In case 
					$(\varphi,\psi)$ is not $F$-length-synchro\-nous, one can effectively compute
			a quadruple $(k,l,k',l')\in\N_{>0}^4$ such that 
			there exist $uv,u'v'\in\Sigma^\triangle$
			and some $F$-reaching idempotent $e\in O$ such that
			$\psi(\ext_{u,v})=\psi(\ext_{u',v'})=e$,
			$\Delta(u)>0$, $\Delta(u')>0$,
					$k=|u|,l=|v|,k'=|u'|$, $l'=|v'|$, and
			$\frac{k}{l}\not=\frac{k'}{l'}$.
				\item $F$-weakly-length-synchronicity for a given $F\subseteq R$.
			\end{enumerate}
\end{enumerate}
\end{prop}
\end{samepage}

\subsection{Proof outline for Theorem~\ref{thm main}}\label{S outline}

Towards proving our main result (Theorem~\ref{thm main}), given a DVPA
$A$, where $L=L(A)$ is a VPL over a visibly pushdown alphabet $\Sigma$, we apply
Proposition~\ref{prop effectivity} and compute its syntactic $\Ext$-algebra
$(R_L,O_L)$ along with its syntactic morphism $(\varphi_L,\psi_L)$ and 
the subset $\varphi_L(L)$.
Then we make the following effective case distinction which immediately
implies Theorem~\ref{thm main}.

\begin{enumerate}
    \item
	If $L$ is not weakly length-synchronous, then $L$ is $\TCO$-hard
		and hence not in $\ACO$
	(Proposition~\ref{prop not wls} in Section~\ref{section lowerbounds}).
		Thus, we can output any $m\geq 2$ since 
		$\MOD_m\reduce\EQ\reduce L$, so 
		$L$ is $\ACCO(m)$-hard for any $m\geq 2$.
    \item
	If $L$ is not quasi-aperiodic, then 
	one can effectively compute some $m\geq 2$ such that
	$\MOD_m\reduce L$
		(Proposition~\ref{prop notAC0} in Section~\ref{section lowerbounds}), so $L$ is $\ACCO(m)$-hard.
    \item
	If $L$ is length-synchronous and $(\varphi_L,\psi_L)$ is
	quasi-aperiodic, then $L\in\ACO$
	(Theorem~\ref{thm:Length-sync_and_quasi-aperiodic} in
	Section~\ref{Section AC0}).
    \item
If $L$ is weakly length-synchronous but not
length-synchronous, and its syntactic morphism
	$(\varphi_L,\psi_L)$ is quasi-aperiodic, 
one can effectively compute vertically visibly pushdown 
	grammars $G_1,\dots,G_m$ generating intermediate VPLs 
	such that $L\eqACO\biguplus_{i\in[1,m]} L(G_i)$
(Theorem~\ref{thm equiv union} in
		Section~\ref{section intermediate}).
	Moreover, already if a VPL $L$ is weakly length-synchronous but not 
		length-synchronous, 
		one can effectively compute 
		$k, l \in \N_{>0}$ with $k \neq l$ such that 
		$\SM_{k, l}\reduce L$
(Proposition~\ref{prop SM lowerbound} in Section~\ref{section intermediate}).
\end{enumerate}

We refer to Section~\ref{S Corollary} for the proof of
Corollary~\ref{corollary OCA}.

\subsection{Lower bounds}\label{section lowerbounds}

The following visibly pushdown languages
are helpful for proving lower bounds.

\begin{defi}
    Let $L \subseteq \Sigma^\triangle$ be a VPL.
    For each $e \in O_L$ and for $\#\not\in\Sigma$ a fresh internal letter
	we define
    \[
	L_e = \set{u \# v \mid (u, v) \in \Con(\Sigma) :
			       \psi_L(\ext_{u, v})=e}
    \]
	and
\begin{align*}
    L_{e \uparrow} & =
    \set{u\#v\mid (u, v) \in \Con(\Sigma) :
		  \Delta(u) > 0, \psi(\ext_{u,v})=e}\\& =
    L_e \cap \set{u\#v\mid (u, v) \in \Con(\Sigma) : \Delta(u) > 0}.
\end{align*}

\end{defi}
	The next lemma shows that 
	both $\varphi_L^{-1}(r)$ and $L_e$ are constant-depth
	reducible to $L$ in case $r\in R_L$ and $e\in O_L$ are
	$\varphi_L(L)$-reaching, respectively.
	\begin{samepage}
\begin{lem}\label{lemma reduction element}
    Let $L \subseteq \Sigma^\triangle$ be a VPL. Then
     \begin{itemize}
	\item $\varphi_L^{-1}(r) \reduce L$ for all $\varphi_L(L)$-reaching 
		$r \in R_L$, and
	\item
	    $L_e \reduce L$ for all $\varphi_L(L)$-reaching $e \in O_L$.
    \end{itemize}
\end{lem}
	\end{samepage}
\begin{proof}
	To show the first point, let us fix some $\varphi_L(L)$-reaching
	$r\in R_L$.
	Thus, there exist $w_r\in\Sigma^\triangle$ and 
	$(u_r,v_r)\in\Con(\Sigma)$
	such that $\varphi_L(w_r)=r$ and 
	$\varphi_L(u_rw_rv_r)\in \varphi_L(L)$.
By definition of the syntactic morphism of $L$ (Definition~\ref{def syntactic}) 
	for all $r_1,r_2\in R_L$ with $r_1\not=r_2$ there exists
	some $e_{r_1,r_2}\in O_L$ such that 
	$e_{r_1,r_2}(r_1)\in \varphi_L(L)\Leftrightarrow e_{r_1,r_2}(r_2)\not\in 
	\varphi_L(L)$.
	For each such $e_{r_1,r_2}\in O_L$ fix
	$(u_{r_1,r_2},v_{r_1,r_2})\in\Con(\Sigma)$
	with $\psi_L(\ext_{u_{r_1,r_2},v_{r_1,r_2}})=e_{r_1,r_2}$.

	Hence, for all $w\in\Sigma^*$ we have
	\[
	    w\in\varphi_L^{-1}(r)\quad\Longleftrightarrow\quad
	    u_rwv_r\in L\ \wedge\
	    \bigwedge_{\substack{r'\in R_L\\r\not=r'}}
	    u_{r,r'}wv_{r,r'}\in
	    L\leftrightarrow u_{r,r'}w_rv_{r,r'}\in L
	    \displaypunct{,}
	\]
	thus showing $\varphi_L^{-1}(r)\reduce L$.

	For the second point, let us fix some $\varphi_L(L)$-reaching 
	$e\in O_L$.
	Fix some $(u_e,v_e)\in\Con(\Sigma)$ such that 
	$\psi_L(\ext_{u_e,v_e})=e$.
	Thus, again, there exist $w_e\in\Sigma^\triangle$ and 
	$(u_e',v_e')\in\Con(\Sigma)$
	such that $\varphi_L(u_e'u_ew_ev_ev_e')\in \varphi_L(L)$.
		Again by definition of the syntactic morphism of $L$
		(Definition~\ref{def syntactic}), 
		for all $e_1,e_2\in O_L$ with $e_1\not=e_2$ there exist some 
		$f_{e_1,e_2}\in O_L$ and some
	$r_{e_1,e_2}\in R_L$ such that 
	$f_{e_1,e_2}(e_1(r_{e_1,e_2}))\in\varphi_L(L)\Leftrightarrow 
	f_{e_1,e_2}(e_2(r_{e_1,e_2}))\not\in\varphi_L(L)$.
	For each such $f_{e_1,e_2}$ and $r_{e_1,e_2}$ fix, respectively,
	$(u_{e_1,e_2},v_{e_1,e_2})\in\Con(\Sigma)$ and
	$w_{e_1,e_2}\in\Sigma^\triangle$ such that 
	$\psi_L(\ext_{u_{e_1,e_2},v_{e_1,e_2}}) = f_{e_1,e_2}$ and
	$\varphi_L(w_{e_1,e_2})=r_{e_1,e_2}$, respectively.
	Hence, for all $u\#v\in\Sigma^*\#\Sigma^*$ we have
	\[
	    u\#v\in L_e\quad\Longleftrightarrow\quad\
	    u_e' u w_e v v_e' \in L\ \wedge\
	    \bigwedge_{\substack{e'\in O_L\\e\not=e'}}
	    u_{e,e'} u w_{e,e'} v v_{e,e'}\in L\leftrightarrow
	    u_{e,e'} u_e w_{e,e'} v_e v_{e,e'} \in L
	    \displaypunct{,}
	\]
	thus showing $L_e\reduce L$.
\end{proof}

The following lower bound has already been sketched in 
Section~\ref{Section Notions}.

\begin{prop}\label{prop not wls}
    If $L$ is not weakly length-synchronous, then $L$ is $\TCO$-hard.
\end{prop}

\begin{proof}
    Recall that $(R_L, O_L)$ is the syntactic $\Ext$-algebra of $L$ and
    $(\varphi_L, \psi_L)\colon
     (\Sigma^\triangle, \ExtMon(\Sigma^\triangle)) \to (R_L, O_L)$
    is its syntactic morphism. Assume 
	that $(\varphi_L, \psi_L)$ is not $\varphi_L(L)$-weakly-length-synchronous.

    Assume first there exist
    $\ext_{u, v}, \ext_{u', v} \in \ExtMon(\Sigma^\triangle)$ satisfying
	that 
    $\psi_L(\ext_{u, v}) = \psi_L(\ext_{u', v})$ that is a
	$\varphi_L(L)$-reaching idempotent such that
	$\Delta(u)$, $\Delta(u') > 0$, but
    $\length{u} \neq \length{u'}$.
    We exploit the fact that $\length{u} \neq \length{u'}$ to reduce 
	$\EQ=\bigset{w\in\set{0,1}^* \bigmid |w|_0=|w|_1}$ to
    $L_{\psi_L(\ext_{u, v})}$. 
	The constant-depth reduction works as follows on input
    $w \in \set{0, 1}^*$:
    \begin{enumerate}
	\item
	    Check if $\length{w} = 2 n$ for some $n \in \N$, reject if it is not
	    the case.
	\item
		Compute $w'=\alpha(w)$, where $\alpha\alpha\set{0,1}^*\rightarrow
		    \Sigma^*$ is the length-multiplying morphism
		    satisfying $\alpha(1)=u^{\length{u'}}$ and 
		    $\alpha(0)=u'^{\length{u}}$.
	\item
	    Accept whenever
	    $w' \# v^{n (\length{u} + \length{u'})} \in L_{\psi(\ext_{u, v})}$.
    \end{enumerate}
	Bearing in mind that $0<\Delta(u)=-\Delta(v)=\Delta(u')$, the latter forms 
	a valid reduction, because given a word $w \in \set{0, 1}^*$ of
    length $2 n$ for an $n \in \N$ that contains $k \in [0, 2 n]$ many
	$1$'s, for
    $w' \# v^{n (\length{u} + \length{u'})}$ to be in $L_{\psi_L(\ext_{u, v})}$,
    it is in particular required that $w' v^{n (\length{u} + \length{u'})}$ is
    well-matched, so it is necessary and sufficient that
	\[
	\begin{array}{lrll}
		&k \cdot \Delta(u) \cdot \length{u'} +
		(2 n - k) \cdot \Delta(u') \cdot \length{u} & =&
		-n \cdot \Delta(v) \cdot (\length{u} + \length{u'})\\
\Longleftrightarrow\qquad&	(k - n) \cdot \Delta(u) \cdot \length{u'} +
		(n - k) \cdot \Delta(u') \cdot \length{u} & = &0\\
		\Longleftrightarrow\qquad&
		(k - n) \cdot \Delta(u) \cdot (\length{u'} - \length{u}) & = &0\\
		\Longleftrightarrow\qquad&k & =& n \displaypunct{.}
    \end{array}
	\]
	Additionally applying Lemma~\ref{lemma reduction element} we
	obtain $\EQ \reduce L_{\psi_L(\ext_{u, v})} \reduce L$.
    Assume now there exist
    $\ext_{u, v}, \ext_{u, v'} \in \ExtMon(\Sigma^\triangle)$ satisfying
	that $\psi_L(\ext_{u, v}) = \psi_L(\ext_{u, v'})$ is an 
	$\varphi_L(L)$-reaching idempotent
    such that $\Delta(u) > 0$ but
    $\length{v} \neq \length{v'}$.
    Symmetrically, one can prove that we also have $\EQ \reduce L$ in this case.

	In conclusion, as $\EQ$ is $\TCO$-complete under 
	constant-depth reductions, it
    follows that $L$ is $\TCO$-hard under constant-depth reductions.
\end{proof}

The following proposition has essentially already been shown
in \cite[Proposition 135]{Ludwigthesis}, yet with some inaccuracies
 that we fix here.
\begin{prop}\label{prop notAC0}
	If $L$ is not quasi-aperiodic, then 
	one can effectively compute some $m\geq 2$ such that
	$\MOD_m\reduce L$.
\end{prop}
\begin{proof}
Since $L$ is not quasi-aperiodic,
	by Point 2 (a) Proposition~\ref{prop effectivity}
	one can effectively compute $k,l\in\N$ such that
	$\psi_L(\ExtMon(\Sigma^\triangle)^{k,l})$
	is not aperiodic. 
	Thus, one
	can compute $m\geq 2$ such that 
$\psi_L(\ExtMon(\Sigma^\triangle)^{k,l})$
	contains the additive group $G=([0,m-1],+,0)$ of $\mathbb{Z}/m\mathbb{Z}$ 
	for some prime number $m$.
	Moreover, there exist 
	$\ext_{u_0,v_0}, \ext_{u_1,v_1}\in\ExtMon(\Sigma^\triangle)^{k,l}$
	such that 
	$\psi_L(\ext_{u_0,v_0})=0_G$ 
	and $\psi_L(\ext_{u_1,v_1})=1_G$.
	Since $G$ is a group
	both $\psi_L(\ext_{u_0,v_0})$ and $\psi_L(\ext_{u_1,v_1})$ are 
	$\varphi_L(L)$-reaching.
	Moreover there exist
	$xy,z\in\Sigma^\triangle$ such that
	$xu_0zv_0y\in L$ if, and only if, $xu_1zv_1y\not\in L$.
	Let us assume without loss of generality that 
	$xu_0zv_0y\not\in L$ and $xu_1zv_1y\in L$
	(the case when $xu_0zv_0y\in L$ and $xu_1zv_1y\not\in L$
	can be proven analogously).
	Let $h_\uparrow,h_\downarrow\colon \set{0,1}^*\rightarrow\Sigma^*$ be the 
	length-multiplying morphisms satisfying $h_\uparrow(i)=u_i$ and
	$h_\downarrow(i)=v_i$ for all $i\in\set{0,1}$.
		We claim that
	\[w\in\text{MOD}_m\Longleftrightarrow
	\bigwedge_{i=1}^{m-1}xh_\uparrow(w)^{i^{m-2}} 
	zh_\downarrow(w^R)^{i^{m-2}}y\not\in L.
	\]
	Let $w_i=xh_\uparrow(w)^{i^{m-2}} 
	zh_\downarrow(w^R)^{i^{m-2}}y$
for all $i\in[1,m-1]$.
Observe that $w_i\in\Sigma^\triangle$ for
	all $i\in[1,m-1]$ directly by definition of 
	the morphisms $h_\uparrow$ and $h_\downarrow$.

To show the above equivalence, let us first assume that $|w|_1$ is 
	divisible by $m$.
	Then we have $\psi_L(\ext_{h^\uparrow(w),h^\downarrow(w^R)})=
	\psi_L(\ext_{u_0,v_0})=0_G$,
	and consequently
	$\psi_L(\ext_{h^\uparrow(w)^{i^{m-2}},h^\downarrow(w)^{i^{m-2}}})=0_G$
	for all $i\in[1,m-1]$. It follows $w_i\not\in L$
	for all $i\in[1,m-1]$, as desired.
	Conversely, assume that $|w|_1$ is not divisible by $m$,
	i.e. $|w|_1\equiv i\text{ mod }m$ for some $i\in[1,m-1]$.
	Hence $\psi_L(\ext_{h^\uparrow(w),h^\uparrow(w^R)})=i_G\not=0_G$
	and thus $\psi_L(\ext_{h^\uparrow(w)^{i^{m-2}},h^\uparrow(w^R)^{i^{m-2}}})=
	(i^{m-1}\text{ mod } m)_G=1_G$ by Fermat's Little Theorem.
	Hence $w_i\in L$ as required. 

	Altogether we obtain $\text{MOD}_m\reduce L$.
\end{proof}

\subsubsection{The non-solvable case}

In this additional section we prove a stronger lower bound, namely
when the syntactic morphism not only is not quasi-aperiodic
but the syntactic $\Ext$-algebra not solvable.
For this we revisit solvable groups and introduce solvable $\Ext$-algebras.

Let $G$ be a finite group. 
The {\em word problem for $G$} is the question, given a word $w_1\cdots w_n$ over $G$,
to decide if their product $w_1\cdots w_n$ in $G$ evaluates to $1_G$.
The {\em commutator} of $g,h\in G$ is $ghg^{-1}h^{-1}\in G$, denoted
by $\lbrack g,h\rbrack$.
The {\em commutator subgroup} $[G,G]$ of $G$ is the subgroup of $G$ that is generated by the commutators of $G$.
We say that $G$ is {\em perfect} if $G=\lbrack G,G\rbrack$.
We say that $G$ is {\em solvable} if in the series of commutator subgroups (a.k.a. derived series)
$G^{(0)},G^{(1)},\ldots$ a trivial group is contained,
where $G^{(0)}=G$ and $G^{(i+1)}=\lbrack G^{(i)},G^{(i)}\rbrack$ for all $i\in\N$.
Thus, note that any non-solvable finite group contains a perfect nontrivial subgroup.

We say the $\Ext$-algebra $(R,O)$ is {\em solvable} if 
all subsets of $R$ or $O$ that are groups
(under the multiplication of $R$, resp.\ of $O$) are solvable. 
We remark that not every 
subset $G\subseteq R$ (resp. $G\subseteq O$) that is a group is necessarily a 
submonoid of $R$ (resp. $O$);
in particular the neutral element of $G$ need not necessarily be the neutral 
element of $O$.
Indeed, for instance assume $R=\set{1,a,b}$ where $1\cdot r=r \cdot 1=r$ for all $r\in R$ and where
$a\cdot b=b\cdot a=b$ and $a\cdot a=b\cdot b=a$; the subset $\set{a,b}$ forms 
the additive group of $\Z/2\Z$ with neutral element $a$.
It is also worth mentioning that since $R$ is (isomorphic to) a submonoid of $O$ we could have equivalently defined an
$\Ext$-algebra to be solvable if all subsets of $O$ that are groups are solvable.
Let us mention additionally that one can prove that if 
$(\varphi,\psi)\colon (\Sigma^\triangle,\ExtMon(\Sigma^\triangle))\rightarrow(R,O)$
is quasi-aperiodic, then $(R,O)$ is solvable. 
In fact, one can prove that
if $(\varphi, \psi)$ is quasi-aperiodic, then $(R, O)$ must contain only Abelian
groups.

Our proof that $L$ is $\NCI$-hard (and thus $\TCO$-hard)
when $(R_L,O_L)$ is not solvable
can be reduced to the case for words~\cite{Barrington89}, by showing that already
$\psi_L(\ExtMon(\Sigma^\triangle)^{k,l})$
contains such a non-solvable group for some fixed $k,l\geq 0$.

\begin{prop}\label{prop notsolvable}
	If $(R_L,O_L)$ is not solvable,
	then $L$ is $\NCI$-hard and thus not in $\ACO$.
\end{prop}

\begin{proof}[Proof of Proposition~\ref{prop notsolvable}]
	Assume $(R_L,O_L)$ is not solvable.  
	Then there exists a subset $G\subseteq O_L$,
	where $G$ is a non-trivial perfect group, i.e. $G=\lbrack G,G\rbrack$.
	Let $\omega$ be the idempotent power of $G$.
	For all $g,h\in G$ there exist $\ext_{u_g,v_g},\ext_{u_h,v_h}\in\ExtMon(\Sigma^\triangle)$
	such that \[\lbrack g,h\rbrack=ghg^{-1}h^{-1}=ghg^{\omega-1}h^{\omega-1}=
	\psi_L\left(\ext_{u_gu_hu_g^{\omega-1}u_h^{\omega-1},v_h^{\omega-1}v_g^{\omega-1}v_hv_g}\right)\]
	and $1_G=g^\omega h^\omega=\psi_L(\ext_{u_g^\omega u_h^\omega,v_h^\omega v_g^\omega})$.
	Therefore, for all $g,h\in G$ we have
	\[
	    \lbrack g,h\rbrack=\psi_L\left(
	    \ext_{u_gu_hu_g^{\omega-1}u_h^{\omega-1},v_h^{\omega-1}v_g^{\omega-1}v_hv_g}\circ\
	    \mathop{\bigcirc}_{\substack{(g',h')\in G^2\\(g',h')\not=(g,h)}}
	    \ext_{u_{g'}^\omega u_{h'}^\omega,v_{h'}^\omega,
	    v_{g'}^\omega}\right)
	    \displaypunct{.}
	\]
	Hence, $\set{\lbrack g,h\rbrack \mid g,h\in G}\subseteq\ExtMon(\Sigma^\triangle)^{k,l}$
	for \[k=\sum_{(g,h)\in G^2}(|u_g|+|u_h|)\cdot\omega\quad\text{and}\quad
	l=\sum_{(g,h)\in G^2}(|v_h|+|v_g|)\cdot\omega\quad.\]
	Since $G=\lbrack G,G\rbrack$ every element of $G$ can be written as the product of at most $|G|$ elements
	in $\set{\lbrack g,h\rbrack\mid g,h\in G}$ and, in fact, even as the product of exactly
	$|G|$ elements in $\set{\lbrack g,h\rbrack\mid g,h\in G}$, since it contains the identity $1_G$.
	Thus, we can conclude that $G\subseteq\psi_L(\ExtMon(\Sigma^\triangle)^{k\cdot |G|,l\cdot|G|}$.
	Since the word problem of any non-solvable finite group is $\NCI$-hard by~\cite{Barrington89}
	and $G\subseteq\psi_L(\ExtMon(\Sigma^\triangle)^{k\cdot|G|,l\cdot|G|}$,
	it follows that the word problem for $G$ is constant-depth reducible to $L$.
	Hence $L$ is $\NCI$-hard and in particular $\TCO$-hard.
\end{proof}

\subsection{In \texorpdfstring{$\ACO$}{AC0}: Length-synchronous and
quasi-aperiodic}\label{Section AC0}

This section is devoted to the following theorem.

\begin{thm}\label{thm:Length-sync_and_quasi-aperiodic}
	If $L$ is length-synchronous and $(\varphi_L,\psi_L)$ is
    quasi-aperiodic, then $L$ is in $\FOplus$ and thus in $\ACO$.
\end{thm}
\noindent
For the rest of this section let us fix 
a VPL $L$, its syntactic $\Ext$-algebra $(R_L,O_L)$,
and its syntactic morphism $(\varphi_L,\psi_L)\colon(\Sigma^\triangle,\ExtMon(\Sigma^\triangle))\rightarrow(R_L,O_L)$.

Before we explain our proof strategy we introduce approximate matchings
and horizontal and vertical evaluation languages.
Approximate matchings generalize the classical matching
relation on well-matched words with respect to our VPL $L$ in the 
sense that they are subsets of the matching relation but
must equal the matching relation on all those
words that are in $L$.
Approximate matchings in the context of visibly pushdown languages
were introduced by Ludwig~\cite{Ludwigthesis}.
We then introduce suitably padded word languages mimicking
the evaluation problem of 
the monoid $R_L$ and the monoid $O_L$,
respectively.

\paragraph{Approximate matchings.}
For any word $w \in \Sigma^*$, we say that two positions
$i, j \in [1, \length{w}]$ in $w$ are \emph{matched} whenever $i < j$,
$w_i \in \Sigma_\call$, $w_j \in \Sigma_\return$ and
$w_{i + 1} \cdots w_{j - 1} \in \Sigma^\triangle$; we also say that $i$ is
{\em matched to} $j$ in $w$. 
Observe that a word $w$ over $\Sigma$ is well-matched if
and only if for each position $i \in [1, \length{w}]$,
\begin{itemize}
    \item
	if $i \in \Sigma_\call$, then there exists a position
	$j \in [1, \length{w}]$ such that $i$ is matched to $j$ in $w$;
    \item
	if $i \in \Sigma_\return$, then there exists a position
	$j \in [1, \length{w}]$ such that $j$ is matched to $i$ in $w$.
\end{itemize}
Given a word $w \in \Sigma^\triangle$, we denote by $\matching{w}$ its
\emph{matching relation} (or \emph{matching}), that is the relation
$
	\set{(i, j) \in [1, \length{w}]^2 \mid \text{$i$ is matched to $j$ in $w$}}.
$
An \emph{approximate matching relative to $L\subseteq\Sigma^\triangle$} is a
function $M\colon \Sigma^* \to {\N_{>0}}^2$ such that $M(w) = \matching{w}$ for
all $w \in L$ and $M(w) \subseteq \matching{w}$ for all
$w \in \Sigma^*\setminus L$.

\paragraph{Horizontal and vertical evaluation languages.}
For all $k \in \N$, we define
\[
    \ExtMon(\Sigma^\triangle)^{k, *} =
    \bigset{\ext_{u, v} \in \ExtMon(\Sigma^\triangle) \bigmid \length{u} = k}
    \quad\text{and}\quad
    \ExtMon(\Sigma^\triangle)^{*, k} =
    \bigset{\ext_{u, v} \in \ExtMon(\Sigma^\triangle) \bigmid \length{v} = k}
    \displaypunct{.}
\]
We also define
$\ExtMon(\Sigma^\triangle)_\uparrow =
 \set{\ext_{u, v} \in \ExtMon(\Sigma^\triangle) \mid \Delta(u) > 0}$
and finally for all $k \in \N$, we define
\[
    \ExtMon(\Sigma^\triangle)^{k, *}_\uparrow =
    \ExtMon(\Sigma^\triangle)^{k, *} \cap \ExtMon(\Sigma^\triangle)_\uparrow
    \quad\text{and}\quad
    \ExtMon(\Sigma^\triangle)^{*, k}_\uparrow =
    \ExtMon(\Sigma^\triangle)^{*, k} \cap \ExtMon(\Sigma^\triangle)_\uparrow
    \displaypunct{.}
\]

Consider the alphabets
$\Gamma_{\varphi_L} =
 \varphi_L(\Sigma^\triangle \setminus \set{\emptyword}) \cup \set{\$}$
and
$\Gamma_{\psi_L}=
 \psi_L\bigl(\ExtMon(\Sigma^\triangle)_\uparrow\bigr) \cup \set{\$}$
for a letter $\$\notin R_L \cup O_L$.
We also define 
\[
	\mathcal{V}_{\varphi_L} =
    \set{\$^k s \mid k \in \N, s \in \varphi_L(\Sigma^{k+1})}^*\ 
     \text{and}\ \mathcal{V}_{\psi_L} =
    \bigset{\$^k f \bigmid
	    k \in \N,
	    f \in
	    \psi_L\bigl(\ExtMon(\Sigma^\triangle)^{k + 1, *}_\uparrow\bigr)}^*
	    \ .
\]
Define the $\varphi_L$-evaluation morphism
$\eval_{\varphi_L}\colon \Gamma_{\varphi_L}^* \to R_L$ by
$\eval_{\varphi_L}(s) = s$ for all
$s \in \varphi_L(\Sigma^\triangle \setminus \set{\emptyword})$ and
$\eval_{\varphi_L}(\$) = 1_R$.
Similarly, define the $\psi_L$-evaluation morphism
$\eval_{\psi_L}\colon \Gamma_{\psi_L}^* \to O_L$ by
$\eval_{\psi_L}(f) = f$ for all
$f \in \psi_L\bigl(\ExtMon(\Sigma^\triangle)_\uparrow\bigr)$ and
$\eval_{\psi_L}(\$) = 1_{O_L}$.
Finally, for all $r \in R_L$, we set
\[
	\mathcal{E}_{\varphi_L, r} = \mathcal{V}_{\varphi_L} \cap 
	\eval_{\varphi_L}^{-1}(r)
\]
and for all $e \in O_L$, we set
\[
    \mathcal{E}_{\psi_L, e} =
    \mathcal{V}_{\psi_L} \cap \eval_{\psi_L}^{-1}(e)
    \displaypunct{.}
\]

\subsubsection{Strategy for the proof of
Theorem~\ref{thm:Length-sync_and_quasi-aperiodic}}
We are now ready to give the proof strategy for 
Theorem~\ref{thm:Length-sync_and_quasi-aperiodic}.
The proof consists of the following steps.
\begin{enumerate}
    \item\label{stp:Quasi-aperiodicity_verif_languages}
	Lemma~\ref{lemma quasi}: $\mathcal{V}_{\varphi_L}$ and
	$\mathcal{V}_{\psi_L}$ are regular languages whose syntactic morphisms
	are quasi-aperiodic.
    \item\label{stp:Quasi-aperiodicity_eval_languages}
	Proposition~\ref{ptn:Quasi-aperiodicity_eval_languages}:
	Let $L$ be a VPL whose  
	syntactic morphism $(\varphi_L, \psi_L)$ is quasi-aperiodic.
	\begin{itemize}
	    \item
		For all $r \in R_L$, the language $\mathcal{E}_{\varphi_L, r}$
		is regular and its syntactic morphism is quasi-aperiodic.
	    \item
		Let $e \in O_L$ such that $\newR_f = \set{(u,v)
		    \in\Con(\Sigma) \mid \psi_L(\ext_{u,v})=f}$ is
		    length-synchronous for each $\varphi_L(L)$-reaching
		    idempotent $f \in O_L$ satisfying $e = g \circ f \circ h$
		    with $g, h \in O_L$.  Then $\mathcal{E}_{\psi_L, e}$ is a
		regular language whose syntactic morphism is quasi-aperiodic.
	\end{itemize}
    \item\label{stp:Approximate_matching_length-sync}
	Proposition~\ref{ptn:Approximate_matching_length-sync}:
	If  $L \subseteq \Sigma^\triangle$ is length-synchronous, then there
	exists an $\FOplus[\Sigma]$-formula $\mu(x, y)$ such that
	$M\colon \Sigma^* \to \powerset{\N_{>0}^2}$ defined by
	$M(w) = \set{(i, j) \in [1, \length{w}]^2 \mid w \models \mu(i, j)}$
	for all $w \in \Sigma^*$ is an approximate matching relative to $L$.
    \item\label{stp:Eval_languages_and_bounded_nesting-depth}
	Proposition~\ref{ptn:Eval_languages_and_bounded_nesting-depth}:
	Assume a VPL $L$ has bounded nesting depth and
	\begin{itemize}
	    \item
		$\mathcal{E}_{\varphi_L, r}$ is a regular language whose
		syntactic morphism is quasi-aperiodic for all
		$\varphi_L(L)$-reaching $r \in R_L$, and
	    \item
		$\mathcal{E}_{\psi_L, e}$ is a regular language whose
		syntactic morphism is quasi-aperiodic for all
		$\varphi_L(L)$-reaching $e \in O_L$.
	\end{itemize}
	Then there exists an $\FOplus[\Sigma, \match]$-sentence $\eta$ such that
	for any approximate matching $M$ relative to $L$, we have $w \in L$ if,
	and only if, $(w, M(w)) \models \eta$ for all $w \in \Sigma^*$.
\end{enumerate}
\noindent
Let us argue that Points~\ref{stp:Quasi-aperiodicity_eval_languages},
\ref{stp:Approximate_matching_length-sync}
and~\ref{stp:Eval_languages_and_bounded_nesting-depth} indeed imply 
Theorem~\ref{thm:Length-sync_and_quasi-aperiodic}
(Point~\ref{stp:Quasi-aperiodicity_verif_languages} will be used in the proof of
Point~\ref{stp:Quasi-aperiodicity_eval_languages}).
Length-synchronicity of $L$ implies weak length-synchronicity of $L$ and thus
bounded nesting depth of $L$ (Proposition~\ref{ptn:Bounded_nesting-depth}).
Point~\ref{stp:Quasi-aperiodicity_eval_languages} implies the other assumptions
of Point~\ref{stp:Eval_languages_and_bounded_nesting-depth}:
length-synchronicity of $L$ means by definition that for each
$\varphi_L(L)$-reaching idempotent $f \in O_L$ we have that
$\newR_f = \set{(u,v) \in \Con(\Sigma) \mid\psi_L(\ext_{u,v})=f}$ 
is length-synchronous, so Point~\ref{stp:Quasi-aperiodicity_eval_languages}
implies that $\mathcal{E}_{\varphi_L,r}$ and $\mathcal{E}_{\psi_L,e}$ are
quasi-aperiodic for all $r\in R_L$ and all $e\in O_L$, respectively.
Finally, combining the $\FOplus[\Sigma, \match]$-sentence of
Point~\ref{stp:Eval_languages_and_bounded_nesting-depth} with the
$\FOplus[\Sigma]$-formula given by
Point~\ref{stp:Approximate_matching_length-sync} that defines an approximate
matching relative to $L$ yields an $\FOplus[\Sigma]$-sentence defining $L$, thus
proving Theorem~\ref{thm:Length-sync_and_quasi-aperiodic}.

\subsubsection{$\mathcal{V}_{\varphi_L}$ and $\mathcal{V}_{\psi_L}$ are quasi-aperiodic
(Proof of Point 1)}
Before proving Point 1 in the proof strategy for
Theorem~\ref{thm:Length-sync_and_quasi-aperiodic}
we require the following auxiliary lemma. 
It provides an important 
periodicity property of $\Ext$-algebra morphisms.
\begin{lem}\label{lem:Finite_Ext-algebra_morphisms_stability}
	The following periodicity holds:
    \begin{enumerate}
	\item\label{it:concatenation_stability}
There exist $t\in \N$ and $p\in \N_{>0}$ such that
	    $\varphi_L(\Sigma^\triangle \cap \Sigma^i) =
	     \varphi_L(\Sigma^\triangle \cap \Sigma^j)$
	    for all $i, j \in \N$ satisfying $i, j \geq t$ and
	    $i \equiv j \pmod{p}$.
	\item\label{it:extension_left-stability}
There exist $t\in \N$ and $p\in \N_{>0}$ such that
	    $\psi_L\bigl(\ExtMon(\Sigma^\triangle)^{i, *}_\uparrow\bigr) =
	     \psi_L\bigl(\ExtMon(\Sigma^\triangle)^{j, *}_\uparrow\bigr)$
	    for all $i, j \in \N$ satisfying $i, j \geq t$ and
	    $i \equiv j \pmod{p}$.
	\item\label{it:extension_right-stability}
There exist $t\in \N$ and $p\in \N_{>0}$ such that
	    $\psi_L\bigl(\ExtMon(\Sigma^\triangle)^{*, i}_\uparrow\bigr) =
	     \psi_L\bigl(\ExtMon(\Sigma^\triangle)^{*, j}_\uparrow\bigr)$
	    for all $i, j \in \N$ satisfying $i, j \geq t$ and
	    $i \equiv j \pmod{p}$.
    \end{enumerate}
\end{lem}
\begin{proof}
	To prove Point 1 recall that $\varphi_L^{-1}(r)$ is a VPL 
	and hence a context-free language for all $r\in R_L$.
	By Parikh's Theorem~\cite[Section 3]{EGKL11} 
	it follows that $S_r=\bigset{|w| \bigmid w\in\Sigma^\triangle, \varphi_L(w)=r}\subseteq\N$ is
	a semilinear set for all $r\in R_L$.
	It follows that 
	for all $U\subseteq R_L$ the set
	$S_U=\bigset{|w| \bigmid w\in\Sigma^\triangle, \varphi_L(w)\in U}\subseteq \N$
	is semilinear since semilinear sets are closed under union.
	Point 1 follows immediately from this observation.
	
	Next we prove Point 2, Point 3 can be proven analogously.
	According to Lemma~\ref{lemma Le} in Section~\ref{Para quasi}
	for $\#\not\in\Sigma$
	the language $L_e=\set{u\#v\mid uv\in\Sigma^\triangle :
	\psi_L(\ext_{u,v})=e}$ is a VPL for all $e\in O_L$.
	As the language $K=\set{u\#v\mid u,v\in\Sigma^\triangle}$ is obviously a VPL, 
	it follows that
	for all $e\in O_L$ 
	the language
	\[L_{e^\uparrow}=L_e\setminus K=\set{u\#v\mid uv\in\Sigma^\triangle :
	\psi_L(\ext_{u,v})=e, \Delta(u)>0}\subseteq L_e
	\] is a VPL as well.
	By Lemma~\ref{lemma semilinear} in Section~\ref{Para quasi} the set
	\[S_e=\set{(k,l)\in\N\times\N\mid
	     \exists u\in\Sigma ^k,
		     v\in^l : u \# v \in L_{e^\uparrow}}\]
is semilinear as well for all $e\in O_L$.
As a consequence we obtain that for all $Y\subseteq O_L$ the set
\[
	S_Y=\set{(k,l)\in\N\times\N\mid \exists u\in\Sigma^k,
	v\in\Sigma^l,e\in Y : u \# v \in L_{e^\uparrow}}\subseteq\N\times\N
\]
is semilinear as well since semilinear sets are closed under union.
	Since for all $Y\subseteq O_L$
	the set $\set{k\in\N\mid \psi_L(\ExtMon(\Sigma^\triangle)_\uparrow^{k,*}=Y}$
	is nothing but the projection of $S_Y$ onto the first component
	and semilinear sets are closed under projection, Point 2 follows.
\end{proof}

The following lemma holds irrespective of whether the syntactic morphism 
$(\varphi_L,\psi_L)$ of $L$ is quasi-aperiodic or not.

\begin{lem}\label{lemma quasi}
	$\mathcal{V}_{\varphi_L}$, $\mathcal{V}_{\psi_L}$ 
	are regular languages whose syntactic
    morphisms are quasi-aperiodic.
\end{lem}
\begin{proof}
    Take $t \in \N$ and $p \in \N_{>0}$ given by
    Lemma~\ref{lem:Finite_Ext-algebra_morphisms_stability} such that
    $\psi_L\bigl(\ExtMon(\Sigma^\triangle)^{i, *}_\uparrow\bigr) =
     \psi_L\bigl(\ExtMon(\Sigma^\triangle)^{j, *}_\uparrow\bigr)$
    for all $i, j \in \N$ satisfying $i, j \geq t$ and $i \equiv j \pmod{p}$.
    Define $\theta_{t, p}\colon \N \to \N$ as
    \[
\theta_{t, p}(n) =
\begin{cases}
   n & \text{if $n < t$}\\
   \min\set{n' \in \N \mid n' \geq t \wedge n' \equiv n \pmod{p}}
   & \text{otherwise}
\end{cases}
    \]
    for all $n \in \N$.
	Take $M$ to be the syntactic monoid of $\mathcal{V}_{\psi_L}$ and
	$h\colon \Gamma_{\psi_L}^* \to M$ to be its syntactic morphism.

    If there exists $f \in \psi_L\bigl(\ExtMon(\Sigma^\triangle)_\uparrow\bigr)$
    and $k \in \N$ such that $\$^k f \notin \mathcal{V}_{\psi_L}$,
    let us fix some $f_\bot$ and $k_\bot$ that satisfy this.
    Observe that for all $k \in \N$, we have that
    $h(\$^k) = h(\$^{\theta_{t, p}(k)})$. Further, for all
    $n \in \N_{>0}$, $k_1, \ldots, k_{n + 1} \in \N$ and
    $f_1, \ldots, f_n \in \psi_L\bigl(\ExtMon(\Sigma^\triangle)_\uparrow\bigr)$,
    we have
    $h(\$^{k_1} f_1 \cdots \$^{k_n} f_n \$^{k_{n + 1}}) =
     h(\$^{\theta_{t, p}(k_1)} f_1 \alpha
 \$^{\theta_{t, p}(k_{n + 1})})$,
    where
    \[
\alpha =
\begin{cases}
   \emptyword
   & \text{if
   $\$^{k_2} f_2 \cdots \$^{k_n} f_n \in
    \mathcal{V}_{\psi_L}$}\\
   \$^{k_\bot} f_\bot & \text{otherwise}
   \displaypunct{.}
\end{cases}
    \]
    Therefore, $M$ is finite and thus $\mathcal{V}_{\psi_L}$ is regular.
    Let $l \in \N_{>0}$ be the stability index of $h$
and take $q \in\N_{>0}$ such that $q\cdot l\geq t$ and $q \cdot l \equiv 0 \pmod{p}$. 
By definition, we
	have $h(\Gamma_{\psi_L}^l) = h(\Gamma_{\psi_L}^{q \cdot l})$. 
    Thus, to show that $h$ is quasi-aperiodic it is sufficient to prove that 
	for all $m \in h(\Gamma_{\psi_L}^{q \cdot l})$, we have $m^2 = m^3$.
	Indeed, given $m \in h(\Gamma_{\psi_L}^{q \cdot l})$, only the following
    three cases can occur.
    \begin{enumerate}
\item
   $m = h(\$^{q \cdot l})$.
   In this case, we have
   \[
m^2 = h(\$^{2 \cdot q \cdot l}) =
h(\$^{\theta_{t, p}(2 \cdot q \cdot l)}) =
h(\$^{\theta_{t, p}(q \cdot l)}) =
h(\$^{q \cdot l}) = m
\displaypunct{,}
   \]
   where the third equality follows from $\theta_{t,p}(2\cdot q\cdot l)=\theta_{t,p}(q\cdot l)$.
\item
   $m = h(\$^{k_1} f \$^{k_\bot} f_\bot \$^{k_2})$
   for $f \in \psi_L\bigl(\ExtMon(\Sigma^\triangle)_\uparrow\bigr)$ and
   $k_1, k_2 \in \N$ satisfying $\theta_{t, p}(k_1) = k_1$ and
   $\theta_{t, p}(k_2) = k_2$.
   In this case, we have
   \[
m^2 =
h(\$^{k_1} f \$^{k_\bot} f_\bot \$^{k_1 + k_2}
    f \$^{k_\bot} f_\bot \$^{k_2}) =
h(\$^{k_1} f \$^{k_\bot} f_\bot \$^{k_2}) = m
\displaypunct{,}
   \]
   where the second equality follows from
   $\$^{k_\bot} f_\bot \$^{k_1 + k_2} f
    \$^{k_\bot} f_\bot \notin \mathcal{V}_{\psi_L}$.
\item
   $m = h(\$^{k_1} f \$^{k_2})$
   for $f \in \psi_L\bigl(\ExtMon(\Sigma^\triangle)_\uparrow\bigr)$ and
   $k_1, k_2 \in \N$ satisfying $\theta_{t, p}(k_1) = k_1$ and
   $\theta_{t, p}(k_2) = k_2$.
   If
   $f \in
    \psi_L\bigr(\ExtMon(\Sigma^\triangle)^{k_1 + k_2 + 1, *}_\uparrow\bigl)$,
   then
   \[
m^2 =
h(\$^{k_1} f \$^{k_1 + k_2} f \$^{k_2}) =
h(\$^{k_1} f \$^{k_2}) = m
   \]
   because $\$^{k_1 + k_2} f \in \mathcal{V}_{\psi_L}$.
   Otherwise,
   $f \notin
    \psi_L\bigl(\ExtMon(\Sigma^\triangle)^{k_1 + k_2 + 1, *}_\uparrow\bigr)$
   and then
   \[
m^2 =
h(\$^{k_1} f \$^{k_1 + k_2} f \$^{k_2}) =
h(\$^{k_1} f \$^{k_\bot} f_\bot \$^{k_2})
   \]
   because $\$^{k_1 + k_2} f \notin \mathcal{V}_{\psi_L}$,
   so
   \[
m^3 =
h(\$^{k_1} f \$^{k_\bot} f_\bot
    \$^{k_1 + k_2} f \$^{k_2}) =
h(\$^{k_1} f \$^{k_\bot} f_\bot \$^{k_2}) =
m^2
   \]
   because
   $\$^{k_\bot} f_\bot \$^{k_1 + k_2} f \notin
    \mathcal{V}_{\psi_L}$.
    \end{enumerate}
    Therefore, $\mathcal{V}_{\psi_L}$ is a regular language whose
    syntactic morphism is quasi-aperiodic.
\end{proof}

\subsubsection{Quasi-aperiodicity of evaluation languages
$\mathcal{E}_{\varphi_L,r}$
and 
$\mathcal{E}_{\psi_L,e}$ (Proof of Point 2)}
One important consequence of Lemma~\ref{lemma quasi} is that for all $r \in R_L$
and $e \in O_L$, the languages $\mathcal{E}_{\varphi_L, r}$ and
$\mathcal{E}_{\psi_L, e}$ are in fact regular languages.
The following proposition gives conditions under which those languages have
moreover quasi-aperiodic syntactic morphisms when the syntactic morphism of $L$
itself is.

\begin{samepage}
\begin{prop}
\label{ptn:Quasi-aperiodicity_eval_languages}
    Let $L$ be a VPL whose syntactic morphism $(\varphi_L, \psi_L)$ is
    quasi-aperiodic.
    \begin{itemize}
	\item
	    For all $r \in R_L$, the language $\mathcal{E}_{\varphi_L, r}$ is
	    regular and its syntactic morphism is quasi-aperiodic.
	\item
	    Let $e \in O_L$ such that $\newR_f = \set{(u,v)
	    \in\Con(\Sigma) \mid \psi_L(\ext_{u,v})=f}$ is
	    length-synchronous for each $\varphi_L(L)$-reaching idempotent
	    $f \in O_L$ satisfying $e = g \circ f \circ h$ with $g, h \in
	    O_L$\footnote{In the terminology of Green's relations, to be
	    introduced in the proof of Proposition~\ref{prop SM lowerbound},
	    this is equivalent to the fact that $e \GreenJleq f$.}. Then
	    $\mathcal{E}_{\psi_L, e}$ is a regular language whose syntactic
	    morphism is quasi-aperiodic.
    \end{itemize}
\end{prop}
\end{samepage}
\begin{proof}
    We already know that for all $r \in R_L$ and $e \in O_L$, the languages
    $\mathcal{E}_{\varphi_L, r}$ and $\mathcal{E}_{\psi_L, e}$ are
    regular. To prove the lemma, we then just have to prove that
    \begin{itemize}
	\item
	    if there exists $r \in R_L$ such that the syntactic morphism of
	    $\mathcal{E}_{\varphi_L, r}$ is not quasi-aperiodic, then there
	    exists $k \in \N$ such that
	    $\varphi(\Sigma^\triangle \cap \Sigma^k)$ contains a semigroup
	    that is not aperiodic;
	\item
	    if there exists $e \in O_L$ such that the syntactic morphism of
	    $\mathcal{E}_{\psi_L, e}$ is not quasi-aperiodic and for each
	    $\varphi_L(L)$-reaching idempotent $f \in O_L$ such that there
	    exist $f', f'' \in O_L$ satisfying $e = f' \circ f \circ f''$ we
	    have that
	    $\newR_f =
	     \set{(u,v) \in \Sigma^* \times \Sigma^* \mid
		  uv\in\Sigma^\triangle, \Delta(u)>0, \psi_L(\ext_{u,v})=f}$ 
	    is length-synchronous, then there exist $k, l \in \N$ such that
	    $\psi_L(\ExtMon(\Sigma^\triangle)^{k, l})$ contains a semigroup that
	    is not aperiodic.
    \end{itemize}
    Indeed, the first point allows to conclude that $(\varphi_L, \psi_L)$ is not
    quasi-aperiodic, since if there exists a non-aperiodic semigroup $S$
    contained in $\varphi_L(\Sigma^\triangle \cap \Sigma^k)$, then
    $\set{\leftmult{s} \mid s \in S}$ is a semigroup contained in
    $\psi_L(\ExtMon(\Sigma^\triangle)^{k, 0})$ (because for each $s \in S$,
    there exists $w \in \Sigma^\triangle \cap \Sigma^k$ satisfying
    $\varphi_L(w) = s$, so that
    $\psi_L(\ext_{w, \emptyword}) = \leftmult{\varphi_L(w)} = \leftmult{s}$).
    But this semigroup is non-aperiodic as well, since as $S$ is non-aperiodic,
    it must be that for all $i \in \N_{>0}$, there exists $s \in S$ such that
    $s^i \neq s^{i + 1}$, so that
    ${\leftmult{s}}^i \neq {\leftmult{s}}^{i + 1}$.

    We only prove the second point, the first point can be proved in a similar
    way by leaving out the last paragraph of the following proof, that is the
    sole place where we need, given an $e \in O_L$, length-synchronicity of
    $\newR_f$ for each $\varphi_L(L)$-reaching idempotent $f \in O_L$
    such that there exist $f', f'' \in O_L$ satisfying
    $e = f' \circ f \circ f''$.

    Take $t \in \N$ and $p \in \N_{>0}$ given by
    Lemma~\ref{lem:Finite_Ext-algebra_morphisms_stability} such that
    $\psi_L\bigl(\ExtMon(\Sigma^\triangle)^{i, *}_\uparrow\bigr) =
     \psi_L\bigl(\ExtMon(\Sigma^\triangle)^{j, *}_\uparrow\bigr)$
    for all $i, j \in \N$ satisfying $i, j \geq t$ and $i \equiv j \pmod{p}$.

    Assume there exists $e \in O_L$ such that the syntactic morphism of
    $\mathcal{E}_{\psi_L, e}$ is not quasi-aperiodic.
    Take $M$ to be the syntactic monoid of $\mathcal{E}_{\psi_L, e}$
    and $h\colon \Gamma_{\psi_L}^* \to M$ to be its syntactic morphism.
    Let $s \in \N_{>0}$ be the stability index of $h$ and let $\omega \geq 2$
    be a multiple both of the idempotent power of $M$ and the idempotent power
    of $O_L$. 
    Non-quasi-aperiodicity of $h$ implies that there exists
    $g \in h(\Gamma_{\psi_L}^s)$ satisfying $g^\omega \neq g^{\omega + 1}$.

    By definition of the stability index, there exists
	$w \in \Gamma_{\psi_L}^{q \cdot s}$ for $q \in \N_{>0}$ such that
    $q \cdot s \geq t$ and $q \cdot s \equiv 0 \pmod{p}$ satisfying
    $h(w) = g$. Since
    $t \leq q \cdot s \leq q \cdot s \cdot \omega \leq
     q \cdot s \cdot (\omega + 1)$
    and
    $q \cdot s \equiv q \cdot s \cdot \omega \equiv q \cdot s \cdot (\omega + 1)
     \pmod{p}$,
    we cannot have $w = \$^{q \cdot s}$, for otherwise we would have
    $g^\omega = h(\$^{q \cdot s \cdot \omega}) =
     h(\$^{q \cdot s \cdot (\omega + 1)}) = g^{\omega + 1}$
    because
    $\psi_L\bigl(
	\ExtMon(\Sigma^\triangle)^{q \cdot s \cdot \omega + k + 1, *}_\uparrow
     \bigr) =
     \psi_L\bigl(
	\ExtMon(\Sigma^\triangle)^{q \cdot s \cdot (\omega + 1) + k + 1,
				   *}_\uparrow
     \bigr)$
    for all $k \in \N$.
    Therefore, we have
    $w = \$^{k_1} f_1 \cdots \$^{k_n} f_n \$^{k_{n + 1}}$ for
    $n \in \N_{>0}$, $k_1, \ldots, k_{n + 1} \in \N$ and
    $f_1, \ldots, f_n \in \psi_L\bigl(\ExtMon(\Sigma^\triangle)_\uparrow\bigr)$.
    Since $g^\omega \neq g^{\omega + 1}$, there exist $x, y \in \Gamma_{\psi_L}^*$
    such that either $x w^\omega y \in \mathcal{E}_{\psi_L, e}$ and
    $x w^{\omega + 1} y \notin \mathcal{E}_{\psi_L, e}$, or
    $x w^\omega y \notin \mathcal{E}_{\psi_L,  e}$ and
    $x w^{\omega + 1} y \in \mathcal{E}_{\psi_L, e}$.
    Assume the first case holds. Then we have $x = x' \$^{k_x}$ and
    $y = \$^{k_y} y'$ with $k_x, k_y \in \N$ and $x', y' \in 
    \Gamma_{\psi_L}^*$
    satisfying
    $x', \$^{k_x + k_1} f_1, \$^{k_2} f_2, \ldots,
     \$^{k_n} f_n, \$^{k_{n + 1} + k_1} f_1,
     \$^{k_{n + 1} + k_y} y' \in \mathcal{V}_{\psi_L}$
    and
    $\eval_{\psi_L}(x')\circ (f_1 \circ \cdots \circ f_n)^\omega \circ
     \eval_{\psi_L}(y') = e$.
    Therefore, we also have
    $x w^{\omega + 1} y \in \mathcal{V}_{\psi_L}$, hence since
    $x w^{\omega + 1} y \notin \mathcal{E}_{\psi_L, e}$ we necessarily
    have
    \begin{align*}
	e
	& = \eval_{\psi_L}(x') \circ
	    (f_1 \circ \cdots \circ f_n)^\omega \circ
	    \eval_{\psi_L}(y')\\
	& \neq \eval_{\psi_L}(x w^{\omega + 1} y) =
	       \eval_{\psi_L}(x') \circ
	       (f_1 \circ \cdots \circ f_n)^{\omega + 1} \circ
	       \eval_{\psi_L}(y')
	\displaypunct{.}
    \end{align*}
    Thus we have
    $(f_1 \circ \cdots \circ f_n)^\omega \neq
     (f_1 \circ \cdots \circ f_n)^{\omega + 1}$
    and
    $\$^{k_{n + 1} + k_1} f_1 \$^{k_2} f_2 \cdots \$^{k_n} f_n
     \in \mathcal{V}_{\psi_L}$.
    This is also true for the case when
    $x w^\omega y \notin \mathcal{E}_{\psi_L, e}$ and
    $x w^{\omega + 1} y \in \mathcal{E}_{\psi_L, e}$.

    Therefore, we have
    $(f_1 \circ \cdots \circ f_n)^\omega \neq
     (f_1 \circ \cdots \circ f_n)^{\omega + 1}$
    with 
    \[(f_1 \circ \cdots \circ f_n)^i \in
     \psi_L\bigl(\ExtMon(\Sigma^\triangle)^{q \cdot s \cdot i, *}_\uparrow\bigr) =
     \psi_L\bigl(\ExtMon(\Sigma^\triangle)^{q \cdot s, *}_\uparrow\bigr)\]
    for each $i \in \N_{>0}$ because
    $k_{n + 1} + k_1 + \cdots + k_n + n = q \cdot s \geq t$ and
    $k_{n + 1} + k_1 + \cdots + k_n + n = q \cdot s \equiv 0 \pmod{p}$.
    But given $\omega'$ the idempotent power of
    $\set{(f_1 \circ \cdots \circ f_n)^i \mid i \in \N_{>0}}$, we have that
    $(f_1 \circ \cdots \circ f_n)^\omega =
     (f_1 \circ \cdots \circ f_n)^{\omega \omega'} =
     (f_1 \circ \cdots \circ f_n)^{\omega'}$,
    so that
    $(f_1 \circ \cdots \circ f_n)^{\omega'} \neq
     (f_1 \circ \cdots \circ f_n)^{\omega' + 1}$,
    hence $\set{(f_1 \circ \cdots \circ f_n)^i \mid i \in \N_{>0}}$ is not
    aperiodic.

    Assume additionally that for each $\varphi_L(L)$-reaching idempotent
    $f \in O_L$ such that there exist $f', f'' \in O_L$ satisfying
    $e = f' \circ f \circ f''$ we have that
    $\newR_f =
     \set{(u,v) \in \Sigma^* \times \Sigma^* \mid
	  uv\in\Sigma^\triangle, \Delta(u)>0, \psi_L(\ext_{u,v})=f}$ 
    is length-synchronous.
    For each $i \in \N_{>0}$, let
    $\ext_{u_i, v_i} \in
     \psi_L\bigl(\ExtMon(\Sigma^\triangle)^{q \cdot s, *}_\uparrow\bigr)$
    such that $\psi_L(\ext_{u_i, v_i}) = (f_1 \circ \cdots \circ f_n)^i$.
    If $(f_1 \circ \cdots \circ f_n)^\omega$ were 
    not $\varphi_L(L)$-reaching, then it
    would imply that $\ext_{u_\omega, v_\omega}$ is not $L$-reaching. 
    This would in turn
    entail that for all $z \in \Sigma^\triangle$ and
    $\ext_{\alpha, \beta} \in \ExtMon(\Sigma^\triangle)$ we have
    \[
	\ext_{\alpha, \beta}(\ext_{u_\omega, v_\omega}(z)) \notin L \wedge
	\ext_{\alpha, \beta}(\ext_{u_1 u_\omega, v_\omega v_1}(z)) =
	\ext_{\alpha u_1, v_1 \beta}(\ext_{u_\omega, v_\omega}(z)) \notin L
	\displaypunct{,}
    \]
    so that it would follow that
    \[(f_1 \circ \cdots \circ f_n)^\omega = \psi_L(\ext_{u_\omega, v_\omega}) =
     \psi_L(\ext_{u_1 u_\omega, v_\omega v_1}) =
     (f_1 \circ \cdots \circ f_n)^{\omega + 1}\displaypunct{,}\]
    a contradiction.
    Hence, since $(f_1 \circ \cdots \circ f_n)^\omega$ is a
    $\varphi_L(L)$-reaching idempotent and
    \[
	e =
	\begin{cases}
	    \eval_{\psi_L}(x') \circ (f_1 \circ \cdots \circ f_n)^\omega \circ
	    \eval_{\psi_L}(y') &
		\text{if $x w^\omega y \in \mathcal{E}_{\psi_L, e}$ and
		      $x w^{\omega + 1} y \notin \mathcal{E}_{\psi_L, e}$}\\
	    \eval_{\psi_L}(x') \circ
	    (f_1 \circ \cdots \circ f_n)^{\omega + 1} \circ \eval_{\psi_L}(y') &
		\text{o/w ($x w^\omega y \notin \mathcal{E}_{\psi_L,  e}$
		      and $x w^{\omega + 1} y \in \mathcal{E}_{\psi_L, e}$)}
	\end{cases}
	\!\text{,}
    \]
    it follows that $\newR_{(f_1 \circ \cdots \circ f_n)^\omega}$ is
    length-synchronous. So for all $i \in \N, i \geq 2$, since
    $\psi_L(\ext_{u_1^\omega, v_1^\omega}) =
     \psi_L(\ext_{u_i^\omega, v_i^\omega}) = 
     (f_1 \circ \cdots \circ f_n)^\omega$
    with $\Delta(u_1^\omega) > 0$ and $\Delta(u_i^\omega) > 0$, since
    $\length{u_1} = \length{u_i}$, we have
    \[
	\frac{\length{u_1^\omega}}{\length{v_1^\omega}} =
	\frac{\length{u_i^\omega}}{\length{v_i^\omega}}
	\Rightarrow
	\frac{\length{u_1}}{\length{v_1}} =
	\frac{\length{u_i}}{\length{v_i}}
	\Rightarrow
	\length{v_1} = \length{v_i}
	\displaypunct{.}
    \]
    To conclude, we obtain that the non-aperiodic semigroup
    $\set{(f_1 \circ \cdots \circ f_n)^i \mid i \in \N_{>0}}$ is contained in
    $\psi_L(\ExtMon(\Sigma^\triangle)^{q \cdot s, \length{v_1}})$.
\end{proof}

The following remark states that the length-synchronicity 
precondition in the second point of 
Proposition~\ref{ptn:Quasi-aperiodicity_eval_languages}
is important. In fact it shows that weak length-synchronicity
is not sufficient.
\begin{samepage}
\begin{rem}\label{remark not weakly}
    For the second point of
    Proposition~\ref{ptn:Quasi-aperiodicity_eval_languages} it is generally not
    sufficient to assume, given $e$, that for each $\varphi_L(L)$-reaching
    idempotent $f \in O_L$ such that there exist $g, h \in O_L$ satisfying
    $e = g \circ f \circ h$ we have that
    $\newR_f =
     \set{(u,v) \in \Sigma^* \times \Sigma^* \mid
	  uv\in\Sigma^\triangle, \Delta(u)>0, \psi_L(\ext_{u,v})=f}$ 
    is weakly length-synchronous.
    Indeed, the VPL $K$ generated by the grammar with rules
    \begin{align*}
	S & \rightarrow a S b_1 \mid ac T b_2 \mid \emptyword\\
	T & \rightarrow a T b_1 \mid ac S b_2
	\displaypunct{.}
    \end{align*}
    using $S$ as start symbol is not length-synchronous, but weakly
    length-synchronous, and has a quasi-aperiodic syntactic morphism. However,
    for the syntactic $\Ext$-algebra $(R_K, O_K)$ and the syntactic morphism
    $(\varphi_K, \psi_K)$ of $K$, we claim that there exists a
    $\varphi_K(K)$-reaching $e \in O_K$ such that $\mathcal{E}_{\psi_K, e}$ is
    a regular language whose syntactic morphism is not quasi-aperiodic while, as
    $K$ is weakly length-synchronous, for each $\varphi_L(L)$-reaching
    idempotent $f \in O_L$ we have that
    $\newR_f =
     \set{(u,v) \in \Sigma^* \times \Sigma^* \mid
	  uv\in\Sigma^\triangle, \Delta(u)>0, \psi_L(\ext_{u,v})=f}$ 
    is weakly length-synchronous.

    Let $\Gamma$ be the visibly pushdown alphabet of $K$.
    Note that we have $K\subset \SM_{1,2}$, where
    $\SM_{1,2}=L(S\rightarrow aSb_1|acSb_2|\varepsilon)$ is the VPL initially
    introduced in Example~\ref{example SM}.
    For all $uv, u'v' \in \SM_{1,2}$ with $u, u' \in \set{a, c}^+$,
    $v, v' \in \set{b_1, b_2}^+$, $\length{u}_c \equiv \length{u'}_c \pmod{2}$
    we have $x u z v y \in K \Leftrightarrow x u' z v' y \in K$ for all
    $x y, z \in \Gamma^\triangle$.
    This implies that if we set $e_0 = \psi_K(\ext_{a, b_1})$ and
    $e_1 = \psi_K(\ext_{a c, b_2})$, we have that for all $u v \in \SM_{1, 2}$
    with $u \in \set{a, c}^+$, $v \in \set{b_1, b_2}^+$, it holds that
    $\psi_K(\ext_{u, v}) = e_{\length{u}_c\!\! \mod 2}$. Therefore, while
    $e_0 \neq e_1$, we have $e_0 \circ e_1 = e_1 \circ e_0 = e_1$ and
    $e_0 \circ e_0 = e_1 \circ e_1 = e_0$.

    Consider the length-multiplying monoid morphism
    $\beta\colon \set{0, 1}^* \to \Gamma_{\psi_K}^*$ such that
    $\beta(0) = e_0 e_0$ and $\beta(1) = \$ e_1$. Then
    $\MOD_2 = \beta^{-1}(\mathcal{E}_{\psi_K, e_0})$, so
    $\mathcal{E}_{\psi_K, e_0}$ cannot have a quasi-aperiodic syntactic
    morphism, for otherwise, by closure of the class of regular languages whose
    syntactic morphism is quasi-aperiodic under inverses of length-multiplying
    morphisms (see~\cite{Straubing-2002}), we would have that $\MOD_2$ has a
    quasi-aperiodic syntactic morphism.
\end{rem}
\end{samepage}

\subsubsection{Approximate matching relation in $\FOplus$ (Proof of Point 3)}
\label{Section approximate}

The following proposition states that 
there is a $\FOplus[\Sigma]$-definable approximate matching
relative to any length-synchronous visibly pushdown language.
\begin{prop}\label{ptn:Approximate_matching_length-sync}
 If  $L \subseteq \Sigma^\triangle$ is length-synchronous, then
    there exists an $\FOplus[\Sigma]$-formula $\eta(x, y)$ such that
	$M\colon \Sigma^* \to \powerset{\N_{>0}^2}$ defined by
    $
	M(w) = \set{(i, j) \in [1, \length{w}]^2 \mid w \models \eta(i, j)}
    $
    for all $w \in \Sigma^*$ is an approximate matching relative to $L$.
\end{prop}

The technical heart of the proof is the following lemma
whose proof is postponed and 
will take most part of this subsubsection.
This lemma realizes the characterization
of length-synchronicity given by 
Proposition~\ref{ptn:Length-synchronicity_equivalences}
via an $\FOplus[\Sigma]$-formula.
\begin{samepage}
    \begin{lem}
    \label{lem:Synchronous_element_matching}
	    Assume that $(\varphi_L,\psi_L)$ is 
	    weakly length-synchronous.
	    Let $e \in O_L$ be $\varphi_L(L)$-reaching
	    and assume that 
	    $\newU_e=\set{(u,v) \in\Con(\Sigma)\mid  
	    e\circ\psi_L(\ext_{u,v})=e}$ 
	    is length-synchronous.
	Then there exists an $\FOplus[\Sigma]$-formula $\pi_e(x, x', y', y)$ such
	that for all $w \in \Sigma^+$ and
	$i, i', j', j \in [1, \length{w}], i \leq i' < j' \leq j$
	    the following holds,
	\begin{itemize}
	    \item
		if $w\models \pi_e(i, i', j', j)$, then
		$w_i \cdots w_{i'} w_{j'} \cdots w_j \in \Sigma^\triangle$
			and 
	    \item
		if $w_i \cdots w_{i'} w_{j'} \cdots w_j \in \Sigma^\triangle$
			and $(w_i\dots w_{i'},w_{j'}\dots w_j)\in
			\newU_e$, 
		then $w\models \pi_e(i, i', j', j)$.
	\end{itemize}
    \end{lem}
\end{samepage}

\paragraph{Building an approximate matching assuming predicates $\pi_e$.}
\label{paragraph approximate}
Let us prove Proposition~\ref{ptn:Approximate_matching_length-sync}
by making use of Lemma~\ref{lem:Synchronous_element_matching}.
\begin{proof}[Proof of Proposition~\ref{ptn:Approximate_matching_length-sync}.]
By assumption $(\varphi_L, \psi_L)$ is
    $\varphi_L(L)$-length-synchronous.
	Thus, the set of contexts 
	$\newU_e$ is length-synchronous
	for all $\varphi_L(L)$-reaching $e\in O_L$
	by Proposition~\ref{prop length synchronous}.
    Moreover, there exists $d_L \in \N$
    bounding the nesting depth of the words in $L$
Proposition~\ref{ptn:Bounded_nesting-depth}.
	For defining our desired formula $\mu$, we will construct 
$\FOplus[\Sigma]$
	formulas $\mu_d$ and $\mu_d^\uparrow$ for all $0\leq d\leq d_L$
	with the following properties:
 for all $w \in \Sigma^+$ and for all 
    $i, j \in [1, \length{w}]$, we have
	\begin{samepage}
    \begin{itemize}
	    \item if $w\models \mu_d^\uparrow(i,j)$ or 
		    $w\models\mu_d(i,j)$,
		    then $w_i \cdots w_j \in \Sigma^\triangle$,
	    \item if $w\in L$, $w_i\dots w_j\in\Sigma^\triangle$,
		    $\ndepth(w_i\dots w_j)\leq d$ and $i$ is matched to $j$ in $w$, then
		    $w\models \mu_d^\uparrow(i,j)$, and  
	    \item if $w\in L$, $\ndepth(w_i\dots w_j)\leq d$ and 
		    $w_i\dots w_j\in\Sigma^\triangle$,
		    then $w\models\mu_d(i,j)$.
    \end{itemize}
	\end{samepage}
	We therefore define $\mu=\mu_{d_L}$.
	The construction of $\mu_d^\uparrow$ and $\mu_d$ is 
	by induction on $d$.
	We set
	\[
	\mu_0(i,j)=\bot
	\text{ and }
	\mu_0(i,j)=\forall z \bigl(x \leq z \leq y \rightarrow \Sigma_\internal(z)\bigr).
	\]
	Let us assume $d>0$. The formula $\mu_d$ is easily
	defined assuming $\mu_d^\uparrow$. We define
	\begin{eqnarray*}
		\mu_d(x,y)&\!\!=\!\!&\forall z\biggl[x\leq z\leq y\rightarrow\\
		&&\quad\Bigl(\Sigma_\internal(z)\vee\exists 
		z'\bigl(
		    \begin{aligned}[t]
			& (\Sigma_\call(z)\wedge\Sigma_\return(z')\wedge
			   \mu_d^\uparrow(z,z'))\vee\\
			& (\Sigma_\call(z')\wedge\Sigma_\return(z)\wedge
			   \mu_d^\uparrow(z',z))
			\bigr)\Bigr)\biggl]
			\displaypunct{.}
		    \end{aligned}
	\end{eqnarray*}
	It remains to define $\mu_d^\uparrow$.
	Let us assume $u=w_i\dots w_j\in\Sigma^\triangle$, 
	that $i$ is matched to $j$ in $w$ and that $\ndepth(u)=d>0$. 
Hence, $u=a_1vb_1\in\Sigma^\triangle$ for some $a_1\in\Sigma_\call$,
	$b_1\in\Sigma_\return$, and $v\in\Sigma^\triangle$.
	We then apply Lemma~\ref{L nesting induction} which states
	that $u$ has a nesting-maximal
	stair factorization 
	$u=\ext_{x_1,y_1}\circ\ext_{a_1,b_1}\circ\dots
	\ext_{x_{k},y_{k}}\circ\ext_{a_{k},b_{k}}(u')$
	such that for some $h\in[1,k]$,
	setting $u_\ell=\ext_{a_\ell,b_\ell}\circ\ext_{x_{\ell+1},y_{\ell+1}}
			\circ\cdots\circ\
			\ext_{a_k,b_k}(u')$
	for all $\ell \in [1, k]$ and $u_{k + 1} = u'$, we have
	\begin{enumerate}
		\item $\ndepth(u)=\ndepth(u_h)=d$,
		\item $\ndepth(u_{h+1})=d-1$,
					and 
		\item $\ndepth(x_1),\ndepth(y_1),\dots,\ndepth(x_h),\ndepth(y_h)<d$.
	\end{enumerate}
	We remark that $x_1=y_1=\varepsilon$. 
Let $i=i_1<\dots<i_h$ and $j_h<\dots<j_1=j$ be the positions
that correspond to the positions of the letters $a_1,\dots,a_h\in\Sigma_\call$ and 
$b_1,\dots, b_h\in\Sigma_\return$ of $u$ in $w$, respectively:
more precisely
$i_\ell=i+\length{x_1\cdots a_{\ell-1}x_\ell}$
	and $j_\ell=\length{x_1a_1\cdots x_ka_ku'b_ky_k\cdots b_{\ell+1}y_{\ell+1}}+1$
	for all $\ell\in[1,h]$.
	The formula $\mu_{d}^\uparrow$
	could guess the positions
	$i=i_1<\dots<i_h$ and $j_h<\dots<j_1=j$
	and verify the following (recalling that $x_1=y_1=\varepsilon$):
	\begin{enumerate}[(a)]
	\item the infix $w_{i_h+1}\cdots w_{j_h-1}=\ext_{x_{h+1},y_{h+1}}(u_{h+1})$ 
	is well-matched, and 
		\item the word $w_{i_1}\cdots w_{i_{h}} w_{j_h}\cdots 
			w_{j_1}$ is well-matched.
				\end{enumerate}
	Point (a) can be realized via the formula $\mu_{d-1}$ by making 
	use of Point 2 from above, whereas
	Point (b) can be realized by the following ad-hoc formula, 
	this
	time making use of Point 3 and from above:
	\[
		\kappa_h(x,x',y',y)=\exists x_1\cdots x_h\exists y_1\cdots y_h
		\biggl(
		\begin{aligned}[t]
		    & x=x_1\wedge x'=x_h\wedge y'=y_h\wedge y=y_h\wedge\\
		    & \bigwedge_{t=1}^h \bigl(\Sigma_\call(x_t)\wedge
					      \Sigma_\return(y_t)\bigr)\wedge\\
		    & \bigwedge_{t=1}^{h-1} \bigl(\mu_{d-1}(x_t+1,x_{t+1}-1)\wedge
						  \mu_{d-1}(y_{t+1}+1,y_t-1)\bigr)
		    \biggr)
		    \text{.}
		\end{aligned}
	\]

	The problem with this approach is that 
	the size of the formula depends on the size of $w$.
	For instance, for $a\in\Sigma_\call$, $b\in\Sigma_\return$,
	$c\in\Sigma_\internal$, and $u=a^ncb^n$ we have
	$\ndepth(u)=\ndepth(acb)=1$ for all $n\geq 1$.
	Hence we would have $h=n-1$, so $h$ would depend on $u$
	which is problematic. 
	Therefore, towards expressing 
	Point (b) by 
a formula whose size only depends on $|O_L|$, let us define,
	for all $\ell,\ell'\in[1,h]$, the product
	\[e_{\ell,\ell'}=\psi_L(\ext_{x_\ell,y_\ell}\circ
	\ext_{a_{\ell},b_{\ell}}\cdots\ext_{x_{\ell'},y_{\ell'}}
	\circ\ext_{a_{\ell'},b_{\ell'}})\quad\text{and}\quad
	e_\ell=e_{1,\ell}.\]
We remark that all $e_{\ell,\ell'}$ are $\varphi_L(L)$-reaching
since $w$ is assumed to be in $L$.
	We say an interval
	$I=[s,t]\subseteq[1,h]$ is {\em repetitive} if $s<t$ and $e_s=e_t$.

\medskip 
	\noindent
\begin{clm}\label{claim pigeon}
	There exist indices $1=t_0\leq s_1<t_1<s_2<t_2<\cdots
	<s_q<t_q\leq s_{q+1}=h$ such that $[s_1,t_1],\dots, [s_q,t_q]$ are
	all repetitive and for
	$D_0=[t_0,s_1],D_1=[t_1,s_2],\dots,D_q=[t_q,s_{q+1}]$ we have
	$q+\sum_{p=0}^q|D_p|\leq 3\card{O_L}$.
\end{clm}

\medskip 
\noindent
	{\em Proof of the Claim. }
For all $z\in[1,h]$ let
$\lambda(z)=\max\set{\ell\in[1,h]\mid e_\ell=e_z}$.
Observe $\lambda(z)\geq z$ for all $z\in[1,h]$ and
that $|\lambda([1,h])|\leq|O_L|$.
We define $t_0=1$. 
Let $p>0$ and assume that we have already defined $t_{p-1}$.
In case $t_{p-1}=h$ we are done and define $q=p-1$ and $s_{q+1}=h$.
So let us assume $t_{p-1}<h$.
In case there exists $z\in[t_{p-1},h]$ such that $z<\lambda(z)$
we define $s_p=\min\set{z\in[t_{p-1},h]\mid z<\lambda(z)}$
and $t_p=\lambda(s_p)$,
otherwise (i.e.\ in case $z=\lambda(z)$ for all $z\in[t_{p-1},h]$)
we are done and define $q=p-1$ and $s_{q+1}=h$.
Immediately by definition we have
$1=t_0\leq s_1<t_1<s_2<t_2<\dots<s_q<t_q\leq s_{q+1}=h$ (because if we had
$t_{p-1} = s_p$ for a $p \in [2, q]$, we would have
$e_{s_{p-1}} = e_{t_{p-1}} = e_{s_p}$,
so $\lambda(s_p) = \lambda(s_{p-1}) = t_{p-1} = s_p < \lambda(s_p)$, a
contradiction) and $e_{s_p}=e_{t_p}$ for all $p \in [1, q]$.
Moreover, the intervals $[s_1,t_1],\dots,[s_q,t_q]$ are indeed all repetitive.
Since moreover $t_p\in\lambda([1,h])$ for all $p\in[1,q]$ and 
$|\lambda([1,h])|\leq|O_L|$
we must have $q\leq|O_L|$.
Now let $D_0=[t_0,s_1],D_1=[t_1,s_2],\dots,D_q=[t_q,s_{q+1}]$.
Clearly, these sets are pairwise disjoint.
Moreover, by construction, the only elements $z \in \bigcup_{p = 0}^q D_p$ such
that $z < \lambda(z)$ are those in $X=\set{s_1, \ldots, s_q}$, so that all
elements $z \in (\bigcup_{p = 0}^q D_p) \setminus X$ satisfy $z=\lambda(z)$,
i.e.\ are elements from $\lambda([1,h])$.
Thus, we obtain
$q+\sum_{p=0}^{q}\card{D_p} = q + \card{\bigcup_{p=0}^q D_p} \leq
 \card{O_L}+\card{X}+\card{\lambda([1,h])}=3\card{O_L}$.
	\qed

	\medskip
	\noindent
Let $1=t_0\leq s_1<t_1<s_2<t_2<\cdots<
	s_q<t_q\leq s_{q+1}=h$ be the indices satisfying 
	Claim~\ref{claim pigeon}
	along with
	$D_0=[t_0,s_1],D_1=[t_1,s_2],\dots,D_q=[t_q,s_{q+1}]$.
	Let $d_p=|D_p|$ for all $p\in[0,q]$.
	Since, for all $p\in[1,q]$, the non-empty interval
$[s_p,t_p]$ is repetitive, 
	we have $e_{s_p}=e_{t_p}$ and thus 
	obtain
	\[e_{s_p}=e_{t_p}=
	e_{s_p}\circ 
	\psi_L(\ext_{x_{s_p+1}\cdots a_{t_p},b_{t_p}\cdots y_{s_p+1}}).
	\]
	Hence, 
	we have
	$w\models\pi_{e_{s_p}}(i_{s_p}+1,i_{t_p},j_{t_p},j_{s_p}-1)$ 
	where $\pi_{e_{s_p}}$ is the formula given by
	Lemma~\ref{lem:Synchronous_element_matching}
	(recall that $e_{s_p}$ is $\varphi_L(L)$-reaching and that
	$\newU_{e_{s_p}}$ is length-synchronous).
	We can therefore use the formula $\pi_{e_{s_p}}$ to 
	witness that 
$w_{i_{s_p}+1}\cdots w_{i_{t_p}}w_{j_{t_p}}\cdots w_{j_{s_p}-1}$ 
is indeed a well-matched word.
	It will thus remain to verify that
	$w_{i_{t_p}}\cdots w_{i_{s_{p+1}}}w_{j_{s_{p+1}}}\cdots w_{i_{t_p}}$
	is well-matched for all $p\in[0,q]$: this can be guaranteed by evaluating
	$\kappa_{d_p}(i_{t_p},i_{s_{p+1}},j_{s_{p+1}},i_{t_p})$.
	We can now define our final formula $\mu_d^\uparrow$:
	\begin{align*}
		\mu_d^\uparrow(x,y) = & 
		\bigvee_{\substack{q\in[0,|O_L|]\\
		d_0,\dots,d_q\geq 1:\\
		q+d_0+\dots+d_q\leq 3|0|}}\exists 
		x_1\dots x_{q+1}\exists x_0'\dots x_q'
		\exists y_1\dots y_{q+1}\exists y_1'\dots y_0'\\
		& \biggl[
		  \begin{aligned}[t]
		    & x_0\leq x_1<x_1'<x_2<\dots<x_q'<y_q'<y_q<\dots<y_1'<y_1\leq y_0'\wedge \\
		    & x_0'=x\wedge y_0'=y\wedge \mu_{d-1}(x_{q+1}+1,y_{q+1}-1)\wedge\\
		    & \bigwedge_{p=1}^q\left(\bigvee_{e\in O_L \text{ $\varphi_L(L)$-reaching}} \pi_e(x_p+1,x_p',y_p',y_p-1)\right)
		    \wedge\\
		    & \bigwedge_{p=0}^{q} \kappa_{d_p}(x_p',x_{p+1},y_{p+1},y_p')
		    \biggr]
		    \displaypunct{.}
		    \hspace{79mm}\qedhere
		  \end{aligned}
	\end{align*}
\end{proof}

The following remark is obvious but will be important in Section~\ref{section intermediate}.
\begin{rem}\label{remark up}
	When constructing our predicate
	$\mu_{d}^\uparrow$, 
	we could have replaced any subset of the predicates $\pi_e$,
	where $e$ is $\varphi_L(L)$-reaching from above,
	by the predicate $\pi_e^{\textsf{exact}}$ expressing that
	for all $w \in \Sigma^+$ and
	$i, i', j', j \in [1, \length{w}], i \leq i' < j' \leq j$
	it holds:
	\begin{eqnarray*}
		w\models\pi_e^{\textsf{exact}}(i,i',j',j)&
		\ \Longleftrightarrow\ &
	w_i\dots w_{i'}w_{j'}\dots w_{j}\in\Sigma^\triangle,
	e\circ\psi_L(\ext_{w_i\cdots w_{i'},w_{j'}\cdots w_j})=e,
		\text{ and }\\&&\Delta(w_i\cdots w_{i'})>0
	\end{eqnarray*}
\end{rem}

	It remains to prove Lemma~\ref{lem:Synchronous_element_matching}.
 
 \subsubsection*{Proof of Lemma~\ref{lem:Synchronous_element_matching}}
	In essence, our proof is inspired by the approach taken
	in~\cite[Proof of Proposition~126]{Ludwigthesis}, which is itself a
	flawed adaptation 
	of the approach taken
	in~\cite[Proof of Lemma~15]{Krebs-Lange-Ludwig-2015}.

	Let $\alpha_e \in \Q_{>0}$, $\beta_e \in \N$ and $\gamma_e \in \N_{>0}$
	given by 
	Proposition~\ref{ptn:Length-synchronicity_equivalences} for $e$. There
	exist unique $n_e, d_e \in \N_{>0}$ that are relatively prime such that
	$\alpha_e = \frac{n_e}{d_e}$.
	We are going to build an $\FOplus$-formula $\pi_e(x, x', y', y)$ such
	that for all $w \in \Sigma^+$ and
	$i, i', j', j \in [1, \length{w}], i \leq i' < j' \leq j$, we have that
	$w \models \pi_e(i, i', j', j)$ if, and
	only if, all of the following conditions are satisfied:
	\begin{enumerate}[(i)]
	    \item\label{ctn:sync_1}
		$\frac{i' - i + 1}{j - j' + 1} = \frac{n_e}{d_e}$;
	    \item\label{ctn:sync_2}
		$-\beta_e \leq
		 \Delta(w_i \cdots w_{i + k \cdot n_e - 1}
			w_{j - k \cdot d_e + 1} \cdots w_j) \leq \beta_e$
		for all $k \in \N_{>0}$ such that $k \leq (j - j' + 1) / d_e$ and
		$\Delta(w_i \cdots w_{i'} w_{j'} \cdots w_j) = 0$;
	    \item\label{ctn:sync_3}
		$\Delta(w_{i + (q - 1) \cdot \gamma_e} \cdots
			w_{i + q \cdot \gamma_e - 1})
		 \geq 1$
		for all $q \in \N_{>0}$ such that $q \cdot \gamma_e \leq i' - i + 1$
		and $\Delta(w_i \cdots w_{i + p - 1}) \geq 0$ for all
		$p \in [1, i' - i + 1]$;
	    \item\label{ctn:sync_4}
		$\Delta(w_{j - q \cdot \gamma_e + 1} \cdots
			w_{j - (q - 1) \cdot \gamma_e})
		 \leq -1$
		for all $q \in \N_{>0}$ such that $q \cdot \gamma_e \leq j - j' + 1$
		and $\Delta(w_{j - p + 1} \cdots w_j) \leq 0$ for all
		$p \in [1, j - j' + 1]$.
	\end{enumerate}
	\noindent
	Let us first prove that these four conditions whose conjunction 
	    the $\FOplus$-formula $\pi_e(x, x', y', y)$ will express,
	indeed imply the two conditions of the lemma.

	If conditions~\eqref{ctn:sync_1} to~\eqref{ctn:sync_4} are satisfied for
	a $w \in \Sigma^+$ and
	$i, i', j', j \in [1, \length{w}], i \leq i' < j' \leq j$, we actually
	have that $w_i \cdots w_{i'} w_{j'} \cdots w_j \in \Sigma^\triangle$.
	Indeed, condition~\eqref{ctn:sync_2} ensures that
	$\Delta(w_i \cdots w_{i'} w_{j'} \cdots w_j) = 0$.
	Conditions~\eqref{ctn:sync_3} and~\eqref{ctn:sync_4} then additionally
	imply that $\Delta(w_i \cdots w_{i + p - 1}) \geq 0$ for all
	$p \in [1, i' - i + 1]$ and
	$\Delta(w_i \cdots w_{i'} w_{j'} \cdots w_{j' + p - 1}) \geq 0$ for all
	$p \in [1, j - j' + 1]$. This is because if there were a
	$p \in [1, j - j' + 1]$ such that
	$\Delta(w_i \cdots w_{i'} w_{j'} \cdots w_{j' + p - 1}) < 0$, then it
	should be that $\Delta(w_{j' + p} \cdots w_j) > 0$ with $p \leq j - j'$
	as we already know that
	$\Delta(w_i \cdots w_{i'} w_{j'} \cdots w_j) = 0$: this would be a
	contradiction to condition~\eqref{ctn:sync_4}. 
	    
	Conversely, let us fix some $w \in \Sigma^+$ and indices
	$i, i', j', j \in [1, \length{w}]$ such that $ i \leq i' < j' \leq j$, 
	$w_i \cdots w_{i'} w_{j'} \cdots w_j \in \Sigma^\triangle$,
	$\Delta(w_i \cdots w_{i'}) > 0$ and
	$e \circ \psi_L(\ext_{w_i \cdots w_{i'}, w_{j'} \cdots w_j}) = e$.
	In the terminology of 
Proposition~\ref{ptn:Length-synchronicity_equivalences},
for $F=\varphi_L(L)$, 
we have $(w_i \cdots w_{i'}, w_{j'} \cdots w_j)\in\newU_e$.
	We claim that Points~\eqref{ctn:sync_1} to~\eqref{ctn:sync_4} are actually
	satisfied. Indeed, 
	recalling that $L$ is length-synchronous by assumption,
	2(a) of 
	Proposition~\ref{ptn:Length-synchronicity_equivalences} for $e$
	in fact states that that Point~\eqref{ctn:sync_1} is satisfied. 
	Next,
	since for all
	$k \in \N_{>0}$ such that $k \leq \frac{j - j' + 1}{d_e} = \frac{i' - i + 1}{n_e}$,
	the word $w_i \cdots w_{i + k \cdot n_e - 1}$ is a prefix of
	$w_i \cdots w_{i'}$ and the word $w_{j - k \cdot d_e + 1} \cdots w_j$ is
	a suffix of $w_{j'} \cdots w_j$ such that
	$\frac{\length{w_i \cdots w_{i + k \cdot n_e - 1}}}
	      {\length{w_{j - k \cdot d_e + 1} \cdots w_j}} =
	 \frac{k \cdot n_e}{k \cdot d_e}=\alpha_e$,
	it must hold that 
	$-\beta_e \leq
	 \Delta(w_i \cdots w_{i + k \cdot n_e - 1}
		w_{j - k \cdot d_e + 1} \cdots w_j) \leq \beta_e$
	by Point 2(b) of 
	Proposition~\ref{ptn:Length-synchronicity_equivalences}.
	We have that 
	$\Delta(w_i \cdots w_{i'} w_{j'} \cdots w_j) = 0$ 
	immediately follows from our assumption 
    $w_i \cdots w_{i'} w_{j'} \cdots w_j \in \Sigma^\triangle$, 
    thus Point~\eqref{ctn:sync_2} holds.
	Another consequence of our assumption $w_i \cdots w_{i'} w_{j'} \cdots w_j \in \Sigma^\triangle$
	is that 
	$\Delta(w_i \cdots w_{i + p - 1}) \geq 0$ for all
	$p \in [1, i' - i + 1]$ and
	$\Delta(w_i \cdots w_{i'} w_{j'} \cdots w_{j' + p - 1}) \geq 0$ for all
	$p \in [1, j - j' + 1]$. 
	This implies that
	$\Delta(w_{j - p + 1} \cdots w_j) \leq 0$ for all
	$p \in [1, j - j' + 1]$, as already argued above.
	Since
	$w_{i + (q - 1) \cdot \gamma_e} \cdots w_{i + q \cdot \gamma_e - 1}$ is
	a factor of $w_i \cdots w_{i'}$ of length $\gamma_e$ for all
	$q \in \N_{>0}$ such that $q \cdot \gamma_e \leq i' - i + 1$ and
	$w_{j - q \cdot \gamma_e + 1} \cdots w_{j - (q - 1) \cdot \gamma_e}$ is
	a factor of $w_{j'} \cdots w_j$ of length $\gamma_e$ for all
	$q \in \N_{>0}$ such that $q \cdot \gamma_e \leq j - j' + 1$,
	by Points 2(c) and 2(d) of 
	Proposition~\ref{ptn:Length-synchronicity_equivalences}, we finally have
	that conditions~\eqref{ctn:sync_3} and~\eqref{ctn:sync_4} are also
	satisfied.

	It now remains to construct the formula $\pi_e(x, x', y', y)$. We set
	\begin{align*}
	    \pi_e(x, x', y', y) =
	    & (x' - x + 1) \cdot d_e = (y - y' + 1) \cdot n_e \wedge\\
	    & \mu_{n_e, d_e, \beta_e}(x, x', y', y) \wedge\\
	    & \nu_{\gamma_e}^+(x, x') \wedge \nu_{\gamma_e}^-(y', y),
	\end{align*}
	where the first line checks condition~\eqref{ctn:sync_1}, the
	$\FOplus$-formula $\mu_{n_e, d_e, \beta_e}(x, x', y', y)$ will check
	condition~\eqref{ctn:sync_2} under the assumption
	condition~\eqref{ctn:sync_1} is satisfied and the $\FOplus$-formulas
	$\nu_{\gamma_e}^+(x, x')$ and $\nu_{\gamma_e}^-(y', y)$ respectively
	will check conditions~\eqref{ctn:sync_3} and~\eqref{ctn:sync_4}.
	We now explain how to build those formulas.

	\noindent
	\paragraph*{Helper formulas.}
	For all $k \in \N_{>0}$ and $h \in \Z$ such that $-k \leq h \leq k$, we let
	\[
	    H_k^h(x) =
	    \bigvee_{\substack{I, J \subseteq [1, k]\\I \cap J = \emptyset\\
			       \card{I} - \card{J} = h}}
	    \Bigl(
		\bigwedge_{p \in I} \Sigma_\call(x + p - 1) \wedge
		\bigwedge_{p \in J} \Sigma_\return(x + p - 1) \wedge
		\bigwedge_{p \in [1, k] \setminus (I \cup J)}
		    \Sigma_\internal(x + p - 1)
	    \Bigr)
	\]
	such that for all $w \in \Sigma^+$ and
	$i \in [1, \length{w}]$ such that $i \leq \length{w} - k + 1$, we have
	$w\models H_k^h(i)$ if, and only if,
	$\Delta(w_i \cdots w_{i + k - 1}) = h$.

	For all $n, d \in \N_{>0}$ relatively prime and
	$h \in \Z, -n - d \leq h \leq n + d$, we define
	\[
	    D^h_{n, d}(x, y, z) =
	    \bigvee_{\substack{-n \leq h_1 \leq n\\-d \leq h_2 \leq d\\
		       h_1 + h_2 = h}}
	    \Bigl(
		H_n^{h_1}\bigl(x + (z - 1) \cdot n\bigr) \wedge
		H_d^{h_2}\bigl(y - z \cdot d + 1\bigr)
	    \Bigr)
	    \displaypunct{,}
	\]
	such that for all $w \in \Sigma^+$ and
	$i, j, k \in [1, \length{w}]$ with $i + k \cdot n - 1 \leq \length{w}$
	and $j - k \cdot d + 1 \geq 1$, we have
	$w\models D^h_{n, d}(i, j, k)$ if, and only if,
	\[
	    \Delta(w_{i + (k - 1) \cdot n} \cdots w_{i + k \cdot n - 1}
		   w_{j - k \cdot d + 1} \cdots w_{j - (k - 1) \cdot d}) = h
		   \displaypunct{.}
	\]

	\noindent
	\paragraph*{Formula $\mu_{n, d, q}(x, x', y', y)$.}
	For each $p\in\N$ let $\Gamma_p = \set{a_{-p}, \ldots, a_{-1}, a_0, a_1, \ldots, a_p}$
	and define $\Delta_p\colon \Gamma_p^* \to \Z$ to be the $p$-height
	monoid morphism satisfying $\Delta_p(a_h) = h$ for all
	$a_h \in \Gamma_p$.
	Consider the language
	\[
	    L_{p, q} =
	    \set{w \in \Gamma_p^* \mid \Delta_p(w) = 0 \wedge
				       \forall i \in [1, \length{w}],
				       -q \leq \Delta_p(w_1 \cdots w_i) \leq q}
	    \displaypunct{.}
	\]
	We claim that this language is recognized by a finite aperiodic monoid.
	This implies, by a theorem by McNaughton and Papert
	(see~\cite[Theorem~VI.1.1]{Books/Straubing-1994}), that there exists an
	$\FOord[\Gamma_{n + d}]$-sentence $\tilde{\mu}_{p, q}$ defining
	$L_{p, q}$.

	Let now $n, d \in \N_{>0}$ relatively prime and $q \in \N$.
	Consider $w \in \Sigma^+$ and
	$i, i', j', j \in [1, \length{w}]$ such that $i \leq i' < j' \leq j$
	and $\frac{i' - i + 1}{j - j' + 1} = \frac{n}{d}$. 
	We want to check whether we have
	\[
	    -q \leq \Delta(w_i \cdots w_{i + k \cdot n - 1}
			   w_{j - k \cdot d + 1} \cdots w_j) \leq q
	\]
	for all $k \in \N_{>0}$ such that $k \leq (j - j' + 1) / d$ and moreover
	$\Delta(w_i \cdots w_{i'} w_{j'} \cdots w_j) = 0$.
	Since $n$ and $d$ are relatively prime, this means that there exists
	$l \in [1, \length{w}]$ such that $i' - i + 1 = l \cdot n$ and
	$j - j' + 1 = l \cdot d$. We can hence decompose $w_i \cdots w_{i'}$
	as $u_1 \cdots u_l$ with $u_1, \ldots, u_l \in \Sigma^n$ and
	$w_{j'} \cdots w_j$ as $v_l \cdots v_1$ with
	$v_1, \ldots, v_l \in \Sigma^d$. 
	Observe that $\Delta(u_iv_i)\in[-n-d,n+d]$ for all $i\in[1,l]$.
	Using this decomposition, we now need
	to check whether
	$-q \leq \Delta(u_1 v_1) + \cdots + \Delta(u_k v_k) \leq q$ for all
	$k \in [1, l]$ and $\Delta(u_1 v_1) + \cdots + \Delta(u_l v_l) = 0$.
	This is equivalent to checking whether the word
	$\tilde{w} = a_{\Delta(u_1 v_1)} \cdots a_{\Delta(u_l v_l)}$ in
	$\Gamma_{n + d}^*$ belongs to $L_{n + d, q}$.

	We thus transform the $\FOord[\Gamma_{n + d}]$-sentence
	$\tilde{\mu}_{n + d, q}$ into an $\FOplus[\Sigma]$-formula
	$\mu_{n, d, q}(x, x', \allowbreak y', y)$ by
	\begin{itemize}
	    \item
		replacing any quantification $\exists z \rho(z)$ by
		$\exists z \bigl(z \leq (y - y' + 1) / d \wedge \rho(z)\bigr)$;
	    \item
		replacing any quantification $\forall z \rho(z)$ by
		$\forall z
		 \bigl(z \leq (y - y' + 1) / d \rightarrow \rho(z)\bigr)$;
	    \item
		replacing any atomic formula of the form $a_h(z)$ for
		$a_h \in \Gamma_{n + d}$ by $D^h_{n, d}(x, y, z)$.
	\end{itemize}
	By this translation for all $w \in \Sigma^+$ and
	$i, i', j', j \in [1, \length{w}]$ with $i \leq i' < j' \leq j$
    and
	$\frac{i' - i + 1}{j - j' + 1} = \frac{n}{d}$
	we have 
	$w\models \mu_{n, d, q}(i, i', j', j)$
	if, and only if,
	$-q \leq
	 \Delta(w_i \cdots w_{i + k \cdot n - 1}
		w_{j - k \cdot d + 1} \cdots w_j) \leq q$
	for all $k \in \N_{>0}, k \leq (j - j' + 1) / d$ and
	$\Delta(w_i \cdots w_{i'} w_{j'} \cdots w_j) = 0$.

	It remains to show that $L_{p, q}$ is recognized by a finite aperiodic
	monoid for all $p, q \in \N$. 	
	Set $Q_q = \set{-q, \ldots, -1, 0, 1, \ldots, q, \bot}$ and consider the
	monoid ${Q_q}^{Q_q}$ with function composition from left to right.
	For each $a_h \in \Gamma_p$, we define the function
	$f_{a_h}\colon Q_q \to Q_q$ to be such that
	\[
	    f_{a_h}(h') =
	    \begin{cases}
		h'+ h & \text{if $h' \neq \bot$ and $-q \leq h' + h \leq q$}\\
		\bot & \text{otherwise}
	    \end{cases}
	\]
	for all $h' \in Q_q$.
	We take $M_{p, q}$ to be the submonoid of ${Q_q}^{Q_q}$ generated by
	$\set{f_{a_h} \mid a_h \in \Gamma_p}$ and define
	$\varphi_{p, q}\colon \Gamma_p^* \to M_{p, q}$ as the unique monoid
	morphism such that $\varphi_{p, q}(a_h) = f_{a_h}$ for all
	$a_h \in \Gamma_p$.

	It is straightforward to show, by induction on the length of $w$, that
	for all $w \in \Gamma_p^*$ and all $h \in Q_q$, we have
	\[
	    \varphi_{p, q}(w)(h) =
	    \begin{cases}
		h + \Delta_p(w) &
		    \text{if $h \neq \bot$ and
			  $-q \leq h + \Delta_p(w_1 \cdots w_i) \leq q$ for all
			  $i \in [1, \length{w}]$}\\
		\bot & \text{otherwise.}
	    \end{cases}
	\]
	Thus
	$L_{p, q} = \varphi_{p, q}^{-1}(\set{f \in M_{p, q} \mid f(0) = 0})$.
	We claim that the monoid $M_{p, q}$ is aperiodic.
	Indeed, take $f \in M_{p, q}$; we claim that
	$f^{2 q + 1} = f^{2 q + 2}$. Since $M_{p, q}$ is generated by
	$\set{f_{a_h} \mid a_h \in \Gamma_p}$, there exists $w \in \Gamma_p^*$
	satisfying $\varphi_{p, q}(w) = f$. There are three subcases to consider.
	\begin{itemize}
	    \item
		If $\Delta_p(w) = 0$, then since
		$h + \Delta_p(w^{n - 1} w_1 \cdots w_i) =
		 h + \Delta_p(w_1 \cdots w_i)$
		for all $h \in \Z, -q \leq h \leq q$, for all $n \in \N_{>0}$
		and $i \in [1, \length{w}]$, we have that $f^n = f$ for all
		$n \in \N_{>0}$.
	    \item
		If $\Delta_p(w) > 0$, then since
		$q < h + \Delta_p(w^{2 q + 1}) \leq h + \Delta_p(w^{2 q + 2})$
		for all $h \in \Z, -q \leq h \leq q$, both
		$f^{2 q + 1}$ and $f^{2 q + 2}$ must be equal to the function
		sending every element to $\bot$.
	    \item
		If $\Delta_p(w) < 0$, then since
		$h + \Delta_p(w^{2 q + 2}) \leq h + \Delta_p(w^{2 q + 1}) < -q$
		for all $h \in \Z, -q \leq h \leq q$, both
		$f^{2 q + 1}$ and $f^{2 q + 2}$ must be equal to the function
		sending every element to $\bot$.
	\end{itemize}

	\noindent
	\paragraph*{Formula $\nu_l^+(x, x')$.}
	For all $l \in \N_{>0}$, we let
	\begin{align*}
	    \nu_l^+(x, x') =
	    & \bigwedge_{p = 1}^{l^2}
	      \Bigl(x' - x + 1 \geq p \rightarrow
		    \bigwedge_{k = 1}^p \bigvee_{h = 0}^k H_k^h(x)\Bigr) 
	      \ \wedge\\
	    & \forall z
	      \Bigl(z \cdot l \leq x' - x + 1 \rightarrow
		    \bigvee_{h = 1}^l H_l^h\bigl(x + (z - 1) \cdot l\bigr)
	      \Bigr)
	    \displaypunct{.}
	\end{align*}

	Fix any $w \in \Sigma^+$ and $i, i' \in [1, \length{w}]$ such that $i \leq i'$. 
    We have $w\models \nu_l^+(i, i')$ if, and only if, 
	$\Delta(w_i \cdots w_{i + p - 1}) \geq 0$ for all
	$p \in [1, \min\set{l^2, i' - i + 1}]$ and
	$\Delta(w_{i + (q - 1) \cdot l} \cdots w_{i + q \cdot l - 1}) \geq 1$
	for all $q \in \N_{>0}$ such that $q \cdot l \leq i' - i + 1$.
	The latter is clearly equivalent to having
    $\Delta(w_{i + (q - 1) \cdot l} \cdots w_{i + q \cdot l - 1}) \geq 1$
	for all $q \in \N_{>0}, q \cdot l \leq i' - i + 1$ and
	$\Delta(w_i \cdots w_{i + p - 1}) \geq 0$ for all
	$p \in [1, i' - i + 1]$, as required.
	
	\noindent
	\paragraph*{Formula $\nu_l^-(y', y)$.}
	For all $l \in \N_{>0}$, we let
	\begin{align*}
	    \nu_l^-(y', y) =
	    & \bigwedge_{p = 1}^{l^2}
	      \Bigl(y - y' + 1 \geq p \rightarrow
		    \bigwedge_{k = 1}^p \bigvee_{h = 0}^k H_k^{-h}(y - k + 1)
	      \Bigr)\ \wedge\\
	    & \forall z
	      \Bigl(z \cdot l \leq y - y' + 1 \rightarrow
		    \bigvee_{h = 1}^l H_l^{-h}\bigl(y - z \cdot l + 1\bigr)
	      \Bigr)
	    \displaypunct{.}
	\end{align*}
	Therefore, analogously as for $\nu_l^+(x, x')$, 
	for all $w \in \Sigma^+$ and $j', j \in [1, \length{w}]$ such that $j' \leq j$, we have
	$w\models \nu_l^-(j', j)$ if, and only if,
	$\Delta(w_{j - q \cdot l + 1} \cdots w_{j - (q - 1) \cdot l}) \leq -1$
	for all $q \in \N_{>0}$ such that $q \cdot l \leq j - j' + 1$ and
	$\Delta(w_{j - p + 1} \cdots w_j) \leq 0$ for all
	$p \in [1, j - j' + 1]$.

\subsubsection{Evaluation in $\FOplus$ (Proof of Point 4)}
\label{Section Evaluation}

The following proposition states a sufficient condition for
a VPL to be
$\FOplus[\Sigma,\match]$-definable assuming that an approximate matching is present as
built-in predicate.
It makes use of bounded nesting depth and quasi-aperiodicity of both the vertical
and horizontal evaluation language.

\begin{samepage}
\begin{prop}
\label{ptn:Eval_languages_and_bounded_nesting-depth}
    Let $L$ be a VPL satisfying the following:
    \begin{itemize}
	\item $L$ has bounded nesting depth,
	\item
	    $\mathcal{E}_{\varphi_L, r}$ is a regular language whose syntactic
	    morphism is quasi-aperiodic for all $\varphi_L(L)$-reaching
	    $r \in R_L$, and
	\item
	    $\mathcal{E}_{\psi_L, e}$ is a regular language whose syntactic
	    morphism is quasi-aperiodic for all $\varphi_L(L)$-reaching
	    $e \in O_L$.
    \end{itemize}
    Then there exists an $\FOplus[\Sigma, \match]$-sentence $\eta$ such that for
    any approximate matching $M$ relative to $L$, we have $w \in L$ if, and only
    if, $(w, M(w)) \models \eta$ for all $w \in \Sigma^*$.
\end{prop}
\end{samepage}

\begin{proof}
    By hypothesis, there exists $d_L \in \N$ bounding the nesting depth of the
    words in $L$.

    By hypothesis also, for each $\varphi_L(L)$-reaching $r \in R_L$, the
    language $\mathcal{E}_{\varphi_L, r}$ is regular and its syntactic morphism
    is quasi-aperiodic.
    This implies, by~\cite[Theorem~VI.4.1]{Books/Straubing-1994}, that for each
    $\varphi_L(L)$-reaching $r \in R_L$, there exists an
    $\FO_{\Gamma_{\varphi_L}}[<, \MODpred]$-sentence $\nu_{\varphi_L, r}$
    defining $\mathcal{E}_{\varphi_L, r}$.

    Finally, by hypothesis, for each $\varphi_L(L)$-reaching $e \in O_L$, the
    language $\mathcal{E}_{\psi_L, e}$ is regular and its syntactic morphism is
    quasi-aperiodic.
    Again, by~\cite[Theorem~VI.4.1]{Books/Straubing-1994}, for each
    $\varphi_L(L)$-reaching $e \in O_L$, there exists an
    $\FO_{\Gamma_{\psi_L}}[<, \MODpred]$-sentence $\nu_{\psi_L, e}$ defining
    $\mathcal{E}_{\psi_L, e}$.

    \subparagraph{Auxiliary formulas.}
    We introduce a few auxiliary formulas that all assume access to the full
    matching relation $M^\triangle(w)$, represented by the relational symbol
    $\match$.

    First let us define a formula $A$ such that for all $w \in \Sigma^\triangle$
    and $i, j, k \in [1, \length{w}]$ satisfying
    $w_i \cdots w_j \in \Sigma^\triangle$, we have that
    $(w, \matching{w}) \models A(i, j, k)$ if, and only if, $i \leq k < j$ and
    $\Delta(w_i \cdots w_k) > 0$.
    We let
    \[
	A(x, y, z) =
	\exists x' \exists y'
	(x \leq x' \leq z < y' \leq y \wedge x' \match y')
	\displaypunct{.}
    \]

    Next, we define a formula $U$ such that for all $w \in \Sigma^\triangle$ and
    $i, i', k \in [1, \length{w}]$, we have that
    $(w, \matching{w}) \models U(i, i', k)$ if, and only if, 
    $i \leq k \leq i'$ and $k$ is matched to some position larger than $i'$ in
    $w$.
    We let
    \[
	U(x, x', z) =
	x \leq z \leq x' \wedge \exists t (z \match t \wedge x' < t)
	\displaypunct{.}
    \]

    The last formulas we introduce are $N_d$ which express that the infix
    $w_i\cdots w_j\in\Sigma^\triangle$ of $w\in\Sigma^\triangle$ has nesting
    depth at least $d\geq 0$.
    More precisely, for all $d \in \N$, we introduce auxiliary formulas $N_d$
    such that for all $w \in \Sigma^\triangle$ and $i, j \in [1, \length{w}]$
    satisfying $w_i \cdots w_j \in \Sigma^\triangle$, we have that
    $(w, \matching{w}) \models N_d(i, j)$ if, and only if,
    $\ndepth(w_i \cdots w_j) \geq d$.
    The case $d=0$ is trivial since we can set $N_0(i,j) = \top$.

    Take $w \in \Sigma^\triangle$ such that $\ndepth(w) = d' \geq 1$.
    Note that then $w$ can be factorized as $w=w_1uw_2$ such that
    $w_1, w_2 \in \Sigma^\triangle$,
    $u\in\Sigma_\call\Sigma^\triangle\Sigma_\return$ and $\ndepth(w)=\ndepth(u) = d'$.
    This means that $u=a_1vb_1$ for $a_1\in\Sigma_\call$, $b_1\in\Sigma_\return$
    and $v\in\Sigma^\triangle$. We then apply
    Lemma~\ref{lem:Maximal_stair_factorization} and
    Lemma~\ref{L nesting induction} implying that $u$ has a nesting-maximal
    stair factorization 
    \[
	u = \ext_{x_1,y_1} \circ \ext_{a_1,b_1} \circ \cdots \circ
	    \ext_{x_k,y_k} \circ \ext_{a_k,b_k}(u')
    \]
    for which there exists $h\in[1,k]$ such that, setting
    $u_i=\ext_{a_i,b_i}\circ\ext_{x_{i+1},y_{i+1}}
		    \circ\cdots\circ\
		    \ext_{a_k,b_k}(u')$
    for all $i \in [1, k]$ and $u_{k + 1} = u'$, we have $\ndepth(u)=\ndepth(u_h) = d'$
    and
    \[\ndepth(\ext_{x_{h+1},y_{h+1}} \circ \ext_{a_{h+1}, b_{h+1}} \circ \cdots \circ
	\ext_{a_k,b_k}(u')) = \ndepth(u_{h+1}) = d'-1\displaypunct{.}\]
    Thus, by definition of the nesting depth of a well-matched word,
    $u_h=a_hz_1z_2b_h$ for $z_1,z_2\in\Sigma^\triangle$ satisfying
    $\ndepth(z_1) = \ndepth(z_2) = d'-1$.

    Hence, we set
    \[
	N_1(x, y) =
	\exists x' \exists y' (x \leq x' < y' \leq y \wedge x' \match y')
    \]
    and for $d\geq 2$ we set 
    \[
	N_d(x, y) =
	\exists x' \exists y' \exists z
	\bigl(
	\begin{aligned}[t]
	    & x \leq x' < z < y' \leq y \wedge x' \match y' \wedge
	      \neg A(x'+1, y'-1, z) \wedge\\
	    & N_{d - 1}(x'+1, z) \wedge N_{d - 1}(z+1, y'-1)
	    \bigr)
	    \displaypunct{.}
	\end{aligned}
    \]

    \paragraph*{Main construction.}
    To build the $\FOplus[\Sigma, \match]$-sentence $\eta$, we build 
    $\FOplus[\Sigma, \match]$-formulas 
    \begin{itemize}
	\item
	    $\eta_{d, r}^\uparrow(x, y)$ for all $d \in \N$ and all
	    $\varphi_L(L)$-reaching $r \in R_L$ and 
	\item
	    $\eta_{d, r}(x, y)$ for all $d\in\N$ and all
	    $\varphi_L(L)$-reaching $r\in R_L$
    \end{itemize}
    that also assume access to the full matching relation $M^\Delta(w)$.
    They will have the following properties
    for all $w\in\Sigma^\triangle$ and all $i,j\in[1,\length{w}]$:
    \begin{itemize}
	\item
	    if $i$ is matched to $j$ in $w$, then
	    $(w, \matching{w}) \models \eta_{d, r}^\uparrow(i, j)$ if, and only
	    if, $\ndepth(w_i \cdots w_j) \leq d$ and $\varphi_L(w_i \cdots w_j) = r$
	    and 
	\item
	    if $w_i \cdots w_j \in \Sigma^\triangle$, then
	    $(w, \matching{w}) \models \eta_{d, r}(i, j)$ if, and only if,
	    $\ndepth(w_i \cdots w_j) \leq d$ and $\varphi_L(w_i \cdots w_j) = r$.
    \end{itemize}
    Let the formula $E$ be defined as
    $\forall x (x \neq x)$ if $\emptyword \in L$ and $\bot=\exists x (x\neq x)$
    otherwise.
Observe that $w\models \forall x (x \neq x)$ if, and only if,
	$w=\varepsilon$.
    Our final formula $\eta$ will then be defined as
    \begin{align*}
	\eta =
	& \forall z \exists t
	  \bigl(
	    (\Sigma_\call(z) \rightarrow z \match t) \wedge
	    (\Sigma_\return(z) \rightarrow t \match z)
	  \bigr) \wedge\\
	& \Bigl(
	    E \vee
	    \exists x \exists y
	    \bigl(
		\neg \exists x' (x' < x) \wedge \neg \exists y' (y < y') \wedge
		\bigvee_{r \in \varphi_L(L)} \eta_{d_L, r}(x, y)
	    \bigr)
	  \Bigr).
    \end{align*}
    
    It now remains to build $\eta_{d, r}^\uparrow(x, y)$ and $\eta_{d, r}(x, y)$
    for all $d \in \N$ and $\varphi_L(L)$-reaching $r \in R_L$.
    The construction is by induction on $d$.
    Let $r \in R_L$ that is $\varphi_L(L)$-reaching.
    We define $\eta_{0,r}^\uparrow(x,y)=\bot$.
    We define $\eta_{0,r}$ as 
    \[
	\eta_{0,r}(x,y) = \neg N_1(x,y) \wedge \tau_0(\nu_{\varphi_L,r})
	\displaypunct{,}
    \]
    where the translation $\tau_0$ is inductively defined as follows:
    \begin{itemize}
	\item $\tau_0(z<z')=z<z'$;
	\item
	    $\tau_0(s(z))=
	     \bigvee_{c \in \varphi_L^{-1}(s) \cap \Sigma_\internal} c(z)$
	    for all
	    $s \in \varphi_L(\Sigma^\triangle \setminus \set{\emptyword})$;
	\item
	    $\tau_0(\MODpred_m(z))=
	     \exists t
	     \bigl((z = x \rightarrow 1 = t \cdot m) \wedge
	     (z \neq x \rightarrow z - x + 1 = t \cdot m)\bigr)$
	    for all $m \in \N_{>0}$;
	\item $\tau_0(\$(z))=\bot$;
	\item
	    $\tau_0(\rho_1(\boldsymbol{z}_1) \wedge \rho_2(\boldsymbol{z}_2))=
	     \tau_0(\rho_1(\boldsymbol{z}_1))\wedge
	     \tau_0(\rho_1(\boldsymbol{z}_2))$;
	\item
	    $\tau_0(\neg\rho(\boldsymbol{z}))=\neg\tau_0(\rho(\boldsymbol{z}))$;
	\item
	    $\tau_0(\exists z \rho(z,\boldsymbol{z}))=
	     \exists z \bigl(x \leq z \leq y \wedge
			     \tau_0(\rho(z,\boldsymbol{z}))\bigr)$.
    \end{itemize}
	\noindent
    Now let $d>0$.
    Let us first define $\eta_{d,r}$ when assuming that we have already defined
    $\eta_{d,r}^\uparrow$.
    Given $w\in\Sigma^\triangle \setminus \set{\emptyword}$ and
    $i,j\in[1,\length{w}]$ such that
    $w_i \cdots w_j \in \Sigma^\triangle \setminus \set{\emptyword}$, note that
    in case $\ndepth(w_i \cdots w_j)\leq d$, then one can factorize $w_i \cdots w_j$
    as $w_i \cdots w_j = u_1 \cdots u_m$ such that
    $u_\ell\in\Sigma_\internal \cup \Sigma_\call\Sigma^\triangle\Sigma_\return$
    and $\ndepth(u_\ell)\leq d$ for all $\ell\in[1,m]$. Note also that if
    $\varphi_L(w_i \cdots w_j) = r$, as $r$ is $\varphi_L(L)$-reaching, then
    $\varphi_L(u_\ell)$ is $\varphi_L(L)$-reaching for all $\ell \in [1, m]$.
    Using these observations we define
    \[
	\eta_{d,r}(x,y) =
	\neg N_{d+1}(x,y) \wedge
	\forall z
	\Bigl(
	    x \leq z \leq y \rightarrow
	    \bigl(
		\tau_1(\$(z)) \vee
		\bigvee_{\substack{s \in \varphi_L(\Sigma^\triangle \setminus
						   \set{\emptyword})\\
			 \text{$\varphi_L(L)$-reaching}}}
		\tau_1(s(z))
	    \bigl)
	\Bigr) \wedge
	\tau_1(\nu_{\varphi_L,r})
	\displaypunct{,}
    \]
    where the translation $\tau_1$ agrees with the above translation $\tau_0$
    (where, as expected, occurrences of $\tau_0$ are 
    replaced by $\tau_1$)
    except for the following kinds of subformulas:
    \begin{itemize}
	\item $\tau_1(\$(z))=A(x,y,z)$;
	\item
	    $\tau_1(s(z)) =
	     \neg A(x, y, z) \wedge
	     \Bigl(\bigvee_{c \in \varphi_L^{-1}(s) \cap \Sigma_\internal} c(z)
		   \vee
		   \exists t
		   \bigl(x \leq t \leq y \wedge t \match z \wedge
			 \eta_{d, s}^\uparrow(t, z)\bigr)
	     \Bigr)$
	    if $s$ is $\varphi_L(L)$-reaching, $\tau_1(s(z)) = \bot$ otherwise.
    \end{itemize}
    It remains to define $\eta_{d,r}^\uparrow$.

    We first construct for all $\varphi_L(L)$-reaching $e\in O_L$ a formula
    $\chi_{d,e}(x,x',y',y)$ such that for all $w \in \Sigma^\triangle$ and
    $i, i', j', j \in [1, \length{w}], i \leq i' < j' \leq j$ we have that if
    $w_i \cdots w_{i'} w_{j'} \cdots w_j \in \Sigma^\triangle$ and $i'$ is
    matched to $j'$ in $w$, then given
    \[
	\ext_{w_i\cdots w_{i'},w_{j'}\cdots w_j}=
	\ext_{x_1,y_1}\circ\ext_{a_1,b_1}\circ\cdots\circ
	\ext_{x_k,y_k}\circ\ext_{a_k,b_k}
    \]
    the stair factorization of $\ext_{w_i\cdots w_{i'},w_{j'}\cdots w_j}$
    provided by Lemma~\ref{lemma ext}, we have
    $(w, \matching{w}) \models \chi_{d,e}(i,i',j',j)$ if, and only if,
    $\ndepth(x_\ell), \ndepth(y_\ell) < d$ for all $\ell \in [1, k]$ and
    $\psi_L(\ext_{w_i\cdots w_{i'},w_{j'}\cdots w_j})=e$.
    Given $w, i, i', j', j$ and the associated stair factorization as above,
    note that whenever $\psi_L(\ext_{w_i\cdots w_{i'},w_{j'}\cdots w_j}) = e$, as $e$
    is $\varphi_L(L)$-reaching, then $\varphi_L(x_\ell)$ and
    $\varphi_L(y_\ell)$ are $\varphi_L(L)$-reaching for all $\ell \in [1, k]$.
    If additionally $\ndepth(x_\ell), \ndepth(y_\ell) < d$ for all $\ell \in [1, k]$, we
    can inductively make use of the formulas
    $\set{\eta_{d-1,r'}\mid r'\in R_L \text{ $\varphi_L(L)$-reaching}}$ in
    order to evaluate
    $\varphi_L(x_1), \varphi_L(y_1), \ldots, \varphi_L(x_k), \varphi_L(y_k)$.
    Let $p_1, \ldots, p_k \in [i, i']$ be the positions in $w_i \cdots w_{i'}$
    where, respectively, $a_1, \ldots, a_k$ in the above factorization of
    $\ext_{w_i \cdots w_{i'}, w_{j'} \cdots w_j}$ appear: the formula
    $\chi_{d,e}$ will verify if
    $\sigma_i\cdots\sigma_{i'}\in\mathcal{E}_{\psi_L,e}$, where
    \[\sigma_q=
     \begin{cases}
	\leftmult{\varphi_L(x_m)}\circ\rightmult{\varphi_L(y_m)}
	\circ\psi_L(\ext_{a_m,b_m}) &
	    \text{if $q = p_m$ for $m \in [1, k]$}\\
	\$ & \text{otherwise}
     \end{cases}\]
    for all $q \in [i, i']$.
    Hence we set
    \[
	\chi_{d, e}(x, x', y', y) =
	\forall z
	\Bigl(
	    x \leq z \leq x' \rightarrow
	    \bigl(
		\tau_2(\$(z)) \vee
		\bigvee_{\substack{f \in
				   \psi_L
				   \bigl(
					\ExtMon(\Sigma^\triangle)_\uparrow
				   \bigr)\\
				   \text{$\varphi_L(L)$-reaching}}}
		\tau_2(f(z))
	    \bigl)
	\Bigr) \wedge
	\tau_2(\nu_{\psi_L,e})
	\displaypunct{,}
    \]
    where the translation $\tau_2$ agrees with translation $\tau_0$ (where, as
    expected, occurrences of $\tau_0$ are replaced by $\tau_2$) with the
    following exceptions:
    \begin{itemize}
	\item
	    $\tau_2(\exists z \rho(z,\boldsymbol{z}))=
	     \exists z
	     \bigl(x \leq z \leq x' \wedge \tau_2(\rho(z,\boldsymbol{z}))\bigr)$
	\item $\tau_2(\$(z))=\neg U(x,x',z)$
	\item
	    $\tau_2(f(z)) =
	     \begin{cases}
		\exists t
		\Bigl(
		\begin{aligned}[t]
		    & U(x, x', z) \wedge z \match t \wedge\\
		    & \bigl(\iota_{d,f}(x,x',y',y,z,t)\vee
			    \zeta_{d,f}(x,x',z,t)\bigr)
		      \Bigr)
		\end{aligned} &
		    \text{if $f$ is $\varphi_L(L)$-reaching}\\
		\bot & \text{otherwise}
	     \end{cases}$
	    for each
	    $f \in \psi_L\bigl(\ExtMon(\Sigma^\triangle)_\uparrow\bigr)$, where 
	    \begin{align*}
		\iota_{d,f}(x,x',y',y,z,t) =
		& \neg \exists z' \bigl(U(x, x', z') \wedge z' < z\bigl)
		  \wedge\\
		& \begin{aligned}[t]
		    & \bigvee_{\substack{a \in \Sigma_\call,
					 b \in \Sigma_\return,
					 r', r'' \in R_L\\
					 f=\leftmult{r'} \circ
					 \rightmult{r''}\circ
					 \psi_L(\ext_{a, b})}}
		      \biggl(
		      \begin{aligned}[t]
			& a(z) \wedge b(t) \wedge\\
			& \Bigl(\bigl(x < z \wedge
				      \eta_{d - 1, r'}(x, z - 1)\bigr) \vee
				F_{r'}(x, z)\Bigr) \wedge\\
			& \Bigl(\bigl(t < y \wedge
				      \eta_{d - 1, r''}(t + 1, y)\bigr) \vee
				F_{r''}(y, t)\Bigr)
			  \biggr)
		      \end{aligned}
		  \end{aligned}
	    \end{align*}
	    with
	    $F_s(x, z) =
	     \begin{cases}
		x = z & \text{if $s = 1_{R_L}$}\\
		\bot & \text{otherwise}
	     \end{cases}$
	    for all $\varphi_L(L)$-reaching $s \in R_L$ and
	    \[
		\zeta_{d,f}(x,x',z,t) =
		\exists z' \exists t'
		\biggl(
		\begin{aligned}[t]
		    & U(x, x', z') \wedge z' < z \wedge
		      \neg \exists z'' \bigl(z' < z'' < z \wedge
					     U(x, x', z'')\bigr) \wedge
		      z' \match t' \wedge\\
		    & \bigvee_{\substack{a \in \Sigma_\call,
					 b \in \Sigma_\return,
					 r', r'' \in R_L\\
					 f=\leftmult{r'} \circ
					 \rightmult{r''}\circ
					 \psi_L(\ext_{a, b})}}
		      \Bigl(
		      \begin{aligned}[t]
			  a(z) \wedge b(t) \wedge \,
		      	& \eta_{d - 1, r'}(z' + 1, z - 1) \wedge\\
			& \eta_{d - 1, r''}(t + 1, t' - 1)
			  \Bigr)\biggr)
			\displaypunct{.}
		      \end{aligned}
		\end{aligned}
	    \]
    \end{itemize}
	\noindent
    We now construct $\eta_{d,r}^\uparrow$ itself.
    Given $w\in\Sigma^\triangle \setminus \set{\emptyword}$ and
    $i,j\in[1,\length{w}]$ such that $i$ is matched to $j$ in $w$, observe first
    that the infix $w_i \cdots w_j$ is of the form
    $w_i \cdots w_j=a_1vb_1\in\Sigma^\triangle$ for some
    $a_1\in\Sigma_\call$, $b_1\in\Sigma_\return$, and $v\in\Sigma^\triangle$.
    As above, we can directly express $\ndepth(w_i \cdots w_j)\leq d$ via the
    formula $\neg N_{d+1}$. Assuming this holds, towards expressing that
    $\varphi_L(w_i \cdots w_j)=r$, we make use of
    Lemma~\ref{lem:Maximal_stair_factorization} and
    Lemma~\ref{L nesting induction}: for the infix $w_i \cdots w_j$ there is a
    nesting-maximal stair factorization 
    \[
	w_i \cdots w_j =
	\ext_{x_1,y_1} \circ \ext_{a_1,b_1} \circ \cdots \circ
	\ext_{x_k,y_k} \circ \ext_{a_k,b_k}(u')
    \]
    such that we have
    \begin{enumerate}
	\item $\ndepth(x_1),\ndepth(y_1),\dots,\ndepth(x_k),\ndepth(y_k)<d$; and
	\item
	    if $\varphi_L(w_i \cdots w_j) = r$, as $r$ is
	    $\varphi_L(L)$-reaching, then
	    $\psi_L(\ext_{x_1,y_1} \circ \ext_{a_1,b_1} \circ \cdots \circ
		    \ext_{x_k,y_k} \circ \allowbreak \ext_{a_k,b_k})$
	    and $\varphi_L(u')$ are $\varphi_L(L)$-reaching.
    \end{enumerate}
    By these points, we can use the formulas
    $\set{\chi_{d,e}(x,x',y',y) \mid
	  e \in O_L \text{ $\varphi_L(L)$-reaching}}$
    to evaluate
    $\psi_L(\ext_{x_1,y_1} \circ \ext_{a_1,b_1} \circ \cdots \circ
	    \ext_{x_k,y_k} \circ \ext_{a_k,b_k})$
    and the formulas
    $\set{\eta_{0,r'}\mid r'\in R_L \text{ $\varphi_L(L)$-reaching}}$ to
    evaluate $\varphi_L(u')$.
    We are now ready to give the formula $\eta_{d,r}^\uparrow$. We set
    \[
	\eta_{d, r}^\uparrow(x, y) =
	\neg N_{d + 1}(x, y) \wedge
	\exists x' \exists y'
	\Bigl(
	\begin{aligned}[t]
	    & x \leq x' < y' \leq y \wedge x' \match y' \wedge\\
	    & \bigvee_{\substack{r' \in R_L, e \in O_L\\e(r') = r}}
	      \bigl(
		\chi_{d, e}(x, x', y', y) \wedge
		\eta_{0, r'}(x' + 1, y' - 1)
	      \bigl)
	\Bigr)
	\displaypunct{.}
	& \qedhere
	\end{aligned}
    \]
\end{proof}

\subsection{The intermediate case}\label{section intermediate}

The following theorem effectively characterizes the remaining case, namely
those VPLs that are weakly length-synchronous but not length-synchronous
and whose
syntactic morphism is quasi-aperiodic:
such VPLs are shown to be constant-depth equivalent to a non-empty disjoint union 
of intermediate VPLs.
The computability of $k,l\in\N$ with $k\not=l$ 
such that $\SM_{k,l}\reduce L$ is subject of Section~\ref{k und l}.

\begin{thm}\label{thm equiv union}
If $L$ is weakly length-synchronous but not
length-synchronous, and its syntactic morphism
	$(\varphi_L,\psi_L)$ is quasi-aperiodic, 
one can effectively compute vertically visibly pushdown 
	grammars $G_1,\dots,G_m$ generating intermediate VPLs 
	such that $L\eqACO\biguplus_{i\in[1,m]} L(G_i)$.
	\end{thm}
\noindent
Before we give the proof of the theorem we need a bit of notation.
Let $L\subseteq\Sigma^\triangle$ be a VPL that is weakly length-synchronous,
not length-synchronous, and whose syntactic
morphism $(\varphi_L,\psi_L)$ is quasi-aperiodic. 
By Proposition~\ref{prop effectivity} one can effectively compute its syntactic
$\Ext$-algebra $(R_L, O_L)$, $(\varphi_L, \psi_L)$ and $\varphi_L(L)$ from (a
given DVPA for) $L$.

\newcommand{\Sleft}{\widetilde{\Sigma}}
\newcommand{\Sright}{\overline{\Sigma}}
\newcommand{\Zn}{\mathcal{Z}}

For all $e\in O_L$ 
recall that 
\[
\newU_e=\set{(u,v)\in\Con(\Sigma)\mid e\circ\psi_L(\ext_{u,v})=e}.
\]
For all $\varphi_L(L)$-reaching $e\in O_L$ 
and some fresh internal letter
$\#\not\in\Sigma$ let 
\[
M_e=\set{u\#v\mid 
(u,v)\in\Con(\Sigma), \Delta(u)>0, e\circ\psi_L(\ext_{u,v})=e}.
\]
Note that since $L$ is assumed to be weakly length-synchronous, by
Proposition~\ref{ptn:Weak_length-synchronicity_equivalence}, 
$\newU_e$ is
weakly length-synchronous for all $\varphi_L(L)$-reaching $e\in O_L$.

We claim that one can effectively compute an $\Ext$-algebra recognizing
$M_e$.
First note that 
$M_e = \bigcup_{f\in O_L \text{ s.t. } e\circ f=e} L_f \cap
       \set{u\#v\mid uv\in\Sigma^\triangle,\Delta(u)>0}$.
Second note that for all languages $L_f$ 
one can effectively compute $\Ext$-algebras recognizing them 
by Lemma~\ref{lemma Le}. 
Third note that one can effectively compute an $\Ext$-algebra recognizing
the language $\set{u\#v\mid uv\in\Sigma^\triangle,\Delta(u)>0}$.
Thus, an $\Ext$-algebra recognizing $M_e$ can be effectively computed since
the languages recognizable by $\Ext$-algebras are effectively
closed under Boolean operations.
Next recall that the set $S_e=\set{(k,l) \in \N^2 \mid \exists (u, v)\in\Sigma^k\times\Sigma^l:u\#v\in M_e}\subseteq\N_{>0}^2$
is hence effectively semilinear by Lemma~\ref{lemma semilinear}.
Note that the word relation $\newU_e$ is length-synchronous
if, and only if, there exists some $\alpha\in\Q_{>0}$ such that
$\frac{k}{l}=\alpha$ for all $(k,l)\in S_e$.
Lemma~\ref{lemma collinear} implies that the latter condition is decidable.
As a consequence one can effectively compute the set
\[\Zn=\set{e\in O_L\mid e\text{ is $\varphi_L(L)$-reaching and 
$\newU_e$ is not length-synchronous}}.\]
Observe that since $L$ is not length-synchronous by assumption, we 
have $\Zn\not=\emptyset$ (Proposition~\ref{prop length synchronous}).

Let us introduce two fresh copies 
$\Sleft=\set{\widetilde{\sigma}\mid \sigma\in\Sigma}$ 
and $\Sright=\set{\overline{\sigma}\mid \sigma\in\Sigma}$
of our alphabet $\Sigma$.
Let $\widetilde{\vartheta}\colon (\Sleft\cup\Sigma)^*\rightarrow\Sleft^*$ 
and $\overline{\vartheta}\colon (\Sright\cup\Sigma)^*\rightarrow\Sright^*$
be the (letter-to-letter and hence length-multiplying) morphisms
satisfying $\widetilde{\vartheta}(\sigma)=
\widetilde{\vartheta}(\widetilde{\sigma})=\widetilde{\sigma}$ and
$\overline{\vartheta}(\sigma)=
\overline{\vartheta}(\overline{\sigma})=\overline{\sigma}$ for all
$\sigma\in\Sigma$.
Conversely, let $\widetilde{\vartheta}^{-1}\colon (\Sleft\cup\Sigma)^*\rightarrow\Sigma^*$
and $\overline{\vartheta}^{-1}\colon (\Sright\cup\Sigma)^*\rightarrow\Sigma^*$
be the morphisms satisfying 
$\widetilde{\vartheta}^{-1}(\widetilde{\sigma})=
\widetilde{\vartheta}^{-1}(\sigma)=\sigma$ and
$\overline{\vartheta}^{-1}(\overline{\sigma})
= \overline{\vartheta}^{-1}(\sigma)=\sigma$ for all $\sigma\in\Sigma$.

We define a new visibly pushdown alphabet
$\Upsilon=\Upsilon_\call\cup\Upsilon_\internal\cup\Upsilon_\return$
where $\Upsilon_\call=\Sigma_\call$,
$\Upsilon_\internal=\Sigma_\internal\cup\Sleft\cup\Sright\cup\set{\#}$,
and $\Upsilon_\return=\Sigma_\return$.

For every word $u\#v\in M_e$ consider the unique factorization
\[
    u\#v=\ext_{x_1,y_1}\circ\ext_{a_1,b_1}\circ\cdots\circ
    \ext_{x_k,y_k}\circ\ext_{a_k,b_k}\circ\ext_{x_{k+1},y_{k+1}}(\#)
\]
where $k\geq 1$, $x_1,\dots,x_{k+1},y_1,\dots,y_{k+1}\in\Sigma^\triangle$,
$a_1,\dots,a_k\in\Sigma_\call$, and $b_1,\dots,b_k\in\Sigma_\return$. (Existence
and unicity of this factorization basically just follows from the unique stair
factorization of $\ext_{u, v}$ given by Lemma~\ref{lemma ext} and from the fact
that $\Delta(u) > 0$.)
For these we define
\[
    (u\#v)^\ddagger=
    \ext_{\widetilde{\vartheta}(x_1),\overline{\vartheta}(y_1)}
    \circ\ext_{a_1,b_1}\circ\cdots\circ
    \ext_{\widetilde{\vartheta}(x_{k}),\overline{\vartheta}(y_{k})}
    \circ\ext_{a_k,b_k}\circ
    \ext_{\widetilde{\vartheta}(x_{k+1}),\overline{\vartheta}(y_{k+1})}(\emptyword)
    \in\Upsilon^\triangle
    \displaypunct{.}
\]
Notably, $(u\#v)^\ddagger$ does not contain the letter $\#$.
Finally, for all $e\in\Zn$ we define the language
\[
    N_e = \set{(u\#v)^\ddagger\in\Upsilon^*\mid u\#v\in M_e} \cup
	  \set{\emptyword}
    \subseteq\Upsilon^\triangle
    \displaypunct{.}
\]

\begin{samepage}
\begin{rem}\label{remark flat}
Let be $n\in\N$ be the constant from Lemma~\ref{ptn:hill_small} for $L$
and let $e\in\Zn$.
When setting $F=\varphi_L(L)$ and $(\varphi,\psi)=(\varphi_L,\psi_L)$,
Lemma~\ref{ptn:hill_small} states that the factorization
\[
    u\#v=\ext_{x_1,y_1}\circ\ext_{a_1,b_1}\circ\cdots\circ
    \ext_{x_k,y_k}\circ\ext_{a_k,b_k}\circ\ext_{x_{k+1},y_{k+1}}(\#)
\]
of every word $u\#v\in M_e$ 
satisfies
$\length{x_1},\dots,\length{x_{k+1}},\length{y_1},\dots,\length{y_{k+1}}\leq n$.
As a consequence, for the corresponding factorization 
\[
    (u\#v)^\ddagger=
    \ext_{\widetilde{\vartheta}(x_1),\overline{\vartheta}(y_1)}
    \circ\ext_{a_1,b_1}\circ\cdots\circ
    \ext_{\widetilde{\vartheta}(x_{k}),\overline{\vartheta}(y_{k})}
    \circ\ext_{a_k,b_k}\circ
    \ext_{\widetilde{\vartheta}(x_{k+1}),\overline{\vartheta}(y_{k+1})}(\emptyword)
\]
of every word $(u\#v)^\ddagger\in N_e \setminus \set{\emptyword}$
we have 
$\length{\widetilde{\vartheta}(x_1)},\dots,\length{\widetilde{\vartheta}(x_{k+1})},
\length{\overline{\vartheta}(y_1)},\dots,\length{\overline{\vartheta}(y_{k+1})}\leq n$.
\end{rem}
\end{samepage}

\subsubsection{Proof strategy}
We are now ready to give the proof strategy for 
Theorem~\ref{thm equiv union}.
The proof consists of the following steps.

\begin{enumerate}
	\item
		$N_{e}$ is an intermediate VPL for all $e\in\Zn$.
		Moreover, one can effectively compute 
		a vertically visibly pushdown grammar $G_e$
		witnessing that $N_e$ is indeed intermediate
		(Lemma~\ref{L intermediate N}).
\item $N_e\reduce M_e$ for all $e\in\Zn$
	(Lemma~\ref{L uncolor}).
\item $M_e\reduce N_e$ for all $e\in\Zn$
	(Lemma~\ref{L color}).
    \item $M_e\reduce \biguplus_{\text{$f\in O_L$ is $\varphi_L(L)$-reaching}} L_f$ 
	for all $e\in\Zn$
		(Lemma~\ref{L to old}).
	    \item $L\reduce \biguplus_{e\in\Zn} M_e$
	(Lemma~\ref{L main intermediate}).
\end{enumerate}
\noindent
Let us argue that Theorem~\ref{thm equiv union}
follows from the above steps. 
By Point 1 for all $e\in\Zn$
we have that $N_e$ is an intermediate VPL, for which moreover
one can effectively compute a
vertically visibly pushdown grammar
$G_e$ witnessing that $N_e$ is indeed intermediate.
Recalling that $\Zn\not=\emptyset$, it remains to argue that
$L\eqACO\biguplus_{e\in\Zn} N_e$.
Before we prove this let us recall some basics
of constant-depth reductions.
For this, let $K,L_1,\dots,L_n,K_1,\dots,K_n$ be languages.
Firstly, observe that if $L_i\reduce K_i$ for all $i\in[1,n]$, then 
$\biguplus_{i\in[1,n]}L_i\reduce\biguplus_{i\in[1,n]}K_i$.
Secondly, if $L_i\reduce K$ for all $i\in[1,n]$, then
$\biguplus_{i\in[1,n]} L_i \reduce K$.

\begin{samepage}
Hence we obtain the following sequence of reductions showing
that $L\eqACO\biguplus_{e\in\Zn} N_e$.
\begin{eqnarray*}
	L&\stackrel{\text{Point 5}}{\reduce}&
	\biguplus_{e\in\Zn} M_e\\[0.2cm]
	&\stackrel{\text{Point 3}}{\reduce}&
	\biguplus_{e\in\Zn} N_e\\[0.2cm]
	&\stackrel{\text{Point 2}}{\reduce}& 
	\biguplus_{e\in\Zn} M_e\\[0.2cm]
	&\stackrel{\text{Point 4}}{\reduce}& 
	\biguplus_{\text{$f\in O_L$ is $\varphi_L(L)$-reaching}} L_f\\[0.2cm]
	&\stackrel{\text{Lemma~\ref{lemma reduction element}}}{\reduce}& 
	L
	\displaypunct{.}
\end{eqnarray*}
\end{samepage}

\begin{lem}\label{L intermediate N}
$N_{e}$ is an intermediate language for all $e\in\Zn$.
		Moreover, one can effectively compute 
		a vertically visibly pushdown grammar 
		witnessing that $N_e$ is indeed an intermediate VPL.
\end{lem}
\begin{proof}
    Let $e\in\Zn$.
	     For showing that $N_e$ is an intermediate VPL we
	     first show that $N_e$ is quasi-counterfree. 
	     To this end we show that the syntactic morphism
	     $(\varphi_{N_e},\psi_{N_e})$ of $N_e$ is quasi-aperiodic,
	     which is equivalent by Proposition~\ref{prop quasi-counterfree}.

	     Assume by contradiction that $(\varphi_{N_e},\psi_{N_e})$
	     is not quasi-aperiodic. Then there exist
	     $k,l\in\N$ such that $\psi_{N_e}(\ExtMon(\Upsilon^\triangle)^{k,l})$
	     contains a non-trivial group, say $G\subseteq O_{N_e}$.
	     Let $g_0$ be the identity of $G$ and let $g_1\in G$ be
	     such that $g_1\not=g_0$.
	     Thus, we have $g_1^{i+1}\not=g_1^i$ for all $\N_{>0}$.
	     We claim that all $g\in G$ are $\varphi_{N_e}(N_e)$-reaching:
	     indeed, if $g\in G$ were not $\varphi_{N_e}(N_e)$-reaching, 
	     then the same would hold for all $g'\in G$ since $g'=g'g^{-1}g$,
	     hence implying that $g'$ is the (one and only) 
	     zero of $O_{N_e}$, 
	     contradicting that $G$ is non-trivial.
	     Fix some 
	     $\ext_{u_0,v_0},\ext_{u_1,v_1}\in\ExtMon(\Upsilon^\triangle)^{k,l}$
	     such that $\psi_{N_e}(\ext_{u_0,v_0})=g_0$ and
	     $\psi_{N_e}(\ext_{u_1,v_1})=g_1$.
	     Note that we must have $k>0$ or $l>0$ for otherwise we would
	     have $u_0=u_1=\varepsilon$ and $v_0=v_1=\varepsilon$,
	     a contradiction to $g_0\not=g_1$.
We moreover claim that 
	     $u_0,u_1\not\in\widetilde{\Sigma}^*$
	     and hence $v_0,v_1\not\in\overline{\Sigma}^*$:
	     indeed, a consequence of Remark~\ref{remark flat}
	     is that 
	     $\ext_{u,v}\in\psi_{N_e}^{-1}(f)\cap\widetilde{\Sigma}^s\times
	     \overline{\Sigma}^t$ implies $s,t\leq n$ for all
	     $\varphi_{N_e}(N_e)$-reaching $f\in O_{N_e}$,
	     yet the fact that $\psi_{N_e}(\ext_{u^i,v^i})$ is in $G$ and
	     thus $\varphi_{N_e}(N_e)$-reaching for all $i\geq 1$
	     contradicts this (recall that $|u_1|>0$ or $|v_1|>0$).
	     In other words, both $u_0$ and $u_1$ contain at least one
	     letter from $\Sigma_\call$ and both $v_0$ and $v_1$
	     contain a letter from $\Sigma_\return$.

	     Next we claim that neither $u_0$ nor $u_1$ contain any
	     letter from $\Sigma_\return$: indeed, without loss of generality
	     if $u_1$ were to contain a letter from $\Sigma_\return$
	     then $\psi_{N_e}(\ext_{u_1^2,v_1^2})$ would be the zero of 
	     $O_{N_e}$,
	     contradicting that $\psi_{N_e}(\ext_{u_1^2,v_1^2})$
	     is in $G$.
	     It follows that 
	     $u_0,u_1\in(\widetilde{\Sigma}^*\Sigma_\call\widetilde{\Sigma}^*)^+$
	     and hence that
	     $v_0,v_1\in(\overline{\Sigma}^*\Sigma_\return\overline{\Sigma}^*)^+$.
	     
	     Let 
	     $u_0'=\widetilde{\vartheta}^{-1}(u_0)$,
	     $u_1'=\widetilde{\vartheta}^{-1}(u_1)$,
	     $v_0'=\overline{\vartheta}^{-1}(v_0)$,
	     $v_1'=\overline{\vartheta}^{-1}(v_1)$
	     and note that $\ext_{u_0',v_0'},\allowbreak\ext_{u_1',v_1'}
	     \in\ExtMon(\Sigma^{k,l})$.
    Since $\powerset{O_L}$ forms a monoid there exists
    $p \in \N_{>0}$ such that
    \[
	\psi_L(\set{\ext_{u_0', v_0'}, \ext_{u_1', v_1'}})^p =
	\psi_L(\set{\ext_{u_0', v_0'}, \ext_{u_1', v_1'}})^{2 p}
	\displaypunct{.}
    \]
    This implies that for all $i \in \N_{>0}$, we have
    \[
	\psi_L(\ext_{u_0'^{p - 1} u_1', v_1' v_0'^{p - 1}})^i \in
	\psi_L(\set{\ext_{u_0', v_0'}, \ext_{u_1', v_1'}})^{i p} =
	\psi_L(\set{\ext_{u_0', v_0'}, \ext_{u_1', v_1'}})^p
	\displaypunct{.}
    \]
    Hence, as $\length{u_0}=\length{u_1}=\length{u_0'} = \length{u_1'} = k$ and
    $\length{v_0}=\length{v_1}=\length{v_0'} = \length{v_1'} = l$, the semigroup
    $\set{\psi_L(\ext_{u_0'^{p - 1} u_1', v_1' v_0'^{p - 1}})^i \mid
	  i \in \N_{>0}}$
    is contained in $\psi_L(\ExtMon(\Sigma^\triangle)^{k p, l p})$.
    We show that the latter semigroup is not aperiodic by
    showing 
    $\psi_L(\ext_{u_0'^{p - 1} u_1', v_1' v_0'^{p - 1}})^i\not=
    \psi_L(\ext_{u_0'^{p - 1} u_1', v_1' v_0'^{p - 1}})^{i+1}$ for all
    $i\in\N_{>0}$.
    Indeed, since 
    \[\psi_{N_e}\left(\left(\ext_{u_0^{p-1}u_1,v_1v_0^{p-1}}\right)^{i+1}\right)=g_1^{i+1}\not=
    g_1^{i+2}=\psi_{N_e}\left(\left(\ext_{u_0^{p-1}u_1,v_1v_0^{p-1}}
    \right)^{i+2}\right)\]
there exist
    $\ext_{x,y}\in\ExtMon(\Upsilon^\triangle)$ and $w\in\Upsilon^\triangle$
    such that, without loss of generality 
    \[
    \left(\ext_{x,y}\circ
    \left(\ext_{u_0^{p-1}u_1,v_1v_0^{p-1}}\right)^{i+1}\right)(w)\in N_e
    \;\:\text{and}\;\:
    \left(\ext_{x,y}\circ
    \left(\ext_{u_0^{p-1}u_1,v_1v_0^{p-1}}\right)^{i+2}\right)(w)\not\in N_e
    \displaypunct{.}
    \]
Now since 
$u_0,u_1\in(\widetilde{\Sigma}^*\Sigma_\call\widetilde{\Sigma}^*)^+$ there exists a factorization
$u_0^{p-1}u_1wv_1v_0^{p-1}=x'y'$
such that 
\[
    \left(\ext_{\widetilde{\vartheta}^{-1}(x),\overline{\vartheta}^{-1}(y)}\circ
    \left(\ext_{\widetilde{\vartheta}^{-1}(u_0^{p-1}u_1),
    \overline{\vartheta}^{-1}(v_1v_0^{p-1})}\right)^{i}\right)(
    \widetilde{\vartheta}^{-1}(x')\#\overline{\vartheta}^{-1}(y'))\in M_e\]
    and
    \[
    \left(\ext_{\widetilde{\vartheta}^{-1}(x),\overline{\vartheta}^{-1}(y)}\circ
    \left(\ext_{\widetilde{\vartheta}^{-1}(u_0^{p-1}u_1),
    \overline{\vartheta}^{-1}(v_1v_0^{p-1})}\right)^{i+1}\right)(
    \widetilde{\vartheta}^{-1}(x')\#\overline{\vartheta}^{-1}(y'))\not\in M_e\]
By definition of $M_e$ we obtain
\begin{eqnarray*}
	e&=&e\circ\psi_L\left(\ext_{\widetilde{\vartheta}^{-1}(x(u_0^{p-1}u_1)^ix'),
	\overline{\vartheta}^{-1}(y'(v_1v_0^{p-1})^iy)}\right)\\
	&=&e\circ\psi_L\left(
	\ext_{\widetilde{\vartheta}^{-1}(x),\overline{\vartheta}^{-1}(y)}
	\right)
	\circ
	\psi_L\left(
	\ext_{\widetilde{\vartheta}^{-1}(u_0^{p-1}u_1),
	\overline{\vartheta}^{-1}(v_1v_0^{p-1})}\right)^i
	\circ
	\psi_L\left(
	\ext_{\widetilde{\vartheta}^{-1}(x'),\overline{\vartheta}^{-1}(y')}
	\right)
\end{eqnarray*}
and 
\begin{eqnarray*}
	e&\not=&e\circ\psi_L\left(\ext_{\widetilde{\vartheta}^{-1}(x(u_0^{p-1}u_1)^{i+1}x'),
	\overline{\vartheta}^{-1}(y'(v_1v_0^{p-1})^{i+1}y)}\right)\\
	&=&e\circ\psi_L\left(
	\ext_{\widetilde{\vartheta}^{-1}(x),\overline{\vartheta}^{-1}(y)}
	\right)
	\circ
	\psi_L\left(
	\ext_{\widetilde{\vartheta}^{-1}(u_0^{p-1}u_1),
	\overline{\vartheta}^{-1}(v_1v_0^{p-1})}\right)^{i+1}
	\circ
	\psi_L\left(
	\ext_{\widetilde{\vartheta}^{-1}(x'),\overline{\vartheta}^{-1}(y')}
	\right)\!.
\end{eqnarray*}
Thus, we must have 
    $\psi_L(\ext_{u_0'^{p - 1} u_1', v_1' v_0'^{p - 1}})^i\not=
    \psi_L(\ext_{u_0'^{p - 1} u_1', v_1' v_0'^{p - 1}})^{i+1}$,
    as required.
    As this is true for each $i \in \N_{>0}$, the semigroup
    $\set{\psi_L(\ext_{u_0'^{p - 1} u_1', v_1' v_0'^{p - 1}})^i \mid
	  i \in \N_{>0}}$
    that is contained in $\psi_L(\ExtMon(\Sigma^\triangle)^{k p, l p})$ is
    not aperiodic, contradicting quasi-aperiodicity of
    $(\varphi_L, \psi_L)$.

It remains to show that one can compute a vertically visibly pushdown grammar
	     $G_e$ with $L(G_e)=N_e$ such that 
	     $\Rn(G_e)$ is weakly length-synchronous but not length-synchronous.
	     By Remark~\ref{remark flat} each non-empty word in 
	     $N_e$ is of the form
\[
    (u\#v)^\ddagger=
    \ext_{\widetilde{\vartheta}(x_1),\overline{\vartheta}(y_1)}
    \circ\ext_{a_1,b_1}\circ\cdots\circ
    \ext_{\widetilde{\vartheta}(x_{k}),\overline{\vartheta}(y_{k})}
    \circ\ext_{a_k,b_k}\circ
    \ext_{\widetilde{\vartheta}(x_{k+1}),\overline{\vartheta}(y_{k+1})}(\emptyword)
\]
for some $k\geq 1$ and some words
$x_1,\dots,x_{k+1},y_1,\dots,y_{k+1}\in\Sigma^\triangle$
all of which have length at most $n$ such that moreover
$
e\circ\psi_L(\ext_{u,v})=e
$.
We construct the grammar $G_e=(V,\Upsilon,P,S_e)$ as follows.
The set of nonterminals is $V=\set{S_f\mid f\in O_L}$,
$S_e\in V$ is the start nonterminal, 
the set of productions consists of the union of 
$\set{S_e\rightarrow_{G_e}\varepsilon}$ and
\begin{eqnarray*}
	\left\{\biggl.S_f\rightarrow_{G_e}\widetilde{\vartheta}(x_1)a\widetilde{\vartheta}(x_2)S_g
\overline{\vartheta}(y_2)b\overline{\vartheta}(y_1)\right.
	\!\!\!&\bigm|&\!\!\!
f,g\in O_L, x_1,x_2,y_1,y_2\in\Sigma^\triangle,
|x_1|,|x_2|,|y_1|,|y_2|\leq n,
\\
	&&\!\!\!
a\in\Sigma_\call,b\in\Sigma_\return,
	\left.\biggl.f\circ\psi_L(\ext_{x_1ax_2,y_2by_1})=g\right\}.
\end{eqnarray*}
As a consequence we obtain
\begin{eqnarray*}
	\Rn(G_e)=\left\{\biggl.\left(\widetilde{\vartheta}(x_1)\dots a_k\widetilde{\vartheta}(x_{x+1}),
\overline{\vartheta}(y_{k+1})b_k\dots
\overline{\vartheta}(y_1)\right)\right.
	\!\!\!&\big|&\!\!\!
	x_1,\dots,x_{k+1},x_1,\dots,y_{k+1}\in\Sigma^\triangle,\\
	&& \!\!\!a_1,\dots,a_{k}\in\Sigma_\call,
	 b_1,\dots,b_{k}\in\Sigma_\return,\\
	 &&\!\!\!
	 \left.e\circ\psi_L(\ext_{a_1\dots a_{k}x_{k+1},y_{k+1}b_k\dots y_1})=e
	 \biggl.\right\}
\end{eqnarray*}
\begin{eqnarray*}
	\phantom{\Rn(G_e)}=\left\{\biggl.\left(\widetilde{\vartheta}(x_1)\dots a_k\widetilde{\vartheta}(x_{x+1}),
\overline{\vartheta}(y_{k+1})b_k\dots
\overline{\vartheta}(y_1)\right)\right.
	\!\!\!&\big|&\!\!\!
	x_1,\dots,x_{k+1},x_1,\dots,y_{k+1}\in\Sigma^\triangle,\\
	&& \!\!\!a_1,\dots,a_{k}\in\Sigma_\call,
	 b_1,\dots,b_{k}\in\Sigma_\return,\\
	 &&\!\!\!
	 \left.\left(x_1\dots a_{k}x_{k+1},y_{k+1}b_k\dots y_1\right)\in
	 \newU_e\biggl.\right\}\!.
\end{eqnarray*}
It follows that $\Rn(G_e)$ is weakly length-synchronous
since $\newU_e$ is: indeed, if $\newR(G_e)$ were not weakly length-synchronous
and without loss of generality there were 
to exist $(u,v),(u',v')\in\newR(G_e)$ 
with $u=u'$ and $|v|\not=|v'|$,
then both
$(\widetilde{\vartheta}^{-1}(u),\overline{\vartheta}^{-1}(v))$
and 
$(\widetilde{\vartheta}^{-1}(u'),\overline{\vartheta}^{-1}(v'))$ would
be in $\newU_e$ by definition.
Yet
$\widetilde{\vartheta}^{-1}(u)= \widetilde{\vartheta}^{-1}(u')$
and $|\overline{\vartheta}^{-1}(v)|=|v|\not=|v'|=\overline{\vartheta}^{-1}(v')$,
so this would contradict that $\newU_e$ is indeed weakly
length-synchronous.
Analogously it follows that $\newR(G_e)$ is not length-synchronous
since $\newU_e$ is not length-synchronous by assumption.
\end{proof}

\begin{lem}\label{L uncolor}
    $N_e\reduce M_e$ for all $e\in\Zn$.
\end{lem}
\begin{proof}
Assume we are given $w\in\Upsilon^*$. To decide if $w\in N_e$ using
an oracle to $M_e$ we do the following constant-depth computation:
	\begin{enumerate}
	    \item
		Accept if $w = \emptyword$, otherwise continue.
		\item Check if $w=uv$ for some 
			$u\in(\Sleft\cup\Sigma_\call)^*$
			and some 
			$v\in(\Sright\cup\Sigma_\return)^*$,
			reject if this is not the case.
		\item Check whether $u$ can be factorized as $u=x_1a_1\cdots x_ka_kx_{k+1}$,
			where $k\geq 1$, $x_1,\dots,x_{k+1}\in
			\set{x \in \Sleft^* \mid
			    \length{x} \leq n \wedge
			\widetilde{\vartheta}^{-1}(x) \in \Sigma^\triangle}$
			and $a_1,\dots,a_k\in\Sigma_\call$
			and whether $v$ can be factorized
			as $v=y_{l+1}b_ly_l\cdots a_1y_1$,
			where $l\geq 1$, $y_1,\dots,y_{l+1}\in
			\set{y \in \Sright^* \mid
			    \length{y} \leq n \wedge
			\overline{\vartheta}^{-1}(y) \in \Sigma^\triangle}$
			and $b_1,\dots,b_l\in\Sigma_\return$.
			Reject if it is not possible.
			(Observe that this is doable 
			by a constant depth and polynomial
			size circuit family since we test membership in finite
		    sets that do not depend on the input.)
		\item Finally accept if, and only if, the word
			\[\widetilde{\vartheta}^{-1}(x_1)a_1\cdots
			\widetilde{\vartheta}^{-1}(x_{k-1})
			a_k
			\widetilde{\vartheta}^{-1}(x_{k+1})
			\#
			\overline{\vartheta}^{-1}(y_{l+1})b_l
\overline{\vartheta}^{-1}(y_{l+1})
			\cdots b_1
		    \overline{\vartheta}^{-1}(y_{1})\] 
	is in $M_e$.
	\qedhere
	\end{enumerate}
\end{proof}

\begin{lem}\label{L color}
	$M_e\leq N_e$ for all $e\in\Zn$.
\end{lem}
\begin{proof}
	Assume we are given $w\in(\Sigma\cup\set{\#})^*$,
	where $w=w_1\cdots w_m$ and where $w_i\in\Sigma\cup\set{\#}$
	for all $i\in[1,m]$.
	To decide if $w\in M_e$ using
an oracle to $N_e$ we do the following constant-depth computation:
	\begin{enumerate}
		\item Check if $w=u\#v$ for some 
			$u\in\Sigma^+$
			and some $v\in\Sigma^+$,
			reject otherwise.
		\item For all return letters $b\in\Sigma_\return$
			and all positions $j$ within $u$ 
			at which $b$ appears, check whether there exists a
			position $i$ within $u$ such that
			$1 \leq j - i \leq n - 1$ and the infix
			$w_i\cdots w_j$ is in
			$\Sigma^\triangle$.
			(As above, this is doable by
			a constant depth and polynomial
			size circuit family since we check well-matchedness of
			at most a fixed number of words that does not depend on
			the input.)
			Reject if it is not the case.
		\item For all call letters $a\in\Sigma_\call$
			and all positions $i$ within $v$
			at which $a$ appears, check whether there exists a
			position $j$ within $v$ such that
			$1 \leq j - i \leq n - 1$ and the infix
			$w_i \cdots w_j$ is in
			$\Sigma^\triangle$.
			Reject if it is not the case.
		\item For each position $i$ within $u$, compute $P_\call(i)$
			where $P_\call$ is the unary predicate defined by
			$w\models P_\call(i)$ if, and only if,
			$i$ is a position within $u$,
			$w_i\in\Sigma_\call$,
			and there does not exist any position $j$ within $u$
			such that $1 \leq j - i \leq n - 1$ and the infix
			$w_i\cdots w_j$ is in
			$\Sigma^\triangle$.
		\item For each position $j$ within $v$, compute $P_\return(j)$ 
			where $P_\return$ is the unary predicate defined by
			$w\models P_\return(j)$ if, and only if,
			$j$ is a position within $v$,
			$w_j\in\Sigma_\return$,
			and there does not exist any position $i$ within $v$
			such that $1 \leq j - i \leq n - 1$ and the infix
			$w_i\cdots w_j$ is in
			$\Sigma^\triangle$.
		\item Let $1\leq i_1<i_2\dots <i_k\leq|u|$ be an 
			enumeration of 
			$\set{i\in[1,|u|]\mid w\models P_\call(i)}$
			and let $|u|+2\leq 
			j_l<j_{l-1}\dots <j_1\leq m$ 
			be an enumeration
			of $\set{j\in[|u|+2,m]\mid w\models P_\return(j)}$.
			Build
			\[
			u'=\widetilde{\vartheta}
			(w_{1}\cdots w_{i_1-1})w_{i_1}
			\cdots 
			\widetilde{\vartheta}(w_{i_{k-1}+1}\cdots
			w_{i_k-1})w_{i_k}
			\widetilde{\vartheta}(w_{i_{k+1}+1}\cdots
		    w_{|u|})\]
			and
			\[
			v'=\overline{\vartheta}(w_{|u|+2}\cdots
			w_{j_{l+1}-1})
			w_{j_l}\cdots
			\overline{\vartheta}
			(w_{j_2}+1\cdots w_{j_1-1})w_{j_1}
			\overline{\vartheta}
			(w_{j_1}+1\cdots w_{m}).
		    \]
		\item Accept if, and only if, the word $u'v'$ is in $N_e$.
		\qedhere
	\end{enumerate}

\end{proof}

\begin{lem}\label{L to old}
    $M_e\reduce \biguplus_{\text{$f\in O_L$ is $\varphi_L(L)$-reaching}} L_f$
    for all $e\in\Zn$.
\end{lem}
\begin{proof}
	Note that the following equivalence holds:
	\[
	    u\#v \in M_e\quad\Longleftrightarrow\quad
	    \text{$\exists f\in O_L$ that is $\varphi_L(L)$-reaching}:
	    e\circ f=e\wedge u\#v\in L_f\wedge \Delta(u)>0
	    \displaypunct{.}
	\]
	This holds because for $\ext_{u, v} \in \ExtMon(\Sigma^\triangle)$
	satisfying $e \circ \psi_L(\ext_{u, v}) = e$, as $e$ is
	$\varphi_L(L)$-reaching, $\psi_L(\ext_{u, v})$ must also be
	$\varphi_L(L)$-reaching.
	Assume we are given $w\in(\Sigma\cup\set{\#})^*$. 
	To decide if $w\in M_e$ we do the following constant-depth computation using oracles to  
	$\biguplus_{\text{$f\in O_L$ is $\varphi_L(L)$-reaching}} L_f$:
	\begin{enumerate}
		\item Check if $w=u\#v$ for some $u,v\in\Sigma^*$, reject otherwise.
		\item Check if $u\#v\in L_f$ for some $\varphi_L(L)$-reaching $f\in O_L$
			satisfying $e\circ f=e$, reject otherwise.
		\item Finally, accept if, and only if
			for all
			$\varphi_L(L)$-reaching $f\in O_L$ 
			we have $u\#\not\in L_f$.
	\end{enumerate}
	If the second check is successful, then $\psi_L(\ext_{u,v})$
	is necessarily $\varphi_L(L)$-reaching, so in that case when
	$\Delta(u)=0$ it holds that $u \in \Sigma^\triangle$ and
	$\psi_L(\ext_{u, \emptyword})$ is $\varphi_L(L)$-reaching.
	Hence, in combination with the second check, the
	third check is successful if, and only if $\Delta(u)>0$.
\end{proof}

\begin{lem}\label{L main intermediate}
	$L\reduce \biguplus_{e\in\Zn} M_e$.
\end{lem}
\begin{proof}
    By assumption $L$ is weakly length-synchronous but not length-synchronous,
    and its syntactic morphism $(\varphi_L,\psi_L)$ is quasi-aperiodic.
    There is a constant $d_L$ such that all words in $L$ have 
    nesting depth at most $d_L$ by Proposition~\ref{ptn:Bounded_nesting-depth}.

    By the first point of
    Proposition~\ref{ptn:Quasi-aperiodicity_eval_languages}
    we may assume that the evaluation language  $\mathcal{E}_{\varphi_L,r}$ is
    regular and its syntactic morphism is quasi-aperiodic for all
    $\varphi_L(L)$-reaching $r\in R_L$.
    This implies, by~\cite[Theorem~VI.4.1]{Books/Straubing-1994}, that for each
    $\varphi_L(L)$-reaching $r \in R_L$, there exists an
    $\FO_{\Gamma_{\varphi_L}}[<, \MODpred]$-sentence $\nu_{\varphi_L, r}$
    defining $\mathcal{E}_{\varphi_L, r}$.

    As $L$ is not length-synchronous we cannot assume analogous sentences for
    the evaluation languages $\mathcal{E}_{\psi_L,e}$ for all
    $\varphi_L(L)$-reaching $e\in O_L$. 
    Indeed, Remark~\ref{remark not weakly} provides an example of a weakly
    length-synchronous but non-length-synchronous VPL whose syntactic morphism
    is quasi-aperiodic but for which some evaluation language
    $\mathcal{E}_{\psi_L,e}$ for $e \in O_L$ that is $\varphi_L(L)$-reaching
    has a non-quasi-aperiodic syntactic morphism.

    However, let $e \in O_L$ be such that
    $\newR_e =
	\set{(u,v) \in \Con(\Sigma) \mid\psi_L(\ext_{u,v})=e}$ 
    is length-synchronous. Take any $\varphi_L(L)$-reaching idempotent
    $f \in O_L$ such that there exist $g, h \in O_L$ satisfying
    $e = g \circ f \circ h$. There exist
    $\ext_{x_g, y_g}, \ext_{x_h, y_h} \in \ExtMon(\Sigma^\triangle)$ such that
    $\psi_L(\ext_{x_g, y_g}) = g$ and $\psi_L(\ext_{x_h, y_h}) = h$.
	Let $(u, v), (u', v') \in\Con(\Sigma)$ such that
    $\Delta(u), \Delta(u') > 0$ and
    $\psi_L(\ext_{u, v}) = \psi_L(\ext_{u', v'}) = f$. Because $f$ is
    idempotent, we have that
    $\psi_L(\ext_{u^{\length{v'}}, v^{\length{v'}}}) =
     \psi_L(\ext_{u'^{\length{v}}, v'^{\length{v}}}) = f$,
    thus
    $\psi_L(\ext_{x_g u^{\length{v'}} x_h, y_h v^{\length{v'}} y_g}) =
     \psi_L(\ext_{x_g u'^{\length{v}} x_h, y_h v'^{\length{v}} y_g}) = e$.
    Therefore, because of length-synchronicity of 
	$\newR_e$, it follows
    that
    \begin{align*}
	\frac{\length{x_g u^{\length{v'}} x_h}}
	     {\length{y_h v^{\length{v'}} y_g}}
	& =
	\frac{\length{x_g u'^{\length{v}} x_h}}
	     {\length{y_h v'^{\length{v}} y_g}}\\
	\frac{\length{x_g} + \length{v'} \cdot \length{u} + \length{x_h}}
	     {\length{y_h} + \length{v'} \cdot \length{v} + \length{y_g}}
	& =
	\frac{\length{x_g} + \length{v} \cdot \length{u'} + \length{x_h}}
	     {\length{y_h} + \length{v} \cdot \length{v'} + \length{y_g}}\\
	\length{v'} \cdot \length{u}
	& =
	\length{v} \cdot \length{u'}\\
	\frac{\length{u}}{\length{v}}
	& =
	\frac{\length{u'}}{\length{v'}}
	\displaypunct{.}
    \end{align*}
    So we can conclude that
    $\newR_f =
     \set{(u,v) \in \Con(\Sigma)\mid\psi_L(\ext_{u,v})=f}$ 
    is length-synchronous.
    Thus, by the second point of
    Proposition~\ref{ptn:Quasi-aperiodicity_eval_languages} we may assume that
    the evaluation language  $\mathcal{E}_{\psi_L,e}$ is regular and its
    syntactic morphism is quasi-aperiodic for all $\varphi_L(L)$-reaching
    $e\in O_L$ with $\newR_e$ length-synchronous.
    This implies again, by~\cite[Theorem~VI.4.1]{Books/Straubing-1994}, that for
    each $\varphi_L(L)$-reaching $e \in O_L$ with $\newR_e$
    length-synchronous, there exists an
    $\FO_{\Gamma_{\psi_L}}[<, \MODpred]$-sentence $\nu_{\psi_L, e}$ defining
    $\mathcal{E}_{\psi_L, e}$.

    For proving $L\reduce\biguplus_{e\in\Zn} M_e$ we must thus make use of the
    oracles to $\biguplus_{e\in\Zn} M_e$.
    All of the following predicates can be computed by a circuit family of
    constant depth and polynomial size with access to these oracles.
    More concretely, by accessing oracles to $\biguplus_{e\in\Zn} M_e$, for all
    $e\in\Zn$ we may assume that we have a predicate $\pi_e^{\mathsf{exact}}$
    such that for all $w \in \Sigma^+$ and
    $i, i', j', j \in [1, \length{w}], i \leq i' < j' \leq j$ the following
    holds:
    \begin{eqnarray}\label{P 1}
	w\models\pi_e^{\mathsf{exact}}(i,i',j',j)&\quad\Longleftrightarrow\quad&
	w_i\cdots w_{i'} w_{j'} \cdots w_j\in\Sigma^\triangle,\\
	&& e\circ\psi_L(\ext_{w_i\cdots w_{i'},w_{j'}\cdots w_j})=e
	    \text{ and }\nonumber\\&&\Delta(w_i\cdots w_{i'})>0\nonumber
    \end{eqnarray}

    For all $\varphi_L(L)$-reaching $e\in O_L$ that are {\em not in $\Zn$} we
    may assume, by Lemma~\ref{lem:Synchronous_element_matching}, that we have
    the $\FOplus$-definable (and hence constant-depth computable) predicate
    $\pi_e$ at hand. It has the following properties: for all $w \in \Sigma^+$
    and $i, i', j', j \in [1, \length{w}], i \leq i' < j' \leq j$:
    \begin{itemize}
	\item
	    if $w\models \pi_e(i, i', j', j)$, then
	    $w_i \cdots w_{i'} w_{j'} \cdots w_j \in \Sigma^\triangle$ and 
	\item
	    if $w_i \cdots w_{i'} w_{j'} \cdots w_j \in \Sigma^\triangle$,
	    $\Delta(w_i \cdots w_{i'}) > 0$ and
	    $e \circ \psi_L(\ext_{w_i \cdots w_{i'}, w_{j'} \cdots w_j}) = e$,
	    then $w\models \pi_e(i, i', j', j)$.
    \end{itemize}
	\noindent
    We can first build an approximate matching $\mu$ relative to $L$. This is
    done totally analogously as done in Section~\ref{paragraph approximate} by
    replacing the there appearing $\pi_e$ for each $e\in\Zn$ by our predicate
    $\pi_e^{\mathsf{exact}}$: indeed, Remark~\ref{remark up} states that the
    predicates $\pi_e$ from of Lemma~\ref{lem:Synchronous_element_matching}
    could have been replaced by the predicate $\pi_e^{\mathsf{exact}}$.

    Thus, as in the proof of
    Proposition~\ref{ptn:Eval_languages_and_bounded_nesting-depth} we may assume
    that we have full access to the matching relation $M^\Delta(w)$ of our input
    word $w$.

    For verifying if a given word $w\in\Sigma^\triangle$ is in $L$ we follow the
    same approach as the main construction in Section~\ref{Section Evaluation}.
    It is however important to stress that this time we cannot assume
    quasi-aperiodicity of the syntactic morphisms of the evaluation languages
    $\mathcal{E}_{\psi_L,e}$ for all $\varphi_L(L)$-reaching $e\in O_L$.
    Still, we build formulas 
    \begin{itemize}
	\item
	    $\overline{\eta}_{d, r}^\uparrow(x, y)$ for all $d\in[0,d_L]$
	    and all $\varphi_L(L)$-reaching $r \in R_L$ and 
	\item
	    $\overline{\eta}_{d, r}(x, y)$ for all $d\in[0,d_L]$ and all
	    $\varphi_L(L)$-reaching $r\in R_L$
    \end{itemize}
    that will have the following properties (as $\eta_{d,r}$ and
    $\eta_{d,r}^\uparrow$) for all $w\in\Sigma^\triangle$ and all
    $i,j\in[1,\length{w}]$:
    \begin{itemize}
	\item
	    if $i$ is matched to $j$ in $w$, then
	    $(w, \matching{w}) \models \overline{\eta}_{d, r}^\uparrow(i, j)$
	    if, and only if, $\ndepth(w_i \cdots w_j) \leq d$ and
	    $\varphi_L(w_i \cdots w_j) = r$;
	\item
	    if $w_i \cdots w_j \in \Sigma^\triangle$, then
	    $(w, \matching{w}) \models \overline{\eta}_{d, r}(i, j)$ if, and
	    only if, $\ndepth(w_i \cdots w_j) \leq d$ and
	    $\varphi_L(w_i \cdots w_j) = r$.
    \end{itemize}
	\noindent
    It remains to define the formulas $\overline{\eta}_{d,r}$ and
    $\overline{\eta}_{d,r}^\uparrow$ for all $d\in[0,d_L]$ and all
    $\varphi_L(L)$-reaching $r\in R_L$.
    For the definition of the $\overline{\eta}_{0,r}^\uparrow$ and the 
    $\overline{\eta}_{0,r}$ we can simply reuse $\eta_{0,r}$ and
    $\eta_{0,r}^\uparrow$ as in the proof of 
    Proposition~\ref{ptn:Eval_languages_and_bounded_nesting-depth} respectively
    ($\eta_{0,r}$ will make use of our sentence $\nu_{\varphi_L,r}$).
    So let us assume $d>0$.

    We first construct for all $\varphi_L(L)$-reaching $e\in O_L$ a formula
    $\overline{\chi}_{d,e}(x,x',y',y)$ such that for all
    $w \in \Sigma^\triangle$ and
    $i, i', j', j \in [1, \length{w}], i \leq i' < j' \leq j$ we have that if
    $w_i \cdots w_{i'} w_{j'} \cdots w_j \in \Sigma^\triangle$ and $i'$ is
    matched to $j'$ in $w$, then given
    \[
	\ext_{w_i\cdots w_{i'},w_{j'}\cdots w_j}=
	\ext_{x_1,y_1}\circ\ext_{a_1,b_1}\circ\cdots\circ
	\ext_{x_k,y_k}\circ\ext_{a_k,b_k}
    \]
    the stair factorization of $\ext_{w_i\cdots w_{i'},w_{j'}\cdots w_j}$
    provided by Lemma~\ref{lemma ext}, we have
    $(w, \matching{w}) \models \overline{\chi}_{d,e}(i,i',j',j)$ if, and only
    if, $\ndepth(x_\ell), \ndepth(y_\ell) < d$ for all $\ell \in [1, k]$ and
    $\psi_L(\ext_{w_i\cdots w_{i'},w_{j'}\cdots w_j})=e$.
    Given $w, i, i', j', j$ and the associated stair factorization as above,
    note that whenever $\psi_L(\ext_{w_i\cdots w_{i'},w_{j'}\cdots w_j}) = e$, as $e$
    is $\varphi_L(L)$-reaching, then $\varphi_L(x_\ell)$ and
    $\varphi_L(y_\ell)$ are $\varphi_L(L)$-reaching for all $\ell \in [1, k]$.
    If additionally $\ndepth(x_\ell), \ndepth(y_\ell) < d$ for all $\ell \in [1, k]$, we
    can inductively make use of the formulas
    $\set{\overline{\eta}_{d-1,r'}\mid
	  r'\in R_L \text{ $\varphi_L(L)$-reaching}}$
    in order to evaluate
    $\varphi_L(x_1), \varphi_L(y_1), \ldots, \varphi_L(x_k), \varphi_L(y_k)$.

    As expected, the problems are, firstly, that we cannot necessarily access
    our evaluation languages $\mathcal{E}_{\psi_L,e}$ and, secondly, that we
    have to build a formula that may not depend on $k$.
    As in Section~\ref{paragraph approximate} we define the product
    \[
	e_{\ell,\ell'}=
	\psi_L(\ext_{x_\ell,y_\ell}\circ\ext_{a_{\ell},b_{\ell}}\circ\cdots\circ
	       \ext_{x_{\ell'},y_{\ell'}}\circ\ext_{a_{\ell'},b_{\ell'}})\quad
	\text{and}\quad
	e_\ell=e_{1,\ell}
    \]
    for all $\ell,\ell'\in[1,k]$.
    We say an interval $I=[s,t]\subseteq[1,k]$ is
    \emph{repetitive} if $s<t$ and $e_s=e_t$.

    By Claim~\ref{claim pigeon} there exist indices
    $1=t_0\leq s_1<t_1<s_2<t_2<\cdots<s_q<t_q\leq s_{q+1}=k$ such that
    $[s_1,t_1],\ldots, [s_q,t_q]$ are all repetitive and for
    $D_0=[t_0,s_1],D_1=[t_1,s_2],\dots,D_q=[t_q,s_{q+1}]$ we have
    $q+\sum_{p=0}^q\card{D_p}\leq 3\card{O_L}$.
    Let $i=i_1<\cdots<i_k$ and $j_k<\cdots<j_1=j$ be the positions that
    correspond to the positions of the letters $a_1,\dots,a_k\in\Sigma_\call$
    and $b_k,\dots, b_1\in\Sigma_\return$ of the factorization of
    $\ext_{w_i \cdots w_{i'}, w_{j'} \cdots w_j}$ in $w$, respectively: more
    precisely $i_\ell=i+\length{x_1\cdots a_{\ell-1}x_\ell}$ and
    $j_\ell=\length{x_1 a_1 \cdots x_k a_k w_{i'+1} \cdots w_{j'-1}
		    b_k y_{k} \cdots b_{\ell+1} y_{\ell+1}}+1$
    for all $\ell\in[1,k]$.
    Since the non-empty interval $[s_p,t_p]$ is repetitive for all $p\in[1,q]$, 
    we have $e_{s_p}=e_{t_p}$ and thus obtain
    \[
	e_{s_p}=e_{t_p}=
	e_{s_p}\circ\psi(\ext_{x_{s_p+1}\cdots a_{t_p},b_{t_p}\cdots y_{s_p+1}})
	\displaypunct{.}
    \]
    Given $p \in [1, q]$, if $e_{s_p} \in \Zn$, we can use the predicate
    $\pi_{e_{s_p}}^{\mathsf{exact}}$ to check the above equality; we set
    $\theta_{d, e_{s_p}}(x, x', y', y) =
     \pi_{e_{s_p}}^{\mathsf{exact}}(x, x', y', y)$.
    If $e_{s_p}\notin\Zn$ and is $\varphi_L(L)$-reaching, then
    $\newU_{e_{s_p}}$ is length-synchronous, so for all $e' \in O_L$ such
    that $e_{s_p} \circ e' = e_{s_p}$, we have that $e'$ is
    $\varphi_L(L)$-reaching and
    $\newR_{e'} = \set{(u,v) \in \Con(\Sigma) \mid \psi_L(\ext_{u,v})=e'}$ 
    is length-synchronous. So to check the above equality, we can use the
    formula $\theta_{d, e_{s_p}}(x, x', y', y)$ built by taking the disjunction
    over all $e' \in O_L$ such that $e_{s_p} \circ e' = e_{s_p}$ of the formulas
    defined inductively completely analogously as $\chi_{d, e'}(x, x', y', x')$
    in Section~\ref{Section Evaluation} (using the sentence $\nu_{\psi_L, e'}$
    defining $\mathcal{E}_{\psi_L, e'}$): we simply replace every occurrence of
    $\eta_{d-1,r}$ by $\overline{\eta}_{d-1,r}$.

    Next, for all $m>0$ and all $\varphi_L(L)$-reaching $f\in O_L$ we will 
    construct a formula $\alpha_{d, m, f}(x,x',y',y)$ such that for all
    $w \in \Sigma^\triangle$ and
    $i, i', j', j \in [1, \length{w}], i \leq i' < j' \leq j$ we have that if
    $w_i \cdots w_{i'} w_{j'} \cdots w_j \in \Sigma^\triangle$ and $i'$ is
    matched to $j'$ in $w$, then given
    \[
	\ext_{w_i\cdots w_{i'},w_{j'}\cdots w_j}=
	\ext_{x_1,y_1}\circ\ext_{a_1,b_1}\circ\cdots\circ
	\ext_{x_k,y_k}\circ\ext_{a_k,b_k}
    \]
    the stair factorization of $\ext_{w_i\cdots w_{i'},w_{j'}\cdots w_j}$
    provided by Lemma~\ref{lemma ext}, we have
    $(w, \matching{w}) \models \alpha_{d,m,f}(i,i',j',j)$ if, and only
    if, $\ndepth(x_\ell), \ndepth(y_\ell) < d$ for all $\ell \in [1, k]$,
    $\Delta(w_i \cdots w_{i'})=-\Delta(w_{j'} \cdots w_j)=m$ and
    $\psi_L(\ext_{w_i\cdots w_{i'},w_{j'}\cdots w_j})=e$.
    For 
    $\bm{\sigma}=(\bm{\sigma}_1,\dots,\bm{\sigma}_m)\in{\Sigma_\call}^m$,
    $\bm{\xi}=(\bm{\xi}_1\dots,\bm{\xi}_m)\in{\Sigma_\return}^m$,
    $\bm{r}=(\bm{r}_1,\dots,\bm{r}_m)\in {R_L}^m$, and 
    $\bm{r}^\dagger=(\bm{r}_1^\dagger,\dots,\bm{r}_m^\dagger)\in {R_L}^m$
    we define
    \[
	\prod(\bm{\sigma},\bm{\xi},\bm{r},\bm{r}^\dagger)=
	\bigcirc_{g=1}^{m}\leftmult{\bm{r}_g}\circ
	\rightmult{\bm{r}_g^\dagger}\circ
	\psi_L(\ext_{\bm{\sigma}_g,\bm{\xi}_g}).
    \]
    The formula $\alpha_{d, m, f}$ can be expressed as follows:
    \begin{align*}
	\alpha_{d,m,f}(x,x',y',y) = &
	\bigvee_{\substack{\bm{\sigma} \in {\Sigma_\call}^m,
			   \bm{\xi}\in {\Sigma_\return}^m\\
			   \bm{r},\bm{r}^\dagger\in {R_L}^m:
			   f=\prod(\bm{\sigma},\bm{\xi},\bm{r},\bm{r}^\dagger)}}
	\exists x_1,\dots, x_m\exists y_1,\dots,y_m\\
	& \biggl(
	  \begin{aligned}[t]
	    & x'=x_m \wedge y'=y_m \wedge
	      x\leq x_1<\cdots<x_m<y_m<\cdots<y_1\leq y\wedge\\
	    & \bigwedge_{g=1}^m
	      \bigl(\bm{\sigma}_g(x_g) \wedge \bm{\xi}_g(y_g) \wedge
		    x_g \match y_g\bigr) \wedge\\
	    & \forall z \bigl((x \leq z \leq x' \wedge
			       \bigwedge_{g=1}^m z \neq x_g) \rightarrow
			      \neg U(x, x', z)\bigr) \wedge\\
	    & \Bigl(\bigl(x < x_1 \wedge
			  \overline{\eta}_{d-1,\bm{r}_1}(x,x_1-1)\bigr) \vee
		    F_{\bm{r}_1}(x, x_1)\Bigr) \wedge\\
	    & \Bigl(\bigl(y_1 < y \wedge
			  \overline{\eta}_{d-1,\bm{r}_1^\dagger}(y_1+1,y)\bigr)
		    \vee
		    F_{\bm{r}_1^\dagger}(y_1, y)\Bigr) \wedge\\
	    & \bigwedge_{g=2}^m \overline{\eta}_{d-1,\bm{r}_g}(x_{g-1}+1,x_g-1)
	      \wedge
	      \bigwedge_{g=2}^m
	      \overline{\eta}_{d-1,\bm{r}_g^\dagger}(y_g+1,y_{g-1}-1)
	    \biggr)
	  \end{aligned}
    \end{align*}
    with
    $F_s(x, z) =
     \begin{cases}
	x = z & \text{if $s = 1_{R_L}$}\\
	\bot & \text{otherwise}
     \end{cases}$
    for all $\varphi_L(L)$-reaching $s \in R_L$.

    We are now ready to define the formula $\overline{\chi}_{d,e}$:
    \begin{eqnarray*}
	\overline{\chi}_{d,e}(x,x',y',x) & = & 
	\bigvee_{\substack{q\in[0,\card{O_L}]\\
			   m_0, \dots, m_{q-1}\geq 1 \wedge m_q \geq 0:\\
			   q+m_0+\dots+m_q\leq 3\card{O_L}}}
	\quad
	\bigvee_{\substack{e_0,f_0,e_1,\dots,f_{q-1},e_q,f_q\in O_L\\
			   e_0=1_{O_L}\wedge\forall j\in[1,q]:e_j=e_{j-1}\circ f_{j-1}\\
			   e=e_q\circ f_q}}\\
	&&
		\exists x_1, \dots, x_{q+1}\exists x_0', \dots, x_q'
		\exists y_1, \dots, y_{q+1}\exists y_1', \dots, y_0'\\
		&&\Biggl(
		\begin{aligned}
		    x_0'\leq x_1<x_1'<x_2<\dots<x_q' \leq x_{q+1}<\\
		    y_{q+1} \leq y_q'<y_q<\dots<y_1'<y_1\leq y_0'
		\end{aligned}\wedge \\
		&&\quad x_0'=x\wedge y_0'=y\wedge x_{q+1}=x'\wedge y_{q+1}=y'\wedge\\
		&&\quad \bigwedge_{p=1}^q (x_p \match y_p \wedge x_p' \match y_p') \wedge\\
		&&\quad
		\forall z \Bigl(\bigl(x \leq z < x' \wedge
					    \bigwedge_{p=1}^q (z \neq x_p \wedge
					z \neq x_p')\bigr) \rightarrow
		       \neg U(x, x', z)\Bigr) \wedge\\
		&&\quad
		\bigwedge_{p=1}^q\theta_{d, e_p}(x_p+1,x_p',y_p',y_p-1)
		\wedge\\
		&&\quad
		\alpha_{d, m_0, f_0}(x_0',x_1,y_1,y_0') \wedge
		\bigwedge_{p=1}^{q-1} 
		\alpha_{d, m_p,f_p}(x_p'+1,x_{p+1},y_{p+1},y_p'-1)
		\wedge\\
		&&\quad
		\Bigl(
		\begin{aligned}[t]
		    \bigl(x_q' < x_{q+1} \wedge
			    \alpha_{d, m_q, f_q}(x_q'+1,x_{q+1},y_{q+1},y_q'-1)\bigr)
		    \vee\\
		    F'_{q, m_q, f_q}(x_q',x_{q+1})\Bigr)
		\Biggr)
		\end{aligned}
    \end{eqnarray*}
    with
    $F'_{q, m, f}(x, z) =
     \begin{cases}
	\top & \text{if $q = 0$}\\
	x = z & \text{if $q > 0$, $m = 0$ and $f = 1_{O_L}$}\\
	\bot & \text{otherwise}
     \end{cases}$
    for all $q \in [0, \card{O_L}]$, $m \in [0, 3\card{O_L}]$ and $f \in O_L$.

    The inductive definition of $\overline{\eta}_{d,r}^\uparrow$ is completely
    analogous to the definition of $\eta_{d,r}$ in
    Section~\ref{Section Evaluation}: we simply replace every occurrence of
    $\eta_{d-1,r}$ by $\overline{\eta}_{d-1,r}$ and every occurrence of
    $\chi_{d,e}$ by $\overline{\chi}_{d,e}$.

    The inductive definition of $\overline{\eta}_{d,r}$ is completely analogous
    to the definition of $\eta_{d,r}$ in Section~\ref{Section Evaluation}: we
    access the horizontal evaluation languages $\mathcal{E}_{\varphi_L,r}$ for
    all $\varphi_L(L)$-reaching $r\in R_L$ by making use of the sentence
    $\nu_{\varphi, r}$ and the already defined $\overline{\eta}_{d,r}^\uparrow$.
\end{proof}

\subsubsection{Computation of $k,l$}\label{k und l}

The following proposition implies the computability of $k,l\in\N$
such that $\SM_{k,l}\reduce L$ already when VPL $L$ is
weakly length-synchronous but not length-synchronous.

\begin{prop}\label{prop SM lowerbound}
    If a VPL $L$ is weakly length-synchronous but not length-synchronous, one can effectively compute $k, l \in \N_{>0}$ with $k \neq l$ such that
$\SM_{k, l}\reduce L$ .
\end{prop}
\begin{proof}
    Let 
    $L \subseteq \Sigma^\triangle$ be a weakly length-synchronous VPL that is not
    length-synchronous. 
	According to Point 2 (b) of Proposition~\ref{prop effectivity} 
	one can effectively
	compute a quadruple $(k_0,l_0,k_0',l_0')\in\N_{>0}^4$ 
	for which there exist 
    $\ext_{u, v}, \ext_{u', v'} \in \ExtMon(\Sigma^\triangle)$ such that
	\begin{itemize} 
		\item $|u|=k_0$, $|v|=l_0$, $|u'|=k_0'$, $|v'|=l_0'$,
		\item $\psi_L(\ext_{u, v}) = \psi_L(\ext_{u', v'})$ is a
    $\varphi_L(L)$-reaching idempotent,
			\item $\Delta(u), \Delta(u') > 0$,
				and
			\item $\frac{k_0}{l_0}=\frac{\length{u}}{\length{v}} \neq 
	\frac{\length{u'}}{\length{v'}}=\frac{k_0'}{l_0'}$.
	\end{itemize}
    We can explicitly compute such $\ext_{u, v}$ and $\ext_{u', v'}$ by just
    doing an exhaustive search.
    This enables us to assume without loss of generality while maintaining
    effective computability that $\Delta(u)=\Delta(u')$: indeed, 
    in case $\Delta(u)\not=\Delta(u')$, we can consider
    $\ext_{u_1,v_1}=\ext_{u^{\Delta(u')},v^{\Delta(u')}}$
    and $\ext_{u_2,v_2}=\ext_{(u')^{\Delta(u)},(v')^{\Delta(u)}}$
    satisfying the desired properties.

    Let us now define Green's relations on $O_L$
    (see~\cite[Chapter~3, Section~1]{Books/Pin-1986}).
    Let us consider two elements $x,y$ of $O_L$.
    \begin{itemize}
	\item
	    We write $x \GreenJleq y$ whenever there are elements $e,f$ of $O_L$
	    such that $x = e \circ y \circ f$. We write $x \GreenJ y$ if
		    $x \GreenJleq y$ and $y \GreenJleq x$. We write
	    $x \GreenJlt y$ if $x \GreenJleq y$ and $x \not\GreenJ y$.
	\item
	    We write $x \GreenRleq y$ whenever there is an element $e$ of $O_L$
	    such that $x = y \circ e$. We write $x \GreenR y$ if
	    $x \GreenRleq y$ and $y \GreenRleq x$.
	\item
	    We write $x \GreenLleq y$ whenever there is an element $e$ of $O_L$
	    such that $x = e \circ y$. We write $x \GreenL y$ if
	    $x \GreenLleq y$ and $y \GreenLleq x$.
	\item
	    We write $x \GreenH y$ if $x \GreenR y$ and $x \GreenL y$.
    \end{itemize}
	\noindent
    Observe that because $\Delta(u) = \Delta(u')$, we have that
    $u v' \in \Sigma^\triangle$ and $u' v \in \Sigma^\triangle$, so that we can
    consider the elements
    $\ext_{u u u, v v' v} \!=\! \ext_{u, v} \circ \ext_{u, v'} \circ \ext_{u, v}$
    and
    $\ext_{u u' u, v v v} \!= \ext_{u, v} \circ \ext_{u', v} \circ \ext_{u, v}$ in
    $\ExtMon(\Sigma^\triangle)$.
    These elements satisfy
    $\psi_L(\ext_{u u u, v v' v}) \GreenJleq \psi_L(\ext_{u, v})$ and
    $\psi_L(\ext_{u u' u, v v v}) \GreenJleq \psi_L(\ext_{u, v})$. We claim that we
    actually have $\psi_L(\ext_{u u u, v v' v}) \GreenJlt \psi_L(\ext_{u, v})$ and
    $\psi_L(\ext_{u u' u, v v v})$ $\GreenJlt \psi_L(\ext_{u, v})$.
    Indeed, assume we had
    $\psi_L(\ext_{u u' u, v v v}) \GreenJ \psi_L(\ext_{u, v})$.
    Set $x = \psi_L(\ext_{u, v})$ and $y = \psi_L(\ext_{u', v})$.
    By a classical property of Green's relations
    (see~\cite[Chapter~3, Proposition~1.4]{Books/Pin-1986}), since it would hold
    that $x \circ y \circ x \GreenRleq x$ and $x \circ y \circ x \GreenJ x$, we
    would have $x \circ y \circ x \GreenR x$ and dually, since it would hold
    that $x \circ y \circ x \GreenLleq x$ and $x \circ y \circ x \GreenJ x$, we
    would have $x \circ y \circ x \GreenL x$.
    Therefore, we would have $x \circ y \circ x \GreenH x$. By another classical
    result on Green's relations~\cite[Chapter~3, Corollary~1.7]{Books/Pin-1986},
    as $x$ is an idempotent, its $\GreenH$-class is a group, hence for
    $\omega \in \N_{>0}$ the idempotent power of $O_L$, we would have
    $(x \circ y \circ x)^\omega = x^\omega = x$ (as the only idempotent element
    in a group is the identity). This would finally entail that
    $\psi_L(\ext_{(u u' u)^\omega, (v v v)^\omega}) =
     \psi_L(\ext_{(u u u)^\omega, (v v v)^\omega})$
    is a $\varphi_L(L)$-reaching idempotent and
    $\Delta((u u' u)^\omega) = \Delta((u u u)^\omega) > 0$ but
    $\length{(u u' u)^\omega} \neq \length{(u u u)^\omega}$, a contradiction to
    the fact that $(\varphi_L, \psi_L)$ is $\varphi_L(L)$-weakly-length-synchronous.
    Symmetrically, we can prove that if we had
    $\psi_L(\ext_{u u u, v v' v}) \GreenJ \psi_L(\ext_{u, v})$, this would
    contradict the fact that $(\varphi_L, \psi_L)$ is
    $\varphi_L(L)$-weakly-length-synchronous.

    We distinguish three cases. In each of these we prove that there exist
    $k, l \in \N_{>0}, k \neq l$ such that
    $\SM_{k, l} \reduce L_{\psi_L(\ext_{u, v})}$, so
    that since $L_{\psi_L(\ext_{u, v})} \reduce L$ (by
    Lemma~\ref{lemma reduction element}) and by transitivity of $\reduce$ we
    have $\SM_{k, l} \reduce L$.

    \noindent
    \paragraph*{Case $\length{v} = \length{v'}$.}
    In that case, we necessarily have $\length{u} \neq \length{u'}$.
    Then, we can exploit the fact that matching $u^3$ with $v v' v$ or $u u' u$
    with $v^3$ makes us fall down to a smaller $\GreenJ$-class to reduce
    $\SM_{3 \length{u}, 2 \length{u} + \length{u'}}$ to $L_{\psi_L(\ext_{u, v})}$.
    The constant-depth reduction works as follows on input $w \in \Sigma^*$:
    \begin{enumerate}
	\item
	    check if $w = x y$ with
	    $x \in (ac^{3 \length{u} - 1} +
		    ac^{2 \length{u} + \length{u'} - 1})^*$
	    and $y \in (b_1 + b_2)^*$,
	    reject if it's not the case;
	\item
	    build $x'$ by sending $ac^{3 \length{u} - 1}$ to $u^3$,
	    $ac^{2 \length{u} + \length{u'} - 1}$ to $u u' u$ and
	    $y'$ by sending $b_1$ to $v^3$ and $b_2$ to $v v' v$;
	\item
	    accept whenever $x' \# y' \in L_{\psi_L(\ext_{u, v})}$.
    \end{enumerate}
    This forms a valid reduction. Indeed, take a word $w = x y$ with
    $x \in (ac^{3 \length{u} - 1} + ac^{2 \length{u} + \length{u'} - 1})^n$ for
    $n \in \N$ and $y \in (b_1 + b_2)^m$ for $m \in \N$ and consider $x' \# y'$
    produced by the reduction with $x' \in (u^3 + u u' u)^n$ and
    $y' \in (v^3 + v v' v)^m$. If
    $w \in \SM_{3 \length{u}, 2 \length{u} + \length{u'}}$, then it easily
    follows that $x' \# y' \in L_{\psi_L(\ext_{u, v})}$. Otherwise, if
    $w \notin \SM_{3 \length{u}, 2 \length{u} + \length{u'}}$, then either
    $n \neq m$ and thus $x' y'$ is not well-matched because
    $\Delta(x') = n \cdot 3 \cdot \Delta(u)$ and
    $\Delta(y') = m \cdot 3 \cdot \Delta(v)$, or $n = m$ and thus $x' y'$ is
    well-matched, so
    $\ext_{x', y'} = \ext_{z_1', t_1'} \circ \cdots \circ \ext_{z_n', t_n'}$
    with $z_1', \ldots, z_n' \in \set{u^3, u u' u}$ and
    $t_1', \ldots, t_n' \in \set{v^3, v v' v}$ such that there exists
    $i \in [1, n]$ satisfying
    $\ext_{z_i', t_i'} \in \set{\ext_{u^3, v v' v}, \ext_{u u' u, v^3}}$,
    thereby implying
    \[
	\psi_L(\ext_{x', y'}) \GreenJleq \psi_L(\ext_{z_i', t_i'}) \GreenJlt
	\psi_L(\ext_{u, v})
	\displaypunct{.}
    \]
Our algorithm therefore outputs the pair $(k,l)=(3k_0,2k_0+k_0')$.
    \noindent
    \paragraph*{Case $\length{u} = \length{u'}$.}
    This case is symmetric to the previous case.
    Our algorithm outputs the pair $(k,l)=(2l_0+l_0',3l_0)$.

    \noindent
    \paragraph*{Case $\length{u} \neq \length{u'}$ and
		       $\length{v} \neq \length{v'}$.}

    Then, we can again exploit the fact that matching $u^3$ with $v v' v$ or
    $u u' u$ with $v^3$ makes us fall down to a smaller $\GreenJ$-class to
    reduce $\SM_{A \cdot B', A' \cdot B}$ where $A = 3 \length{u}=3k_0$,
    $A' = 2 \length{u} + \length{u'}=2k_0+k_0'$, $B = 3 \length{v}=3l_0$ and
    $B' = 2 \length{v} + \length{v'}=2l_0+l_0'$ to $L_{\psi_L(\ext_{u, v})}$.
    Indeed, we have $A \cdot B' \neq A' \cdot B$ because otherwise we would have
    \begin{align*}
	3 \length{u} \cdot (2 \length{v} + \length{v'}) & =
	(2 \length{u} + \length{u'}) \cdot 3 \length{v}\\
	6 \length{u} \length{v} + 3 \length{u} \length{v'} & =
	6 \length{u} \length{v} + 3 \length{u'} \length{v}\\
	\length{u} \length{v'} & = \length{u'} \length{v}
	\displaypunct{.}
    \end{align*}
    The constant-depth reduction works as follows on input $w \in \Sigma^*$:
    \begin{enumerate}
	\item
	    check if $w = x y$ with
	    $x \in (ac^{A \cdot B' - 1} + ac^{A' \cdot B - 1})^*$ and
	    $y \in (b_1 + b_2)^*$, reject if it is not the case;
	\item
	    build $x'$ by sending $ac^{A \cdot B' - 1}$ to $(u^3)^{B'}$,
	    $ac^{A' \cdot B - 1}$ to $(u u' u)^B$ and $y'$ by sending $b_1$ to
	    $(v^3)^{B'}$ and $b_2$ to $(v v' v)^B$;
	\item
	    accept whenever $x' \# y' \in L_{\psi_L(\ext_{u, v})}$.
    \end{enumerate}
    This forms a valid reduction. Indeed, take a word $w = x y$ with
    $x = z_1 \cdots z_n$ where $n \in \N$ and
    $z_1, \ldots, z_n \in \set{ac^{A \cdot B' - 1}, ac^{A' \cdot B - 1}}$
    and $y = t_1 \cdots t_m$ where $m \in \N$ and
    $t_1, \ldots, t_m \in \set{b_1, b_2}$. Consider $x' \# y'$ produced by the
    reduction with $x' = z_1' \cdots z_n'$ where
    $z_1', \ldots, z_n' \in \set{(u^3)^{B'}, (u u' u)^B}$ and
    $y' = t_1' \cdots t_m'$ where
    $t_1', \ldots, t_m' \in \set{(v^3)^{B'}, (v v' v)^B}$.
    If $w \in \SM_{A \cdot B', A' \cdot B}$,
    then it easily follows that $x' \# y' \in L_{\psi_L(\ext_{u, v})}$. Otherwise,
    if $w \notin \SM_{A \cdot B', A' \cdot B}$, three situations can occur.
    \begin{itemize}
	\item
	    There exists $i \in [1, \min\set{n, m}]$ such that
	    $z_1 \cdots z_{i - 1} t_{i - 1} \cdots t_1 \in
	     \SM_{A \cdot B', A' \cdot B}$
	    but it holds that
	    $(z_i, t_i) \in
	     \set{(ac^{A \cdot B' - 1}, b_2), (ac^{A' \cdot B - 1}, b_1)}$.
	    Assume first $(z_i, t_i) = (ac^{A' \cdot B - 1}, b_1)$.
	    In this case, let
	    $\tilde{x}' = (u u' u)^{B - 1} z_{i + 1}' \cdots z_n'$ and 
	    $\tilde{y}' = t_n' \cdots t_{i + 1}' (v^3)^{B' - 1}$.
	    If $\Delta(\tilde{x}' \tilde{y}') \neq 0$, then
	    \[
		\Delta(x' y') =
		\Delta(z_1' \cdots z_{i - 1}' (u u' u) v^3
		       t_{i - 1}' \cdots t_1') +
		\Delta(\tilde{x}' \tilde{y}') =
		\Delta(\tilde{x}' \tilde{y}') \neq 0
		\displaypunct{,}
	    \]
	    thus $x' y'$ is not well-matched.
	    Otherwise, if $\Delta(\tilde{x}' \tilde{y}') = 0$, we can show that
	    $\tilde{x}' \tilde{y}'$ is well-matched. Indeed, since
	    $u v \in \Sigma^\triangle$, for all $j \in [1, \length{u}]$, we have
	    $\Delta(u_1 \cdots u_j) \geq 0$ and for all $j \in [1, \length{v}]$,
	    we have
	    $\Delta(v_j \cdots v_{\length{v}}) =
	     -\Delta(u v_1 \cdots v_{j - 1}) \leq 0$.
	    Similarly, since $u' v' \in \Sigma^\triangle$, for all
	    $j \in [1, \length{u'}]$, we have $\Delta(u'_1 \cdots u'_j) \geq 0$
	    and for all $j \in [1, \length{v'}]$, we have
	    $\Delta(v'_j \cdots v'_{\length{v'}}) =
	     -\Delta(u' v'_1 \cdots v'_{j - 1}) \leq 0$.
	    This implies that for all $j \in [1, \length{\tilde{x}'}]$, we have
	    $\Delta(\tilde{x}'_1 \cdots \tilde{x}'_j) \geq 0$
	    and for all $j \in [1, \length{\tilde{y}'}]$, we have
	    $\Delta(\tilde{x}' \tilde{y}'_1 \cdots \tilde{y}'_{j - 1}) =
	     -\Delta(\tilde{y}'_j \cdots \tilde{y}'_{\length{\tilde{y}'}}) \geq
	     0$.
	    Therefore, $\tilde{x}' \tilde{y}' \in \Sigma^\triangle$.
	    Hence $x' y'$ is well-matched and
	    \[
		\ext_{x', y'} =
		\ext_{z_1' \cdots z_{i - 1}', t_{i - 1}' \cdots t_1'} \circ
		\ext_{z_i', t_i'} \circ \ext_{\tilde{x}', \tilde{y}'}
		\displaypunct{,}
	    \]
	    so that
	    \[
		\psi_L(\ext_{x', y'}) \GreenJleq \psi_L(\ext_{z_i', t_i'}) \GreenJlt
		\psi_L(\ext_{u, v})
		\displaypunct{.}
	    \]

	    If we assume that $(z_i, t_i) = (ac^{A \cdot B' - 1}, b_2)$, then we
	    prove in the same way that either $x' y'$ is not well-matched or it
	    is well-matched and
	    $\psi_L(\ext_{x', y'}) \GreenJlt \psi_L(\ext_{u, v})$.
	\item
	    It holds that $n < m$ and
	    $z_1 \cdots z_n t_n \cdots t_1 \in \SM_{A \cdot B', A' \cdot B}$.
	    This entails that
	    $\Delta(x' t_n' \cdots t_1') =
	    \Delta(z_1' \cdots z_n' t_n' \cdots t_1') = 0$,
	    so that
	    \[
		\Delta(x' y') =
		\Delta(x' t_n' \cdots t_1') + \Delta(t_m' \cdots t_{n + 1}') =
		\Delta(t_m' \cdots t_{n + 1}') < 0
	    \]
	    because $m > n$ and $\Delta(v) < 0$ as well as $\Delta(v') < 0$.
	    Therefore, $x' y'$ is not well-matched.
	\item
	    It holds that $n > m$ and
	    $z_1 \cdots z_m t_m \cdots t_1 \in \SM_{A \cdot B', A' \cdot B}$.
	    Symmetrically to the previous case, we can also show that then,
	    $x' y'$ is not well-matched.
    \end{itemize}
    Hence, our algorithm outputs 
    the pair 
    $(k,l)=(A\cdot B',A'\cdot B)=(3k_0(2l_0+l_0'),(2k_0+k_0')3l_0)$
    in this last case.
	    \qedhere
\end{proof}

\subsection{Proof of Corollary~\ref{corollary OCA}}\label{S Corollary}
	Let $A=(Q,\Sigma,q_0,F,\delta_0,\dots,\delta_m)$ be a $m$-VCA 
	and let $L=L(A)$.
	One easily computes from $A'$ a DVPA such that $L(A')=L$.
	Details of this standard translation are omitted.
	It will be sufficient to prove that $L$ is weakly length-synchronous
	if, and only if, $L$ is length-synchronous: indeed,
	one can simply perform the case distinction of Section~\ref{S outline}
	and observe that, under the assumption that weak length-synchronicity
	and length synchronicity coincide, the algorithm for 
	Theorem~\ref{thm main} 
	will either output that $L$ is in $\ACO$ or some
	$m\geq 2$ such that $\MOD_m\reduce L$.

	\newcommand{\Mn}{\mathcal{M}}
	It thus suffices to prove that if $L$ is not length-synchronous, then
	$L$ is not weakly length-synchronous.
	Let $(R_L,O_L)$ be the syntactic $\Ext$-algebra of $L$ along with
	with its syntactic morphism 
	$(\varphi_L,\psi_L)\colon (\Sigma^\triangle,\ExtMon(\Sigma^\triangle))
	\rightarrow(R_L,O_L)$.

	Let $\Mn=[0,m]\times (Q^Q)^{[0,m]}\times (Q^Q)^{[0,m]}$.
	The behavior of the $m$-VCA can be described 
	as follows. 
	To each $\ext_{u,v}\in\ExtMon(\Sigma^\triangle)$ we assign
	the triple $\zeta(\ext_{u,v})=(j,(f_i)_{i\in[0,m]},(g_i)_{i\in[0,m]})
	\in\Mn$, where 
	\begin{itemize}
		\item $j=\min(\Delta(u),m)\in[0,m]$,
		\item $f_i(q)=q'$ where $q'\in Q$ is the unique state
			such that $q(i)\xrightarrow{u}_Aq'(i+j)$ for all $q\in Q$ and 
			all $i\in[0,m]$, and
		\item $g_i(q)=q'$, where $q'\in Q$ is the unique state
			such that $q'(i+j)\xrightarrow{v}_A q'(i)$ for all $q\in Q$ and all $i\in[0,m]$.
	\end{itemize}
	Over $\Mn$ we define the product 
	\[(j,(f_i)_{i\in[0,m]},(g_i)_{i\in[0,m]})\odot_{\Mn}
(j',(f_i')_{i\in[0,m]},(g_i')_{i\in[0,m]})\]
as
\[
(\min(j+j',m),(f_{\min(i+j,m)}'\circ f_i)_{i\in[0,m]},
(g_{\min(i+j,m)}\circ g_{\min(i+j+j',m)}')_{i\in[0,m]}).
\]
We claim $\odot_\Mn$ is associative.
For this, let us fix 
\begin{itemize}
\item $m=(j,(f_i)_{i\in[0,m]},(g_i)_{i\in[0,m]})\in\Mn$,
\item $m'=(j',(f_i')_{i\in[0,m]},(g_i')_{i\in[0,m]})\in\Mn$, and 
\item $m''=(j'',(f_i'')_{i\in[0,m]},(g_i'')_{i\in[0,m]})\in\Mn$.
\end{itemize}
Let $\widetilde{m}=m\odot_\Mn m'=
(\widetilde{j},(\widetilde{f_i})_{i\in[0,m]},
(\widetilde{g_i})_{i\in[0,m]})$
and 
$\overline{m}=m'\odot_\Mn m''=
(\overline{j},(\overline{f_i})_{i\in[0,m]},
(\overline{g_i})_{i\in[0,m]})$.
We need to prove $\widetilde{m}\odot_\Mn m''=m\odot_\Mn\overline{m}$.

Since $j,j',j'',m\geq 0$ we have
$\min(\widetilde{j}+j'',m)=\min(\min(j+j',m)+j'',m)=\min(j+j'+j'',m)=
\min(j+\min(j'+j'',m),m))=\min(j+\overline{j},m)$
associativity holds on the first component.
For each $i\in[0,m]$ the $i^{\text{th}}$ component of the second
component of $\widetilde{m}\odot_\Mn m''$ is 
\begin{eqnarray*}
	f_{\min(i+\widetilde{j},m)}''\circ \widetilde{f}_i
	&=&f_{\min(i+\widetilde{j},m)}''\circ\left(f_{\min(i+j,m)}'\circ f_i\right)\\
	&=&f_{\min(i+\min(j+j',m),m)}''\circ\left(f_{\min(i+j,m)}'\circ f_i\right)\\
	&=&f_{\min(i+j+j',m)}''\circ\left(f_{\min(i+j,m)}'\circ f_i\right)\\
	&=&\left(f_{\min(i+j'+j'',m)}''\circ f_{\min(i+j,m)}'\right)\circ f_i\\
	&=&\left(f_{\min(\min(i+j,m)+j',m)}''\circ f_{\min(i+j,m)}'\right)\circ f_i\\
	&=&\overline{f}_{\min(i+j,m)}\circ f_i
\end{eqnarray*}
the latter of which is the $i^{\text{th}}$ component of the second
component of $m\odot_\Mn \overline{m}$, as required.
For each $i\in[0,m]$ the $i^{\text{th}}$ component of the third
component of $m\odot_\Mn\overline{m}$ is 
\begin{eqnarray*}
	g_{\min(i+j,m)}\circ\overline{g}_{\min(i+j+\overline{j},m))}
	&=& g_{\min(i+j,m)}\circ\overline{g}_{\min(i+j+j'+j'',m)}\\
	&=& g_{\min(i+j,m)}\circ\overline{g}_{\min(\min(i+j,m)+j'+j'',m)}\\
	&=& g_{\min(i+j,m)}\circ\left(
	g_{\min(\min(i+j,m)+j')}'\circ g_{\min(\min(i+j,m),j'+j'',m)}''
	\right)\\
	&=& \left(g_{\min(i+j,m)}\circ g_{\min(\min(i+j,m)+j')}'\right)\circ
	g_{\min(\min(i+j,m),j'+j'',m)}''\\
	&=& \left(g_{\min(i+j,m)}\circ g_{\min(i+j+j',m)}'\right)\circ
	g_{\min(i+\min(j+j',m)+j'',m)}''\\
	&=& \widetilde{g}_i\circ
	g_{\min(i+\widetilde{j}+j'',m)}''
\end{eqnarray*}
the latter of which is the $i^{\text{th}}$ component of the third
component of $\widetilde{m}\odot_\Mn m''$, as required.
Clearly $(0,(\id_Q)^{[0,m]},(\id_Q)^{[0,m]})$ is the identity of $\Mn$
with respect to $\odot_\Mn$, hence $(\Mn,\odot_\Mn)$ is a monoid.
	The following points can easily be verified.
	\begin{enumerate}
		\item The function $\zeta\colon \ExtMon(\Sigma^\triangle)\rightarrow\Mn$
			is a monoid morphism.
		\item For all $\ext_{u,v},\ext_{u',v'}\in\ExtMon(\Sigma^\triangle)$
			with $\Delta(u)=\Delta(u')$
			we have 
			\[\zeta(\ext_{u,v})=\zeta(\ext_{u',v'})
			\quad\Longrightarrow\quad
			\zeta(\ext_{u,v})=\zeta(\ext_{u',v})=
			\zeta(\ext_{u,v'})=\zeta(\ext_{u',v'}).
			\]
		\item For all $\ext_{u,v},\ext_{u',v'}\in\ExtMon(\Sigma^\triangle)$
		we have
			\[
			\zeta(\ext_{u,v})=\zeta(\ext_{u',v'})
			\quad\Longrightarrow\quad
			\psi_L(\ext_{u,v})=\psi_L(\ext_{u',v'}).
			\]
	\end{enumerate}
Now assume that $L$ is not length-synchronous. We will prove
that $L$ is not length-synchronous.
By assumption there exist a $\varphi_L(L)$-reaching idempotent $e\in O_L$
and $\ext_{u,v},\ext_{u',v'}\in\ExtMon(\Sigma^\triangle)$
such that $\Delta(u),\Delta(u')>0$, $\frac{|u|}{|v|}\not=\frac{|u'|}{|v'|}$,
and $\psi_L(\ext_{u,v})=\psi_L(\ext_{u',v'})=e$.
Without loss of generality we may assume $\Delta(u)=\Delta(u')$.
Let $\omega$ denote the idempotent power of $\Mn$.
Consider the elements
\[
\ext_{x,y}=(\ext_{u,v}^{2\omega}\circ\ext_{u',v'}^\omega)^\omega
\text{ and }
\ext_{x',y'}=(\ext_{u,v}^{\omega}\circ\ext_{u',v'}^{2\omega})^\omega.
\]
By definition we have 
$\zeta(\ext_{x,y})=\zeta(\ext_{x',y'})$,
and since $\Delta(x)=\Delta(x')$ we obtain
$\zeta(\ext_{x,y})=\zeta(\ext_{x',y})=\zeta(\ext_{x,y'})=\zeta(\ext_{x',y'})$
by Point 2.
Hence,
\begin{eqnarray*}
\psi_L(\ext_{x,y})=\psi_L(\ext_{x',y})=\psi_L(\ext_{x,y'})=\psi_L(\ext_{x',y'})=e
\end{eqnarray*}
by Point 3.

We finally make a case distinction on whether $|u|=|u'|$ or not.

First, assume $|u|\not=|u'|$. 
Then $|x|\not=|x'|$ by construction. 
Since
$\psi_L(\ext_{x,y})=\psi_L(\ext_{x',y})=e$, which
is an idempotent of $O_L$, we obtain that $L$ is not weakly length-synchronous.

Now assume $|u|=|u'|$.
Since $\frac{|u|}{|v|}\not=\frac{|u'|}{|v'|}$ by assumption, 
we conclude that $|v|\not=|v'|$.
By construction, the latter implies $|y|\not=|y'|$.
Since $\psi_L(\ext_{x,y})=\psi_L(\ext_{x,y'})=e$,
again we obtain that $L$ is not weakly length-synchronous.

\section{Proof of Proposition~\ref{prop effectivity}}\label{Section Effectiveness}

We will prove the different statements appearing in
Proposition~\ref{prop effectivity} in the following subsections.

\subparagraph{Computability of the syntactic $\Ext$-algebra.}

This paragraph will be devoted to proving Point 1 of
Proposition~\ref{prop effectivity}, rephrased in the following proposition.
\begin{prop}\label{prop ext alg}
    Given a DVPA $A$ with $L=L(A)$, one can compute the syntactic $\Ext$-algebra
    $(R_L,O_L)$ of $L$, its syntactic morphism $(\varphi_L,\psi_L)$ and
    $\varphi_L(L)$.
\end{prop}
\noindent
	We require a bit of notation.
For each visibly pushdown alphabet $\Sigma$ and each finite $\Ext$-algebra
$(R,O)$ it follows from Proposition~\ref{ptn:Freeness_Ext-algebras} that
each morphism
$(\varphi,\psi)\colon (\Sigma^\triangle,\ExtMon(\Sigma^\triangle))\rightarrow (R,O)$
has a unique finite presentation: it is given by the tuples 
\[
    (\varphi(c))_{c\in\Sigma_\internal}\quad
    \text{and}\quad (\psi(\ext_{a,b}))_{(a,b)\in\Sigma_\call\times\Sigma_\return}\quad.
\]
The syntactic $\Ext$-algebra $(R_L, O_L)$ of a VPL $L$ over a visibly pushdown
alphabet $\Sigma$ can be represented by any $\Ext$-algebra $(R, O)$ such that
$R$ has $[1, \card{R_L}]$ as base set and such that there exists a bijective
morphism $(\alpha, \beta)\colon (R, O) \to (R_L, O_L)$. Indeed, in that case we
have
\begin{enumerate}
    \item
	$x y = z \Leftrightarrow \alpha(x) \alpha(y) = \alpha(z)$ for all
	$x, y, z \in R$;
    \item
	$x' y' = z' \Leftrightarrow
	 \alpha^{-1}(x') \alpha^{-1}(y') = \alpha^{-1}(z')$
	for all $x', y', z' \in R_L$;
    \item
	$f(x) = y \Leftrightarrow \beta(f)(\alpha(x)) = \alpha(y)$ for all
	$f \in O$ and all $x, y \in R$; and
    \item
	$f'(x') = y' \Leftrightarrow
	 \beta^{-1}(f')(\alpha^{-1}(x')) = \alpha^{-1}(y')$
	for all $f' \in O_L$ and all $x', y' \in R_L$.
\end{enumerate}
\noindent
For the following claim we avoid the tedious standard algebraic constructions on $\Ext$-algebras
to show decidability of the equivalence problem, since
the latter decidability has already been established in~\cite{AlurM04}.
\begin{clm}
    There is an algorithm that decides, given two morphisms into finite
    $\Ext$-algebras
    $(\varphi_1,\psi_1)\colon(\Sigma^\triangle,\ExtMon(\Sigma^\triangle))\rightarrow (R_1,O_1)$
    and 
    $(\varphi_2,\psi_2)\colon(\Sigma^\triangle,\ExtMon(\Sigma^\triangle))\rightarrow (R_1,O_1)$
    for $\Sigma$ a visibly pushdown alphabet and subsets $F_1\subseteq R_1$ and
    $F_2\subseteq R_2$, whether $\varphi_1^{-1}(F_1)=\varphi_2^{-1}(F_2)$.
\end{clm}
\begin{proof}[Proof of the Claim.]
    The proof of Theorem~\ref{thm:vpl:ext} shows that one can effectively
    compute DVPAs $A_1$ and $A_2$ such that $L(A_1)=\varphi_1^{-1}(F_1)$
    and $L(A_2)=\varphi_2^{-1}(F_2)$.
    By~\cite{AlurM04} one can effectively decide if $L(A_1)=L(A_2)$
    by deciding $L(A_1)\subseteq L(A_2)$ and $L(A_2)\subseteq L(A_1)$.
\end{proof}

\begin{proof}[Proof of Proposition~\ref{prop ext alg}]
    By Theorem~\ref{thm:vpl:ext} we first compute from our DVPA $A$ on the
    visibly pushdown alphabet $\Sigma$ an $\Ext$-algebra $(R_A,O_A)$, a morphism
    $(\varphi_A,\psi_A)\colon(\Sigma^\triangle, \ExtMon(\Sigma^\triangle))\to
     (R_A,O_A)$,
    and a subset $F_A \subseteq R_A$ such that $L(A) = \varphi_A^{-1}(F_A)$.
    For an $\Ext$-algebra $(R,O)$ define $\#(R,O)=(\card{R},\card{O})$.
    Let $\prec\subseteq(\N\times\N)^2$ be the lexicographic order on $\N\times\N$, 
    i.e.\ $(i,j)\prec(k,l)$ if, and only if either $i<k$, or $i=k$ and $j<l$.

    Observe that since $(R_A, O_A)$ recognizes $L$, we have that the syntactic
    $\Ext$-algebra $(R_L, O_L)$ of $L$ divides $(R_A, O_A)$ by
    Proposition~\ref{prop:syntactic:recognizes}, so that
    $\#(R_L, O_L) \leq \#(R_A, O_A)$.
    In fact, we have that any $\Ext$-algebra $(R, O)$ having $[1, i]$ for
    $i \in [1, \card{R_L}]$ as base set, satisfying $\#(R, O) \leq \#(R_L, O_L)$
    and recognizing $L$ via a morphism
    $(\varphi, \psi)\colon (\Sigma^\triangle, \ExtMon(\Sigma^\triangle)) \to
     (R, O)$
    is a presentation of $(R_L, O_L)$ with $(\varphi, \psi)$ and $F$
    presentations of, respectively, $(\varphi_L, \psi_L)$ and $\varphi_L(L)$.
    Indeed, since such an $\Ext$-algebra recognizes $L$, by
    Proposition~\ref{prop:syntactic:recognizes} it is divided by $(R_L, O_L)$:
    this implies that $\#(R_L, O_L) \leq \#(R, O)$, but as also
    $\#(R, O) \leq \#(R_L, O_L)$, we have $\#(R, O) = \#(R_L, O_L)$.
    The morphism $(\varphi, \psi)$ must be surjective, otherwise, by
    Lemma~\ref{lem:Surjective_recognition},
    $\bigl(\varphi(\Sigma^\triangle),
	   \restr{\psi(\ExtMon(\Sigma^\triangle))}{\varphi(\Sigma^\triangle)}
     \bigr)$
    would be a sub-$\Ext$-algebra of $(R, O)$ recognizing $L$ such that
    $\#\bigl(\varphi(\Sigma^\triangle),
	     \restr{\psi(\ExtMon(\Sigma^\triangle))}{\varphi(\Sigma^\triangle)}
       \bigr) <
       \#(R, O) = \#(R_L, O_L)$
    while $(R_L, O_L)$ divides
    $\bigl(\varphi(\Sigma^\triangle),
	   \restr{\psi(\ExtMon(\Sigma^\triangle))}{\varphi(\Sigma^\triangle)}
     \bigr)$,
    which is contradictory.
    Therefore, by Lemma~\ref{lem:Factorization_surjective_recognition}, there is
    a surjective morphism $(\alpha, \beta)\colon (R, O) \to (R_L, O_L)$, that
    must be bijective, such that $\varphi_L = \alpha \circ \varphi$, so that
    $(R, O)$ is a presentation of $(R_L, O_L)$ with $(\varphi, \psi)$ and $F$
    presentations of, respectively, $(\varphi_L, \psi_L)$ and $\varphi_L(L)$.

    Under the assumption that such an $\Ext$-algebra exists, we compute
    $(R_L, O_L)$, $(\varphi_L, \psi_L)$ and $\varphi_L(L)$ by enumerating all
    the finitely many triples made of a finite $\Ext$-algebra $(R, O)$, a
    morphism
    $(\varphi,\psi)\colon (\Sigma^\triangle, \ExtMon(\Sigma^\triangle))\to(R,O)$
    and a subset $F \subseteq R$ such that $R$ has $[1, i]$ for
    $i \in [1, \card{R_A}]$ as base set and $\#(R, O) \leq \#(R_A, O_A)$.
    For each of these we test whether $\varphi^{-1}(F)=\varphi_A^{-1}(F_A)$,
    which is possible by the above claim and take $(R, O)$, $(\varphi, \psi)$
    and $F$ from a triple validating this test with $\#(R, O)$ minimal with
    respect to $\prec$.

    It remains to prove that an $\Ext$-algebra $(R, O)$ having $[1, i]$ for
    $i \in [1, \card{R_L}]$ as base set, satisfying $\#(R, O) \leq \#(R_L, O_L)$
    and recognizing $L$ exists.
    Take any bijection $\alpha\colon R_L \to [1, \card{R_L}]$. We define $R$ to
    be the monoid with base set $[1, \card{R_L}]$ and operation defined by
    $x \cdot y = \alpha\bigl(\alpha^{-1}(x) \alpha^{-1}(y)\bigr)$ for all
    $x, y \in [1, \card{R_L}]$. This is a monoid because
    \begin{itemize}
	\item
	    $x \cdot \alpha(1_R) =
	     \alpha\bigl(\alpha^{-1}(x) \alpha^{-1}(\alpha(1_R))\bigr) =
	     \alpha(\alpha^{-1}(x)) =
	     \alpha\bigl(\alpha^{-1}(\alpha(1_R)) \alpha^{-1}(x)\bigr) =
	     \alpha(1_R) \cdot x$
	    for all $x \in R$; and
	\item
	    for all $x, y, z \in R$, we have
	    \begin{align*}
		x \cdot (y \cdot z)
		& = \alpha
		    \biggl(
			\alpha^{-1}(x)
			\alpha^{-1}\Bigl(\alpha\bigl(\alpha^{-1}(y)
						     \alpha^{-1}(z)\bigr)\Bigr)
		    \biggr)\\
		& = \alpha\bigl(\alpha^{-1}(x) \alpha^{-1}(y) \alpha^{-1}(z)\bigr)\\
		& = \alpha
		    \biggl(
			\alpha^{-1}\Bigl(\alpha\bigl(\alpha^{-1}(x)
						     \alpha^{-1}(y)\bigr)\Bigr)
			\alpha^{-1}(z)
		    \biggr)\\
		& = (x \cdot y) \cdot z
		\displaypunct{.}
	    \end{align*}
    \end{itemize}
    Define the function $\beta\colon O_L \to R^R$ by
    $\beta(f')(x) = \alpha\bigl(f'(\alpha^{-1}(x))\bigr)$ for all $f' \in O_L$
    and $x \in R$. Set $O$ to be the monoid with base set $\beta(O_L)$ and with
    composition as operation. This is a monoid because
    \begin{itemize}
	\item
	    $\beta(1_O)(x) =
	     \alpha\bigl(1_O(\alpha^{-1}(x))\bigr) = x = \id_R(x)$
	    for all $x \in R$; and
	\item
	    for all $f', g' \in O_L$,
	    \[
		\beta(f') \circ \beta(g')(x) =
		\alpha
		\Biggl(
		    f'
		    \biggl(
			\alpha^{-1}
			\Bigl(
			    \alpha\bigl(g'(\alpha^{-1}(x))\bigr)
			\Bigr)
		    \biggr)
		\Biggr) =
		\alpha\bigl(f' \circ g'(\alpha^{-1}(x))\bigr) =
		\beta(f' \circ g')(x)
	    \]
	    for all $x \in R$, so that $\beta(f') \circ \beta(g') \in O$.
    \end{itemize}
    Then $(R, O)$ is an $\Ext$-algebra, because for all $r' \in R_L$, we have
    \[
	\beta(\leftmult{r'})(x) =
	\alpha\bigl(\leftmult{r'}(\alpha^{-1}(x))\bigr) =
	\alpha\bigl(\alpha^{-1}(\alpha(r')) \alpha^{-1}(x)\bigr) =
	\alpha(r') \cdot x =
	\leftmult{\alpha(r')}(x)
    \]
    for all $x \in R$ and $\beta(\rightmult{r'})(x) = \rightmult{\alpha(r')}(x)$
    for all $x \in R$, so that $\leftmult{r}, \rightmult{r} \in O$ for all
    $r \in R$ by surjectivity of $\alpha$.

    We now define
    $(\varphi, \psi)\colon (\Sigma^\triangle, \ExtMon(\Sigma^\triangle)) \to (R, O)$
    as the unique morphism satisfying $\varphi(c) = \alpha(\varphi_L(c))$ for all
    $c \in \Sigma_\internal$ and $\psi(\ext_{a, b}) = \beta(\psi_L(\ext_{a, b}))$ for
    all $a \in \Sigma_\call$, $b \in \Sigma_\return$ given by
    Proposition~\ref{ptn:Freeness_Ext-algebras}.
    It is easy to show that then, $\varphi(w) = \alpha(\varphi_L(w))$ for all
    $w \in \Sigma^\triangle$ by structural induction on $w$.
    Hence, by injectivity of $\alpha$, we have
    \begin{align*}
	\varphi^{-1}\bigl(\alpha(\varphi_L(L))\bigr)
	& = \set{w \in \Sigma^\triangle \mid
		 \alpha(\varphi_L(w)) \in \alpha(\varphi_L(L))}\\
	& = \set{w \in \Sigma^\triangle \mid \varphi_L(w) \in \varphi_L(L)} =
	\varphi_L^{-1}(\varphi_L(L)) = L
	\displaypunct{,}
    \end{align*}
    thus $(R, O)$ recognizes $L$.
\end{proof}

\subparagraph{Decidability of quasi-aperiodicity.}\label{Para quasi}
This paragraph is devoted to proving Point 2 (a) 
of Proposition~\ref{prop effectivity},
rephrased in the following proposition.
\begin{prop}\label{prop quasi}
    Given a morphism
    $(\varphi,\psi)\colon (\Sigma^\triangle,\ExtMon(\Sigma^\triangle))\to(R,O)$
    for $\Sigma$ a visibly pushdown alphabet and $(R, O)$ a finite
    $\Ext$-algebra, it is decidable if $(\varphi,\psi)$ is quasi-aperiodic.
If $(\varphi,\psi)$ is not quasi-aperiodic, one can effectively
compute $k,l\in\N$ such that 
			$\psi(\ExtMon(\Sigma^\triangle)^{k,l})$ is
			not aperiodic.
\end{prop}
\noindent
For the rest of this paragraph, let us fix a morphism 
$(\varphi,\psi)\colon(\Sigma^\triangle,\ExtMon(\Sigma^\triangle))\to(R,O)$,
where $\Sigma$ is a visibly pushdown alphabet and $(R,O)$ is some finite
$\Ext$-algebra that is the input to our problem.
We first have the following lemma.
\begin{lem}{\label{lemma Le}}
    For all $e\in O$ one can effectively compute a finite $\Ext$-algebra
    recognizing
    $L_e=
     \set{u\#v\mid uv \in \Sigma^\triangle : \psi(\ext_{u,v})=e}$,
    where $\#$ is a fresh internal letter that does not appear in $\Sigma$,
    along with an associated morphism and subset.
\end{lem}
\begin{proof}
    Let $\Sigma'$ be the alphabet that emerges from $\Sigma$ by
    additionally declaring $\#$ as an internal letter.
    We will construct an $\Ext$-algebra $(R',O')$ and a morphism 
    $(\varphi',\psi')\colon(\Sigma'^\triangle,\ExtMon(\Sigma')^\triangle)\to(R',O')$
    such that for some element $r'\in R'$ we have $L_e=\varphi'^{-1}(r')$.

    We define $R'=R\cup O\cup\set{\bot}$, for some fresh zero $\bot$,
    where multiplication between two elements in $R'$ is 
    defined as follows:
    \begin{itemize}
	\item
	    multiplication between two elements in $R$ is inherited from the
	    monoid $R$;
	\item
	    $r\cdot f=\leftmult{r}\circ f$ and $f\cdot r=\rightmult{r}\circ f$
	    for all $r\in R$ and all $f\in O$;
	\item
	    $\bot$ acts as a zero, i.e.\ $\bot\cdot r'=r'\cdot\bot=\bot$ for all
	    $r'\in R'$;
	\item
	    $f\cdot g=\bot$ for all $f,g\in O$.
    \end{itemize}
    Clearly the identity of $R$ is the identity of $R'$.
    Associativity is immediate except for products of the form
    $r_1\cdot f\cdot r_2, r_1\cdot r_2\cdot f$, and $f\cdot r_1\cdot r_2$,
    where $f\in O$ and $r_1,r_2\in R$. 
    In the first case we have
    \begin{align*}
	(r_1\cdot f)\cdot r_2
	& = (\leftmult{r_1}\circ f)\cdot r_2
	  = \rightmult{r_2}\circ(\leftmult{r_1}\circ f)
	  = (\rightmult{r_2}\circ\leftmult{r_1})\circ f\\
	& = (\leftmult{r_1}\circ\rightmult{r_2})\circ f
	  = r_1\cdot (\rightmult{r_2}\circ f)
	  = r_1\cdot (f\cdot r_2)
	  \displaypunct{.}
    \end{align*}
    In the second case we have
    \[
	(r_1\cdot r_2)\cdot f
	= \leftmult{r_1 r_2}\circ f
	= (\leftmult{r_1}\circ\leftmult{r_2})\circ f
	= \leftmult{r_1}\circ(\leftmult{r_2}\circ f)
	= r_1\cdot(r_2\cdot f)
    \]
    and in the third case we have
    \[
	f\cdot (r_1\cdot r_2)=\rightmult{r_1r_2}\circ f=
	(\rightmult{r_2}\circ\rightmult{r_1})\circ f=
	\rightmult{r_2}\circ(\rightmult{r_1}\circ f)=
	(f\cdot r_1)\cdot r_2
	\displaypunct{.}
    \]
    We define $O'=(R')^{R'}$ which is clearly a monoid for composition and thus
    directly get that $(R', O')$ is an $\Ext$-algebra.
    Applying Proposition~\ref{ptn:Freeness_Ext-algebras}, we define the morphism
    \[(\varphi',\psi')\colon
      (\Sigma'^\triangle,\ExtMon(\Sigma')^\triangle)\to(R',O')\]
    as the unique one satisfying $\varphi'(c)=\varphi(c)$ for all
    $c\in\Sigma_\internal$, $\varphi'(\#)=\id_O$
    and where for all $a \in \Sigma_\call$, $b \in \Sigma_\return$, we have
    \[
	\psi'(\ext_{a,b})(x)=
	\begin{cases}
	    \psi(\ext_{a,b})(x) & \text{if $x\in R$}\\
	    \psi(\ext_{a,b})\circ x & \text{if $x\in O$}\\
	    \bot & \text{otherwise (i.e.\ if $x=\bot$)}
	\end{cases}
    \]
    for all $x\in R'$.
    It suffices to prove the following claim, which directly implies the desired
    equality
    $\varphi'^{-1}(e)=
     \set{u\#v\mid uv \in \Sigma^\triangle \text{ s.t.\ } \psi(\ext_{u,v})=e}$.
    For all $w\in\Sigma'^\triangle$ we have
    \[
	\varphi'(w)=
	\begin{cases}
	    \varphi(w)& \text{ if $w\in\Sigma^\triangle$}\\
	    \psi(\ext_{u,v}) & \text{ if $w=u\#v$ for some $uv\in\Sigma^\triangle$}\\
	    \bot & \text{ otherwise.}
	\end{cases}
    \]
    We prove it by structural induction on $w$. The cases when $w = \emptyword$
    or $w=c\in\Sigma_\internal$ follow immediately from the definition of
    $\varphi'$.
    In case $w=\#=\emptyword\#\emptyword$, we have
    $\varphi'(w)=\id_O=\psi(\ext_{\emptyword,\emptyword})$.

    For the inductive step first assume
    $w=aw'b$ for some $w'\in\Sigma'^\triangle$.
    If $w'$ is neither in $\Sigma^\triangle$ nor of the form
    $u\#v$ with $uv \in \Sigma^\triangle$, then $\varphi'(w') = \bot$ by
    induction hypothesis and thus
    $\varphi'(w) = \psi'(\ext_{a,b})(\varphi'(w'))=\psi'(\ext_{a,b})(\bot) =
     \bot$
    as required.
    If $w'\in\Sigma^\triangle$, then $\varphi'(w')=\varphi(w')\in R$ by
    induction hypothesis, and hence
    $\varphi'(w)=\psi'(\ext_{a,b})(\varphi'(w'))=\psi'(\ext_{a,b})(\varphi(w'))=
     \psi(\ext_{a,b})(\varphi(w'))=\varphi(w)$
    as required.
    If $w'=u\#v$ with $uv\in\Sigma^\triangle$, i.e.\ $w = a u\#v b$, then 
    $\varphi'(w')=\psi(\ext_{u,v})\in O$ by induction hypothesis.
    Hence, we have $\varphi'(w)=\psi'(\ext_{a,b})(\varphi'(w'))=
    \psi'(\ext_{a,b})(\psi(\ext_{u,v}))=\psi(\ext_{a,b})\circ\psi(\ext_{u,v})=
    \psi(\ext_{au,vb})$.

    Finally assume $w=xy$ for some $x,y\in\Sigma'^\triangle\setminus\set{\emptyword}$.
    The case when $x$ or $y$ is neither in $\Sigma^\triangle$ nor of the form
    $u\#v$ with $uv \in \Sigma^\triangle$ is easily handled by applying the
    induction hypothesis and observing that $\bot$ is a zero in $R'$.
    Two other immediate cases are when both $x$ and $y$ are in
    $\Sigma^\triangle$ and when both $x$ and $y$ are of the form $u\#v$ with
    $uv \in \Sigma^\triangle$.
    Consider the case when $x\in\Sigma^\triangle\setminus\set{\emptyword}$ 
    and $y=u\#v$ with $uv \in \Sigma^\triangle$, hence $w=xu\#v$.
    The induction hypothesis yields $\varphi'(x)=\varphi(x)\in R$ and
    $\varphi'(y)=\psi(\ext_{u,v})\in O$.
    We obtain 
    \[
	\varphi'(xy)=\varphi'(x)\cdot\varphi'(y)=
	\varphi(x)\cdot \psi(\ext_{u,v})=
	\leftmult{\varphi(x)}\circ\psi(\ext_{u,v})=\psi(\ext_{xu,v})
    \]
    as required.
    Finally, let us treat the case when $x=u\#v$ with
    $uv\in\Sigma^\triangle$ and
    $y\in\Sigma^\triangle\setminus\set{\emptyword}$, i.e.\ $w=u\#vy$.
    The induction hypothesis yields $\varphi'(x)=\psi(\ext_{u,v})\in O$ and
    $\varphi'(y)=\varphi(y)\in R$.
    We obtain
    \[
	\varphi'(xy)=\varphi'(x)\cdot\varphi'(y)=
	\psi(\ext_{u,v})\cdot\varphi(y)=
	\rightmult{\varphi(y)}\circ\psi(\ext_{u,v})=\psi(\ext_{u,vy})
    \]
    as required.
\end{proof}

The next goal will be to prove that the set of pairs of word
lengths $(\length{u},\length{v})$ of words $u\#v\in L_e$ is effectively
semilinear for each $e\in O$.

A {\em (realtime) pushdown automaton (PDA for short)} is a tuple
$A=(Q,\Sigma,\Gamma,\Omega,q_0,F,\bot)$,
where $Q$ is a finite set of {\em states},
$\Sigma$ is a finite {\em input alphabet},
$\Gamma$ is a finite {\em stack alphabet},
$q_0\in Q$ is an {\em initial state},
$F\subseteq Q$ is the set of {\em final states},
$\bot\in\Gamma\setminus\Sigma$ is the {\em bottom-of-stack symbol},
and $\Omega\subseteq Q\times\Sigma\times\Gamma\times Q\times\Gamma^*$ is a
finite {\em transition relation} such that for all $(p,a,X,q,\alpha)\in\Omega$
we have $\alpha\in\Gamma^*\bot$ if $X=\bot$ and
$\alpha\in(\Gamma\setminus\set{\bot})^*$ otherwise.
The relation $\Omega$ is naturally extended to 
the relation $\Omega^*\subseteq Q\times\Sigma^*\times\Gamma^*\bot\times Q\times\Gamma^*\bot$,
namely as the smallest relation containing the set
$\set{(p,\emptyword,\alpha,p,\alpha)\mid p\in Q,\alpha\in\Gamma^*\bot}$
and such that moreover, if $(p,a,X,q,\alpha)\in\Omega$ and
$(q,w,\alpha\beta,r,\gamma)\in\Omega^*$, then
$(p,aw,X\beta,r,\gamma)\in\Omega^*$.
The {\em language} of $A$ is
$L(A)=\set{w\in\Sigma^*\mid
	   \exists \alpha\in\Gamma^*\bot, \exists q\in F :
	   (q_0,w,\bot,q,\alpha)\in\Omega^*}$.
Hence it is clear that one can compute 
a PDA $A'$ such that $L(A')=L(A)$.

\begin{lem}\label{lemma semilinear}
    Let $A$ be a DVPA that accepts a language over a visibly pushdown alphabet
    $\Sigma'$ such that
    $L(A) \subseteq
     \set{u \# v \mid
	  uv \in \Sigma'^\triangle \setminus \Sigma'^* \# \Sigma'^*}$
    and $\# \in \Sigma'_\internal$.
    Then the set
    \[
	P(L(A))=
	\set{(k,l)\in\N\times\N\mid
	     \exists u\in(\Sigma' \setminus \set{\#})^k,
		     v\in(\Sigma' \setminus \set{\#})^l : u \# v \in L(A)}
    \]
    is effectively semilinear.
\end{lem}
\begin{proof}
    We first compute a PDA $A'$ accepting the same language as $A$, i.e.\
    $L(A')=L(A)$.
    Let us assume without loss of generality that $0,1\not\in\Sigma'$.
    We claim that from $A'=(Q,\Sigma',\Gamma,\Omega,q_0,F,\bot)$ one can 
    compute a PDA $A''$ such that
    \[
	L(A'')=\set{0^{|u|}\#1^{|v|}\mid u\#v\in L(A')}
	\displaypunct{.}
    \]
    Indeed, the PDA $A''$ can simply be computed as follows:
    we set
    \[
	A''=(Q\times\set{0,1},\set{0,1,\#},\Gamma,\Omega',\langle q_0,0\rangle,F\times\set{1},\bot)
	\displaypunct{,}
    \]
    where $\Omega'$ is the union of 
    $\bigset{(\langle p,i\rangle,i,X,\langle q,i\rangle,\alpha) \bigmid
	     i\in\set{0,1}, \exists c\in\Sigma' \setminus \set{\#} :
	     (p,c,X,q,\alpha)\in\Omega}$
    and $\set{(\langle p,0\rangle,\#,X,\langle q,1\rangle,\alpha)\mid
    (p,\#,X,q,\alpha)\in\Omega}$.
    Finally, we apply Parikh's Theorem, cf.~\cite[Section 3]{EGKL11}, which
    implies that the set $\set{(m,n)\in\N\times\N\mid 0^m\#1^n\in L(A'')}=P(L(A))$
    is effectively semilinear.
\end{proof}
We are now ready to prove Proposition~\ref{prop quasi}.
\begin{proof}[Proof of Proposition~\ref{prop quasi}]
    Let $e\in O$.
    By Lemma~\ref{lemma Le} we first compute a finite $\Ext$-algebra
    recoginizing $L_e$, along with an associated morphism and subset. From the
    latter we can compute (by Theorem~\ref{thm:vpl:ext}) a DVPA $A_e$ accepting
    $L_e$. We then use Lemma~\ref{lemma semilinear} to conclude that the set
    \[
	P(L_e) =
	\set{(k,l)\in\N\times\N\mid
	     \exists u\in\Sigma^k,v\in\Sigma^l :
	     uv\in\Sigma^\triangle, \psi(\ext_{u,v})=e}
    \]
    is effectively semilinear, and this holds for all $e \in O$.

    We make use of the folklore fact that semilinear sets are effectively closed
    under Boolean operations, cf.~\cite{CH16} for a recent study.
    To decide whether $(\varphi, \psi)$ is quasi-aperiodic, we go through all
    possible subsets $U \subseteq O$: if it is a subsemigroup of $O$ that is a
    non-trivial group, we compute the set $\bigcap_{e\in U} P(L_e)$ and reject
    if it is non-empty (which is easy to check given a semilinear presentation
    of the set), otherwise we continue. If we were able to go through
    all those subsets without rejecting, we accept.

	Thus, if $(\varphi,\psi)$ is not quasi-aperiodic we can
	find a subset $U\subseteq O$ that contains a non-trivial group and
	output a pair $(k,l)\in \bigcap_{e\in U} P(L_e)$;
	it witnesses that $\psi(\ExtMon(\Sigma^\triangle)^{k,l})$
	is not aperiodic.
\end{proof}

\subparagraph{Decidability of length-synchronicity.}
This paragraph is devoted to proving Point 2 (b)
of Proposition~\ref{prop effectivity},
rephrased in the following proposition.
\begin{prop}\label{prop length effectivity}
    Given a morphism
    $(\varphi,\psi)\colon (\Sigma^\triangle,\ExtMon(\Sigma^\triangle))\to(R,O)$,
    for $\Sigma$ a visibly pushdown alphabet and $(R, O)$ a finite
    $\Ext$-algebra, and some $F\subseteq R$, it is decidable if $(\varphi,\psi)$
    is $F$-length-synchronous.
If $(\varphi,\psi)$ is not length-synchronous, one
	can effectively compute a tuple $(k,l,k',l')\in\N_{>0}^4$
	such that that there exist $uv,u'v'\in\Sigma^\triangle$
	and some $F$-reaching idempotent $e\in O$
	such that $\psi(\ext_{u,v})=\psi(\ext_{u',v'})=e$,
	$k=|u|$, $l=|v|$, $k'=|u'|$, $l'=|v'|$ and  
	$\frac{k}{l}\not=\frac{k'}{l'}$.
\end{prop}
\noindent
Before proving the proposition we need a technical lemma
characterizing when a two-dimensional semilinear set 
contains only vectors with the same slope.
We say two vectors $\vec{x},\vec{y}\in\N^2$ are {\em co-linear} if
$\vec{y}=\alpha\cdot\vec{x}$ for some $\alpha\in\Q_{>0}$
\begin{lem}\label{lemma collinear}
	Let $S=\bigcup_{i\in I}\left(\vec{x}_{i,0}+\sum_{j=1}^{t_i}\N\vec{x}_{i,j}\right)
	\subseteq \N_{>0}^2$ be a non-empty semilinear set,
	where
	$\vec{x}_{i,j}\not=(0,0)$ for all $i\in I$ and all $j\in[0,t_i]$.
	Then,
	\[\left|\left\{\frac{k}{l}\Big| (k,l)\in S\right\}\right|=1\quad\Longleftrightarrow\quad
	\forall i,i'\in I\ \forall j\in[0,t_i]\ \forall j'\in [0,t_{i'}]:
	\vec{x}_{i,j}\text{ and }\vec{x}_{i',j'}\text{ are co-linear}.\]
\end{lem}
\begin{proof}
First assume that 
	$\vec{x}_{i,j}$ and $\vec{x}_{i',j'}$ are co-linear for all 
	$i,i'\in I$, $j\in[0,t_i]$, and $j'\in [0,t_{i'}]$.
	Let $(k,l),(k',l')\in S$.
	That is, $(k,l)=\vec{x}_{i,0}+ n_1\vec{x}_{i,1}+\dots+n_{t_i}\vec{x}_{i,t_i}$
	and $(k',l')=\vec{x}_{i',0}+n_1'\vec{x}_{i',1}+\dots+n_{t_{i'}}'\vec{x}_{i',t_{i'}}$
	for some $i,i'\in I$ and some $n_1,\dots,n_{t_i},n_1',\dots,n_{t_{i'}}'\in
	\N$.
	But due to pairwise co-linearity there exist
	$\alpha,\alpha'\in\Q_{>0}$ such that $(k,l)=\alpha\vec{x}_{i,0}$
	and $(k',l')=\alpha'\vec{x}_{i,0}$, thus implying
	$\frac{k}{l}=\frac{k'}{l'}$.

	Conversely assume that there exist two vectors $(k,l)=\vec{x}_{i,j}$
	and $(k',l')=\vec{x}_{i',j'}$ that are not co-linear.
	In case this is possible when $i\not= i'$ and $j=j'=0$ we are done,
	since then $(k,l),(k',l')\in S$ and thus $\frac{k}{l}\not=\frac{k'}{l'}$.
	Otherwise $\vec{x}_{i,0}$ and $\vec{x}_{i',0}$ are co-linear
	for all $i,i'\in I$, so there must exist 
	$i\in I$ and $j\in[0,t_i]$
	such that $\vec{x}_{i,0}$ and $\vec{x}_{i,j}$ are not co-linear.
	Then $\vec{x}_{i,0}$ and $\vec{x}_{i,0}+\vec{x}_{i,j}$ are in $S$
	but also not co-linear: indeed, if 
	$\alpha\vec{x}_{i,0}=\vec{x}_{i,0}+\vec{x}_{i,j}$ for some $\alpha\in\Q_{>0}$,
	then $\vec{x}_{i,j}=(\alpha-1)\vec{x}_{i,0}$ with $\alpha-1>0$
	due to $\vec{x}_{i,0},\vec{x}_{i,j}\in\N^2\setminus\set{(0,0)}$,
	a contradiction.
	Hence there exist $(k,l),(k',l')\in S$ that are not co-linear, 
	and therefore $\frac{k}{l}\not=\frac{k'}{l'}$.
\end{proof}

\begin{proof}[Proof of Proposition~\ref{prop length effectivity}]
Let us fix the $\Ext$-algebra morphism
$(\varphi,\psi)\colon (\Sigma^\triangle,\ExtMon(\Sigma^\triangle))\to(R,O)$,
where $(R, O)$ is a finite $\Ext$-algebra and where $F\subseteq R$.

Recall that over the alphabet $\Sigma'$, 
	obtained from $\Sigma$ by declaring a fresh
letter $\#$ as internal, the language
\[
    L_{e \uparrow} =
    \set{u\#v\mid uv \in \Sigma^\triangle :
		  \Delta(u) > 0, \psi(\ext_{u,v})=e} =
    L_e \cap \set{u\#v\mid uv \in \Sigma^\triangle : \Delta(u) > 0}
\]
is given for all $e \in O$.
The language $\set{u\#v\mid uv \in \Sigma^\triangle : \Delta(u) > 0}$ is a clearly
	a VPL.
Thus, for all $e \in O$, we have that the set
\[
    P(L_{e \uparrow}) =
    \left\{(k,l)\in\N\times\N\mid
	 \exists u\in\Sigma^k,v\in\Sigma^l :
	 uv\in\Sigma^\triangle, \Delta(u) > 0, \psi(\ext_{u,v})=e\right\}
\]
is effectively semilinear: indeed, given $e \in O$, using Lemma~\ref{lemma Le}
	and Theorem~\ref{thm:vpl:ext}, we can as in the proof of Point 2 (a)
	of Proposition~\ref{prop effectivity} compute a DVPA $A_e$ accepting
$L_e$; we then compute a DVPA $A_{e \uparrow}$ accepting
$L_{e \uparrow} = L(A_e) \cap L(A)$ by using the effective construction given
in~\cite{AlurM04} and finally use Lemma~\ref{lemma semilinear} to conclude.

	Observe that $(\varphi,\psi)$ is $F$-length-synchronous
	if, and only if, for each $F$-reaching idempotent $e \in O$
	for which $P(L_{e\uparrow})$ is non-empty we have
	$|\set{\frac{k}{l}\mid (k,l)\in P(L_{e \uparrow})}|=1.$
	The latter condition is easily seen to be decidable
	by the characterization provided in Lemma~\ref{lemma collinear}.
	Hence, for deciding if $(\varphi,\psi)$ is length-synchronous
	our algorithm verifies if for all
	$F$-reaching $e\in O$ for which $P(L_{e\uparrow})$ is non-empty
	we have 
	$|\set{\frac{k}{l}\mid (k,l)\in P(L_{e \uparrow})}|=1.$
	On the other hand, if this verification fails, i.e.
	in case $(\varphi,\psi)$ is not $F$-length-synchronous,
		our algorithm outputs,
again using the characterization of Lemma~\ref{lemma collinear},
		a quadruple 
	$(k,l,k',l')\in\N_{>0}^4$ such that
	for some $F$-reaching idempotent $e\in O$ we have
	$(k,l),(k',l')\in P(L_{e\uparrow})$ and
	$\frac{k}{l}\not=\frac{k'}{l'}$.
\end{proof}

\subparagraph{Decidability of weak length-synchronicity.}
This paragraph is devoted to proving Point 2 (c) 
of Proposition~\ref{prop effectivity},
rephrased in the following proposition.
\begin{prop}\label{prop weakly length effectivity}
    Given a morphism
    $(\varphi,\psi)\colon (\Sigma^\triangle,\ExtMon(\Sigma^\triangle))\to(R,O)$,
    for $\Sigma$ a visibly pushdown alphabet and $(R, O)$ a finite
    $\Ext$-algebra, and some $F\subseteq R$, it is decidable if $(\varphi,\psi)$
    is $F$-weakly-length-synchronous.
\end{prop}
\noindent
Let us fix the morphism
$(\varphi,\psi)\colon (\Sigma^\triangle,\ExtMon(\Sigma^\triangle))\to(R,O)$,
for $\Sigma$ a visibly pushdown alphabet and $(R, O)$ a finite $\Ext$-algebra,
and some $F\subseteq R$.

Define the new visibly pushdown alphabet $\overline{\Sigma}$ by
$\overline{\Sigma}_\call = \set{\overline{b} \mid b \in \Sigma_\return}$,
$\overline{\Sigma}_\internal = \set{\overline{c} \mid c \in \Sigma_\internal}$
and $\overline{\Sigma}_\return = \set{\overline{a} \mid a \in \Sigma_\call}$.
For all $w \in \Sigma^*$, we define
\[
    \overline{w} =
    \begin{cases}
	\emptyword & \text{if $w = \emptyword$}\\
	\overline{w_n} \cdots \overline{w_1} &
	    \text{if $w = w_1 \cdots w_n$ for $n \in \N_{>0}$ and
		  $w_1, \ldots w_n \in \Sigma$}
	\displaypunct{.}
    \end{cases}
\]
We have the following lemma, that we prove later on.
\begin{lem}{\label{lemma Ke}}
    For all $e\in O$ one can effectively compute a finite $\Ext$-algebra
    recognizing the language of well-matched words
    $K_e=
     \set{u\#\overline{u'}\mid
	  u, u' \in \Sigma^*, \exists v \in \Sigma^* :
	  uv \in \Sigma^\triangle, u'v \in \Sigma^\triangle,
	  \psi(\ext_{u,v})=\psi(\ext_{u',v})=e}$,
    where $\#$ is a fresh internal letter that does not appear in
    $\Sigma \cup \overline{\Sigma}$, along with an associated morphism and
    subset.
\end{lem}

Over the alphabet $\Sigma'$ obtained from $\Sigma \cup \overline{\Sigma}$ by
declaring the fresh letter $\#$ as internal, we define
\begin{align*}
    K_{e \uparrow}
    & = \biggset{u\#\overline{u'}\biggmid
		 \begin{aligned}
		   & u, u' \in \Sigma^*, \exists v \in \Sigma^* :\\
		   & uv \in \Sigma^\triangle, u'v \in \Sigma^\triangle,
		     \Delta(u) > 0, \psi(\ext_{u,v})=\psi(\ext_{u',v})=e
		 \end{aligned}}\\
    & = K_e \cap
	\set{u\#u'\mid
	     uu' \in (\Sigma \cup \overline{\Sigma})^\triangle : \Delta(u) > 0}
\end{align*}
for all $e \in O$.
As in the proof of Point (2) of the second statement of
Proposition~\ref{prop effectivity}, we can prove that the language
$\set{u\#u'\mid
      uu' \in (\Sigma \cup \overline{\Sigma})^\triangle : \Delta(u) > 0}$
is a VPL and thus conclude that for all $e \in O$, the set
\[
    P(K_{e \uparrow}) =
    \biggset{(k,l)\in\N\times\N\biggmid
	     \begin{aligned}
	       & \exists u\in\Sigma^k, u'\in\Sigma^l, v \in \Sigma^*:\\
	       & uv\in\Sigma^\triangle, u'v \in \Sigma^\triangle, \Delta(u) > 0,
		 \psi(\ext_{u,v})=\psi(\ext_{u',v})=e
	     \end{aligned}}
\]
is effectively semilinear.

It is clear that $(\varphi, \psi)$ is $F$-weakly-length-synchronous if and only
if for each idempotent $e \in O$ that is $F$-reaching, there does not exist any
$(x_1, x_2) \in P(K_{e \uparrow})$ such that $x_1 \neq x_2$.
Therefore, to decide whether $(\varphi, \psi)$ is $F$-weakly-length-synchronous,
we go through all $e \in O$: if $e$ is an idempotent that is $F$-reaching, we
compute the set $P(K_{e \uparrow})$ and reject if it contains a vector
$(x_1, x_2)$ such that $x_1 \neq x_2$ (which is easy to check given a semilinear
presentation of the set), otherwise we continue. Finally, if we were able to go
through all those elements without rejecting, we accept.

\begin{proof}[Proof of Lemma~\ref{lemma Ke}]
    Let $\Sigma'$ be the alphabet that emerges from
    $\Sigma \cup \overline{\Sigma}$ by additionally declaring $\#$ as an
    internal letter.
    We will construct an $\Ext$-algebra $(R',O')$ and a morphism
    $(\varphi',\psi')$ from $(\Sigma'^\triangle,\ExtMon(\Sigma')^\triangle)$ to
    $(R',O')$ such that for some subset $F \subseteq R'$ we have
    $K_e=\varphi'^{-1}(F)$.

    Let $\overline{R} = \set{\overline{r} \mid r \in R}$.
    We define
    $R' = R \cup \overline{R} \cup \powerset{O^2} \setminus \emptyset \cup
	  \set{\bot, 1}$,
    for some fresh zero $\bot$ and identity $1$, where multiplication between two elements in $R'$ is 
    defined as follows:
    \begin{itemize}
	\item
	    for all $r_1, r_2 \in R$,
	    \begin{align*}
		r_1 \cdot r_2 & = r_1 r_2 & \overline{r_1} \cdot r_2 & = \bot\\
		\overline{r_1} \cdot \overline{r_2} & = \overline{r_2 r_1} &
		    r_1 \cdot \overline{r_2} & = \bot
		\displaypunct{;}
	    \end{align*}
	\item
	    for all $r \in R$ and $E \in \powerset{O^2} \setminus \emptyset$,
	    \begin{align*}
		r \cdot E & = \set{(\leftmult{r} \circ e_1, e_2) \mid
				   (e_1, e_2) \in E} &
		    E \cdot r & = \bot\\
		E \cdot \overline{r} & = \set{(e_1, \leftmult{r} \circ e_2)
					      \mid (e_1, e_2) \in E} &
		    \overline{r} \cdot E & = \bot
		\displaypunct{;}
	    \end{align*}
	\item
	    for all $E_1, E_2 \in \powerset{O^2} \setminus \emptyset$, we have
	    $E_1 \cdot E_2 = \bot$;
	\item
	    $\bot$ acts as a zero, i.e.\ $\bot\cdot r'=r'\cdot\bot=\bot$ for all
	    $r'\in R'$;
	\item
	    $1$ acts as an identity, i.e.\ $1 \cdot r' = r' \cdot 1 = r'$ for
	    all $r'\in R'$.
    \end{itemize}
    Associativity is immediate except for products of the form
    $\overline{r_1} \cdot \overline{r_2} \cdot \overline{r_3}$,
    $r_1 \cdot E \cdot \overline{r_2}$, $r_1 \cdot r_2 \cdot E$ and
    $E \cdot \overline{r_1} \cdot \overline{r_2}$,
    where $E \in \powerset{O^2} \setminus \emptyset$ and $r_1, r_2, r_3 \in R$. 
    In the first case we have
    \[
	(\overline{r_1} \cdot \overline{r_2}) \cdot \overline{r_3} =
	\overline{r_2 r_1} \cdot \overline{r_3} =
	\overline{r_3 r_2 r_1} =
	\overline{r_1} \cdot \overline{r_3 r_2} =
	\overline{r_1} \cdot (\overline{r_2} \cdot \overline{r_3})
	\displaypunct{.}
    \]
    In the second case we have
    \begin{align*}
	(r_1 \cdot E) \cdot \overline{r_2}
	& = \set{(\leftmult{r_1} \circ e_1, e_2) \mid (e_1, e_2) \in E} \cdot
	    \overline{r_2}\\
	& = \set{(\leftmult{r_1} \circ e_1, \leftmult{r_2} \circ e_2) \mid
		 (e_1, e_2) \in E}\\
	& = r_1 \cdot
	    \set{(e_1, \leftmult{r_2} \circ e_2) \mid (e_1, e_2) \in E}
	  = r_1 \cdot (E \cdot \overline{r_2})
	\displaypunct{.}
    \end{align*}
    In the third case we have
    \begin{align*}
	(r_1 \cdot r_2) \cdot E
	& = \set{(\leftmult{r_1 r_2} \circ e_1, e_2) \mid (e_1, e_2) \in E}\\
	& = \set{(\leftmult{r_1} \circ \leftmult{r_2} \circ e_1, e_2) \mid
		 (e_1, e_2) \in E}\\
	& = r_1 \cdot
	    \set{(\leftmult{r_2} \circ e_1, e_2) \mid (e_1, e_2) \in E}
	  = r_1 \cdot (r_2 \cdot E)
    \end{align*}
    and in the fourth case we have
    \begin{align*}
	E \cdot (\overline{r_1} \cdot \overline{r_2})
	& = \set{(e_1, \leftmult{r_2 r_1} \circ e_2) \mid
		 (e_1, e_2) \in E}\\
	& = \set{(e_1, \leftmult{r_2} \circ \leftmult{r_1} \circ e_2) \mid
		 (e_1, e_2) \in E}\\
	& = \set{(e_1, \rightmult{r_1} \circ e_2) \mid (e_1, e_2) \in E} \cdot
	    \overline{r_2}
	  = (E \cdot \overline{r_1}) \cdot \overline{r_2}
	\displaypunct{.}
    \end{align*}
    We define $O'=(R')^{R'}$ which is clearly a monoid for composition and thus
    directly get that $(R', O')$ is a finite $\Ext$-algebra.
    Applying Proposition~\ref{ptn:Freeness_Ext-algebras}, we define the morphism
    $(\varphi',\psi')\colon
     (\Sigma'^\triangle,\ExtMon(\Sigma')^\triangle)\to(R',O')$
    as the unique one satisfying $\varphi'(c)=\varphi(c)$ and
    $\varphi'(\overline{c}) = \overline{\varphi(c)}$ for all
    $c\in\Sigma_\internal$,
    $\varphi'(\#)=
     \set{(\psi(\ext_{\emptyword, v}), \psi(\ext_{\emptyword, v})) \mid
	  v \in \Sigma^\triangle}$
    and where for all $a, a' \in \Sigma_\call$, $b, b' \in \Sigma_\return$, we
    have
    \begin{align*}
	\psi'(\ext_{a,b})(x) & =
	\begin{cases}
	    \psi(\ext_{a,b})(1_R) & \text{if $x=1$}\\
	    \psi(\ext_{a,b})(x) & \text{if $x\in R$}\\
	    \bot & \text{otherwise}
	\end{cases}\\
	\psi'(\ext_{\overline{b'},\overline{a'}})(x) & =
	\begin{cases}
	    \overline{\psi(\ext_{a',b'})(1_R)} & \text{if $x=1$}\\
	    \overline{\psi(\ext_{a',b'})(x')} &
		\text{if $x=\overline{x'}$ for $x'\in R$}\\
	    \bot & \text{otherwise}
	\end{cases}\\
	\psi'(\ext_{a,\overline{a'}})(x) & =
	\begin{cases}
	    \bigcup_{\substack{b \in \Sigma_\return\\z \in \Sigma^\triangle\\
			       (e_1, e_2) \in x}}
	    \set{(\psi(\ext_{a,bz}) \circ e_1, \psi(\ext_{a',bz}) \circ e_2)} &
		\text{if $x \in \powerset{O^2} \setminus \emptyset$}\\
	    \bot & \text{otherwise}
	\end{cases}\\
	\psi'(\ext_{\overline{b'},b})(x) & = \bot
    \end{align*}
    for all $x\in R'$.
    Note that $(\varphi', \psi')$ is computable because
    \begin{align*}
	\set{(\psi(\ext_{\emptyword, v}), \psi(\ext_{\emptyword, v})) \mid
	     v \in \Sigma^\triangle}
	& = \set{(\rightmult{\varphi(v)}, \rightmult{\varphi(v)}) \mid
		 v \in \Sigma^\triangle}\\
	& = \set{(\rightmult{r}, \rightmult{r}) \mid r \in R}
    \end{align*}
    and
    \begin{align*}
	& \bigcup_{\substack{b \in \Sigma_\return\\z \in \Sigma^\triangle\\
			     (e_1, e_2) \in x}}
	  \set{(\psi(\ext_{a,bz}) \circ e_1, \psi(\ext_{a',bz}) \circ e_2)}\\
	= & \bigcup_{\substack{b \in \Sigma_\return\\z \in \Sigma^\triangle\\
		     (e_1, e_2) \in x}}
	    \set{(\rightmult{\varphi(z)} \circ \psi(\ext_{a,b}) \circ e_1,
		  \rightmult{\varphi(z)} \circ \psi(\ext_{a',b}) \circ e_2)}\\
	= & \bigcup_{\substack{b \in \Sigma_\return\\r \in R\\
		     (e_1, e_2) \in x}}
	    \set{(\rightmult{r} \circ \psi(\ext_{a,b}) \circ e_1,
		  \rightmult{r} \circ \psi(\ext_{a',b}) \circ e_2)}
    \end{align*}
    for all $x \in \powerset{O^2} \setminus \emptyset$ and
    $a, a' \in \Sigma_\call$.

    Now define the set of pairs
    $P = \set{(u, u') \in \Sigma^* \times \Sigma^* \mid
	      \exists v \in \Sigma^* :
	      uv \in \Sigma^\triangle, u'v \in \Sigma^\triangle}$;
    it is not difficult to check that for all $w \in \Sigma'^*$,
    $w \in \Sigma'^\triangle \cap \Sigma^* \# \overline{\Sigma}^*$ if and only
    if $w = u\#\overline{u'}$ for $(u, u') \in P$.
    It suffices to prove the following claim, which directly implies the desired
    equality
    \begin{align*}
	& \varphi'^{-1}(\set{E \in \powerset{O^2} \setminus \emptyset \mid
			   (e, e) \in E})\\
	= & \set{u\#\overline{u'} \mid
		 (u, u') \in P, \exists v \in \Sigma^* :
		 uv \in \Sigma^\triangle, u'v \in \Sigma^\triangle,
	     \psi(\ext_{u, v}) = \psi(\ext_{u', v}) = e}\\
	= & \set{u\#\overline{u'} \mid
		 u, u' \in \Sigma^*, \exists v \in \Sigma^* :
		 uv \in \Sigma^\triangle, u'v \in \Sigma^\triangle,
		 \psi(\ext_{u,v})=\psi(\ext_{u',v})=e}
	\displaypunct{.}
    \end{align*}
    For all $w\in\Sigma'^\triangle$ we have
    \[
	\varphi'(w)=
	\begin{cases}
	    1 & \text{if $w = \emptyword$}\\
	    \varphi(w) & \text{if $w\in\Sigma^\triangle\setminus\set{\emptyword}$}\\
	    \overline{\varphi(w')} & \text{if $w = \overline{w'}$ for $w' \in
	    \Sigma^\triangle\setminus\set{\emptyword}$}\\
	    \set{(\psi(\ext_{u,v}), \psi(\ext_{u', v})) \mid
		 v \in \Sigma^*, uv, u'v \in \Sigma^\triangle} &
		\text{if $w=u\#\overline{u'}$ for $(u, u') \in P$}\\
	    \bot & \text{otherwise.}
	\end{cases}
    \]
    We prove it by structural induction on $w$. The cases when $w = \emptyword$
    or $w=c\in\Sigma_\internal$ or
    $w=\overline{c}\in\overline{\Sigma}_\internal$ follow immediately from the
    definition of $\varphi'$.
    In case $w=\#=\emptyword\#\emptyword$, we have
    \[
	\varphi'(w)=
	\set{(\psi(\ext_{\emptyword, v}), \psi(\ext_{\emptyword, v})) \mid
	     v \in \Sigma^\triangle} =
	\set{(\psi(\ext_{\emptyword, v}), \psi(\ext_{\emptyword, v})) \mid
	     v \in \Sigma^*, \emptyword v \in \Sigma^\triangle}
    \]
    as required.

    For the inductive step first assume
    $w=\alpha w' \beta$ for $w'\in\Sigma'^\triangle$,
    $\alpha \in \Sigma_\call \cup \overline{\Sigma}_\call$ and
    $\beta \in \Sigma_\return \cup \overline{\Sigma}_\return$.
    If $w'$ is neither in $\Sigma^\triangle \cup \overline{\Sigma}^\triangle$
    nor of the form $u\#\overline{u'}$ with $(u, u') \in P$, then
    $\varphi'(w') = \bot$ by induction hypothesis and thus
    $\varphi'(w) = \psi'(\ext_{\alpha,\beta})(\varphi'(w')) =
     \psi'(\ext_{\alpha,\beta})(\bot) = \bot$
    as required, since $w$ is also neither in
    $\Sigma^\triangle \cup \overline{\Sigma}^\triangle$ nor of the form
    $u\#\overline{u'}$ with $(u, u') \in P$.
    If $w' = \emptyword$, then $\varphi'(w') = 1$ and hence
    \begin{align*}
	\varphi'(w)
	& = \psi'(\ext_{\alpha, \beta})(\varphi'(w'))\\
	& = \psi'(\ext_{\alpha, \beta})(1)\\
	& =
	\begin{cases}
	    \psi(\ext_{\alpha, \beta})(1_R) = \varphi(w) &
		\text{if $\alpha \in \Sigma_\call$ and
		      $\beta \in \Sigma_\return$}\\
	    \overline{\psi(\ext_{a, b})(1_R)} = \overline{\varphi(a b)} &
		\text{if $\alpha = \overline{b} \in \overline{\Sigma}_\call$ and
		      $\beta = \overline{a} \in \overline{\Sigma}_\return$}\\
	    \bot & \text{otherwise}
	\end{cases}
    \end{align*}
    as required, because $w = \overline{b} \, \overline{a} = \overline{a b}$ in
    the second case.
    If $w' \in \Sigma^\triangle \setminus \set{\emptyword}$, then
    $\varphi'(w') = \varphi(w') \in R$ by induction hypothesis, and hence
    \begin{align*}
	\varphi'(w) = \psi'(\ext_{\alpha, \beta})(\varphi'(w'))
	& = \psi'(\ext_{\alpha, \beta})(\varphi(w'))\\
	& =
	\begin{cases}
	    \psi(\ext_{\alpha, \beta})(\varphi(w')) = \varphi(w) &
		\text{if $\alpha \in \Sigma_\call$ and
		      $\beta \in \Sigma_\return$}\\
	    \bot & \text{otherwise}
	\end{cases}
    \end{align*}
    as required.
    If $w' \in \overline{\Sigma}^\triangle \setminus \set{\emptyword}$, then
    $w' = \overline{w''}$ for
    $w'' \in \Sigma^\triangle \setminus \set{\emptyword}$, so
    $\varphi'(w') = \overline{\varphi(w'')} \in \overline{R}$ by induction
    hypothesis, and hence
    \begin{align*}
	\varphi'(w)
	& = \psi'(\ext_{\alpha, \beta})(\varphi'(w'))\\
	& = \psi'(\ext_{\alpha, \beta})(\overline{\varphi(w'')})\\
	& =
	\begin{cases}
	    \overline{\psi(\ext_{a, b})(\varphi(w''))} =
	    \overline{\varphi(a w'' b)} &
		\text{if $\alpha = \overline{b} \in \overline{\Sigma}_\call$ and
		      $\beta = \overline{a} \in \overline{\Sigma}_\return$}\\
	    \bot & \text{otherwise}
	\end{cases}
    \end{align*}
    as required, because
    $w = \overline{b} \, \overline{w''} \, \overline{a} = \overline{a w'' b}$ in
    the first case.
    If $w'=u\#u'$ with $(u, u') \in P$, i.e.\ $w = \alpha u\#u' \beta$,
    then
    \[
	\varphi'(w') =
	\set{(\psi(\ext_{u,v'}), \psi(\ext_{u', v'})) \mid
	     v' \in \Sigma^*, uv', u'v' \in \Sigma^\triangle} \in
	\powerset{O^2} \setminus \emptyset
    \]
    by induction hypothesis.
    Hence, we have
    \begin{align*}
	\varphi'(w)
	& = \psi'(\ext_{\alpha, \beta})(\varphi'(w'))\\
	& = \psi'(\ext_{\alpha, \beta})
	    (\set{(\psi(\ext_{u,v'}), \psi(\ext_{u', v'})) \mid
		  v' \in \Sigma^*, uv', u'v' \in \Sigma^\triangle})\\
	& = \bigcup_{\substack{b \in \Sigma_\return\\z \in \Sigma^\triangle}}
	    \set{(\psi(\ext_{au,v'bz}), \psi(\ext_{a'u',v'bz})) \mid
		 v' \in \Sigma^*, uv', u'v' \in \Sigma^\triangle}\\
	& = \set{(\psi(\ext_{au,v}), \psi(\ext_{a'u', v})) \mid
		 v \in \Sigma^*, auv, a'u'v \in \Sigma^\triangle}
    \end{align*}
    if $\alpha = a \in \Sigma_\call$ and
    $\beta = \overline{a'} \in \overline{\Sigma}_\return$ (where the last
    inclusion from right to left follows by considering the unique stair
    factorizations given by Lemma~\ref{lemma ext} for the elements of each pair)
    and $\varphi'(w) = \bot$ otherwise, as required.

    Finally assume $w=xy$ for some
    $x, y \in \Sigma'^\triangle \setminus \set{\emptyword}$.
    The case when $x$ or $y$ is neither in
    $\Sigma^\triangle \cup \overline{\Sigma}^\triangle$ nor of the form
    $u\#\overline{u'}$ with $(u, u') \in P$ is easily handled by applying the
    induction hypothesis and observing that $\bot$ is a zero in $R'$.
    Four other immediate cases are when both $x$ and $y$ are in
    $\Sigma^\triangle$, when $x$ is in $\Sigma^\triangle$ and $y$ in
    $\overline{\Sigma}^\triangle$, when $x$ is in $\overline{\Sigma}^\triangle$
    and $y$ in $\Sigma^\triangle$ and when both $x$ and $y$ are of the form
    $u\#u'$ with $(u, u') \in P$.
    For the case when both $x$ and $y$ are in $\overline{\Sigma}^\triangle$, we
    have that $x = \overline{x'}$ and $y = \overline{y'}$ for
    $x', y' \in \Sigma^\triangle$, so that
    $\varphi'(x) = \overline{\varphi(x')} \in \overline{R}$ and
    $\varphi'(y) = \overline{\varphi(y')} \in \overline{R}$ by induction
    hypothesis, hence
    \[
	\varphi'(x y) = \varphi'(x) \cdot \varphi'(y) =
	\overline{\varphi(x')} \cdot \overline{\varphi(y')} =
	\overline{\varphi(y') \varphi(x')} = \overline{\varphi(y' x')}
    \]
    as required, because
    $x y = \overline{x'} \, \overline{y'} = \overline{y' x'}$.
    Consider the case when
    $x \in (\Sigma^\triangle \cup \overline{\Sigma}^\triangle) \setminus
	   \set{\emptyword}$ 
    and $y = u\#\overline{u'}$ with $(u, u') \in P$, hence
    $w = x u\#\overline{u'}$.
    The induction hypothesis yields
    \[
	\varphi'(x) =
	\begin{cases}
	    \varphi(x) \in R &
		\text{if $x \in \Sigma^\triangle \setminus \set{\emptyword}$}\\
	    \overline{\varphi(x')} \in \overline{R} &
		\text{if $x = \overline{x'}$ for
		      $x' \in \Sigma^\triangle \setminus \set{\emptyword}$}
	\end{cases}
    \]
    and
    $\varphi'(y) =
     \set{(\psi(\ext_{u,v}), \psi(\ext_{u', v})) \mid
	  v \in \Sigma^*, uv, u'v \in \Sigma^\triangle} \in
     \powerset{O^2} \setminus \emptyset$.
    We obtain 
    \begin{align*}
	\varphi'(w)
	& = \varphi'(x) \cdot \varphi'(y)\\
	& = \varphi(x) \cdot
	    \set{(\psi(\ext_{u,v}), \psi(\ext_{u', v})) \mid
		 v \in \Sigma^*, uv, u'v \in \Sigma^\triangle}\\
	& = \set{(\leftmult{\varphi(x)} \circ \psi(\ext_{u,v}),
		  \psi(\ext_{u', v})) \mid
		 v \in \Sigma^*, uv, u'v \in \Sigma^\triangle}\\
	& = \set{(\psi(\ext_{x u,v}), \psi(\ext_{u', v})) \mid
		 v \in \Sigma^*, xuv, u'v \in \Sigma^\triangle}
    \end{align*}
    if $x \in \Sigma^\triangle \setminus \set{\emptyword}$ and
    $\varphi'(w) =
     \overline{\varphi(x')} \cdot
     \set{(\psi(\ext_{u,v}), \psi(\ext_{u', v})) \mid
	  v \in \Sigma^*, uv, u'v \in \Sigma^\triangle} = \bot$ 
    if $x = \overline{x'}$ for
    $x' \in \Sigma^\triangle \setminus \set{\emptyword}$, as required.
    Eventually, let us treat the case when $x = u\#u'$ with $(u, u') \in P$ and
    $y \in (\Sigma^\triangle \cup \overline{\Sigma}^\triangle) \setminus
	   \set{\emptyword}$,
    hence $w = u\#\overline{u'} y$.
    The induction hypothesis yields
    $\varphi'(x) =
     \set{(\psi(\ext_{u,v}), \psi(\ext_{u', v})) \mid
	  v \in \Sigma^*, uv, u'v \in \Sigma^\triangle} \in
     \powerset{O^2} \setminus \emptyset$
    and
    $\varphi'(y) =
     \begin{cases}
	 \varphi(y) \in R &
	     \text{if $y \in \Sigma^\triangle \setminus \set{\emptyword}$}\\
	 \overline{\varphi(y')} \in \overline{R} &
	     \text{if $y = \overline{y'}$ for
		   $y' \in \Sigma^\triangle \setminus \set{\emptyword}$}
     \end{cases}$.
    We obtain 
    \begin{align*}
	\varphi'(w)
	& = \varphi'(x) \cdot \varphi'(y)\\
	& = \set{(\psi(\ext_{u,v}), \psi(\ext_{u', v})) \mid
		 v \in \Sigma^*, uv, u'v \in \Sigma^\triangle} \cdot
	    \overline{\varphi(y')}\\
	& = \set{(\psi(\ext_{u,v}),
		  \leftmult{\varphi(y')} \circ \psi(\ext_{u', v})) \mid
		 v \in \Sigma^*, uv, u'v \in \Sigma^\triangle}\\
	& = \set{(\psi(\ext_{u,v}), \psi(\ext_{y' u', v})) \mid
		 v \in \Sigma^*, uv, y'u'v \in \Sigma^\triangle}
    \end{align*}
    if $y = \overline{y'}$ for
    $y' \in \Sigma^\triangle \setminus \set{\emptyword}$ and
    $\varphi'(w) =
     \set{(\psi(\ext_{u,v}), \psi(\ext_{u', v})) \mid
	  v \in \Sigma^*, uv, u'v \in \Sigma^\triangle} \cdot \varphi(y) = \bot$
    if $x \in \Sigma^\triangle \setminus \set{\emptyword}$, as required, because
    $\overline{y' u'} = \overline{u'} \, \overline{y'}$ in the first case.
\end{proof}

\section{Conclusion}\label{Conclusion}
In this paper we have studied the question of
which visibly pushdown languages lie in the complexity class $\ACO$.

We have introduced the notions of 
length-synchronicity, weak length-synchronicity and quasi-counterfreeness.
We have introduced intermediate VPLs: 
these are quasi-counterfree VPLs
generated by context-free grammars $G$ involving the production
$S\rightarrow_G\varepsilon$ for the start nonterminal $S$ and whose
further productions are all of the form $T\rightarrow_G uT'v$,
where $uv$ is well-matched,
$u\in(\Sigma_\internal^*\Sigma_\call\Sigma_\internal^*)^+$,
$v\in(\Sigma_\internal^*\Sigma_\return\Sigma_\internal^*)^+$,
and the set of contexts $\set{(u,v)\in\Con(\Sigma)\mid S\Rightarrow_G^*uSv}$
is weakly length-synchronous but not length-synchronous.
To the best of our knowledge our community is 
unaware of whether at all there is an intermediate VPL 
that is provably in $\ACO$ (even in $\ACCO$) or 
provably not in $\ACO$. 
In fact we conjecture that every intermediate VPL is neither in $\ACCO$ nor $\TCO$-hard.

Our main result states that there is an algorithm
that, given a visibly pushdown language $L$, 
outputs if $L$ surely lies in $\ACO$, surely does not lie in
$\ACO$ (by providing some $m\geq 2$ such that $L(A)$ is $\ACCO(m)$-hard), 
or outputs a disjoint finite union of intermediate
VPLs that $L$ is constant-depth equivalant to.
In the latter case one can moreover compute distinct $k,l\in\N_{>0}$ 
such that already $\SM_{k,l}=L(S\rightarrow \varepsilon\mid a c^{k-1} S b_1\mid ac^{l-1}Sb_2)$ is constant-depth reducible to $L$.
	
We conjecture that due to the particular nature of intermediate VPLs,
either all of them are in $\ACO$ or all are not:
this conjecture together with our main result indeed implies that there 
is an algorithm that decides if a given visibly pushdown language
is in $\ACO$.

As main tools we carefully revisited $\Ext$-algebras, 
introduced by Czarnetzki et al.~\cite{Czarnetzki-Krebs-Lange-2018}, 
being closely related
to forest algebras, introduced by Boja{\'n}czyk and Walukiewicz~\cite{BW08}.
For the reductions from $\SM_{k,l}$ we made use of Green's relations.

Natural questions arise.
Is there any \emph{concrete} intermediate VPL that
is provably in $\ACCO$, provably not in $\ACO$, or
hard for $\TCO$?
Another exciting question is whether one can
effectively compute those visibly pushdown languages
that lie in the complexity class $\TCO$.
Is there is a $\TCO$/$\NCI$ complexity dichotomy?
For these questions new techniques seem to be necessary.
In this context it is already interesting to mention
there is an $\NCI$-complete visibly pushdown language
whose syntactic $\Ext$-algebra is aperiodic.
Another exciting question is to give an algebraic
characterization of the visibly counter languages.

\bibliography{main}

\end{document}